\documentclass[12pt,english,american]{article}
\usepackage[T1]{fontenc}
\usepackage[latin9]{inputenc}
\usepackage{geometry}
\usepackage{color}
\usepackage{babel}
\usepackage{amsthm}
\usepackage{amstext}
\usepackage{amssymb}
\usepackage{amsmath}
\usepackage{cancel}     
\usepackage[unicode=true,pdfusetitle,
 bookmarks=true,bookmarksnumbered=false,bookmarksopen=false,
 breaklinks=true,pdfborder={0 0 0},backref=false,colorlinks=true]
 {hyperref}
\hypersetup{
 linkcolor=blue,citecolor=blue, urlcolor=blue}
\makeatletter

\newcommand{\noun}[1]{\textsc{#1}}
\newcommand{\red}{black}

\newcommand{\tiQ}{Q}
\newcommand{\mring}{\mathring}
\newcommand{\bmcA}{\mcA}
\newcommand{\mrm}{\mathrm}
\renewcommand{\Re}{\mathrm{Re}}
\newcommand{\tigam}{\ti\ga}

\newcommand{\F}{\mcF}
\newcommand{\res}{\upharpoonleft}
\newcommand{\mcL}{\mathcal{L}}
\newcommand{\veps}{\epsilon}
\newcommand{\mcF}{\mathcal{F}}

\newcommand{\bint}{\,\,\bar{\!\!\int}}
\newcommand{\hint}{\,\,\hat{\!\!\int}}
\newcommand{\q}{i'}
\newcommand{\hhi}{\hat i}
\newcommand{\Q}{\bar Q}
\newcommand{\cphi}{\check{\phi}}
\newcommand{\mm}{q}
\newcommand{\m}{m}
\newcommand{\mr}{\mathrm}
\newcommand{\cpsi}{ \check{\psi} }
\newcommand{\calF}{ {\cal F} }

\newcommand{\tic}{c}

\newcommand{\PS}{S}
\newcommand{\maxi}{\mathrm{max}}
\newcommand{\fib}{\mathrm{fi}}
\newcommand{\free}{\mathrm{free}}

\newcommand{\mcA}{\mathcal A}
\newcommand{\mcB}{\mathcal B}

\newcommand{\kas}{\kappa_{*}}

\newcommand{\hk}{\hat k }    

\newcommand{\pho}{\mathrm{f}}
\newcommand{\alf}{\overline{\al}}

\newcommand{\vv}{v_{\alf}}

\newcommand{\g}{\la }

\newcommand{\wt}{\widetilde}

\newcommand{\ti}{\tilde}

\newcommand{\un}{\underline}

\newcommand{\Om}{\Omega}
\newcommand{\ga}{\gamma}

\newcommand{\La}{\Lambda}

\newcommand{\ka}{\kappa}

\newcommand{\be}{\beta}

\newcommand{\pa}{\partial}

\newcommand{\ov}{\overline}
\newcommand{\vp}{\varphi}
\newcommand{\mfh}{\mathfrak{h}}

\newcommand{\eps}{\varepsilon}
\newcommand{\de}{\delta}

\newcommand{\De}{\Delta}
\newcommand{\e}{e}

\newcommand{\nin}{\noindent}
\newcommand{\si}{\sigma}
\newcommand{\ph}{\phantom}
\newcommand{\h}{\fr{1}{2}}
\newcommand{\nat}{\mathbb{N}}
\newcommand{\hil}{\mathcal{H}}
\newcommand{\om}{\omega}

\newcommand{\mco}{\mathcal{O}}

\newcommand{\supp}{\mathrm{supp}}
\newcommand{\fr}[2]{\frac{#1}{#2}}
\newcommand{\al}{\alpha}
\newcommand{\real}{\mathbb{R}}
\newcommand{\complex}{\mathbb{C}}
\newcommand{\la}{\lambda}
\newcommand{\non}{\nonumber}
\newcommand{\Ga}{\Gamma}
\newcommand{\half}{\fr{1}{2}}
\newcommand{\lan}{\langle}
\newcommand{\ran}{\rangle}
\newcommand{\bc}{ \color{black} }
\newcommand{\bcc}{\color{black}}
\newcommand{\bcb}{\color{black}}

\def\proof{\noindent{\bf Proof. }}
\def\qed{$\Box$\medskip}

\newcommand{\beq}{\begin{equation}}
\newcommand{\eeq}{\end{equation}}
\newcommand{\beqa}{\begin{eqnarray}}
\newcommand{\eeqa}{\end{eqnarray}}
\newcommand{\ben}{\begin{arabicenumerate}}
\newcommand{\een}{\end{arabicenumerate}}

\newcommand{\sib}{\boldsymbol{\si}}

\def\bel{\begin{lem} } 
\def\eel{\end{lem} }
\def\bet{\begin{thm}}
\def\eet{\end{thm}}
\def\bed{\begin{defn}}
\def\eed{\end{defn} }

\def\bec{\begin{cor}}
\def\eec{\end{cor}}
\def\ber{\begin{rem}}
\def\eer{\end{rem}}

\usepackage{enumitem}		
\theoremstyle{plain}
\newtheorem{thm}{\protect\theoremname}[section]
\theoremstyle{definition}
\newtheorem{defn}[thm]{\protect\definitionname}
\theoremstyle{plain}
\newtheorem{prop}[thm]{\protect\propositionname}
\theoremstyle{plain}
\theoremstyle{remark}
\newtheorem{rem}[thm]{\protect\remarkname}
\theoremstyle{plain}
\newtheorem{lem}[thm]{\protect\lemmaname}
\theoremstyle{plain}
\newtheorem{cor}[thm]{\protect\corollaryname}

\usepackage{txfonts,refstyle,xcolor}

\newref{con}{name = Conjecture\ }
\newref{prop}{name = Proposition\ }
\newref{def}{name = Definition\ }
\newref{sec}{name = Section\ }
\newref{sub}{name = Section\ }
\newref{thm}{name = Theorem\ }
\newref{lem}{name = Lemma\ }
\newref{cor}{name = Corollary\ }
\newref{fig}{name = Figure\ }
\newref{rem}{name =  Remark\ }

\usepackage{bbm}

\usepackage[all]{xy}

\newcommand{\xyR}[1]{%
     \makeatletter
     \xydef@\xymatrixrowsep@{#1}
     \makeatother
}

\newcommand{\xyC}[1]{%
     \makeatletter
     \xydef@\xymatrixcolsep@{#1}
     \makeatother
}

\newcommand{\ko}{\color{red}}

\newcommand{\ncol}[1]{\color{normalcolor}}

\makeatother

\addto\captionsamerican{\renewcommand{\corollaryname}{Corollary}}
\addto\captionsamerican{\renewcommand{\definitionname}{Definition}}
\addto\captionsamerican{\renewcommand{\lemmaname}{Lemma}}
\addto\captionsamerican{\renewcommand{\propositionname}{Proposition}}
\addto\captionsamerican{\renewcommand{\remarkname}{Remark}}
\addto\captionsamerican{\renewcommand{\theoremname}{Theorem}}
\addto\captionsenglish{\renewcommand{\corollaryname}{Corollary}}
\addto\captionsenglish{\renewcommand{\definitionname}{Definition}}
\addto\captionsenglish{\renewcommand{\lemmaname}{Lemma}}
\addto\captionsenglish{\renewcommand{\propositionname}{Proposition}}
\addto\captionsenglish{\renewcommand{\remarkname}{Remark}}
\addto\captionsenglish{\renewcommand{\theoremname}{Theorem}}
\providecommand{\corollaryname}{Corollary}
\providecommand{\definitionname}{Definition}
\providecommand{\lemmaname}{Lemma}
\providecommand{\propositionname}{Proposition}
\providecommand{\remarkname}{Remark}
\providecommand{\theoremname}{Theorem}
\renewcommand{\1}{\!\!\!}

\begin{document}
\title{Coulomb scattering in the massless Nelson model II. \\ Regularity of ground states} 
\author{
{\bf Wojciech Dybalski}\\
Zentrum Mathematik, Technische Universit\"at M\"unchen,\\
E-mail: {\tt dybalski@ma.tum.de}
\and
{\bf Alessandro Pizzo}\\
Dipartimento di Matematica, Universit\`a di Roma ``Tor
Vergata''\\ 
E-mail: {\tt pizzo@axp.mat.uniroma2.it}}

\date{}

\maketitle

\abstract 
For the massless Nelson model we provide detailed information about the dependence of the normalized ground 
states $\check{\psi}_{P,\sigma}$ of the fiber single-electron  Hamiltonians $H_{P,\sigma}$
on the total momentum $P$ and the infrared cut-off $\sigma$. This information is obtained with the help of the  iterative analytic perturbation theory.
In particular, we derive bounds of the form
\[
\|\partial_{P_i}\check{\psi}_{P,\sigma}\|,\  \| \partial_{P_i} \partial_{P_j}\check{\psi}_{P,\sigma}  \|\leq \frac{c}{\sigma^{\delta_{\lambda_0}}}, 
\] 
for  some constant $c$ and a function of the maximal admissible coupling  constant $\lambda_0\mapsto \delta_{\lambda_0}$ s.t. $\lim_{\lambda_0\to 0}\de_{\lambda_0}=0$. These results hold both in the infrared-regular and 
infrared-singular case.
They are exploited in part I of this series to construct the two-electron scattering states in the infrared-regular massless Nelson model (in the absence of an infrared cut-off)  along the lines of Haag-Ruelle scattering theory. They should also be relevant for the problem of scattering of two infraparticles in the infrared-singular Nelson model, whose solution is the goal of this series of papers.
 {\bcc Although a part of a larger
investigation, the present work is written in a self-contained fashion.} 


\section{Introduction}
\setcounter{equation}{0}

In spite of  tremendous progress  in the mathematical description of scattering of light  
and matter in the framework of non-relativistic QED \cite{CFP07,DG04, Pi05, FGS04, Sp97, DK11}, processes involving 
several massive particles (`electrons') remained untreated. This important gap in our understanding can be traced
back to serious conceptual and technical difficulties arising at the multi-electron level. On the conceptual side,
it is essential to follow the lines of relativistic scattering theory \cite{Ha58, Ru62, Dy05}, in spite of the fact that the `bare electron', appearing
in the model Hamiltonian, is a non-relativistic quantum-mechanical particle.  In the presence of a fixed infrared cut-off \cite{Fr73.1}  
and for massive  photons \cite{Al73} a suitable variant of Haag-Ruelle scattering theory was found long time ago. 
However, only in our recent paper \cite{DP12.1}, this framework was  generalized  to  theories with non-trivial 
infrared structure. In \cite{DP12.1} we constructed scattering states of two electrons in the infrared-regular massless Nelson
model (without an infrared cut-off) and proved their tensor product structure. This result relies on detailed spectral properties of 
the ground states of the single-particle fiber Hamiltonians which encode localization of physical electrons in space. In the present 
paper {\bcc and in \cite{DP16} we prove}  these properties using  iterative analytic perturbation theory \cite{Pi03, KM12, CFP09, BBP13}. {\bcb  Hence, our spectral results here derived are the basis for the rigorous scattering theory of non-relativistic quantum particles interacting with a quantized massless boson field.  
While the interaction mediated by low energy (massless) bosons still  requires some regularization in \cite{DP12.1}, no infrared cut-off is present.}

Let $H$ be the Hamiltonian of the massless Nelson model and $H_P$ be the usual fiber Hamiltonians at fixed total momentum $P$ 
{\bcc defined in  (\ref{Hamiltonian})  below}.
The analysis of the spectrum of $H_{P}$ was initiated in \cite{Fr73,Fr74.1} and advanced in \cite{Pi03},   
{\bcb where the iterative analytic perturbation theory was developed}. Interesting results on the spectrum of the  Nelson model with a slightly different
form factor were also obtained in \cite{AH12}  by means of different methods. The results of the present paper go much beyond the existing applications of the iterative analytic perturbation theory. In order  to elucidate these improvements let us now discuss our
 findings in non-technical terms: Let $H_{P,\si}$ be the fiber Hamiltonians with an infrared cut-off $\si>0$ 
which have normalized ground state vectors $\cpsi_{P,\si}$ {\bcb (with phases fixed in Definition~\ref{abuse-definition})} corresponding to isolated simple eigenvalues $E_{P,\si}$.  {\bcb We recall that in the case of the physical electrodynamic interaction (morally corresponding to the form factor (\ref{form-factor-formula}) with $\bar{\alpha}=0$)  the vector $\cpsi_{P,\si}$ tends weakly to zero as $\sigma\to 0$.}
In all cases $\bar{\alpha}\geq 0$, the behaviour of the derivatives $\pa^{\be}_P\cpsi_{P,\si}$ in the limit $\si\to 0$ carries information about the localization
of the electron in space and therefore is of central importance for the problem of scattering of several particles. In more
technical terms, these derivatives enter the proof of existence of scattering states via the non-stationary phase method  \cite{DP12.1}.

The analysis of these derivatives, given in the present paper, proceeds as follows: With the help of the unitary dressing
transformation $W_{P,\si}$, defined  in (\ref{W-definition}) below, we obtain the following formula 
\beqa
\pa_{P_i} \cpsi_{P,\si}=W_{P,\si}^*\frac{1}{H_{P,\si}^{W}-E_{P,\si}}(\Gamma_{P,\si})_{i}\check{\phi}_{P,\si}, \label{intro-first-derivative}
\eeqa
where $H_{P,\si}^{W}=W_{P,\si} H_{P,\si}  W_{P,\si}^*$ and $\check{\phi}_{P,\si}=W_{P,\si} \cpsi_{P,\si}$. 
The expression $\Gamma_{P,\si}$,
defined by the formula
\beqa
\Ga_{P,\si}=\nabla E_{P,\si}-W_{P,\si}(P-P_{\pho})W_{P,\si}^*, \label{Gamma-def-intro}
\eeqa
{\bcb where $P_{\pho}$ is the free photon momentum operator}, has the property $\lan \cphi_{P,\si}, (\Gamma_{P,\si})_{i} \cphi_{P,\si}\ran=0$ {\bcc due to the {\color{\red}Hellman}-Feynman formula}. The behaviour
of the r.h.s. of (\ref{intro-first-derivative})  as a function of $\si$ can be controlled with the help of  iterative analytic perturbation theory, 
and we obtain in Corollary~\ref{technical-corollary} that
\beqa
\|\frac{1}{H_{P,\si}^{W}-E_{P,\si}}(\Gamma_{P,\si})_{i}\check{\phi}_{P,\si}\|\leq  \fr{c}{\si^{\de_{\g_0}}}, \label{analytic-perturbation-theory}
\eeqa 
where $c$ is independent of $\si$ and  $\g_0\mapsto \de_{\g_0}$ {\bcc tends to zero as a function of the maximal admissible coupling constant.} 
Clearly, we have an analogous bound on the first derivative of the vector
\beqa
\|\pa_{P_i} \cpsi_{P,\si}\|\leq  \fr{c}{\si^{\de_{\g_0}}}. \label{first-derivative-estimate-intro}
\eeqa
This preliminary result is within reach of existing {\color{\red}applications} of iterative analytic perturbation theory (see e.g. \cite{KM12, FP10}  for estimates
comparable to (\ref{analytic-perturbation-theory}) in different models). However, it is not sufficient for constructing scattering
states of two electrons. For this purpose the second derivative is needed.

The analysis of the second derivative of  $\cpsi_{P,\si}$ is considerably more difficult. It starts with a derivation of a formula
which has the following form
\beqa
\pa_{P_i}\pa_{P_{i'}} \cpsi_{P,\si} \1 &=& \1 W_{P,\si}^*\bigg( \frac{Q_{P,\si}^{\perp}}{H_{P,\si}^{W}-E_{P,\si}}(\Gamma_{P,\si})_{i'}\frac{1}{H_{P,\si}^{W}-E_{P,\si}}(\Gamma_{P,\si})_{i}\check{\phi}_{P,\si}+\{i\leftrightarrow i'\}\bigg) \non\\
&  &-W_{P,\si}^*\check{\phi}_{P,\si}\langle \check{\phi}_{P,\si},(\Gamma_{P,\si})_{i'}\frac{1}{(H_{P,\si}^{W}-E_{P,\si})^{2}}(\Gamma_{P,\si})_{i}\check{\phi}_{P,\si}\rangle,
\label{second-deriv-form}
\eeqa
where $Q_{P,\sigma}=|\cphi_{P,\si}\ran\lan \cphi_{P,\si}|$. The last term on the r.h.s. of (\ref{second-deriv-form}) can be estimated 
using (\ref{analytic-perturbation-theory}). To control the first two terms we use iterative analytic perturbation theory to prove the 
following bound (cf. Corollary~\ref{technical-corollary})
\beqa
& &\|\frac{Q_{P,\sigma}^{\perp}}{H_{P,\sigma}^{W}-E_{P,\sigma}}(\Gamma_{P,\sigma})_{i}\frac{1}{H_{P,\sigma}^{W}-E_{P,\sigma}}(\Gamma_{P,\sigma})_{i'}\check{\phi}_{P,\sigma}\|\leq \frac{c}{\sigma^{2\delta_{\g_0}}},\label{eq:inequality-cont-1-intro}
\eeqa
and consequently we obtain:
\beqa
\|\pa_{P_i}\pa_{P_{i'}} \cpsi_{P,\si}\|\leq \frac{c}{\sigma^{2\delta_{\g_0}}}. \label{second-derivative-intro-x}
\eeqa 
We emphasize that the proof of (\ref{eq:inequality-cont-1-intro}), which is the main technical result of this paper, is
much more involved than the proof of (\ref{analytic-perturbation-theory}) and other applications of the iterative analytic 
perturbation theory available in the literature. It requires extensive use of
direct integral representations (cf. Subsection~\ref{direct-subsection}) and application of novel maximal
modulus principle arguments (cf. Subsection~\ref{maximal-mod-sub}).  Another complication comes from unexpected
singular terms, which prevent simple power-counting arguments (we refer  e.g. to  $B^*(G|^{n-1}_n)\phi_{P,n-1}$ and 
$b^*\big(\mathcal{I}|_{n}^{n-1}\cdot g|^{n-1}_n \big)\phi_{P,n-1}$ in (\ref{unexp})). A more thorough discussion of these 
new ingredients is presented in Subsection~\ref{novel-subsection}.  

 We stress that our main spectral results  reported in (\ref{first-derivative-estimate-intro}) and (\ref{second-derivative-intro-x}) hold also when $\bar{\alpha}=0$ in the form factor that models the interaction (see (\ref{form-factor-formula})). These results are surely needed  for the construction of scattering of several infraparticles in the infrared-singular massless
Nelson model, which is the goal of this series of papers. However, differently from the control of the effective dynamics up to some time scale related to the coupling constant (cf \cite{BCFFS13}), the construction of scattering states for infraparticles requires a nontrivial control of the clouds of asymptotic real photons "attached" to the charged particles emerging out of the scattering; see \cite{Pi05}, \cite{CFP09} for the case of a single infraparticle. This control poses new conceptual  problems in the case of many electrons. An overview of them can be found in \cite{Dy17} along with clear-cut mathematical conjectures, with the aim to  provide a consistent LSZ scattering theory that corrects the accepted (in the physics community) but incomplete ansatz provided by Fadeev and Kulish and based on a modified asymptotic dynamics \emph{a la Dollard}.  

We hope  that the {\color{\red} progress}  from the present paper will enable other interesting investigations of regularity of the
ground state vectors in models of non-relativistic QED.  For example, 
in order to improve the control  of the $P-$ dependence  of $\check{\phi}_{P,\sigma}$, uniformly in $\sigma$, our results naturally come into play. We recall that the H\"older regularity in $P$, {\bcb uniformly in $\sigma$}, of $\check{\phi}_{P,\sigma}$ is a crucial ingredient in the analysis of the effective dynamics of an electron coupled to an external potential and interacting with the quantized radiation; see \cite{BCFFS13}. 
{\bcb In fact, by expressing the l.h.s of (\ref{intro-first-derivative}) in terms of $\check{\phi}_{P,\si}$, we obtain 
\beqa \label{dev-phi}
\partial_{P_i} \check{\phi}_{P,\si}=\frac{1}{H^{W}_{P,\si}-E_{P,\si}} (\Gamma_{P,\si})_i\check{\phi}_{P,\si}+\Big(\int d^3k\,  f^{(i)}_{P,\si}(k) b(k)- h.c.\Big)\check{\phi}_{P,\si},
\eeqa
where $b(k)$ is the (improper) annihilation operator and $f^{(i)}_{P,\si}$ is a concrete function which behaves as $|k|^{-3/2}$ for  $\si\leq |k|\leq 1$. We conjecture that this latter singularity
is compensated by the the first term on the r.h.s. of (\ref{dev-phi}) and altogether $\|\partial_{P_i} \check{\phi}_{P,\si} \|\leq c$ holds uniformly in $\si$. Such a bound would require
certain refinements of estimate~(\ref{analytic-perturbation-theory}), with more explicit control of the singular behaviour. We will not pursue this direction here, but we believe that it is within reach
of the methods developed in this paper}. 


Our paper is organized as follows: In Section~\ref{Preliminaries-and-results} we state our main result about regularity
of the ground states and discuss a corollary about regularity of wave functions, proven in \cite{DP16}. In Section~\ref{Prel-sec}
we include some preliminaries on Hamiltonians and their ground states, the shift of the infrared cut-off, standard resolvent bounds,
direct integral representations and the maximal modulus principle. Section~\ref{section-on-norm-bounds} is the core of the 
paper, it includes the inductive proof of estimate~(\ref{eq:inequality-cont-1-intro}).  In Section~\ref{derivatives-section}  we derive
formulas for the derivatives of ground states and ground state energies, including (\ref{intro-first-derivative}), (\ref{second-deriv-form}),
and conclude the proof of our main result (Theorem~\ref{preliminaries-on-spectrum}), including estimates (\ref{first-derivative-estimate-intro}), (\ref{second-derivative-intro-x}). In the main part of the paper, outlined above, the focus is on new results. In the appendices we typically reestablish
(under our assumptions and for the present model) relevant results  from the existing literature.  Thereby the paper is essentially self-contained.

\vspace{0.5cm}

\noindent{\bf Acknowledgment:} A.P. thanks the Hausdorff Research Institute of Mathematics, Bonn, for hospitality. A.P. was supported by the NSF grant
\#DMS-0905988. A.P. acknowledges the MIUR Excellence Department Project awarded
to the Department of Mathematics, University of Rome Tor Vergata, CUP
E83C18000100006.

W.D.   thanks the University of California Davis and the Hausdorff  Research Institute for  Mathematics, Bonn,  for hospitality. 
W.D. is  supported by the German Research Foundation (DFG) within the
grants SP181/25--2,  DY107/1--1, DY107/2--1. Moreover, he would like to acknowledge the support of 
the Danish Council for Independent Research, grant no. 09-065927 "Mathematical Physics", and of the Lundbeck Foundation. Last
but not least, W.D. would like to thank his wife Jasmin for her continuing support in the course of this long-term project.

\section{Results}\label{Preliminaries-and-results} 
\setcounter{equation}{0}
In this section we state the main results of this paper which were announced already in Section~1.2 of \cite{DP12.1}.

Let $\mfh_{\fib}=L^2(\real^{3}, d^3k)$ be the single-photon subspace in the fiber picture.
Let $\mcF:=\Ga(\mfh_{\fib})$ be the symmetric Fock space over $\mfh_{\fib}$ and let us denote the
corresponding (improper) creation and annihilation operators  by $b^*(k)$ and $b(k)$. {\bcc The free
fiber Hamiltonian  of the massless Nelson model at a fixed total momentum $P$ has the form
\beqa
H_{P,\free}:=\h(P-P_{\pho})^2+H_{\pho}, \ \textrm{ where } \  H_{\pho}:=\int d^3k\, |k|b^*(k)b(k), \quad P_{\pho}:=\int d^3k\, k b^*(k)b(k), \label{free-Hamiltonian}
\eeqa
whereas} the interacting fiber Hamiltonian is given by
\beqa
H_{P}:=\h(P-P_{\pho})^2+H_{\pho}+\int d^3k\, \vv(k)\, ( b(k)+b^*(k) ). \label{Hamiltonian}
\eeqa
To define  the form factor $\vv$ we need to introduce some notation: Let $\mcB_r$ be the open ball of radius $r>0$ centered at zero.
Let $\chi_{r}\in C_0^{\infty}(\real^3)$ be a function which is rotationally invariant, non-increasing in the radial direction, supported in $\mcB_r$ 
and equal to one on $\mcB_{(1-\eps_0)r}$, for $0<\eps_0<1$.  Let $0<|\g|\leq 1$ be the coupling constant  
and $\ka>0$  the  ultraviolet cut-off which is kept fixed in our investigation. For simplicity of the presentation we will set $\ka=1$. The form-factor is given by
\beqa 
\vv(k):=\g\fr{  \chi_{\ka}(k) |k|^{\alf} }{(2|k|)^\half}, \label{form-factor-formula}
\eeqa
where $\h\geq \alf\geq 0$ is a parameter which controls the infrared behaviour of the system. (For $\alf=0$ we say that the model is infrared
singular, for $\alf>0$ that it is infrared regular). This parameter is kept fixed.

 An important role in the spectral analysis of the Hamiltonians $H_{P}$ is played by their counterparts  $H_{P,\si}$
with an infrared cut-off $\si\in (0,\ka]$. They are  given by
\beqa
H_{P,\si}:=\h(P-P_{\pho})^2+H_{\pho}+\int d^3k\, \vv^{\si}(k)\, ( b(k)+b^*(k) ). \label{infrared-cut-off-Hamiltonian}
\eeqa  
The form-factor $\vv^{\si}$, which carries information both about the (sharp) infrared cut-off $\si$ and the (smooth) ultraviolet cut-off $\ka$, is defined as
\beqa
\vv^{\si}(k):=\g\fr{ \chi_{[\si,\ka)}(k)    |k|^{\alf} }{(2|k|)^\half}, \label{IR-cut-off-propagator}
\eeqa
where $\chi_{[\si,\ka)}(k):=\mathbf{1}_{  \mcB'_{\si}   }(k)\chi_{\ka}(k)$, $\mcB_r:=\{\, k\in\mathbb{R}^{3} \,  | \, |k|<r \, \}$ and 
$\mcB_r'=\real^3\backslash \mcB_r$. {\bcc By the Kato-Rellich  theorem, $H_P$ and $H_{P,\si}$ are self-adjoint operators on 
$D(H_{P,\free})$. }

As our analysis concerns the  bottom of the spectrum of the fiber Hamiltonians, let us define
\beqa
 E_{P}:=\inf \sib(H_{P}), \quad  E_{P,\si}:=\inf \sib(H_{P,\si}), \label{original-definition}
\eeqa
where $\sib$ denotes the spectrum.
Since the model is non-relativistic, we are interested in small values of the total momentum $P$ for which the electron 
moves slower than the photons. For this reason we consider $P$ from the set 
\beqa
\PS:=\{\, P\in \real^3\,|\, |P|< P_{\maxi}\,\}
\eeqa
for  $P_{\maxi}=1/3$. Since we  work in the weak coupling regime, we  fix some sufficiently small $\g_0>0$, specified in Theorem~\ref{preliminaries-on-spectrum},
and restrict attention to $|\g|\in (0, \g_0]$.

In the following theorem, which is our main result,  we collect the results concerning $E_P, E_{P,\si}$  and the corresponding ground state vector $\cpsi_{P,\si}$ 
which are  needed in the scattering theory of two electrons in \cite{DP12.1}. 
\begin{thm}\label{preliminaries-on-spectrum} Fix $0\leq \alf \leq 1/2$ and let $P_{\maxi}=1/3$. Then there exists $\g_0>0$ and $\ka \geq \ka_{\la_0}>0$ s.t. for all $|\g|\in (0, \g_0]$ and $P\in S:=\mcB_{P_{\maxi}}$
  the following statements hold:
\begin{enumerate}[label = \textup{(\alph*)}, ref =\textup{(\alph*)},leftmargin=*]
\item\label{cut-off-part} For $\si\in (0,\ka_{\la_0}]$, $E_{P,\si}$ is a simple eigenvalue corresponding to a normalized eigenvector $\cpsi_{P,\si}$,
whose phase is fixed in Definition~\ref{abuse-definition} below. $S\ni P\mapsto E_{P,\si}$ is analytic and strictly convex, {\color{\red}  for all} $\si\in (0,\ka_{\la_0}]$.
Moreover,  for some $0<\de_{\la_0}<1/4$, specified below
\beqa
& &|\pa_{P}^{\be_1}E_{P,\si}|\leq c,\quad  |\pa_{P}^{\be_2}E_{P,\si}|\leq c, \quad  |\pa_{P}^{\be_3}E_{P,\si}|\leq c/\si^{\de_{\la_0}} \label{velocity-boundedness}\\
& &\|\pa^{\be}_P\cpsi_{P,\si}\|\leq  c /\si^{\de_{\g_0}} \label{state-bound}
\eeqa
for  multiindices $\be_i$, $\be$ s.t. $|\be_i|=i$ and $0<|\be|\leq 2$. 
\item\label{energy-part} For $\si\in (0,\ka_{\la_0}]$ the estimate 
\begin{equation}
|E_{P}-E_{P,\si}|\leq c\si \label{energy-convergence-bound}
\end{equation}
holds true. Moreover,  $S\ni P\mapsto E_P$ is twice continuously differentiable and strictly convex.
\item\label{eigenvectors-convergence} For $\alf>0$, $E_P$ is an eigenvalue corresponding to a normalized eigenvector $\cpsi_{P}$.
Moreover, for a suitable choice of the phase of $\cpsi_{P}$ and  $\si\in (0,\ka_{\la_0}]$
\beqa
\|\cpsi_P-\cpsi_{P,\si}\|\leq c {\bcb (\alf)^{-1} }   \si^{\alf}.
\eeqa
\end{enumerate}
The constant $c$ above is independent of $\si$, $P$, $\g$, $\alf$ within the assumed restrictions. Clearly, all statements above
remain true after replacing $\g_0$ by some $\ti\g_0\in (0,\g_0]$. The resulting function
  $\ti\g_0\mapsto \de_{\ti\g_0}$ can be chosen s.t. $\lim_{\ti\g_0\to 0} \de_{\ti\g_0}=0$.
\end{thm}
We give the proof of this theorem in the last part of Section~\ref{derivatives-section}. The most innovative statements are  
the bound on the third derivative of $E_{P,\si}$ in (\ref{velocity-boundedness}) and on the first and second derivative
of $\cpsi_{P,\si}$ in (\ref{state-bound}).
Other items  have already been established  in the
Nelson model or in similar models: The fact that $S\ni P\mapsto E_P$, $S\ni P\mapsto E_{P,\si}$ are twice continuously differentiable and convex has been
shown  in non-relativistic and semi-relativistic QED in \cite{FP10, KM12, BCFS07} and in the
Nelson model with a slightly different form-factor in  \cite{AH12}.  
The bound in (\ref{energy-convergence-bound})
can be extracted from \cite{Pi03}. The first statement in \ref{cut-off-part}  has been established already in \cite{Fr74.1}.
Part \ref{eigenvectors-convergence} is implicit in \cite{Pi03}.

It turns out that the regularity of the vector $\cpsi_{P,\si}$, established in Theorem~\ref{preliminaries-on-spectrum},
is not sufficient for scattering theory of two electrons. One also needs analogous bounds for the wave functions  
$\{f^{\mm}_{P,\si} \}_{\mm\in \nat_0}$ of $\cpsi_{P,\si}$. Although these estimates are thoroughly discussed and proven in
a companion paper \cite{DP16}, we state them here as they constitute an interesting corollary of Theorem~\ref{preliminaries-on-spectrum}.

Clearly, each $f^{\mm}_{P,\si}$   is a square-integrable function symmetric in $\mm$ variables from $\real^3$. We introduce the following auxiliary functions:
\beqa
 g^{\mm}_{\si}(k_1,\ldots, k_{\mm}):=\prod_{i=1}^{\mm}\fr{ c\g  \chi_{[\si,\kas)}(k_i) |k_i|^{\alf} }{|k_i|^{3/2}},\quad \kas:=(1-\eps_0)^{-1}\ka, \label{def-g-ks}
\eeqa
{\color{\red}for some $0<\eps_0<1$}, where $\tic$ is a positive constant independent of $\mm,\si, P$ and $\g$ within the restrictions specified above. 
Next, we introduce the notation
\beqa
\mcA_{r_1,r_2}:=\{\, k\in \real^3 \,|\, r_1<|k|<r_2 \,\}, \label{A-set}
\eeqa
where $0\leq r_1< r_2$. 
Now we are ready to state the required properties of the functions $f^{\mm}_{P,\si}$:

\begin{thm}\label{main-theorem-spectral} 
Fix $0\leq \alf \leq 1/2$  and let $P_{\maxi}=1/3$. Then there exists  $\lambda_0>0$ and $\ka> \ka_{\la_0}>0$ s.t. for all $P \in S=\mathcal{B}_{P_{\maxi}}$,
$\lambda\in (0, \lambda_0]$ and $\si\in (0, \ka_{\la_0}]$ there holds:
\begin{enumerate}[label = \textup{(\alph*)}, ref =\textup{(\alph*)},leftmargin=*]

\item \label{f-m-support} Let  $\{f^q_{P,\sigma}\}_{q\in\nat_0}$ be the $q$-particle components of $\cpsi_{P,\sigma}$
and let $\overline{\mathcal{A}}_{\sigma,\kappa}^{\times q}$ be defined  as the Cartesian product of $q$ copies of the closure of the set $\mathcal{A}_{\sigma,\kappa}$ introduced in (\ref{A-set}). Then,
  for any $P\in S$, the function $f^q_{P,\sigma}$ is supported in $\overline{\mathcal{A}}_{\sigma,\kappa}^{\times q}$.

\item \label{f-m-smoothness} The  function
\begin{equation}
S\times \mathcal{A}_{\sigma,\infty}^{\times q}\ni (P; k_1,\dots, k_q) \mapsto f^q_{P,\sigma}(k_1,\dots, k_q) \label{momentum-wave-functions}
\end{equation}
is twice continuously differentiable and extends by continuity, together with its derivatives, to the set 
$S\times \overline{\mathcal{A}}_{\sigma,\infty}^{\times q}$. 

\item\label{derivatives-bounds} For any multiindex $\beta$, $0\leq |\beta|\leq 2$, the function (\ref{momentum-wave-functions}) satisfies
\begin{eqnarray}
|\partial^{\beta}_{k_l}f^q_{P,\sigma}(k_1,\dots, k_q)|  \1&\leq&  \1 \frac{1}{\sqrt{q!}} |k_l|^{-|\beta|}g^q_{\sigma}(k_1,\dots, k_q), \label{simple-spectral-bound}\\
|\partial^{\beta}_{P}f^q_{P,\sigma}(k_1,\dots, k_q)|  \1 &\leq&  \1  \frac{1}{\sqrt{q!}} \bigg( \frac{1}{\sigma^{\delta_{\lambda_0}}}\bigg)^{| \beta |}  g^q_{\sigma}(k_1,\dots, k_q), \label{infrared-spectral-bound}\\
|\partial_{P^{i'}}\partial_{k_l^i}f^q_{P,\sigma}(k_1,\dots, k_q)|  \1 &\leq&  \1 \frac{1}{\sqrt{q!}}\frac{1}{\sigma^{\delta_{\lambda_0}}}|k_l|^{-1}g^q_{\sigma}(k_1,\dots, k_q), \label{mixed-spectral-bound}
\end{eqnarray}
where the function $\tilde{\lambda}_0 \mapsto \delta_{\tilde{\lambda_0}}$ has the properties specified in Theorem~\ref{preliminaries-on-spectrum} and $g^q_{\sigma}$ is defined in (\ref{def-g-ks}).
\end{enumerate}
\end{thm}
We give the proof of Theorem~\ref{main-theorem-spectral} {\bcc in  a companion paper \cite{DP16}}.  Parts \ref{f-m-support}, \ref{f-m-smoothness}   and   estimate~(\ref{simple-spectral-bound}) in 
\ref{derivatives-bounds}  can be extracted from \cite{Fr73,Fr74.1,Fr73.1} or proven using the methods from these papers. 
Estimate~(\ref{infrared-spectral-bound}) for $|\be|=1$ and (\ref{mixed-spectral-bound}) rely on the standard formula for the ground state
wave functions from \cite{Fr73,Fr74.1} combined with the spectral ingredient (\ref{analytic-perturbation-theory}). This latter ingredient is relatively well
established in models of non-relativistic QED \cite{FP10,KM12} and proven for the Nelson model in the present paper (see Theorem~\ref{thm:induction-convergence} below). The key new result  in Theorem~\ref{main-theorem-spectral} is the bound (\ref{infrared-spectral-bound}) for $|\be|=2$ which 
requires a novel, `infrared safe'  formula for the wave functions, derived in \cite{DP16} by solving a non-commutative recurrence relation.
This new formula is then combined with the spectral ingredient (\ref{eq:inequality-cont-1-intro}) which is the main technical result of the present paper (see Theorem~\ref{main-technical-result} below). 
  
  The infrared regular case (i.e. $0<\alf\leq 1/2$) is meanwhile relatively well understood (see e.g. \cite{GH09,FFS14}). It is likely that in this situation    the results  from Theorems~\ref{preliminaries-on-spectrum}, \ref{main-theorem-spectral}  could be strengthened and
  their proofs simplified. While this would suffice for scattering of two electrons in the infrared regular Nelson model, described
  in \cite{DP12.1}, it would not help to control collisions of two infraparticles, which is the goal of this project. In this context the
  iterative analytic perturbation theory, which we use and further develop in this paper, appears to be the optimal approach.

\vspace{1.2cm}

\noindent \bf Standing assumptions and conventions\rm: 
\begin{enumerate}

\item We  denote by $c,c',c''$ numerical constants which may change from line to line.  These constants are universal in the sense that are independent of $\si$, $P$, $\g$, $\alf$, $\epsilon$, within the assumed restrictions.

\item Upper or lower indices $i,i',\hhi,\hhi'=1,2,3$ denote components of  vectors in $\real^3$.

\item We denote by $\lan \psi_1, \psi_2\ran$ the scalar product of two vectors $\psi_1,\psi_2$ in a Hilbert space.

\item {\color{\red}The contour integrals are oriented counterclockwise.} Moreover, we use the convention
\[
-\frac{1}{2\pi i}\oint_{\gamma}\equiv\oint_{\gamma}
\]
for a contour $\gamma$. With this convention $\oint_{\gamma} dz/z=-1$ if $\gamma$ is a circle centered at zero.

\item The symbol $\mathcal{B}_{r}$ denotes the set $\{\, k\in\mathbb{R}^{3}\,  : \,|k|<r \, \}$.

\item $\chi_{r}\in C_0^{\infty}(\real^3)$ is a function which is rotationally invariant, non-increasing in the radial direction, supported in $\mcB_r$ 
and equal to one on $\mcB_{(1-\eps_0)r}$, for some fixed $0<\eps_0<1$. 

\item {\bcb The symbol $\int_{r_1}^{r_2}   d^3k$ stands for $\int_{\mcB_{r_2}\backslash \mcB_{r_1}  } d^3k$.}

\item The symbol $\hint_{\!\! r_1}^{r_2}   d^3k$ stands for $\int_{\mcB_{r_2}\backslash \mcB_{r_1}  } d^3k \, \chi_{\ka}(k)|k|^{\alf}$, where $\kappa=1$ is
the UV cut-off. 

\item The symbol $\bint_{\!\! r_1}^{r_2}   d^3k$ stands for $\int_{\mcB_{r_2}\backslash \mcB_{r_1}  } d^3k \, \chi_{\ka}(k)^2|k|^{2\alf}  $.

\item We set $\sigma_{n}:=\kappa\epsilon^{n}$, where  $n\in \nat_0$ and $0<\epsilon\leq\frac{1}{2}$, $\ka=1$.

\item We set $\mcA|^{n-1}_n:=\mcA_{\si_n, \si_{n-1}}=\{\, k\in \real^3 \,|\, \si_n<|k|<\si_{n-1} \,\}$.

\item The symbol $\hint_{\!\!\!\!(\bmcA|^{n-1}_n)^{\times m}}d^{3m}k$ stands for $\int_{\mcB_{\si_{{\color{\red}n-1}}}\backslash \mcB_{\si_{{\color{\red}n}}}  } d^3k_1 \, \ldots \int_{\mcB_{\si_{{\color{\red}n-1}}}\backslash \mcB_{\si_{{\color{\red}n}}}  } d^3k_m\, \chi_{\ka}(k_1)|k_1|^{\alf}\ldots \chi_{\ka}(k_m)|k_m|^{\alf}$.

\item The symbol  $\bint_{\!\!\!\!(\bmcA|^{n-1}_n)^{\times m}}d^{{\color{\red}3m}}k$ stands for  $\int_{\mcB_{\si_{{\color{\red}n-1}}}\backslash \mcB_{\si_{{\color{\red}n}}}  } d^3k_1 \, \ldots \int_{\mcB_{\si_{{\color{\red}n-1}}}\backslash \mcB_{\si_{{\color{\red}n}}}  } d^3k_m\, \chi_{\ka}(k_1)^2|k_1|^{2\alf}\ldots \chi_{\ka}(k_m)^2|k_m|^{2\alf}$.

\item  If $X$ is an element of a Banach space  and $\|X\|\leq cY$ for some $Y\in \real_+$ then we write $X=\mco(Y)$.
If, furthermore, $Y\leq c'Z$ for some $Z\in \real_+$ we will write $X=\mco(Y)=\mco(Z)$.

\item The symbol $\mathbb{N}_0$ stands for $\mathbb{N}\cup\{0\}$.

\end{enumerate}

\section{Preliminaries}\label{Prel-sec}
\subsection{Hamiltonians and their ground states} \label{first-lemma-app}
\setcounter{equation}{0}

{\bcc This subsection concerns the ground states of physical, transformed and intermediate Hamiltonian{\color{\red}s}, defined by 
formulas~(\ref{H-physical}), (\ref{transformed-hamiltonian}) and (\ref{intermediate-Ham}), respectively. One reason for
introducing the transformed Hamiltonians is that their ground states exist in the limit $\si\to 0$ (cf. Corollary~\ref{vector-existence} below.)}

First, we introduce a sequence of infrared cut-offs 
\beqa
\sigma_{n}:=\kappa\epsilon^{n},\quad n\in \nat_0, \quad \quad0<\epsilon\leq\frac{1}{2}
\eeqa
and the corresponding cut-off Hamiltonians at fixed total momentum and fixed ultraviolet
cut-off $\kappa=1$, i.e.,
\beqa
H_{P,\si_n}=\frac{(P-P_{\pho})^{2}}{2}+\lambda\hint_{\!\!\!\si_n}^{\ka}d^3k\frac{1}{\sqrt{2|k|} }\{b(k)+b^{*}(k)\}+H_{\pho}.
\label{H-physical}
\eeqa
From now on, we will also write $H_{P,n}:=H_{P,\si_n}$ and $E_{P,n}:=E_{P,\si_n}$. \begin{rem} Since we work in the low-coupling regime, we
will always assume that $|\la|\in (0, \la_0]$, where $\lambda_0$ is assumed to be sufficiently small though we will often omit this specification.  
Furthermore, we assume that $\veps\in (0,\veps_0]$ for some $0<\veps_0\leq 1/2$. 
The maximal values $\la_0, \veps_0$ will be reduced in the course of the argument, but  will remain non-zero
and depend only on universal constants. (Until the end of Section~\ref{Prel-sec}
we will have $\veps_0=1/2$, then this parameter will be decreased in 
Theorems~\ref{thm:induction-convergence} and \ref{main-technical-result}). 
At any stage of the paper it is  assumed that $\la_0$ and $\veps_0$ are  chosen {\color{\red}small enough} so that all the preceding results hold. 
\end{rem}

For $n\in\mathbb{N}_0$, we define the bosonic Fock spaces for $ m \leq n$
\begin{eqnarray}
{\cal F}\!\!\!&:=&\!\!\!{\Gamma}\left(L^{2}\left(\mathbb{R}^{3},d^3k\right)\right),\label{Fock1}\\
{\cal F}_{n}\!\!\!&:=&\!\!\!{\Gamma}\left(L^{2}\left(\mathbb{R}^{3}\setminus{\cal B}_{\sigma_{n}}, d^3k\right)\right),\qquad\label{Fock2}\\
{\cal F}|^{m}_{n}\!\!\!&:=&\!\!\!{\Gamma}\left(L^{2}\left({\cal B}_{\sigma_{m}}\setminus{\cal B}_{\sigma_n}, d^3k\right)\right).\label{Fock3}
\end{eqnarray}
In all these Fock spaces we shall use the same symbol $\Omega$ to
denote the vacuum vector. Clearly, $\calF_n=\calF_{n-1}\otimes \calF|_n^{n-1}$.
For a vector $\psi$ in $\mathcal{F}_{n-1}$ and
an operator $O$ on $\mathcal{F}_{n-1}$ we shall use the same symbol
to denote the vector $\psi\otimes\Omega$ in $\mathcal{F}_{n}$ and
the operator $O\otimes 1_{\mathcal{F}|_{n}^{n-1} }$ on $\mathcal{F}_{n}$,
respectively, where $1_{\mathcal{F}|_{n}^{n-1}}$ is the identity
operator on ${\cal F}|_{n}^{n-1}$ (e.g., 
$\hint_{\!\!\kappa\epsilon^{n-1}}^{\kappa}d^3k\, b(k)\,
\upharpoonleft{\cal F}_{n}\equiv\hint_{\!\!\kappa\epsilon^{n-1}}^{\kappa}d^3k\, b(k)\otimes 1_{\mathcal{F}|_{n}^{n-1}}$).  
{\color{\red} For a given operator $A$ on $\mcF$ we set $\|A\|_{\mathcal{F}_n}:=\|A \res \mathcal{F}_n \|$. }
The Hamiltonians $H_{P,n}$ act on the Hilbert space $\cal F$. 

{\bcc Proposition~\ref{prop:Hamiltonian-results}, stated below, provides a basis for our investigation. This result is
standard, but  {\color{\red}in the available literature (see \cite{Pi03, BDP12, CFP09, KM12}}) there are differences in terms of models and ranges of parameters. Thus to make our work accessible {\color{\red}to} non-experts we provide a rather detailed proof in 
Appendix~\ref{proof-of-basic-lemma}.} Some intermediate results from Appendix~\ref{proof-of-basic-lemma} will also be used at later stages of our discussion (e.g., estimate (\ref{Gap-bound}) enters into the proof of the resolvent bound (\ref{L-standard-estimates}) below).
\begin{prop}
\label{prop:Hamiltonian-results} 
Let   $|P|\leq P_{\max}=1/3$,  $|\la|\in (0,\la_0]$, $\veps\in (0,1/2]$ and $|\la|\leq \epsilon^2$.  {\bcc  Then, for all $n\in \nat_0$ 
the resolvent $(H_{P,n}-z)^{-1}\res \mcF_n$ is well-defined in 
$(1/16)\si_n\leq |E_{P,n-1}-z|\leq (1/3)\si_n$. Furthermore,
the Hamiltonian $H_{P,n}\res \mcF_n$
has a unique (unnormalized) ground state vector $\psi_{P,n}$, and the corresponding
ground state energy coincides with  $E_{P,n}:=\inf\sib(H_{P,n}\res\mcF)$. $\psi_{P,n}$ is defined iteratively
from $\psi_{P,0}\equiv\Omega$ by the following relation valid for $n\geq1$, }
\beqa
\psi_{P,n}:=\oint_{\gamma_{P,n}}\frac{dw_{n}}{H_{P,n}-w_{n}}\psi_{P,n-1},\quad
\gamma_{P,n}:=\left\{w_{n}\in\mathbb{C}\,\, \big| \,\, |w_{n}-E_{P,n-1}|=\frac{\sigma_{n}}{3}\right\}.\label{first def psi}
\eeqa
The following properties hold true: The function $S\ni P\mapsto E_{P,n}$ is analytic for any fixed  $n\in\nat_0$. 
Moreover,
\beqa
& &E_{P-k,n}-E_{P,n}\geq-\bigg(\frac{1}{3}+c|\lambda|\bigg)|k|,\quad\quad k\in\mathbb{R}^{3}, \label{eq:difference-bound} \\
& &|\nabla E_{P,n}|\leq\frac{1}{3}+c|\lambda|,\label{eq:gradient-bound}\\
& &E_{P,n}\leq E_{P,n-1},\label{eq:monotonicity}\\
& &|E_{P,n-1}-E_{P,n}|\leq c_{\Delta E}\lambda^{2}\sigma_{n-1}, \label{eq:energy-shift}\\
& &\mathrm{Gap}\, (H_{P,n}\upharpoonleft\mathcal{F}_{n}):=\inf_{\phi\perp\psi_{P,n},\phi\in\mathcal{F}_{n}}\lan\frac{\phi}{\|\phi\|},(H_{P,n}-E_{P,n})\frac{\phi}{\|\phi\|}\ran\geq\frac{\sigma_{n}}{3},\label{eq:gap-estimate} \\
& &{\bcc \mathrm{Gap}\, (H_{P,n-1}\res \mathcal{F}_{n}) \geq \fr{\si_n}{2}}, \label{last-gap-estimate}
\eeqa
where $c$, $c_{\Delta E}$ are universal constants. 
\end{prop}
{\bcc \begin{rem} It is clear from the above proposition that formula (\ref{first def psi}) {\color{black} and the estimate in (\ref{eq:energy-shift}) imply}
\beqa
\psi_{P,n}=\oint_{\gamma_{P,n+1}}\frac{dw_{n+1}}{H_{P,n}-w_{n+1}}\psi_{P,n-1}, \label{second def psi}
\eeqa
as we modified the contour within the region of holomorphy. This variant will  be more convenient for applications
in this paper.
\end{rem}
}

Next, we introduce the transformed Hamiltonians and some related formulas
that will be needed in Lemma~\ref{lem:Basic-estimates} and for the
proof of Theorem \ref{thm:induction-convergence}. These definitions have been introduced in  \cite{Pi03}.
\begin{defn}\label{def:transformed-hamiltonian-results} Let us introduce the Weyl operators:
\beqa
& &W_{P,n}:=\exp(-\lambda\hint^{\kappa}_{\!\!\!\sigma_n}\frac{d^3k}
{\sqrt{2}|k|^{\frac{3}{2}}\alpha_{P,n}(\hat{k} ) }\{  b(k)-b^{*}(k) \}), \label{W-definition} \\
& &\widetilde{W}_{P,n}:=\exp(-\lambda\hint^{\kappa}_{\!\!\!\sigma_n}\frac{d^3k}{\sqrt{2}|k|^{\frac{3}{2}}\alpha_{P,n-1}(\hat{k})}\{b(k)-b^{*}(k)\})\,, \label{intermediate}
\eeqa
where $\alpha_{P,n}(\hat{k}):=(1-\hat{k}\cdot\nabla E_{P,n})$. The \emph{transformed} Hamiltonians and the \emph{intermediate} Hamiltonians 
are given by
\beqa
& &H_{P,n}^{W}:=W_{P,n}H_{P,n}W_{P,n}^{*}, \label{transformed-hamiltonian}\\
& &\hat{H}_{P,n}^{W}:=\widetilde{W}_{P,n}H_{P,n}\widetilde{W}_{P,n}^{*}=
\widetilde{W}_{P,n}W_{P,n}^{*}H_{P,n}^{W}W_{P,n}\widetilde{W}_{P,n}^{*}\,,
\label{intermediate-Ham}
\eeqa
respectively. The equalities are meant on $D(H_{P,n})=D(H_{P,\free})$ which is invariant under $W_{P,n}^{*}, \widetilde{W}_{P,n}^{*}$ \cite{Fr77}.
(Equalities of unbounded operators in the remaining part of this section are meant on $D(H_{P,\free})$ unless stated otherwise).
\end{defn}
Now we recall an iterative construction of ground states of the modified and transformed Hamiltonians from~\cite{Pi03}.
\begin{cor} \label{phi-vectors-corollary}{\bcc Let   $|P|\leq P_{\max}=1/3$,  $|\la|\in (0,\la_0]$, $\veps\in (0,1/2]$ and $|\la|\leq \epsilon^2$. }
The ground state vectors $\phi_{P,n}$
and $\hat{\phi}_{P,n}$ of the Hamiltonians $H_{P,n}^{W}$ and $\hat{H}_{P,n}^{W}$,
respectively, are iteratively defined by 
\beqa
{\bcc \hat{\phi}_{P,n}:=\oint_{\gamma_{P,n+1}}\frac{dw_{n+1}}{\hat{H}_{P,n}^{W}-w_{n+1}}\phi_{P,n-1}},
\quad \phi_{P,n}:=W_{P,n}\widetilde{W}_{P,n}^{*}\hat{\phi}_{P,n},\quad\phi_{P,0}:=\Omega. \label{phi-definitions}
\eeqa
{\bcc They satisfy }
\beqa
|\nabla E_{P,n-1}-\nabla E_{P,n}|\leq c_{1}[\lambda^{2}\sigma_{n-1}+\|\frac{\hat{\phi}_{P,n}}{\|\hat{\phi}_{P,n}\|}-\frac{\phi_{P,n-1}}{\|\phi_{P,n-1}\|}\|]. \label{grad-difference}
\eeqa
\end{cor}
\proof{\bcc {\color{\red}The contour integral in expression}~(\ref{phi-definitions}) is well defined by Proposition~\ref{prop:Hamiltonian-results} and (\ref{second def psi}). 
Estimate~(\ref{grad-difference}), which follows from the proof of Lemma A.2 of \cite{Pi03}, is  verified in Lemma~\ref{gradient-shift-lemma}}. \qed
\subsection{{\color{\red} Shift} of the cut-off from $\si_{n-1}$ to $\si_{n}$} \label{shift-of-the-cut-off-subsection}
{\bcc In Subsection~\ref{first-lemma-app} we studied how the ground state energy $E_{P,n}$ and the ground state
vectors $\psi_{P,n}, \phi_{P,n}, \hat\phi_{P,n}$ behave under the change of the {\color{\red}infrared} cut-off by one step. In this subsection
we determine the behaviour of various relevant operator quantities. The most important relations are (\ref{eq:resolvent-expansion}),
(\ref{shift-of-gamma}), (\ref{shift-of-Q}). We start with a simple computational result. }
\bel\label{transformed-Ham}\cite{Pi03} The transformed Hamiltonian has the form
\beqa \label{def-H^W}
H_{P,n}^{W}=\frac{\Gamma_{P,n}^{2}}{2}+\int d^3k\,\alpha_{P,n}(k)|k|b^{*}(k)b(k)+c_P^{\sigma_n},
\eeqa
where 
\beqa
\Gamma_{P,n}\!\!\!&:=&\!\!\!\Pi_{P,n}-\frac{\lan W_{P,n}\psi_{P,n},\,\Pi_{P,n}W_{P,n}\psi_{P,n}\ran}{\lan \psi_{P,n},\,\psi_{P,n}\ran}
={\bcc \nabla E_{P,n}-W_{P,n}(P-P_{\pho})W_{P,n}^*}, 
\label{Ga-formula} \\
\Pi_{P,n}\!\!\!&:=&\!\!\!P_{\pho}-\lambda\hint^{\kappa}_{\!\!\!\sigma_n}\frac{d^3k\, k}{\sqrt{2}|k|^{\frac{3}{2}}\alpha_{P,n}(\hat{k}) }\{b(k)+b^{*}(k)\},\label{Pi-formula}\\
c_P^{\sigma_n}\!\!\!&:=&\!\!\!\frac{P^2}{2}-\frac{1}{2}(P-\nabla E_{P,n})^2-\lambda^2{\bint}^{\kappa}_{\!\!\!\sigma_n} \frac{d^3k}{2|k|^2\alpha_{P,n}(\hat{k})}.
\label{c-P-formula}
\eeqa
{\bcc 
Furthermore, $\lan \phi_{P,n}, \Ga_{P,n}\phi_{P,n}\ran=0$}.
\eel
{\bcc\nin 
In Lemma \ref{resolvent-expansion-one} we will consider a new contour, namely
\beqa\label{def-gammatilde-0}
{\bcc \tigam_{P,n}:=\{\, z_n\in\complex\,|\, \Re\, z_n=E_{P,n-1}+\fr{\si_n}{3}\,\}, }
\eeqa
which appears in the statements of Theorems~\ref{thm:induction-convergence} and \ref{main-technical-result}. {\color{\red}We stress that while} the former
theorem could  be proven with the standard contour $\ga_{P,n}$, this does not seem to be the case for the latter theorem.
The following expansion is used in the first step of the proof of Theorems~\ref{thm:induction-convergence} and \ref{main-technical-result} .
\bel\label{resolvent-expansion-one}\cite{Pi03}   
For   $|P|\leq P_{\max}=1/3$,  $|\la|\in (0,\la_0]$, $\veps\in (0,1/2]$ and $|\la|\leq \epsilon^4$ 
the resolvent of the intermediate Hamiltonian~(\ref{intermediate-Ham}) satisfies 
\beqa
\frac{1}{\hat{H}_{P,n}^{W}-{\bcc z_{n+1}}}=
\frac{1}{H_{P,n-1}^{W}+\Delta c_{P}|_{n}^{n-1}-z_{n+1}}
\sum_{j=0}^{\infty}\{-H_I^{W}|_{n}^{n-1}\frac{1}{H_{P,n-1}^{W}+\Delta c_{P}|_{n}^{n-1}-z_{n+1}}\}^{j},
\label{eq:resolvent-expansion}
\eeqa
{\bcc for $z_{n+1}\in \gamma_{P,n+1}\cup \ti\gamma_{P,n+1}$, where the equality is understood on $\mcF_n$} and
\beqa
H_I^{W}|_{n}^{n-1} \1&:=& \1\frac{1}{2}(\Gamma_{P,n-1}\cdot(\mathcal{L}|_{n}^{n-1}+\mathcal{I}|_{n}^{n-1})+\mathrm{h.c.})
+\frac{1}{2}(\mathcal{L}|_{n}^{n-1}+\mathcal{I}|_{n}^{n-1})^{2},\\
\mathcal{L}|_{n}^{n-1} \1&:=& \1-\lambda\hint_{\!\!\sigma_{n}}^{\sigma_{n-1}}d^{3}k\frac{k(b(k)+b^{*}(k))}{\sqrt{2}|k|^{3/2}\alpha_{P,n-1}(\hat{k})},
\label{Ln-definitions}\\
\mathcal{I}|_{n}^{n-1} \1&:=& \1 \lambda^{2} 
{\bint}_{\!\!\!\sigma_{n}}^{\sigma_{n-1}}  d^{3}k\frac{k}{2|k|^{3}[\alpha_{P,n-1}(\hat{k})]^{2}},
\label{I-definitions}\\
\Delta c_{P}|_{n}^{n-1} \1&:=& \1-\lambda^2{\bint}^{\sigma_{n-1}}_{\!\!\!\sigma_{n}}\frac{d^3k}{2|k|^2\alpha_{P,n}(\hat k)}. \label{Delta-cp}
\eeqa
\eel
\ber
We warn the reader that $H_I^{W}|_{n}^{n-1}$ depends on $P$, although this is not reflected by our notation.
\eer
\proof {\bcc By a standard computation we obtain $\hat{H}_{P,n}^{W}=H_{P,n-1}^W+\Delta c_P|_n^{n-1}+H_I^{W}|_{n}^{n-1}$.  The convergence  of the {\color{\red}series}  follows from estimate (\ref{H-I-standard-estimates}) below. Also the restriction $|\la|\leq \veps^4$ enters via this estimate. \qed}\\
\nin The following two lemmas are obtained again by standard computations. We remark that formula~(\ref{shift-of-Pi}) below enters into the proof
of estimate~(\ref{grad-difference}) in Lemma~\ref{gradient-shift-lemma}}.
\bel\label{Pi-shifting}\cite{Pi03} We recall that $\Pi_{P,n}$ is given by (\ref{Pi-formula}) and define
\beqa
\hat{\Pi}_{P,n}:=\widetilde{W}_{P,n}W_{P,n}^{*}\Pi_{P,n}W_{P,n}\widetilde{W}_{P,n}^{*}.
\eeqa
There holds
\beqa
\hat{\Pi}_{P,n}-\Pi_{P,n-1}
=\mathcal{L}|_{n}^{n-1}+\la^2{\bint}_{\!\!\!\si_{n}}^{\si_{n-1}} d^3k\, \fr{k(\al_{P,n}^2(\hk)-\al_{P,n-1}^2(\hk))  }{2|k|^3 \al_{P,n-1}^2(\hk)\al_{P,n}^2(\hk)}.
\label{shift-of-Pi}
\eeqa
\eel
\bel\label{Gamma-shifting}\cite{Pi03} We recall that $\Ga_{P,n}$ is given by (\ref{Ga-formula}). We define
\beqa
\hat{\Gamma}_{P,n} \1&:=& \1\widetilde{W}_{P,n}W_{P,n}^{*}\Gamma_{P,n}W_{P,n}\widetilde{W}_{P,n}^{*},\\
\Delta\Gamma_{P}|_{n}^{n-1}  \1&:=& \1 \hat{\Gamma}_{P,n}-\Gamma_{P,n-1}\label{eq:delta-gamma},\\
\Delta'\Gamma_{P}|_{n}^{n-1} \1&:=& \1-\nabla E_{P,n-1}+\nabla E_{P,n}+\mathcal{L}|_{n}^{n-1}.\label{eq:delta'-gamma}
\eeqa
There holds
\beqa\label{def-deltagamma}
\Delta\Gamma_{P}|_{n}^{n-1}=\Delta'\Gamma_{P}|_{n}^{n-1}+\mathcal{I}|_{n}^{n-1},
 \eeqa
or, in other words,
\beqa
\hat{\Gamma}_{P,n}-\Gamma_{P,n-1}=-\nabla E_{P,n-1}+\nabla E_{P,n}+\mathcal{L}|_{n}^{n-1}+\mathcal{I}|_{n}^{n-1}. 
\label{shift-of-gamma}
\eeqa
\eel
{\bcc \nin The reader should note that the difference $\hat{\Gamma}_{P,n}-\Gamma_{P,n-1}$ in formula~(\ref{shift-of-gamma}) is 
controlled by the difference $-\nabla E_{P,n-1}+\nabla E_{P,n}$, which in turn depends on $\hat\phi_{P,n}-\phi_{P,n-1}$
via estimate~(\ref{grad-difference}). These relations will be used in the proof of Theorems~\ref{thm:induction-convergence} and \ref{main-technical-result}.

In the last lemma of this subsection we analyse the ground state projections. This information will be used in the first step of the proof of Theorem~\ref{main-technical-result}. }
\bel\label{Q-shifting}  Recall from Remark~\ref{energy-consistency} that 
the ground state eigenvector of $H_{P,n}$ (defined on $\mcF$) is
$\psi^{(\infty)}_{P,n}{\color{black}:}=\psi_{P,n}\otimes \Om$ in $\mcF=\mcF_n\otimes \mcF|^n_{\infty}$. {\bcb Consistently with (\ref{second def psi}), we define the following projections on $\mcF$
for $n\geq 1$ }
\beqa
& &\Q_{P,n}:={\bcb \oint_{\ga_{P,n+1}}\fr{dw_{n+1}}{H_{P,n}-w_{n+1}},}
\quad\quad\quad \Q_{P,n}^{\bot}{\color{black}:}=1-  \Q_{P,n},     \label{Q-def-first}\\
& &Q_{P,n}:=W_{P,n} \Q_{P,n} W_{P,n}^*,  \quad\quad\quad\quad\quad   Q_{P,n}^{\bot}{\color{black}:}=1-Q_{P,n},     \label{3.37}\\
&    &\hat{Q}_{P,n}:=\widetilde{W}_{P,n}W_{P,n}^{*}{\color{black}Q_{P,n}}W_{P,n}\widetilde{W}_{P,n}^{*},
\quad \   \hat{Q}_{P,n}^{\bot}{\color{black}:}=1-\hat{Q}_{P,n}.
\label{Q-def-last}
\eeqa
$\Q_{P,n}, Q_{P,n}, \hat{Q}_{P,n}$  leave $\mcF_m$ invariant for $m\geq n$.
Furthermore, for   $|P|\leq P_{\max}=1/3$,  $|\la|\in (0,\la_0]$, $\veps\in (0,1/2]$ and $|\la|\leq \epsilon^4$  
there holds the following equality on $\mcF_n$
\beqa
\hat{Q}_{P,n}^{\perp} 
  =  Q_{P,n-1}^{\perp}
    -\oint_{{\bcc \gamma_{P,n+1} }}{\bcc dw_{n+1}}\frac{1}{H_{P,n-1}^{W}+\Delta c_{P}|_{n}^{n-1}-w_{n+1}}\sum_{j=1}^{\infty}\{-H_I^{W}|_{n}^{n-1}\frac{1}{H_{P,n-1}^{W}+\Delta c_{P}|_{n}^{n-1}-w_{n+1}}\}^{j}. \label{shift-of-Q}
\eeqa
\eel
\proof {\bcc {\color{\red}Making use of Lemma~\ref{resolvent-expansion-one}, we write on $\mcF_n$}
\beqa
\hat{Q}_{P,n}^{\perp} \1&:=& \1\widetilde{W}_{P,n}W_{P,n}^{*}{\color{black}Q_{P,n}^{\perp}}W_{P,n}\widetilde{W}_{P,n}^{*}\non\\
 \1&=& \1 1-\oint_{\ga_{P,n+1}}\fr{dw_{n+1}}{ \hat H_{P,n}^W  -w_{n+1} } \non\\
 \1 &=& \1 1-\oint_{\ga_{P,n+1}} dw_{n+1}\frac{1}{H_{P,n-1}^{W}+\Delta c_{P}|_{n}^{n-1}-w_{n+1}}
\sum_{j=0}^{\infty}\{-H_I^{W}|_{n}^{n-1}\frac{1}{H_{P,n-1}^{W}+\Delta c_{P}|_{n}^{n-1}-w_{n+1}}\}^{j}.
\eeqa
We conclude the proof by noting that on $\mcF_n$
\beqa
1-\oint_{\ga_{P,n+1}} dw_{n+1}\frac{1}{H_{P,n-1}^{W}+\Delta c_{P}|_{n}^{n-1}-w_{n+1}}=1-{\bcb \fr{(|\phi_{P,n-1}\ran\otimes |\Om\ran)(\lan \phi_{P,n-1}|\otimes \lan \Om|)  }{ \| \phi_{P,n-1}\otimes \Om \|^2} }=Q_{P,n-1}^{\perp}.
\label{problematic}
\eeqa
Here we need that the resolvent is well defined {\color{\red} and the contour $-\Delta c_{P}|_{n}^{n-1}+\ga_{P,n+1}$ encloses no spectral point of $ H_{P,n-1}^{W}$ except for the ground state energy.} 
Since  (cf. Proposition \ref{prop:Hamiltonian-results} {\bcb and formula (\ref{Delta-cp})})   $|\Delta c_{P}|_{n}^{n-1}|\leq c\la^2\si_{n-1}$, {\color{\red}$|E_{P,n-1}-E_{P,n}|\leq c_{\Delta E}\lambda^{2}\sigma_{n-1}$, $|w_{n+1}-E_{P,n}|=\frac{\sigma_{n+1}}{3}$},  and {\color{\red}$ \mathrm{Gap}\, (H_{P,n-1}\res \mathcal{F}_{n}) \geq \fr{\si_n}{2}$}, 
this can be achieved, uniformly in $n$, by choosing $\la_0$ sufficiently small. } Now the first equality in  (\ref{problematic}) follows from  {\bcb Proposition~\ref{prop:Hamiltonian-results}}. \qed

\subsection{Standard resolvent bounds}
{\bcc In this subsection we collect some standard bounds on the quantities introduced above which will be heavily used in the
proofs of Theorems~\ref{thm:induction-convergence} and \ref{main-technical-result}.}  
The proof of the following lemma is given in Appendix~\ref{standard-resolvent-bounds-section}.
\begin{lem} \label{lem:Basic-estimates} Define 
\beqa
H_{P,n-1}^{W+}\1 &:=&\1H_{P,n-1}^{W}+\Delta c_{P}|_{n}^{n-1},\label{def-H+}\\
(H_{I}^{W}|_{n}^{n-1})_{\mr{mix}} \1 & := &\1 \h\big( \Gamma_{P,n-1}\cdot(\mathcal{L}|_{n}^{n-1}+\mathcal{I}|_{n}^{n-1})+\mr{h.c.}\big),\\
\Delta(H_{I}^{W}|_{n}^{n-1})_{\mr{mix}}\1 & := & \1(H_{I}^{W}|_{n}^{n-1})_{\mr{mix}}-
(\mathcal{L}|_{n}^{n-1}+\mathcal{I}|_{n}^{n-1})\cdot\Gamma_{P,n-1}\non\\
\1&=&\1\sum_{i=1}^3\h[(\Ga_{P,n-1})_i,  (\mathcal{L}|_{n}^{n-1})_i],\\
(H_{I}^{W}|_{n}^{n-1})_{\mr{quad}} \1& := &\1 H_{I}^{W}|_{n}^{n-1}-(H_{I}^{W}|_{n}^{n-1})_{\mr{mix}}=
\h(\mathcal{L}|_{n}^{n-1}+ \mathcal{I}|_{n}^{n-1} )^2.
\label{quad-def}
\eeqa
Let  $|P|\leq P_{\max}=1/3$,  $|\la|\in (0,\la_0]$, $\veps\in (0,1/2]$ and $|\la|\leq \epsilon^4$. 
Then,  for $z_{n+1}\in\tigam_{P,n+1}\cup \ga_{P,n+1}$, the following
estimates hold true for universal constants $c_{0}, c_{2}, c_{3}, c_4$: 
\beqa
& &\|\frac{1}{H_{P,n-1}^{W+}-z_{n+1}}\|_{\mathcal{F}_{n}}\leq \frac{c_{0}}{\sigma_{n+1}}, \label{first}\\
& &\|\frac{1}{H_{P,n-1}^{W+}-z_{n+1}}(\Gamma_{P,n-1})_{i}\|_{\mathcal{F}_{n}}\leq\frac{c_{0}}{\sigma_{n+1}},\label{3.11-Gamma}  \\
& &\|\frac{1}{H_{P,n-1}^{W+}-z_{n+1}} (\mathcal{I}|_{n}^{n-1})_{i} \|_{\mathcal{F}_{n}}\leq c_{2}|\lambda|^{\frac{1}{2}}, \label{I-standard-estimate}\\ 
& &\|\frac{1}{H_{P,n-1}^{W+}-z_{n+1}}(\mathcal{L}|_{n}^{n-1})_{i}\|_{\mathcal{F}_{n}}\leq c_{2}|\lambda|^{\frac{1}{2}}, \label{L-standard-estimates} \\
& & \|\frac{1}{H_{P,n-1}^{W+}-z_{n+1}}(H_{I}^{W}|_{n}^{n-1})\|_{\mathcal{F}_{n}}\leq c_{2}|\lambda|^{\frac{1}{2}}, 
\label{H-I-standard-estimates}
\\
& & \|\frac{1}{H_{P,n-1}^{W+}-z_{n+1}}(H_{I}^{W}|_{n}^{n-1})_{\mr{quad}}\|_{\mathcal{F}_{n}}\leq c_{3}|\lambda|\sigma_{n-1},
\label{3.53}\\
& & \|\frac{1}{H_{P,n-1}^{W+}-z_{n+1}}\Delta(H_{I}^{W}|_{n}^{n-1})_{\mr{mix}}\|_{\mathcal{F}_{n}}\leq c_{4}|\lambda|^{\frac{1}{2}}\sigma_{n-1}.\label{3.54}
\eeqa
All estimates are uniform in $z_{n+1}$ within the specified restrictions. The estimate in (\ref{L-standard-estimates}) still
holds after replacing $(\mathcal{L}|_{n}^{n-1})_{i}$ with its creation or annihilation parts $(\mathcal{L}|_{n}^{n-1})_{i}^{(\pm)}$. The estimate
in (\ref{I-standard-estimate}) can readily be improved to $c_2|\la|^{3/2}$ (cf. (\ref{I-estimate-x})), but it will be convenient to have the same bound
in (\ref{I-standard-estimate})--(\ref{H-I-standard-estimates}). 
\end{lem}
\begin{rem} \label{remark-intro} {\bcc
It will be convenient to assume in the following that 
\beqa
|E_{P,n-1}-E_{P,n}|\!\!\!&\leq&\!\!\! \fr{1}{20}\si_{n+1}, \label{conv-assumption-E}\\
|\Delta c_{P}|_{n}^{n-1}|\!\!\!&\leq&\!\!\! \fr{1}{20}\si_{n+1}. \label{conv-assumption-c}
\eeqa
By (\ref{Delta-cp}) and (\ref{eq:energy-shift}) this can be ensured by reducing $\la_0$ and using $|\la|\leq \epsilon^4$. 
Furthermore, we require that 
\beqa
c_{2}|\lambda_0|^{1/2}\leq 1/2, \quad  c_0c_2 |\la_0|^{1/4}\leq 1/2, \label{new-la-restrictions}
\eeqa
where $c_0,c_2$ are universal constants appearing in Lemma~\ref{lem:Basic-estimates}.
This is needed, in particular, in (\ref{n=0-verification}) below. }
\end{rem}
\subsection{Direct integral representations}\label{direct-subsection}

In this subsection we introduce direct integral representations of transformed Hamiltonians which 
simplify the proof of Theorem~\ref{thm:induction-convergence} and are crucial for the proof of Theorem~\ref{main-technical-result}. The relevance of such representations for iterative analytic perturbation theory was noticed in \cite{KM12} (see e.g.
Step 1 of the proof of Lemma 6.1 of this reference).

Let us first introduce the following auxiliary Hamiltonians, acting on the Hilbert space $\mcF$,  for $k_1, \ldots, k_m$ s.t. $\si_n \leq |k_{\ell}|\leq \si_{n-1}$ and $n\geq 1$:
\beqa
& &[H_{P,n-1}^{W+}]_{k_1,\ldots, k_m}:=H_{P,n-1}^{W+}+\sum_{\ell=1}^m (k_{\ell}\cdot \Gamma_{P,n-1}+\alpha_{P, n-1}(\hat k_{\ell})|k_{\ell}|)+\frac{( \sum_{\ell=1}^m k_{\ell} )^2}{2}\label{first-k-k-k-Hamiltonian} \\
& &\ph{444444444444}=W_{P,n-1}H_{P-\sum_{\ell=1}^mk_{\ell} ,n-1} W_{P,n-1}^*+\De c_P|^{n-1}_n+\sum_{\ell=1}^m|k_{\ell}|,
\label{second-k-k-k-Hamiltonian}
\eeqa
{\color{\red}where  the step from (\ref{first-k-k-k-Hamiltonian}) to (\ref{second-k-k-k-Hamiltonian})  uses (\ref{transformed-hamiltonian}), (\ref{Ga-formula}), and (\ref{def-H+}).}

To exhibit the geometric meaning of these Hamiltonians we recall that $\mcF_n=\mcF_{n-1}\otimes \mcF|^{n-1}_n$
and $\mcF|^{n-1}_n$ is the direct sum of $m$-particle subspaces $(\mcF|^{n-1}_n)^{(m)}$, $m\geq 0$. We have
$(\mcF|^{n-1}_n)^{(1)}:=L^2(\bmcA|^{n-1}_n, d^3k)$ and $(\mcF|^{n-1}_n)^{(m)}:=
L^2_{\mathrm{s}}(  (\bmcA|^{n-1}_n)^{\times m}, d^{3m}k)$, where the subscript $\mathrm{s}$ denotes
symmetric subspace.
Thus we have the standard identifications
\beqa
& &\mcF_{n-1}\otimes (\mcF|^{n-1}_n)^{(1)}\simeq \int^{\oplus}_{\bmcA|^{n-1}_n} \, d^3k\, \mcF_{n-1}, \\
& &\mcF_{n-1}\otimes (\mcF|^{n-1}_n)^{(m)}\simeq \bigg[\int^{\oplus}_{(\bmcA|^{n-1}_n)^{\times m} } \, d^{3m}k \, \mcF_{n-1}\bigg]_{\mathrm{s}},
\eeqa
where $[\ldots ]_{\mathrm{s}}$ denotes the subspace of symmetric sections. In these terms we can write (e.g. in the sense of functional calculus)
\beqa
& &H^{W+}_{P,n-1}\res (\mcF_{n-1}\otimes (\mcF|^{n-1}_n)^{(m)})\simeq \int^{\oplus}_{ (\bmcA|^{n-1}_n)^{\times m}  }  d^{3m}k\,\, \big([H^{W+}_{P,n-1}]_{k_1,\ldots,k_m} 
\res  \mcF_{n-1}\big),  \,
\label{first-direct-just}\\
& &\Ga_{P,n-1}\res (\mcF_{n-1}\otimes (\mcF|^{n-1}_n)^{(m)})\simeq \int^{\oplus}_{(\bmcA|^{n-1}_n)^{\times m}}  d^{3m}k\, \big( (\Ga_{P,n-1}+\sum_{\ell=1}^{m}k_{\ell}) \res  \mcF_{n-1}\big). \label{second-direct} 
\eeqa
To justify (\ref{first-direct-just}) for $m=1$ it suffices to write
\beqa
& &H^{W+}_{P,n-1}\res (\mcF_{n-1}\otimes (\mcF|^{n-1}_n)^{(1)})\non\\
& &=W_{P,n-1}\bigg(\h\big(P-P_{\pho}|^0_{n-1}\otimes 1-1\otimes  k |^{n-1}_n \big)^2 +H_{\pho}|^0_{n-1}\otimes 1
+1\otimes |k|\,|^{n-1}_n+\la\Phi|^0_{n-1}{\color{\red}\otimes 1}+{\color{\red} 1\otimes}\De c_P|^{n-1}_n\bigg)W_{P,n-1}^*, \non\\
\label{tensor-product-representation}
\eeqa
where $k |^{n-1}_n, |k|\,|^{n-1}_n$ are multiplication operators on $L^2(\bmcA|^{n-1}_n, d^3k)$ and $\Phi|^0_{n-1}, H_{\pho}|^0_{n-1}, P_{\pho}|^0_{n-1}$ are defined in (\ref{Phi})-(\ref{P-pho}). The case $m>1$ is treated analogously. \\\

\nin\textbf{Example}. To illustrate the above definitions we write the direct integral representation
of the following expression which  appears in  Lemma~\ref{middle-subsection-auxiliary-lemma} below:
\beqa
& &\frac{1}{H_{P,n-1}^{W+}-z_{n+1}} 
(\Gamma_{P,n-1})_i\frac{1}{H_{P,n-1}^{W+}-z'_{n+1}} (\mathcal{L}|_{n}^{n-1})^{(+)}_{i'} \phi_{P,n-1}
\label{example-first-expression}\\
& &\simeq-\la\hint^{\oplus}_{\!\!\!\! \bmcA|^{n-1}_n}  \fr{d^3k}{\sqrt{2}|k|^{3/2}} 
\frac{k_{i'}}{\alpha_{P,n-1}(\hat{k})}\frac{1}{[H_{P,n-1}^{W+}]_k-z_{n+1}}(\Gamma_{P,n-1}+k)_{i} \frac{1}{[H_{P,n-1}^{W+}]_k-z'_{n+1}}
\phi_{P,n-1},\label{example-expression}
\eeqa
where $z_{n+1}, z'_{n+1}\in \ti\ga_{P,n+1}$ and we used definition~(\ref{Ln-definitions}): 
\beqa
(\mathcal{L}|_{n}^{n-1})_{i'}^{(+)} \1&:=& \1-\lambda\hint_{\!\! \bmcA|^{n-1}_n   }  
\fr{d^{3}k}{\sqrt{2}|k|^{3/2}}   \frac{k_{i'}}{\alpha_{P,n-1}(\hat{k})}b^{*}(k).
\label{Ln-definitions-example}
\eeqa
By standard properties of direct integrals, we also have
\beqa
\|(\ref{example-first-expression})\|=|\la|\bigg( \bint_{\!\!\!\! \bmcA|^{n-1}_n}\fr{d^3k}{2|k|^3} 
\fr{k_{i'}^2}{\alpha_{P,n-1}(\hat{k})^2}
\|\frac{1}{[H_{P,n-1}^{W+}]_k-z_{n+1}}(\Gamma_{P,n-1}+k)_{i} \frac{1}{[H_{P,n-1}^{W+}]_k-z'_{n+1}}\phi_{P,n-1}\|^2 \bigg)^{1/2}. \label{norm-example}
\eeqa
There is a different way to do the above computations, which we only sketch, as it will not be used
in the following. Considering that for $\si_{n}\leq |k|\leq \si_{n-1}$
\beqa
\frac{1}{H_{P,n-1}^{W+}-z_{n+1}}b^*(k)=b^*(k)\frac{1}{[H_{P,n-1}^{W+}]_k-z_{n+1}}, \quad
\Ga_{P,n-1}b^*(k)=b^*(k)(\Ga_{P,n-1}+k),
\eeqa
and referring to definition~(\ref{Ln-definitions-example}), we can simply commute  
$(\mathcal{L}|_{n}^{n-1})^{(+)}_{i'}$ to the left in (\ref{example-first-expression}), obtaining
\beqa
(\ref{example-first-expression})=-\la\hint_{\!\!\!\! \bmcA|^{n-1}_n}  \fr{d^3k}{\sqrt{2}|k|^{3/2}} 
\frac{k_{i'}}{\alpha_{P,n-1}(\hat{k})} b^*(k)\frac{1}{[H_{P,n-1}^{W+}]_k-z_{n+1}}(\Gamma_{P,n-1}+k)_{i} \frac{1}{[H_{P,n-1}^{W+}]_k-z'_{n+1}}
\phi_{P,n-1}.  \label{non-rigorous-example}
\eeqa
This gives (\ref{norm-example}) by an obvious computation using $[b(k), b^*(k')]=\de(k-k')$ 
and $b(k)\phi_{P,n-1}=0$ for $\si_{n}\leq |k|\leq \si_{n-1}$.  We remark that (\ref{example-expression}) can be seen as a rigorous implementation of (\ref{non-rigorous-example}). \\\

In the following  lemma we collect the relevant properties of $[H_{P,n-1}^{W+}]_{k_1,\ldots, k_m}$. We recall that they are understood as operators on $\mcF$.
\bel\label{pull-through-resolvents} {\bcc Suppose that  $|P|\leq P_{\max}=1/3$, $|\la|\in (0,\la_0]$, $\veps\in (0,1/2]$ and $|\la|\leq \veps^4$. Then, for  
$\sigma_{n}\leq   |k_{1}|, \ldots, |k_m| \leq \sigma_{n-1}$,
 $n\in \nat$ and $z_{n+1}\in  \ga_{P,n+1}\cup \ti \ga_{P,n+1} $}
\beqa
& &\|\frac{1}{[H_{P,n-1}^{W+}]_{k_1, \ldots, k_m}-z_{n+1}}\|    \leq  \frac{c}{|k_1|+\cdots+|k_m|},  \label{res-est}\\ 
& &\|\frac{1}{[H_{P,n-1}^{W+}]_{k_1,\ldots, k_m} -z_{n+1}}(\Gamma_{P,n-1})_i\| \leq  \frac{cm}{|k_1|+\cdots+|k_m|}.  \label{res-gamma-est}
\eeqa
Furthermore,  for any  $\phi\in \mcF_{n}$
\beqa
\|\frac{1}{[H_{P,n-1}^{W+}]_{k_1,\ldots, k_m}-z_{n+1}}Q_{P,n-1}^{\bot}(\Gamma_{P,n-1})_i\phi\|\leq c\,{\color{\red}m} \| Q_{P,n-1}^{\bot}  \frac{1}{H^{W+}_{P,n-1}-z_{n+1}}(\Gamma_{P,n-1})_i\phi\|, \label{estimate-Q}
\eeqa
where the r.h.s. is well defined by (\ref{3.11-Gamma}).
\eel
\newcommand{\nko}{{\ko n-1}}
\proof  We treat only the case of $z_{n+1}\in  \ti\ga_{P,n+1}$ as the case of circle contours is analogous and simpler. 
{\color{\red} Making use of  representation~(\ref{second-k-k-k-Hamiltonian}), we} estimate
\beqa
[H_{P,n-1}^{W+}]_{k_1,\ldots,  k_m}-\mathrm{Re}\,z_{n+1}\1&=&\1 W_{P,n-1}(H_{P-\sum_{\ell=1}^m k_{\ell},n-1}-E_{P-\sum_{\ell=1}^mk_{\ell},n-1}) W_{P,n-1}^*\non\\
& &+(E_{P-\sum_{\ell=1}^mk_{\ell},n-1}-E_{P,n-1}) +(E_{P,n-1}-E_{P,n}) \non\\
& &+(E_{P,n}-\mathrm{Re}\,z_{n+1})+\De c_P|^{n-1}_n+\sum_{\ell=1}^{m}|k_{\ell}|\non\\
\1 &\geq & \1  -\big(1/3+c|\la|\big)\big|\sum_{\ell=1}^{m}k_{\ell}\big|-(1/3+1/10)\si_{n+1}+\sum_{\ell=1}^{m}|k_\ell| \non\\
\1 &\geq & \1  -\big(1/3+c|\la|\big)\sum_{\ell=1}^m|k_{\ell}|+(1/2)\sum_{\ell=1}^m|k_{\ell}|  \non\\
& &+(1/2)(\sum_{\ell=1}^m|k_{\ell}|)-(1/2)(1/3+1/10)(1/m)
(\sum_{\ell=1}^m|k_\ell|)
\geq c'(\sum_{\ell=1}^m|k_{\ell}|),  \label{positivity-computation-x}
\eeqa
for $c'>0$ and $\la_0$ sufficiently small, both independent of $m$.  Here we made use of Remark~\ref{energy-consistency},
which tells us that $E_{P-\sum_{\ell=1}^mk_{\ell},n-1}$ is the infimum of the spectrum of $H_{P-\sum_{\ell=1}^m k_{\ell},n-1}$
over the full Fock space $\mcF$. (Therefore the norms in (\ref{res-est}), (\ref{res-gamma-est}) can be taken over $\mcF$).
Furthermore,  in (\ref{positivity-computation-x}) we used
 Proposition~\ref{prop:Hamiltonian-results}, restrictions (\ref{conv-assumption-E}),  (\ref{conv-assumption-c}) and 
$\si_{n+1}\leq (1/2)\si_n\leq (1/2)(1/m)\sum_{\ell=1}^m|k_{\ell}|$. 
This gives the estimate in~(\ref{res-est}).

Now we consider the bound in (\ref{res-gamma-est}). {\color{\red} We recall that by assumption $\si_n \leq |k_{\ell}|\leq \si_{n-1}\leq 1$. Then, making} use of 
(\ref{second-k-k-k-Hamiltonian}), (\ref{res-est}) and $|\nabla E_{P,n}|\leq \fr{1}{3}+c|\la|$ we have, setting $A(z_{n+1}):=H_{P-\sum_{\ell=1}^m k_{\ell},n-1}+\De c_P|^{n-1}_n+\sum_{\ell=1}^m|k_{\ell}|-z_{n+1}$,
\beqa
\big\|\frac{1}{[H_{P,n-1}^{W+}]_k-z_{n+1}}(\Gamma_{P,n-1})_i\big\|
\1 &=&\1 \big\| A(z_{n+1})^{-1}  (-(P-P_{\pho})+\nabla E_{P,n}  )_i   \big\|\non\\
\1 &\leq& \1\big\|A(z_{n+1})^{-1} (P-\sum_{\ell=1}^mk_{\ell}-P_{\pho})_i   \big\|
+O\big(m \big(\sum_{\ell=1}^m|k_{\ell}|\big)^{-1}\big)\non\\
\1 &\leq& \1 c\big\|A(z_{n+1})^{-1} H_{P-\sum_{\ell=1}^mk_{\ell},\free}  A(\bar z_{n+1})^{-1} \big\|^{1/2}+O\big(m \big(\sum_{\ell=1}^m|k_{\ell}|\big)^{-1}\big)  \non\\
\1 &\leq &\1 c\big\|A(z_{n+1})^{-1} H_{P-\sum_{\ell=1}^m k_{\ell},n-1}  A(\bar z_{n+1})^{-1} \big\|^{1/2}+O\big(m \big(\sum_{\ell=1}^m|k_{\ell}|\big)^{-1}\big), 
\eeqa
where in the last step we made use of Lemma~\ref{energy-bound-zero} and again of (\ref{res-est}). By writing
$H_{P-\sum_{\ell=1}^m k_{\ell},n-1}=A(z_{n+1})-(\De c_P|^{n-1}_n+\sum_{\ell=1}^m|k_{\ell}|-z_{n+1})$, 
considering separately the case $|\mrm{Im}\,z_{n+1}|\leq 1$ and
$|\mrm{Im}\, z_{n+1}|\geq 1$, and making use again of (\ref{res-est}) we conclude the
proof of (\ref{res-gamma-est}). 

To show estimate~(\ref{estimate-Q}), 
we apply the resolvent expansion truncated at first
order using representation~(\ref{first-k-k-k-Hamiltonian}) of the Hamiltonian
\begin{eqnarray}
\frac{1}{[H_{P,n-1}^{W+}]_{k_1,\ldots, k_m}-z_{n+1}}\1&=&\1\frac{1}{H_{P,n-1}^{W+}-z_{n+1}}\non\\
& &\1+\frac{1}{[H_{P,n-1}^{W+}]_{k_1,\ldots,k_m}-z_{n+1}}\bigg[-\sum_{\ell=1}^m \big(k_{\ell}\cdot \Gamma_{P,n-1}
+\al_{P,n-1}(\hat k_{\ell})|k_{\ell}|\big)-\h \big(\sum_{\ell=1}^{m}k_{\ell}\big)^2 \bigg]  \frac{1}{H_{P,n-1}^{W+}-z_{n+1}}.\non\\
\label{first-trunc-fin}
\end{eqnarray}
Now, using the bounds (\ref{res-est}), (\ref{res-gamma-est}) {\color{\red} and the constraint $\si_n \leq |k_{\ell}|\leq \si_{n-1}\leq 1$},
we obtain
\beqa
\|\frac{1}{[H_{P,n-1}^{W+}]_{k_1,\ldots, k_m}-z_{n+1}}Q_{P,n-1}^{\bot}(\Gamma_{P,n-1})_i\phi\|\leq cm\| Q_{P,n-1}^{\bot}  \frac{1}{H^{W+}_{P,n-1}-z_{n+1}}(\Gamma_{P,n-1})_i\phi\|. \label{estimate-Q-x}
\eeqa
This concludes the proof of (\ref{estimate-Q}) and of the lemma. \qed

\subsection{Maximal modulus principle arguments}\label{maximal-mod-sub}

We recall the standard maximal modulus principle:
\begin{thm}\label{maximal-modulus} Let $f$ be a holomorphic function of $m$ complex variables with 
a bounded {\bcb connected} region of holomorphy $\mco\subset \complex^m$, which  extends by continuity to the boundary $\pa\mco$. Then
\beqa
\sup_{z\in \mco}|f(z)|=\sup_{z\in \pa\mco}|f(z)|. \label{max-mod}
\eeqa
\end{thm}
\nin Now we state a simple application: 
\bel\label{maximal-modulus-appl-one} Let $F$ be a Hilbert space valued function
of $m$ complex variables, holomorphic in
\beqa
\ti\ga_{P,n}^-:=\{\,\un{z}:=( z_1, \ldots, z_m)\in\complex^m\,|\, \mathrm{Re}\,z_i<E_{P,n-1}+\si_n/3, \ i=1, \ldots, m\,\},  
\eeqa
which extends by continuity to the closure of this region. Suppose furthermore that
\beqa
\lim_{R\to\infty}\sup_{\un{z} \in C_{R,n}^{\times m}}\|F(\un{z})\|=0, \label{limiting-R}
\eeqa
where $C_{R,n}:=\{\, z\in \complex\,|\,z= E_{P,n-1}+\si_n/3+Re^{i{\color{\red}\phi}}, {\color{\red}\phi}\in [\pi/2,3\pi/2]   \,  \}$.
Then, for any $\un{z}\in \ti\ga_{P,n}^-$,
\beqa
\|F(\un{z})\|\leq \sup_{\un{z'}\in \ti\ga_{P,n}^{\times m}}\|F(\un{z'})\|,
\label{appl-maximal-modulus}
\eeqa
{\color{\red}where $\ti\ga_{P,n}$ is defined in (\ref{def-gammatilde-0}).}
\eel
\proof For any $\psi\in \hil$ {\color{\red} with $\|\psi\|=1$} consider $f_{\psi}(\un{z}):=\lan \psi, F(\un{z})\ran$. We note that we cannot apply
Theorem~\ref{maximal-modulus} directly to $f_{\psi}$, because the region $\ti\ga_{P,n}^-$ is unbounded.
Let $\mco_{R,n}\subset \complex$ be the bounded region limited by the line 
$\ti\ga_{P,n}$ and the half-circle $C_{R,n}$.
Given $\un{z}\in \ti\ga_{P,n}^-$ we choose $R$ large enough so that $\un{z}\in \mco_{R,n}^{\times m}$. Then, by (\ref{max-mod})
\beqa
|f_{\psi}(\un{z})|\leq \sup_{\un{z'}\in \pa\mco_{R,n}^{\times m}}|f_{\psi}(\un{z'})|\leq \max\bigg\{ \sup_{\un{z'}\in \ti\ga_{P,n}^{\times m}}|f_{\psi}(\un{z'})|, \ \sup_{\un{z'}\in C_{R,n}^{\times m} }|f_{\psi}(\un{z'})|\,\bigg\}=\sup_{\un{z}'\in \ti\ga_{P,n}^{\times m}}|f_{\psi}(\un{z'})|\leq \sup_{\un{z'}\in \ti\ga_{P,n}^{\times m}}\|F(\un{z'})\|,\,\,\,
\eeqa
where in the third step we took $R$ sufficiently large and used (\ref{limiting-R}). Now (\ref{appl-maximal-modulus}) follows
by taking {\color{black}the} supremum over $\psi$. \qed

\section{Main technical result}\label{section-on-norm-bounds}
\setcounter{equation}{0}
In this section we state and prove our main technical result, which is Theorem~\ref{main-technical-result} below.
To our knowledge this result goes beyond the existing applications of iterative analytic perturbation theory.
The proof relies on Theorem~\ref{thm:induction-convergence} stated below, which is proven in Appendix~\ref{app-induction-convergence}. We recall that results comparable to Theorem~\ref{thm:induction-convergence} were obtained in \cite{FP10,KM12} for different models but the case of the Nelson
model was left aside. 
\begin{thm} \label{thm:induction-convergence} {\bcc Let $|P|\leq P_{\max}=1/3$. 
Then, for any  $\delta$,  with $\frac{1}{4}>\delta>0$, 
there exist $\epsilon(\delta)>0$ and $\lambda_{0}\equiv\lambda_{0}(\epsilon(\delta))>0$ with the following property:
For any  $\veps\in (0, \veps(\de)]$ and $|\la|\in (0, \la_0 ]$ s.t. $|\la|\leq \veps^8$ we have {\bcb for $n\geq 0$}:}\\
i) 
\beqa
\sup_{z_{n+1}\in{\bcc \ti\gamma_{P,n+1}  }}\|\frac{1}{H_{P,n}^{W}-z_{n+1}}(\Gamma_{P,n})_{i}\phi_{P,n}\,\|\leq\frac{1}{\sigma_{n}^{\delta}},\quad\quad i=1,2,3,
\eeqa
ii) 
\beqa
\|\hat{\phi}_{P,n+1}-\phi_{P,n}\|\leq|\lambda|^{\frac{1}{4}}\sigma_{n}^{1-\delta},
\eeqa
iii) 
\beqa
\|\phi_{P,n+1}\|\geq1-2{\bcc \sum_{j=0}^{n}}\{|\lambda|^{\frac{1}{4}} {\bcb  \sigma_{j}^{\frac{1}{2}} }\}\geq\frac{1}{2}.
\eeqa
\end{thm}
\begin{rem}\label{max-mod-1} Making use of the auxiliary bounds (\ref{conv-assumption-E}), (\ref{conv-assumption-c}), it is easy to see that
for $z_{n+1}\in \ti\ga_{P,n+1}, z_{n+2}\in \ti\ga_{P,n+2}$ we have $\mathrm{Re}\,z_{n+2}-\De c_P|^{n}_{n+1}\leq \mathrm{Re}\, z_{n+1}$. Given this,
it is an immediate consequence of claim i) of Theorem~\ref{thm:induction-convergence} and the maximal modulus principle (Lemma~\ref{maximal-modulus-appl-one}) that the first estimate below holds
\beqa
\sup_{z_{n+2}\in{  \ti\gamma_{P,n+2} }}\|\frac{1}{H_{P,n}^{W+}-z_{n+2}}(\Gamma_{P,n})_{i}\phi_{P,n}\|\leq\frac{1}{\sigma_{n}^{\delta}},   \quad 
\sup_{z_{n+1}\in{  \ti\gamma_{P,n+1} }}\|\frac{1}{H_{P,n-1}^{W+}-z_{n+1}}(\Gamma_{P,n-1})_{i}\phi_{P,n-1}\|\leq\frac{1}{\sigma_{n-1}^{\delta}}.
\label{alternative-claim-i}
\eeqa
Here the gap estimate (\ref{eq:gap-estimate}) and the fact  that $\Ga_{P,n}\phi_{P,n}$ is orthogonal to $\phi_{P,n}$ ensured the holomorphy in the relevant region. The second inequality  above follows analogously (for  $n\geq 1$) from claim i) for  $n\to n-1$ and thus it can be used in the inductive proof of
Theorem~\ref{thm:induction-convergence}. (It is also useful in the proof of  Theorem~\ref{main-technical-result} below). Alluding again to the maximal modulus principle, we can replace $\ti\ga_{P,n+2}, \ti \ga_{P,n+1}$ with $\ga_{P,n+2},  \ga_{P,n+1}$ in (\ref{alternative-claim-i}). 
\end{rem}


It may be demanding for the reader to scrutinize the proof of the following theorem.
We suggest that he/she opens two copies of the paper on a large computer screen to
have both the main line of the argument and the relevant auxiliary lemmas and definitions 
simultaneously  in front of his/her eyes.  
\begin{thm}\label{main-technical-result} {\bcc Let $|P|\leq P_{\max}=1/3$. 
Then, for any  $\delta$,  with $\frac{1}{4}>\delta>0$, 
there exist $\epsilon^*(\delta)>0$ and $\lambda^*_{0}\equiv\lambda^*_{0}(\epsilon^*(\delta))>0$ with the following property:
For any  {\color{\red}$\veps\in (0, \veps^*(\de)]$ and $|\la|\in (0, \la^*_0 ]$} s.t. $|\la|\leq \veps^8$ we have  {\bcb for $n\geq 0$}:}\\
\begin{equation}
\sup_{z_{n+1},z'_{n+1}\in{\bcc \ti\gamma_{P,n+1} } } \|\frac{Q_{P,n}^{\perp}}{H_{P,n}^{W}-z_{n+1}}(\Gamma_{P,n})_{i}\frac{1}{H_{P,n}^{W}-z'_{n+1}}(\Gamma_{P,n})_{i'}\phi_{P,n}\|\leq\frac{1}{\sigma_{n}^{2\delta}}.\label{eq:statement-3}
\end{equation}
\end{thm}
\begin{rem} {\bcc Obviously, we have $\la^*_0\in (0,\la_0]$ and $\veps^*(\de)\in (0,\veps(\de)]$, where
$\la_0$ and $\veps(\de)$ are the values fixed in Theorem~\ref{thm:induction-convergence}}.
\end{rem}
The remaining part of this section is devoted to the proof of Theorem~\ref{main-technical-result} which is divided into several subsections.  The inequality in  (\ref{eq:statement-3}) is
manifestly true for $n=0$ because $(\Gamma_{P,0})_{i}\phi_{P,0}=0$. Therefore we can assume that (\ref{eq:statement-3}) is fulfilled for $n-1$ ($n\geq1$) and  prove that, consequently, it holds for $n$.  Our induction hypothesis has the form
\beqa
\sup_{z_{n},z'_{n}\in  \ti\gamma_{P,n}  } \|\frac{Q_{P,n-1}^{\perp}}{H_{P,n-1}^{W}-z_{n}}(\Gamma_{P,n-1})_{i}\frac{1}{H_{P,n-1}^{W}-z'_{n}}(\Gamma_{P,n-1})_{i'}\phi_{P,n-1}\|\leq\frac{1}{\sigma_{n-1}^{2\delta}}. \label{main-theorem-inductive-hypothesis}
\eeqa
\begin{rem}\label{max-mod-2}
{\color{\red}Similarly as in (\ref{alternative-claim-i}), with the help of the maximal modulus principle (Lemma~\ref{maximal-modulus-appl-one}), 
inequality~(\ref{main-theorem-inductive-hypothesis})  implies an estimate which is more convenient in applications:
\beqa
\sup_{z_{n+1},z'_{n+1}\in \gamma_{P,n+1} \cup \ti\gamma_{P,n+1}   } \|\frac{Q_{P,n-1}^{\perp}}{ {\bcb H_{P,n-1}^{W+}}-z_{n+1}}(\Gamma_{P,n-1})_{i}\frac{1}{ {\bcb H_{P,n-1}^{W+}}-z'_{n+1}}(\Gamma_{P,n-1})_{i'}\phi_{P,n-1}\|\leq\frac{1}{\sigma_{n-1}^{2\delta}}. \label{alternative-inductive-hypothesis}
\eeqa}
We also remark that in the  following discussion $z_{n+1}, z'_{n+1}$ belong to $\ti\gamma_{P,n+1}$ (or to $\gamma_{P,n+1} \cup \ti\gamma_{P,n+1}$ if explicitly specified), whereas $w_{n+1},w'_{n+1}$ belong to  $ \gamma_{P,n+1} $. 
\end{rem}

Proceeding to the inductive proof, we use the unitary operator $\widetilde{W}_{P,n}W_{P,n}^{*}$ to switch in (\ref{eq:statement-3})
from the given expression to the corresponding expression with `hats'
\begin{equation}\label{transf-res-gamma-res-gamma}
\|\,\frac{Q_{P,n}^{\perp}}{H_{P,n}^{W}-z_{n+1}}(\Gamma_{P,n})_{i}\frac{1}{H_{P,n}^{W}-z'_{n+1}}(\Gamma_{P,n})_{i'}\phi_{P,n}\|=\|\,\frac{\hat{Q}_{P,n}^{\perp}}{\hat{H}_{P,n}^{W}-z_{n+1}}(\hat{\Gamma}_{P,n})_{i}\frac{1}{\hat{H}_{P,n}^{W}-z'_{n+1}}(\hat{\Gamma}_{P,n})_{i'}\hat{\phi}_{P,n}\|,
\end{equation}
{\color{\red}where  $z_{n+1},z'_{n+1}\in  \ti\gamma_{P,n+1}$.}
{\color{\red}Using the definition in (\ref{def-H+}) and formulae (\ref{eq:resolvent-expansion}), (\ref{eq:delta-gamma}), and (\ref{shift-of-Q}), } we proceed with the full expansion of $\hat{Q}_{P,n}^{\perp},\hat{\phi}_{P,n}$,
$\frac{1}{\hat{H}_{P,n}^{W}-z_{n+1}}$, and $(\hat{\Gamma}_{P,n})_{i}$:
\begin{eqnarray}
& &\hat{Q}_{P,n}^{\perp}\frac{1}{\hat{H}_{P,n}^{W}-z_{n+1}}(\hat{\Gamma}_{P,n})_{i}\frac{1}{\hat{H}_{P,n}^{W}-z'_{n+1}}(\hat{\Gamma}_{P,n})_{i'}\hat{\phi}_{P,n}
\label{main-expression-splitting}\\
& &=  \Big\{\tiQ_{P,n-1}^{\perp} -\sum_{r=1}^{\infty}\oint_{\gamma_{P,n+1}}dw'_{n+1}\{\frac{1}{H_{P,n-1}^{W+}-w'_{n+1}}(-)H_{I}^{W}|_{n}^{n-1}\}^{r}\frac{1}{H_{P,n-1}^{W+}-w'_{n+1}}\Big\}\times \label{uno} {\color{red} }\\
& &\times \sum_{l=0}^{\infty}   
\{\frac{1}{H_{P,n-1}^{W+}-z_{n+1}}(-)H_{I}^{W}|_{n}^{n-1}\}^{l}\frac{1}{H_{P,n-1}^{W+}-z_{n+1}} \times   \label{un mezzo}\\
& &\ph{4}\times (\Gamma_{P,n-1}+\Delta\Gamma_{P}|_{n}^{n-1})_{i}  \times\label{due}\\
& &\ph{4}\times\sum_{l'=0}^{\infty}\{\frac{1}{H_{P,n-1}^{W+}-z'_{n+1}}  (-)H_{I}^{W}|_{n}^{n-1} \}^{l'}\frac{1}{H_{P,n-1}^{W+}-z'_{n+1}} \times \label{tre} \\
& &\ph{4}\times (\Gamma_{P,n-1}+\Delta\Gamma_{P}|_{n}^{n-1})_{i'} \times \label{quattro}\\
& &\ph{4}\times
\sum_{l''=0}^{\infty}\oint_{\gamma_{P,n+1}}dw_{n+1}\{\frac{1}{H_{P,n-1}^{W+}-w_{n+1}}(-)H_{I}^{W}|_{n}^{n-1}\}^{l''}\frac{1}{H_{P,n-1}^{W+}-w_{n+1}}\phi_{P,n-1}\label{cinque}. 
\end{eqnarray}
To manipulate this long expression we introduce some short-hand notations. We suppress the dependence on $P$ and $n$
since these parameters do not change within the relevant part of the arguments, and set
\beqa
R\1&:=&\1\frac{1}{H_{P,n-1}^{W+}-z_{n+1}}, \quad R':=\frac{1}{H_{P,n-1}^{W+}-z'_{n+1}},\quad  R'':=\frac{1}{H_{P,n-1}^{W+}-w_{n+1}},\label{R-R'-def}
\\
R'''\1&:=&\1\frac{1}{H_{P,n-1}^{W+}-w'_{n+1}}, \quad V:=(-)H_{I}^{W}|_{n}^{n-1}.\label{R'''-V-def}
\eeqa
We also set
\begin{eqnarray}
(\tiQ_{P,n-1}^{\perp})^q=\left \{\begin{array}{ll} \tiQ_{P,n-1}^{\bot}& \textrm{ for $q=0$, } \\
   \sum_{r=1}^{\infty}\oint_{\gamma_{P,n+1}}dw'_{n+1}\{ R'''V\}^{r}R''' &   \textrm{ for $q=1$. }
\end{array} \right.
\end{eqnarray}
Furthermore, we define for $q=0,1$
\beqa
I^q \1&:=&\1 (\tiQ_{P,n-1}^{\perp})^q\sum_{l=0}^{\infty}   
\{ RV \}^{l}  R
 (\Gamma_{P,n-1}+\Delta\Gamma_{P}|_{n}^{n-1})_{i}  
\sum_{l'=0}^{\infty}\{  R'V  \}^{l'}  R' 
 (\Gamma_{P,n-1}+\Delta\Gamma_{P}|_{n}^{n-1})_{i'} \non\\
& &\ph{4444444444444444444444444444}\times
\sum_{l''=0}^{\infty}\oint_{\gamma_{P,n+1}}dw_{n+1}\{ R''V \}^{l''}  R'' \phi_{P,n-1},  \label{I-q-definition}
\eeqa
and note that $(\ref{main-expression-splitting})=I^0-I^1$. Thus to conclude the inductive argument, it suffices
to show that $\| I^q\|\leq c \si_{n-1}^{-2\de}=(c\veps^{2\de})\si_{n}^{-2\de}$ and then set $\veps(\de)$ sufficiently
small.   We divide the argument into three cases studied in Subsections~\ref{Main-result-first-subsection}, \ref{Main-result-second-subsection},
\ref{Main-result-third-subsection}. This lengthy discussion is preceded  by Subsection~\ref{novel-subsection}, which explains briefly the main novel ingredients of the proof.  {\color{\red}Without further notice, in the estimates below we assume the constraints in Remark \ref{remark-intro} and use the results of Section \ref{Prel-sec}. }

\subsection{Main novel ingredients of the proof}\label{novel-subsection}

The overall strategy of the inductive proof of Theorem~\ref{main-technical-result} is similar to proving claim i)  in Theorem~\ref{thm:induction-convergence}, but there are several additional complications, which we would like to briefly explain in this subsection.
To demonstrate various pitfalls of the present proof, we will state several plausible looking relations decorated by `?'. We believe that
all these relations are incorrect (although we did not attempt to formally disprove them).    We hope
that the remarks below will convince the reader that the proof of 
Theorem~\ref{main-technical-result}  is more than just a tedious application of an existing method. {\bcb\emph{Concerning notation, we warn the reader that from now on the symbol $\mathcal{O}(...)$ is also referred to operators and means that their norm $\|\ldots \|_{\mcF_n}$ is $\mathcal{O}(...)$.}}

\subsubsection{Singular terms $b^*(g_i|^{n-1}_n)\phi_{P,n-1}$, $B^*(G|^{n-1}_n)\phi_{P,n-1}$   } 
\label{singular-terms-subsubsection}

Let us consider the following contribution to (\ref{I-q-definition}), which will be studied systematically in the second part of the proof, in Subsection~\ref{Main-result-second-subsection},
\beqa
I^q_{\nat_0,\nat, 0}\1&:=&\1 (\tiQ_{P,n-1}^{\perp})^q\sum_{l=0}^{\infty}  \{ RV \}^{l} 
 R (\Gamma_{P,n-1}+\Delta\Gamma_{P}|_{n}^{n-1})_{i}  \sum_{l'=0}^{\infty}\{  R'V  \}^{l'} (R'V) R' 
 (\Gamma_{P,n-1}+\Delta\Gamma_{P}|_{n}^{n-1})_{i'}
 \phi_{P,n-1}.
\eeqa
More specifically, we would like to look at the following part
\beqa
I^q_{\nat_0,\nat, 0}\1&\ni&\1 (\tiQ_{P,n-1}^{\perp})^q\sum_{l=0}^{\infty}  \{ RV \}^{l} 
 R (\Gamma_{P,n-1} )_{i}  \sum_{l'=0}^{\infty}\{  R'V  \}^{l'} (R'V) R'  (\Gamma_{P,n-1})_{i'}  \phi_{P,n-1}\non\\
\1&\ni&\1 (\tiQ_{P,n-1}^{\perp})^q\sum_{l=0}^{\infty}  \{ RV \}^{l} 
 R (\Gamma_{P,n-1} )_{i}  \sum_{l'=0}^{\infty}\{  R'V  \}^{l'}\tiQ_{P,n-1}^{\perp}(R'V) R'  (\Gamma_{P,n-1})_{i'}  \phi_{P,n-1}, \label{informal-discussion}
 \eeqa
where $\ni$ means that we simply dropped the terms which are irrelevant for the present informal discussion.

By Lemma~\ref{lem:Basic-estimates} 
(cf. also (\ref{simple-sum-over-l}) below) we have
\beqa
\sum_{l=0}^{\infty}  \{ RV \}^{l}=\mco(1),  \quad  R (\Gamma_{P,n-1} )_{i}=\mco(\si_{n+1}^{-1}), \quad   \sum_{l'=0}^{\infty}\{  R'V  \}^{l'}=\mco(1).
\label{informal-estimates}
\eeqa   
Thus in order to obtain the desired bound, namely $(\ref{informal-discussion})=\mco(\si_{n-1}^{^{-2\de}})$, we have to control the factor $R (\Gamma_{P,n-1} )_{i}$, for example by compensating
the singularity $\si_{n+1}^{-1}$ appearing in the second estimate in (\ref{informal-estimates}). One might try to achieve this by establishing the following bound
\beqa
\tiQ_{P,n-1}^{\perp}(R'V) R'  (\Gamma_{P,n-1})_{i'}  \phi_{P,n-1}=^{\!\!\!\!?}\mco(|\la|^{1/2}\si_{n-1}^{1-2\de}), \label{wrong}
\eeqa
and then using $|\la|\leq \veps^8$ to estimate $|\la|^{1/2}\si_{n+1}^{-1}\si_{n-1}^{1-2\de}=|\la|^{1/2}\veps^2 \si_{n-1}^{-2\de}\leq \si_{n-1}^{-2\de}$.
At  first sight  (\ref{wrong}) may seem plausible: By Theorem~\ref{thm:induction-convergence} we have 
$R'  (\Gamma_{P,n-1})_{i'}  \phi_{P,n-1}=\mco(\si_{n-1}^{-\de})$.  Furthermore, by (\ref{quad-def}), 
\beqa
\label{form-H_I-new-informal}
{\color{black}-}V:=H_I^{W}|_{n}^{n-1}=(\mathcal{L}|_{n}^{n-1}+\mathcal{I}|_{n}^{n-1})\cdot\Gamma_{P,n-1}
+\Delta(H_{I}^{W}|_{n}^{n-1})_{\mr{mix}}+(H_{I}^{W}|_{n}^{n-1})_{\mr{quad}}, \label{V-informal}
\eeqa
where by  Lemma~\ref{lem:Basic-estimates} and definition~(\ref{I-definitions})  
\beqa
R'\Delta(H_{I}^{W}|_{n}^{n-1})_{\mr{mix}}=\mco(|\la|^{1/2}\si_{n-1}), \quad  R'(H_{I}^{W}|_{n}^{n-1})_{\mr{quad}}=\mco(|\la|\si_{n-1}), \quad  |\mathcal{I}|_{n}^{n-1}|\leq c_{I}\la^2\si_{n-1}. 
\eeqa
Given this information we are seemingly close to establishing (\ref{wrong}), since we can write
\beqa
\tiQ_{P,n-1}^{\perp}(R'V)  R'  (\Gamma_{P,n-1})_{i'}  \phi_{P,n-1}=\tiQ_{P,n-1}^{\perp}R'(\mathcal{L}|_{n}^{n-1}\cdot\Gamma_{P,n-1} )  R'  (\Gamma_{P,n-1})_{i'}  \phi_{P,n-1}+\mco(|\la|^{1/2}\si_{n-1}^{1-2\de}), \label{almost-wrong}
\eeqa
where we also used the induction hypothesis (\ref{main-theorem-inductive-hypothesis}) to treat the term involving $\mathcal{I}|_{n}^{n-1}$.

Let us now pay close attention to the first term on the r.h.s. of (\ref{almost-wrong}), involving $\mathcal{L}|_{n}^{n-1}\cdot\Gamma_{P,n-1}$. 
First, we recall the definition of $\mathcal{L}|_{n}^{n-1}$ 
\beqa
\mathcal{L}|_{n}^{n-1}:=-\lambda\hint_{\!\!\! {\color{black}\mcA |_n^{n-1}} }d^{3}k\frac{k(b(k)+b^{*}(k))}{\sqrt{2}|k|^{3/2}\alpha_{P,n-1}(\hat{k})},
\eeqa
which easily gives $\mathcal{L}|_{n}^{n-1} \phi_{P,n-1}=\mco(|\la|^{1/2}\si_{n-1})$. One could hope that by combining this property with the inductive
hypothesis (\ref{main-theorem-inductive-hypothesis}) the desired bound follows, i.e., that the following implication holds
\beqa
& &\bigg\{\, Q_{P,n-1}^{\perp}R'(\Gamma_{P,n-1})_{i}R' (\Gamma_{P,n-1})_{i'}\phi_{P,n-1}=\mco( \sigma_{n-1}^{-2\delta} ), \quad 
\mathcal{L}|_{n}^{n-1} \phi_{P,n-1}=\mco(|\la|^{1/2}\si_{n-1})
\bigg\} \label{implication-wrong-1}\\
& &\ph{444444444444444444444444444}\Downarrow?\non\\
& &\ph{4444444444}\tiQ_{P,n-1}^{\perp}R'(\mathcal{L}|_{n}^{n-1}\cdot\Gamma_{P,n-1} )  R'  (\Gamma_{P,n-1})_{i'}  \phi_{P,n-1}=^{\!\!\!\!?}\mco(|\la|^{1/2}\si_{n-1}^{1-2\de}).
\label{implication-wrong-2}
\eeqa
It turns out, however, that both  the implication in (\ref{implication-wrong-1})-(\ref{implication-wrong-2}) and the bounds in (\ref{implication-wrong-2}), (\ref{wrong})
resist verification. The  infrared regularity encoded in the (correct) estimates in (\ref{implication-wrong-1}) does not  carry over by simple `power counting' to the expression on the l.h.s. of (\ref{implication-wrong-2}). 

Let us now outline a correct treatment of the l.h.s. of (\ref{implication-wrong-2}),  whose  complete discussion is given in  Lemma~\ref{b-star-lemma} below.  
First, we note that  only the creation part $(\mathcal{L}|_{n}^{n-1})^{(+)}$ has a non-zero contribution to (\ref{implication-wrong-2}), i.e.
we can write
\beqa
\tiQ_{P,n-1}^{\perp}R'(\mathcal{L}|_{n}^{n-1}\cdot\Gamma_{P,n-1}) R'  (\Gamma_{P,n-1})_{i'}  \phi_{P,n-1}=
R'((\mathcal{L}|_{n}^{n-1})^{(+)}\cdot\Gamma_{P,n-1}) R'  (\Gamma_{P,n-1})_{i'}  \phi_{P,n-1}, \label{dropping-projection}
\eeqa
where we also observed that the vector on the r.h.s. of (\ref{dropping-projection}) is automatically in the range of $\tiQ_{P,n-1}^{\perp}$. 
In contrast to (\ref{dropping-projection}), for  the induction hypothesis in (\ref{implication-wrong-1})  the projection $\tiQ_{P,n-1}^{\perp}$ is essential.  Thus 
one could doubt the implication in (\ref{implication-wrong-1})-(\ref{implication-wrong-2}) already at the present stage.  To understand better why this implication is problematic, we compute, 
using the direct integral representations from Subsection~\ref{direct-subsection},
\beqa
& &R'((\mathcal{L}|_{n}^{n-1})^{(+)}\cdot\Gamma_{P,n-1}) R'  (\Gamma_{P,n-1})_{i'}  \phi_{P,n-1}\non\\
& &\ph{444444444444}:=\frac{1}{H_{P,n-1}^{W+}-z'_{n+1}} ( (\mathcal{L}|_{n}^{n-1})^{(+)}\cdot\Gamma_{P,n-1}  ) \frac{1}{H_{P,n-1}^{W+}-z'_{n+1}} (\Gamma_{P,n-1})_{i'}\phi_{P,n-1}\label{R-LG-R-G-phi-informal}\\
& &\ph{4444444444444}{\color{\red}\simeq }-\la\hint^{\oplus}_{\!\!\!\bmcA|^{n-1}_n}  \fr{d^3k}{\sqrt{2}|k|^{3/2}} 
\frac{k_{i}}{\alpha_{P,n-1}(\hat{k})}\times \non\\
& &\ph{44444444444444444}\times\frac{1}{[H_{P,n-1}^{W+}]_k-z'_{n+1}}Q_{P,n-1}^{\perp} (\Gamma_{P,n-1})_i 
\frac{1}{H_{P,n-1}^{W+}-z'_{n+1}} (\Gamma_{P,n-1})_{i'}\phi_{P,n-1}\quad\quad
\label{second-case-second-est-0-informal} \\
& &\ph{444444444444444} -\la\hint^{\oplus}_{\!\!\!\!\bmcA|^{n-1}_n}  \fr{d^3k}{\sqrt{2}|k|^{3/2}} \frac{k_{i}}{\alpha_{P,n-1}(\hat{k})}\times\non\\
& &\ph{44444444444444444}\times \frac{1}{[H_{P,n-1}^{W+}]_k-z'_{n+1}} Q_{P,n-1}(\Gamma_{P,n-1})_i \frac{1}{H_{P,n-1}^{W+}-z'_{n+1}} (\Gamma_{P,n-1})_{i'}\phi_{P,n-1},\quad\quad
\label{second-case-second-est-1-informal}
\eeqa
where in the second step we  inserted $1=Q_{P,n-1}+Q_{P,n-1}^{\perp}$ inside the direct integral.  The term in (\ref{second-case-second-est-0-informal})  can
indeed  be treated using the induction hypothesis and the structure of $\mathcal{L}|_{n}^{n-1}$ (as we hoped to treat the whole expression (\ref{R-LG-R-G-phi-informal})
in (\ref{implication-wrong-1})-(\ref{implication-wrong-2})). Using  estimate~(\ref{estimate-Q}),
$|k_i|\leq \si_{n-1}$ and $|\la|^{1/2}|\ln \, \veps|^{1/2}\leq c$ we conclude that $(\ref{second-case-second-est-0-informal})=\mco(|\la|^{1/2}\si_{n-1}^{1-2\de})$.
But we are still left with (\ref{second-case-second-est-1-informal}).  A careful analysis in Lemma~\ref{b-star-lemma} shows that there is a function $g_{i'}|^{n-1}_n$
with support in $\bar{\bmcA}|^{n-1}_n$ s.t.
\beqa
(\ref{second-case-second-est-1-informal})=b^*\big(g_{i'}|^{n-1}_n \big)\phi_{P,n-1}+\mco(|\la|^{1/2}\si_{n-1}^{1-2\de}) \quad\textrm{ and }\quad  |g_{i'}(k)|^{n-1}_n|\leq |\la|\fr{c}{|k|^{3/2}}\si_{n-1}^{-\de}, \label{last-step-informal}
\eeqa
where $b^*(g_{i'}|^{n-1}_n):=\int d^3k\,g_{i'}(k)|^{n-1}_nb^*(k)$. As the estimate in (\ref{last-step-informal}) only gives 
$b^*\big(g_{i'}|^{n-1}_n \big)\phi_{P,n-1}=\mco(|\la|^{1/2}\si_{n-1}^{-\de})$, we see no hope for establishing (\ref{implication-wrong-2}), (\ref{wrong}).
Instead, the first relation in (\ref{last-step-informal}) should be substituted back to (\ref{informal-discussion}) and treated using separate arguments 
(see  Lemmas~\ref{interaction-x}, \ref{R-Ga-g-phi} below) so that the desired estimate $(\ref{informal-discussion})=\mco(\si_{n-1}^{^{-2\de}})$ is eventually established.

To conclude this discussion we remark that in the last part of the proof, in Subsection~\ref{Main-result-third-subsection}, also `two-photon' singular terms of the following form
appear
\beqa
B^*(G|^{n-1}_n)\phi_{P,n-1}:=\int d^3k_1d^3k_2\, G(k_1,k_2)|^{n-1}_n b^*(k_1)b^*(k_2)\phi_{P,n-1}. \label{B-star-informal}
\eeqa
Here  $G|^{n-1}_n$ is a function with support in $\bar\mcA|^{n-1}_n \times  \bar\mcA|^{n-1}_n$ which satisfies the pointwise bound
\beqa
|G(k_1,k_2)|^{n-1}_n|\leq c|\la|^2|k_1|^{-3/2} |k_2|^{-3/2} \si_{n-1}^{1-\de}.
\eeqa
This bound gives $B^*(G|^{n-1}_n)\phi_{P,n-1}=\mco(|\la| \si_{n-1}^{1-2\de})$ which is less regular than the error term $\mco(|\la|^{1/2}\si_{n-1}^{2-2\de})$
in relation~(\ref{unexp}) below, where (\ref{B-star-informal}) enters.   The origin of (\ref{B-star-informal}), which will be briefly discussed in  Subsection~\ref{maximal-modulus-sub},
is analogous to the origin of $b^*\big(g_{i'}|^{n-1}_n \big)\phi_{P,n-1}$ above.

\subsubsection{Straight-line contours $\ti\ga_{P,n}$ and the maximal modulus principle} \label{maximal-modulus-sub}
 Let us now consider the following contribution to (\ref{I-q-definition}), which will be studied systematically 
 in the last part of the proof, in Subsection~\ref{Main-result-third-subsection}
 \beqa
 I^q_{\nat_0,\nat_0,\nat}\1&:=&\1 (\tiQ_{P,n-1}^{\perp})^q\sum_{l=0}^{\infty}   
\{ RV \}^{l}  R
 (\Gamma_{P,n-1}+\Delta\Gamma_{P}|_{n}^{n-1})_{i}  
\sum_{l'=0}^{\infty}\{  R'V  \}^{l'}  R' 
 (\Gamma_{P,n-1}+\Delta\Gamma_{P}|_{n}^{n-1})_{i'}\times\non\\
& &\ph{444444444444444444444444}\times
\sum_{l''=0}^{\infty}\oint_{\gamma_{P,n+1}}dw_{n+1}\{ R''V \}^{l''} R''V  R''  \phi_{P,n-1}\label{informal-second-section}\\
\1&\ni &\1 (\tiQ_{P,n-1}^{\perp})^q\sum_{l=0}^{\infty}   
\{ RV \}^{l}  R (\Gamma_{P,n-1})_{i}  
\sum_{l'=0}^{\infty}\{  R'V  \}^{l'}  R'  (\Gamma_{P,n-1})_{i'}\times\non\\
& &\ph{4444444444444}\times
\sum_{l''=0}^{\infty}\oint_{\gamma_{P,n+1}}dw_{n+1}f(w_{n+1}) \{ R''V \}^{l''} Q_{P,n-1}^{\bot} R''V R''V   \phi_{P,n-1}. \label{informal-integral-equation-second-section}
\eeqa
 Here in the second step we dropped the terms proportional to $\Delta\Gamma_{P}|_{n}^{n-1}$ and the $l''=0$ term in (\ref{informal-second-section}) as they will not
 be relevant for the present discussion. We also noticed that $R''\phi_{P,n-1}=f(w_{n+1})\phi_{P,n-1}$, where the numerical function $f(w_{n+1})=\mco(\si_{n+1}^{-1})$ will be compensated
 by the length of the integration contour $\gamma_{P,n+1}$ in (\ref{informal-integral-equation-second-section}).  We recall from (\ref{informal-estimates})  that 
 simple-minded norm estimates of the factors $R (\Gamma_{P,n-1})_{i}$, $R'  (\Gamma_{P,n-1})_{i'}$  yield together a singularity  $\mco(\si_{n+1}^{-2})$, which has
 to be tamed by a proper treatment of  $Q_{P,n-1}^{\bot} R''V R''V   \phi_{P,n-1}$.  Lemma~\ref{B-star-lemma} gives the following estimate
  \beqa
Q_{P,n-1}^{\bot}R''VR''V\phi_{P,n-1}
=B^*(G|^{n-1}_n)\phi_{P,n-1}+(\mathcal{I}|_{n}^{n-1})_{i'} b^*\big( g_{i'}|^{n-1}_n \big)\phi_{P,n-1}+\mco(|\la|^{1/2}\si_{n-1}^{2-2\de}), \label{unexp-informal}
\eeqa
where the summation over $i'=1,2,3$ is understood and the singular terms $B^*(G|^{n-1}_n)\phi_{P,n-1}$, 
$b^*\big(g_{i'}|^{n-1}_n \big)\phi_{P,n-1}$ were discussed in Subsection~\ref{singular-terms-subsubsection}. The error term  
$\mco(|\la|^{1/2}\si_{n-1}^{2-2\de})$  compensates the $\mco(\si_{n+1}^{-2})$ singularity directly, while the singular terms are substituted back to 
(\ref{informal-integral-equation-second-section}) and treated by separate arguments, in particular  in Lemmas~\ref{B-star-lemma-one}, \ref{B-star-lemma-three}.  Eventually, the desired bound
$(\ref{informal-integral-equation-second-section})=\mco(\si_{n-1}^{-2\de})$ is established.

In the remaining part of this subsection we will have a closer look at one particular contribution
to the expression $Q_{P,n-1}^{\bot}R''VR''V\phi_{P,n-1}$, namely (cf. definition~(\ref{V-informal}))
\beqa
R'' \big( (\mathcal{L}|_{n}^{n-1})^{(+)}\cdot \Ga_{P,n-1} \big) R''
\big( (\mathcal{L}|_{n}^{n-1})^{(+)}\cdot \Ga_{P,n-1} \big) \phi_{P,n-1},
\label{informal-contribution-second-subsection}
\eeqa
where we dropped the projection $Q_{P,n-1}^{\bot}$ as this vector is manifestly in its range.
Making use of the direct integral representations from  Subsection~\ref{direct-subsection}, and recalling that $R''=(H^{W+}_{P,n-1}-w_{n+1})^{-1}$, we obtain
\beqa
(\ref{informal-contribution-second-subsection})
\1&\simeq&\1 \fr{\la^2}{\sqrt{2}}\hint^{\oplus}_{\!\!\!\!(\bmcA|^{n-1}_n)^{\times 2}}d^3k_1d^3k_2 
\bigg(\fr{1}{\sqrt{2}|k_1|^{3/2}} \frac{k_{1,j}}{\alpha_{P,n-1}(\hat{k}_1)}  \fr{1}{\sqrt{2}|k_2|^{3/2}} \frac{k_{2,j'}}{\alpha_{P,n-1}(\hat{k}_2)}\times\non\\
& &\times\frac{1}{[H_{P,n-1}^{W+}]_{k_1,k_2}-w_{n+1}}(\Ga_{P,n-1}+k_2)_j \frac{1}{[H_{P,n-1}^{W+}]_{k_2}-w_{n+1}}(\Ga_{P,n-1})_{j'}\phi_{P,n-1}+\{1\leftrightarrow 2\}\bigg)\quad\quad\quad
\label{L-G-L-G-1-informal}\\
\1&\ni &\1 \fr{\la^2}{\sqrt{2}}\hint^{\oplus}_{\!\!\!\!(\bmcA|^{n-1}_n)^{\times 2} }d^3k_1d^3k_2 
\bigg(\fr{1}{\sqrt{2}|k_1|^{3/2}} \frac{k_{1,j}}{\alpha_{P,n-1}(\hat{k}_1)}  \fr{1}{\sqrt{2}|k_2|^{3/2}} \frac{k_{2,j'}}{\alpha_{P,n-1}(\hat{k}_2)}\times\non\\
& &\ph{4}\times\frac{1}{[H_{P,n-1}^{W+}]_{k_1,k_2}-w_{n+1}} Q_{P,n-1}^{\bot}(\Ga_{P,n-1})_j \frac{1}{[H_{P,n-1}^{W+}]_{k_2}-w_{n+1}}(\Ga_{P,n-1})_{j'}\phi_{P,n-1}+\{1\leftrightarrow 2\}\bigg)\quad\quad\quad\label{L-G-L-G-2-informal}\\
& &+ \fr{\la^2}{\sqrt{2}}\hint^{\oplus}_{\!\!\!\!(\bmcA|^{n-1}_n)^{\times 2} }d^3k_1d^3k_2 
\bigg(\fr{1}{\sqrt{2}|k_1|^{3/2}} \frac{k_{1,j}}{\alpha_{P,n-1}(\hat{k}_1)}  \fr{1}{\sqrt{2}|k_2|^{3/2}} \frac{k_{2,j'}}{\alpha_{P,n-1}(\hat{k}_2)}\times\non\\
& &\ph{4}\times\frac{1}{[H_{P,n-1}^{W+}]_{k_1,k_2}-w_{n+1}} Q_{P,n-1}(\Ga_{P,n-1})_j \frac{1}{[H_{P,n-1}^{W+}]_{k_2}-w_{n+1}}(\Ga_{P,n-1})_{j'}\phi_{P,n-1}+\{1\leftrightarrow 2\}\bigg),\quad\quad\quad\label{term-which-complicates-informal} 
\eeqa
where in the second step we skipped the term proportional to $k_{2,j}$ from $(\Ga_{P,n-1}+k_2)_j$
(as it will not be relevant for the present discussion) and inserted $1=Q_{P,n-1}+Q_{P,n-1}^{\bot}$
similarly as in (\ref{second-case-second-est-0-informal})-(\ref{second-case-second-est-1-informal}).
As one might expect from this latter computation, (\ref{term-which-complicates-informal}) gives
rise to the singular term $B^*(G|^{n-1}_n)\phi_{P,n-1}$ in (\ref{unexp-informal}). Our main concern
here is, however, the term in (\ref{L-G-L-G-2-informal}), which will eventually contribute to 
the error term $\mco(|\la|^{1/2}\si_{n-1}^{2-2\de})$ in  (\ref{unexp-informal}).

Our analysis of (\ref{L-G-L-G-2-informal}) starts from a declaration that the factor 
$\la^2 k_{1,j}k_{2,j'}$, which satisfies $\la^2 |k_{1,j}|\ |k_{2,j}|\leq \la^2 \si^2_{n-1}$, should contribute
the  $|\la|^{1/2}\si^2_{n-1}$--part of the anticipated estimate $\mco(|\la|^{1/2}\si_{n-1}^{2-2\de})$.
Thus for the remaining part of (\ref{L-G-L-G-2-informal}) we need to establish a bound of order
$\si^{-2\de}$, possibly multiplied by some inverse power of $\veps$. Given the form of this bound and the structure of  this remaining part, it is natural  to try to relate it to the inductive hypothesis (\ref{alternative-inductive-hypothesis}).

To this end, we first use estimate 
(\ref{estimate-Q}) with $\phi:=([H_{P,n-1}^{W+}]_{k_2}-w_{n+1})^{-1}(\Gamma_{P,n-1})_{j'}\phi_{P,n-1}$, which gives for $\si_n\leq |k_1|, |k_2|\leq \si_{n-1}$
\beqa
& &\| \frac{1}{[H_{P,n-1}^{W+}]_{k_1,k_2}-w_{n+1}}Q_{P,n-1}^{\perp} (\Gamma_{P,n-1})_j   \frac{1}{[H_{P,n-1}^{W+}]_{k_2}-w_{n+1}} (\Gamma_{P,n-1})_{j'}\phi_{P,n-1}\|\non\\
& &\ph{4444444444444}\leq 
c\| Q_{P,n-1}^{\perp}\frac{1}{H_{P,n-1}^{W+}-w_{n+1}} (\Gamma_{P,n-1})_j
\frac{1}{[H_{P,n-1}^{W+}]_{k_2}-w_{n+1}} (\Gamma_{P,n-1})_{j'}\phi_{P,n-1}\|. 
\label{informal-towards-maximal-modulus}
\eeqa
In Lemma~\ref{last-case-additional-lemma} below we  essentially replace
$[H_{P,n-1}^{W+}]_{k_2}$  with $H^{W+}_{P,n-1}$ by a suitable expansion of the
resolvent.
To explain this argument, we first recall that
\beqa
[H_{P,n-1}^{W+}]_{k}:=H_{P,n-1}^{W+}+k\cdot \Gamma_{P,n-1}+\frac{|k|^2}{2}+\alpha_{P, n-1}(\hat k)|k|.
\eeqa
Next, we expand the resolvent of $[H_{P,n-1}^{W+}]_{k}$ in 
(\ref{informal-towards-maximal-modulus})
to the second order:
\beqa
\frac{1}{[H_{P,n-1}^{W+}]_{k}-w_{n+1}}\1&=&\1\frac{1}{H_{P,n-1}^{W+}+\alpha_{P, n-1}(\hat k)|k|-w_{n+1}} \label{0-informal}\\
& &+\frac{1}{H_{P,n-1}^{W+}+\alpha_{P, n-1}(\hat k)|k|-w_{n+1}}
[-k\cdot \Gamma_{P,n-1}-\frac{|k|^2}{2}]\frac{1}{H_{P,n-1}^{W+}+\alpha_{P, n-1}(\hat k)|k|-w_{n+1}}\quad\quad \label{1-informal}\\
& &+\frac{1}{[H_{P,n-1}^{W+}]_{k}-w_{n+1}}\bigg\{[-k\cdot \Gamma_{P,n-1}-\frac{|k|^2}{2}]\frac{1}{H_{P,n-1}^{W+}+\alpha_{P, n-1}(\hat k)|k|-w_{n+1}}\bigg\}^2.\quad\label{3-informal}
\eeqa
We set $\ti w_{n+1}:=w_{n+1}-\alpha_{P, n-1}(\hat k)|k|$ and  consider the contribution of (\ref{0-informal}) to (\ref{informal-towards-maximal-modulus}). We have
\beqa
& &\|Q_{P,n-1}^{\bot}\frac{1}{H_{P,n-1}^{W+}-w_{n+1}} (\Ga_{P,n-1})_j \frac{1}{H_{P,n-1}^{W+}-\ti{w}_{n+1}}(\Ga_{P,n-1})_{j'}\phi_{P,n-1}\|\non\\
& &\leq \sup_{z'_{n+1}\in \ti\ga_{P,n+1} }\|Q_{P,n-1}^{\bot}\frac{1}{H_{P,n-1}^{W+}-w_{n+1}} (\Ga_{P,n-1})_j \frac{1}{H_{P,n-1}^{W+}-z'_{n+1}}(\Ga_{P,n-1})_{j'}\phi_{P,n-1}\|\leq \si_{n-1}^{-2\de},
\label{maximal-modulus-principle-estimate} 
\eeqa
where we used that $\alpha_{P, n-1}(\hat k)|k|\geq 0$,  applied the maximal modulus principle and the inductive hypothesis via relation~(\ref{alternative-inductive-hypothesis}). We refrain from a complete analysis of the higher order terms~(\ref{1-informal}), (\ref{3-informal}) here, which give
contributions to (\ref{informal-towards-maximal-modulus}) of order $\mco(\veps^{-4}\si_{n-1}^{-2\de})$ and eventually lead to 
$(\ref{L-G-L-G-2-informal})=\mco(|\la|^{1/2}\si_{n-1}^{2-2\de})$. 
However, we would like to remark for future reference that an alternative expansion, related to (\ref{0-informal})-(\ref{3-informal}) by
\beqa
\fr{1}{H_{P,n-1}^{W+}+\alpha_{P, n-1}(\hat k)|k|-w_{n+1}} &\to& \fr{1}{H_{P,n-1}^{W+}-w_{n+1}}, \label{shift-informal-one}\\
\,[-k\cdot \Gamma_{P,n-1}-\frac{|k|^2}{2}] &\to& [-k\cdot \Gamma_{P,n-1}- \alpha_{P, n-1}(\hat k)|k| -\frac{|k|^2}{2}], \label{shift-informal-two}
\eeqa
generates  higher order terms which, substituted to (\ref{informal-towards-maximal-modulus}),  do not appear to behave as $\si_{n-1}^{-2\de}$. 
Thus there is no obvious alternative to  expansion (\ref{0-informal})-(\ref{3-informal}) and 
the use of  variable $\ti w_{n+1}:=w_{n+1}-\alpha_{P, n-1}(\hat k)|k|$ introduced above.

We would like to stress that the use of the straight-line contour $\ti \ga_{P,n+1}$ in the
induction hypothesis (\ref{alternative-inductive-hypothesis}) was essential for estimate (\ref{maximal-modulus-principle-estimate}). If instead of (\ref{alternative-inductive-hypothesis})
we only had
\beqa
\sup_{w'_{n+1},w''_{n+1}\in \gamma_{P,n+1}    } \|\frac{Q_{P,n-1}^{\perp}}{H_{P,n-1}^{W}-w'_{n+1}}(\Gamma_{P,n-1})_{i}\frac{1}{H_{P,n-1}^{W}-w''_{n+1}}(\Gamma_{P,n-1})_{i'}\phi_{P,n-1}\|
\leq \sigma_{n-1}^{-2\delta}, \label{alternative-inductive-hypothesis-informal}
\eeqa
we could not conclude.  Indeed, for $w_{n+1}\in \ga_{P,n+1}$ the resulting variable 
$\ti w_{n+1}:=w_{n+1}-\alpha_{P, n-1}(\hat k)|k|$ is always outside of the circle $\ga_{P,n+1}$
and the maximal modulus principle does not apply.  (Using the larger circle $\ga_{P,n}$, in analogy to (\ref{main-theorem-inductive-hypothesis}), does not help).  A direct application of the Cauchy integral formula, which is a potential tool to handle such problems, does not apply for the same reason. Also a shift of the resolvent as in (\ref{shift-informal-one}), (\ref{shift-informal-two})
generates problematic error terms as mentioned above.  

Finally we remark that for the induction hypothesis of Theorem~\ref{thm:induction-convergence} straight-line contours $\ti \ga_{P,n+1}$ were not essential. If instead of the induction hypothesis
(\ref{alternative-claim-i}) we had
\beqa
\sup_{w_{n+1}\in{  \gamma_{P,n+1} }}\|\frac{1}{H_{P,n-1}^{W+}-w_{n+1}}(\Gamma_{P,n-1})_{i}\phi_{P,n-1}\|\leq   \sigma_{n-1}^{-\delta},
\label{alternative-claim-i-informal}
\eeqa
the induction would still close, yielding a weaker variant of Theorem~\ref{thm:induction-convergence}  with $\ti\ga_{P,n+1}$ replaced with $\ga_{P,n+1}$. (This variant 
would not suffice to prove Theorem~\ref{main-technical-result}, however).

\subsection{Contribution to $I^q$ with $l\in \nat_0$, $l'=0$, $l''=0$} \label{Main-result-first-subsection}
{\color{black}The} contribution we analyse in this subsection is stated below. {\bcb In  (\ref{4.17})-(\ref{4.19}) below we insert $Q_{P,n-1}^{\perp}+Q_{P,n-1}=1$ and  use  $\sum_{l=0}^{\infty} \{ RV \}^{l} =1+\sum_{l=0}^{\infty}\{ RV \}^{l} RV$. }
\beqa
I^q_{\nat_0,0,0}\1&:=&\1(\tiQ_{P,n-1}^{\perp})^q
\sum_{l=0}^{\infty}   
\{ RV \}^{l}  R  (\Gamma_{P,n-1}+\Delta\Gamma_{P}|_{n}^{n-1})_{i}  
 R'  (\Gamma_{P,n-1}+\Delta\Gamma_{P}|_{n}^{n-1})_{i'}\phi_{P,n-1}\non\\
\1&=&\1(\tiQ_{P,n-1}^{\perp})^q\sum_{l=0}^{\infty}   
\{ RV \}^{l}   Q_{P,n-1}^{\perp} R
 (\Gamma_{P,n-1})_{i}  
 R'  (\Gamma_{P,n-1})_{i'}\phi_{P,n-1} \label{4.17}\\
&&+\de_{q,1}(\tiQ_{P,n-1}^{\perp})^qQ_{P,n-1} R
 (\Gamma_{P,n-1})_{i}  
 R'  (\Gamma_{P,n-1})_{i'}\phi_{P,n-1}\label{4.18}\\
&&+(\tiQ_{P,n-1}^{\perp})^q\sum_{l=0}^{\infty}   
\{ RV \}^{l} RVQ_{P,n-1} R
 (\Gamma_{P,n-1})_{i}  
 R'  (\Gamma_{P,n-1})_{i'}\phi_{P,n-1}\label{4.19} \\
&&+(\tiQ_{P,n-1}^{\perp})^q\sum_{l=0}^{\infty}   
\{ RV \}^{l}  R
 (\Delta\Gamma_{P}|_{n}^{n-1})_{i}  
 R'  (\Gamma_{P,n-1})_{i'}\phi_{P,n-1} \label{4.20}\\
&&+(\tiQ_{P,n-1}^{\perp})^q\sum_{l=0}^{\infty}   
\{ RV \}^{l}  R
 (\Gamma_{P,n-1})_{i}  
 R'  (\Delta\Gamma_{P}|_{n}^{n-1})_{i'}\phi_{P,n-1}\label{cont-here}\\
&&+(\tiQ_{P,n-1}^{\perp})^q\sum_{l=0}^{\infty}   
\{ RV \}^{l}  R
 (\Delta\Gamma_{P}|_{n}^{n-1})_{i}  
 R'  (\Delta\Gamma_{P}|_{n}^{n-1})_{i'}\phi_{P,n-1}{\color{\red}=\mco(\si_{n-1}^{-2\de}).} \label{final-estimate}
\eeqa
To justify the last estimate we first recall that {\color{\red}by (\ref{H-I-standard-estimates})} in Lemma~\ref{lem:Basic-estimates}
\beqa
\|\sum_{l=0}^{\infty}\{ RV \}^{l}\|_{\mcF_n}:=\| \sum_{l=0}^{\infty}\big\{\frac{1}{H_{P,n-1}^{W+}-z_{n+1}}(-)H_{I}^{W}|_{n}^{n-1}\big\}^{l}\|_{\mcF_n}
\leq \fr{1}{1-|\la|^{1/2}c_2}. \label{simple-sum-over-l}
\eeqa
Now (\ref{final-estimate}) follows from Lemmas~\ref{ind-hypothesis}-- \ref{De-Ga-De-Ga} below.
\bel\label{ind-hypothesis} {\color{\red}(Control of (\ref{4.17}))} Under the assumptions of Theorem~\ref{main-technical-result} and the inductive hypothesis
\beqa
& &Q_{P,n-1}^{\perp} R
 (\Gamma_{P,n-1})_{i}  
 R'  (\Gamma_{P,n-1})_{i'}\phi_{P,n-1}\non\\
& &:=
Q_{P,n-1}^{\perp} \frac{1}{H_{P,n-1}^{W+}-z_{n+1}}
 (\Gamma_{P,n-1})_{i}  
  \frac{1}{H_{P,n-1}^{W+}-z'_{n+1}} (\Gamma_{P,n-1})_{i'}\phi_{P,n-1}=\mco(\si_{n-1}^{-2\de}).
\eeqa
\eel
\proof This is a consequence of the inductive hypothesis stated in (\ref{alternative-inductive-hypothesis}). \qed
\bel {\color{\red}(Control of (\ref{4.18}))} Under the assumptions of Theorem~\ref{main-technical-result} and for $q=1$
\beqa
& &\de_{q,1}(\tiQ_{P,n-1}^{\perp})^qQ_{P,n-1}\non\\
& &:=\sum_{r=1}^{\infty}\oint_{\gamma_{P,n+1}}dw'_{n+1}\{\frac{1}{H_{P,n-1}^{W+}-w'_{n+1}}(-)H_{I}^{W}|_{n}^{n-1}\}^{r}
\frac{1}{H_{P,n-1}^{W+}-w'_{n+1}}Q_{P,n-1}=\mco(|\la|^{1/2}\si_{n-1}^{1-\de}).
\eeqa
\eel
\proof We first note that
\beqa
\frac{1}{H_{P,n-1}^{W+}-w'_{n+1}}Q_{P,n-1}=f(w'_{n+1})Q_{P,n-1},
\eeqa
where $f(w'_{n+1})=\mco(\si_{n+1}^{-1})$. Furthermore, by Lemma~\ref{first-case-aux-lemma} below
\beqa
\frac{1}{H_{P,n-1}^{W+}-w'_{n+1}}(-)H_{I}^{W}|_{n}^{n-1}Q_{P,n-1}=\mco(|\la|^{1/2}\si_{n-1}^{1-\de}).
\eeqa
Thus taking into account that by Lemma~\ref{lem:Basic-estimates}
\beqa
\|\frac{1}{H_{P,n-1}^{W+}-w'_{n+1}}(-)H_{I}^{W}|_{n}^{n-1}\|_{\mcF_n}=\mco(|\la|^{1/2}),
\eeqa
and that the length of the region of integration is proportional to $\si_{n+1}$, we have
\beqa
\de_{q,1}(\tiQ_{P,n-1}^{\perp})^qQ_{P,n-1}=\mco(|\la|^{1/2}\si_{n-1}^{1-\de}).
\eeqa
This concludes the proof. \qed
\bel {\color{\red}(Control of (\ref{4.18}) and (\ref{4.19}))} Under the assumptions of Theorem~\ref{main-technical-result} 
\beqa
& &Q_{P,n-1} R (\Gamma_{P,n-1})_{i}   R'  (\Gamma_{P,n-1})_{i'}\phi_{P,n-1}\non\\
& &:=Q_{P,n-1} \frac{1}{H_{P,n-1}^{W+}-z_{n+1}} (\Gamma_{P,n-1})_{i}   
\frac{1}{H_{P,n-1}^{W+}-z'_{n+1}}  (\Gamma_{P,n-1})_{i'}\phi_{P,n-1}=\mco(\si_{n+1}^{-1}\si_{n-1}^{-\de})\,.
\eeqa
\eel
\proof Clearly, the expression from the statement of the lemma is bounded in norm by
\beqa
& &\fr{c_0}{\si_{n+1}}\|Q_{P,n-1}(\Gamma_{P,n-1})_{i}  \|_{\mcF_{n-1}}  \big\|\frac{1}{H_{P,n-1}^{W+}-z'_{n+1}}  (\Gamma_{P,n-1})_{i'}\phi_{P,n-1}\big\|     
\leq\fr{c_0}{\si_{n+1}}\|Q_{P,n-1}(\Gamma_{P,n-1})_{i}  \|_{\mcF_{n-1}} \fr{1}{\si_{n-1}^{\de}}, \label{Caucha-arg-main-proof}
\eeqa
where we applied estimate (\ref{first}) of Lemma~\ref{lem:Basic-estimates} and  Theorem~\ref{thm:induction-convergence} (more precisely second formula
in (\ref{alternative-claim-i})).
To conclude the proof we note that (cf.  formulas~(\ref{Ga-formula}), (\ref{3.37}), and (\ref{eq:gradient-bound}))
\beqa
\|Q_{P,n-1}(\Gamma_{P,n-1})_{i}  \|_{\mcF_{n-1}}\leq c+\|\bar Q_{P,n-1}(P-P_{\pho})_i\|_{\mcF_{n-1}}\leq c+c' \|\bar Q_{P,n-1}H_{P,\free}\bar Q_{P,n-1}\|^{1/2}_{\mcF_{n-1}},  \label{Ga-Q-bound}
\eeqa
which is bounded uniformly in $n$ by Lemma~\ref{energy-bound-zero}. \qed
\begin{rem}
{\color{black}Without further notice, in the next lemmas we  repeatedly make use of the constraint $|\la|^{1/2}\leq \veps^{4}$.}
\end{rem}
\bel\label{first-case-aux-lemma} {\color{\red}(Control of (\ref{4.19}))} Under the assumptions of Theorem~\ref{main-technical-result} 
\beqa
RVQ_{P,n-1}:=\frac{1}{H_{P,n-1}^{W+}-z_{n+1}}(-)H_{I}^{W}|_{n}^{n-1}Q_{P,n-1}=\mco(|\la|^{1/2}\si_{n-1}^{1-\de}).
\eeqa
The same bound is true for $z_{n+1}\in \ga_{P,n+1}$.
\eel
\proof We recall from (\ref{quad-def}) that
\begin{equation}\label{form-H_I-new}
H_I^{W}|_{n}^{n-1}:=(\mathcal{L}|_{n}^{n-1}+\mathcal{I}|_{n}^{n-1})\cdot\Gamma_{P,n-1}
+\Delta(H_{I}^{W}|_{n}^{n-1})_{\mr{mix}}+(H_{I}^{W}|_{n}^{n-1})_{\mr{quad}}.
\end{equation}
We consider the contributions from the respective terms  on the r.h.s. of (\ref{form-H_I-new}) to the expression in the statement of the lemma. First, making use of Lemma~\ref{final-subsection-auxiliary-lemma-one}, we obtain
\beqa
\frac{1}{H_{P,n-1}^{W+}-z_{n+1}}\mathcal{L}|_{n}^{n-1}\cdot \Gamma_{P,n-1} Q_{P,n-1}=\mco(|\la|{\color{\red}|\ln\, \veps|^{1/2}}\si_{n-1}^{1-\de})\,.
\eeqa
Next, we have by (\ref{I-definitions})
\beqa
|\mathcal{I}|_{n}^{n-1}|\leq c_I |\lambda|^{2}\sigma_{n-1}.
\eeqa
Thus we can write,  applying Theorem~\ref{thm:induction-convergence},
\beqa
\frac{1}{H_{P,n-1}^{W+}-z_{n+1}}\mathcal{I}|_{n}^{n-1}\cdot \Gamma_{P,n-1} Q_{P,n-1}=\mco(|\lambda|^{2}\sigma_{n-1} \sigma_{n-1}^{-\de}).
\eeqa
Finally, by Lemma~\ref{lem:Basic-estimates},
\beqa
\frac{1}{H_{P,n-1}^{W+}-z_{n+1}} (\Delta(H_{I}^{W}|_{n}^{n-1})_{\mr{mix}}+(H_{I}^{W}|_{n}^{n-1})_{\mr{quad}} )Q_{P,n-1}=
\mco(|\la|^{1/2}\si_{n-1}). \textrm{\qed}
\eeqa
\bel\label{DeGa-Ga} {\color{\red}(Control of (\ref{4.20}))} Under the assumptions of Theorem~\ref{main-technical-result} 
\beqa
R(\Delta\Gamma_{P}|_{n}^{n-1})_{i} R'(\Gamma_{P,n-1})_{i'}\phi_{P,n-1}
=\frac{1}{H_{P,n-1}^{W+}-z_{n+1}}(\Delta\Gamma_{P}|_{n}^{n-1})_{i} \frac{1}{H_{P,n-1}^{W+}-z'_{n+1}}(\Gamma_{P,n-1})_{i'}\phi_{P,n-1}=\mco(\si_{n-1}^{-2\de}).
\eeqa
\eel
\proof We recall that
\beqa  \label{De-E-I-0}
\Delta\Gamma_{P}|_{n}^{n-1}:=
-\nabla E_{P,n-1}+\nabla E_{P,n}+\mathcal{I}|_{n}^{n-1}
+\mathcal{L}|_{n}^{n-1}.
\eeqa
Making use of the facts that (cf. (\ref{grad-difference}), claims ii), iii) of Theorem~\ref{thm:induction-convergence} and  (\ref{I-definitions}))
\beqa
|\nabla E_{P,n-1}-\nabla E_{P,n}|\leq c_1[\la^2\si_{n-1}+4|\la|^{1/4}\si_{n-1}^{1-\de}],\quad |\mathcal{I}|_{n}^{n-1}|\leq c_I |\lambda|^{2}\sigma_{n-1}, \label{remiander-one-new}
\eeqa 
we have
\beqa
|-\nabla E_{P,n-1}+\nabla E_{P,n}+\mathcal{I}|_{n}^{n-1}|\leq c |\la|^{1/4}\si_{n-1}^{1-\de}. \label{De-E-I}
\eeqa
Now, by   Theorem~\ref{thm:induction-convergence}, we can write
\beqa
& &\frac{1}{H_{P,n-1}^{W+}-z_{n+1}}( -\nabla E_{P,n-1}+\nabla E_{P,n}+\mathcal{I}|_{n}^{n-1}  )_{i} \frac{1}{H_{P,n-1}^{W+}-z'_{n+1}}(\Gamma_{P,n-1})_{i'}\phi_{P,n-1}\non\\
& &\ph{4444444444444444444444444}=\mco(|\la|^{1/4}\si_{n+1}^{-1}\si_{n-1}^{1-\de}\si_{n-1}^{-\de})=\mco(\si_{n-1}^{-2\de}).
\eeqa
Arguing similarly and making use of {\color{\red}(\ref{L-standard-estimates})} in Lemma~\ref{lem:Basic-estimates}, we get
\beqa
\frac{1}{H_{P,n-1}^{W+}-z_{n+1}}(  \mathcal{L}|_{n}^{n-1}   )_{i} \frac{1}{H_{P,n-1}^{W+}-z'_{n+1}}(\Gamma_{P,n-1})_{i'}\phi_{P,n-1}
=\mco(|\la|^{1/2}\si_{n-1}^{-\de}).
\eeqa
This concludes the proof. \qed
\bel {\color{\red}(Control of (\ref{cont-here}))} Under the assumptions of Theorem~\ref{main-technical-result} 
\beqa
R (\Gamma_{P,n-1})_{i}  R'  (\Delta\Gamma_{P}|_{n}^{n-1})_{i'}\phi_{P,n-1}
:=\frac{1}{H_{P,n-1}^{W+}-z_{n+1}}(\Gamma_{P,n-1} )_{i} \frac{1}{H_{P,n-1}^{W+}-z'_{n+1}}
(\De\Gamma_{P}|_n^{n-1})_{i'}\phi_{P,n-1}=\mco(\si_{n-1}^{-2\de}).
\eeqa
\eel
\proof We have $\Delta\Gamma_{P}|_{n}^{n-1}:=-\nabla E_{P,n-1}+\nabla E_{P,n}+\mathcal{I}|_{n}^{n-1}
+\mathcal{L}|_{n}^{n-1}$ and we consider first the contribution of the first three terms. Arguing as in
the proof of Lemma~\ref{DeGa-Ga}, we have
\beqa
& &\frac{1}{H_{P,n-1}^{W+}-z_{n+1}}(\Gamma_{P,n-1})_{i} \frac{1}{H_{P,n-1}^{W+}-z'_{n+1}}(  -\nabla E_{P,n-1}+\nabla E_{P,n}+\mathcal{I}|_{n}^{n-1}  )_{i'}\phi_{P,n-1}\non\\
& &\ph{444444444444444444444444444}=\mco(|\la|^{1/4}\si_{n-1}^{1-\de}\si_{n+1}^{-1}\si_{n-1}^{-\de} )=
\mco(\si_{n-1}^{-2\de}).
\eeqa
Now we consider the contribution {\color{\red}proportional to} $\mathcal{L}|_{n}^{n-1}$.  By Lemma~\ref{middle-subsection-auxiliary-lemma} we
have
\beqa
\frac{1}{H_{P,n-1}^{W+}-z_{n+1}}(\Gamma_{P,n-1})_{i} \frac{1}{H_{P,n-1}^{W+}-z'_{n+1}}
(  \mathcal{L}|_{n}^{n-1}  )_{i'}\phi_{P,n-1}=\mco(|\la|^{1/2}\si_{n-1}^{-\de}),
\eeqa
which concludes the proof. \qed

\bel\label{De-Ga-De-Ga} {\color{\red}(Control of l.h.s.(\ref{final-estimate}))} Under the assumptions of Theorem~\ref{main-technical-result} 
\beqa
R (\Delta\Gamma_{P}|_{n}^{n-1})_{i}R'  (\Delta\Gamma_{P}|_{n}^{n-1})_{i'}\phi_{P,n-1}   
:=\frac{1}{H_{P,n-1}^{W+}-z_{n+1}}(\Delta\Gamma_{P}|_{n}^{n-1})_{i}\frac{1}{H_{P,n-1}^{W+}-z'_{n+1}}(\Delta\Gamma_{P}|_{n}^{n-1})_{i'}\phi_{P,n-1}=\mco(\si_{n-1}^{-2\de}).\label{yet-another-term}
\eeqa
\eel
\proof Recalling from (\ref{De-E-I}) that $\Delta\Gamma_{P}|_{n}^{n-1}=X|^{n-1}_n
+\mathcal{L}|_{n}^{n-1}$, where $X|^{n-1}_n:=-\nabla E_{P,n-1}+\nabla E_{P,n}+\mathcal{I}|_{n}^{n-1}$ is a vector in $\real^3$
satisfying
\beqa
|X|^{n-1}_n|\leq  
c|\la|^{1/4}\si_{n-1}^{1-\de}, \label{standard-bound-X}
\eeqa
we write 
\beqa
\mathrm{l.h.s.}(\ref{yet-another-term})\1&=&\1\frac{1}{H_{P,n-1}^{W+}-z_{n+1}}  ( X|^{n-1}_n)_i\frac{1}{H_{P,n-1}^{W+}-z'_{n+1}}
(X|^{n-1}_n)_{i'}\phi_{P,n-1} \label{first-contr}\\
& &+\frac{1}{H_{P,n-1}^{W+}-z_{n+1}}  ( X|^{n-1}_n)_i\frac{1}{H_{P,n-1}^{W+}-z'_{n+1}}
(\mathcal{L}|_{n}^{n-1})_{i'}\phi_{P,n-1}\label{second-contr}\\
& &+\frac{1}{H_{P,n-1}^{W+}-z_{n+1}} ( \mathcal{L}|_{n}^{n-1})_{i} \frac{1}{H_{P,n-1}^{W+}-z'_{n+1}}
( X|^{n-1}_n )_{i'}\phi_{P,n-1}\label{third-contr}\\
& &+\frac{1}{H_{P,n-1}^{W+}-z_{n+1}}  (\mathcal{L}|_{n}^{n-1})^{(+)}_i
\frac{1}{H_{P,n-1}^{W+}-z'_{n+1}}
(\mathcal{L}|_{n}^{n-1})_{i'}\phi_{P,n-1}. \label{fourth-contr}\\
& &+\frac{1}{H_{P,n-1}^{W+}-z_{n+1}}  (\mathcal{L}|_{n}^{n-1})^{(-)}_i
\frac{1}{H_{P,n-1}^{W+}-z'_{n+1}}
(\mathcal{L}|_{n}^{n-1})_{i'}\phi_{P,n-1}, \label{fifth-contr}
\eeqa
where $(\mathcal{L}|_{n}^{n-1})^{(\pm)}_i$ are the creation/annihilation parts of ${\color{\red}(\mathcal{L}|_{n}^{n-1})_i}$.
We have by (\ref{standard-bound-X}) and Lemma~\ref{lem:Basic-estimates}
\beqa
\|(\ref{first-contr})\| \leq c|\la|^{1/2} \bigg(\fr{\si_{n-1} }{\si_{n+1}} \bigg)^2\fr{1}{\si_{n-1}^{2\de}}\leq \fr{c'}{\si_{n-1}^{2\de}}.
\eeqa
Now we analyse the terms involving $\mathcal{L}|_{n}^{n-1}$. Clearly, 
\beqa
(\ref{second-contr})=-\la\hint^{\oplus}_{\!\!\!\! \bmcA|^{n-1}_n}  \fr{d^3k}{\sqrt{2}|k|^{3/2}} 
\frac{k_{i'}}{\alpha_{P,n-1}(\hat{k})}\frac{1}{[H_{P,n-1}^{W+}]_k-z_{n+1}}
\frac{( X|^{n-1}_n)_i}{[H_{P,n-1}^{W+}]_k-z'_{n+1}}\phi_{P,n-1},\quad
\eeqa
and therefore, by Lemma~\ref{pull-through-resolvents}, {\bcb (which gives $|k_{i'}| \,\|([H_{P,n-1}^{W+}]_k-z_{n+1})^{-1}\|_{\mcF_n} 
\|([H_{P,n-1}^{W+}]_k-z'_{n+1})^{-1}\|_{\mcF_n}  \leq c|k|^{-1}$)}, $|k|^{-1}\leq \si_n^{-1}$ and estimate~(\ref{standard-bound-X}), we have
\beqa  \label{second-third-contr}
\|(\ref{second-contr})\|\leq c|\la|^{5/4}\si_{n-1}^{1-\de}|\ln\,\veps|^{1/2}\si_n^{-1}\leq 
c|\la|^{5/4} \fr{\si_{n-1}}{\si_n} |\ln\,\veps|^{1/2}\fr{1}{\si_{n-1}^{\de} }\leq \fr{c'}{\si_{n-1}^{\de}}.
\eeqa
Now we consider (\ref{third-contr}). In the direct integral notation we have
\beqa
(\ref{third-contr})\simeq -\la    \hint^{\oplus}_{\1 \bmcA|^{n-1}_n}  \fr{d^3k}{\sqrt{2}|k|^{3/2}} 
\frac{k_{i}}{\alpha_{P,n-1}(\hat{k})}\frac{1}{[H_{P,n-1}^{W+}]_k-z_{n+1}}\frac{( X|^{n-1}_n )_{i'}}{H_{P,n-1}^{W+}-z'_{n+1}}
\phi_{P,n-1}.
\eeqa
by Lemma~\ref{pull-through-resolvents} and (\ref{standard-bound-X})  we obtain
\beqa
\|(\ref{third-contr})\|\leq c|\la|^{5/4} \fr{\si_{n-1}}{\si_{n+1}} |\ln\,\veps|^{{1/2}}\fr{1}{\si_{n-1}^{\de} }\leq \fr{c'}{\si_{n-1}^{\de}}.
\eeqa

Next, we consider (\ref{fourth-contr}). In the direct integral representation it has the form  
\beqa
(\ref{fourth-contr})\1 &\simeq& \1 \fr{\la^2}{\sqrt{2}} {\hint}^{\oplus}_{\1(\bmcA|^{n-1}_n)^{\times 2}}d^3k_1d^3k_2 \, \frac{1}{[H_{P,n-1}^{W+}]_{k_1,k_2}-z_{n+1}} \times\non\\
& &\times \bigg\{ \fr{1}{\sqrt{2}|k_1|^{3/2}} \frac{k_{1,i}}{\alpha_{P,n-1}(\hat{k}_1)}
\fr{1}{\sqrt{2}|k_2|^{3/2}} \frac{k_{2,i'}}{\alpha_{P,n-1}(\hat{k}_2)}\frac{1}{[H_{P,n-1}^{W+}]_{k_2}-z_{n+1}}\phi_{P,n-1}+\{1\leftrightarrow 2\}\bigg\}.
\eeqa 
Now making use of Lemma~\ref{pull-through-resolvents} we get 
$\|(\ref{fourth-contr})\|\leq c|\la|^2|\ln\,\veps|\leq c'$. Finally we estimate (\ref{fifth-contr}). It is easy to see that
\beqa
& &(\ref{fifth-contr})=\la^2\frac{1}{H_{P,n-1}^{W+}-z_{n+1}}  {\color{\red}\bint_{\!\!\!\!\bmcA|^{n-1}_n}} d^3k  \fr{1}{2|k|^{3}} \frac{k_{i}k_{i'}  }{\alpha_{P,n-1}(\hat{k})^2}
\frac{1}{[H_{P,n-1}^{W+}]_k-z'_{n+1}}
\phi_{P,n-1}\non\\  
& &\ph{4444444444444444444444444444444444}=\mco(\la^2 |{\color{\red}\ln\,\veps|} \si_{n-1}\si_{n+1}^{-1})=\mco(|\la|),
\eeqa
where we made use of Lemma~\ref{lem:Basic-estimates} and (\ref{res-est}).  This concludes the proof. \qed
\begin{rem}
{\bcb  In the estimate of (\ref{second-third-contr})  and other direct integral estimates in the proof of Lemma~\ref{De-Ga-De-Ga}, we follow a procedure that is not optimal but is enough for our purposes and will be applied (without further notice) to similar quantities studied in the rest of this paper.  Namely, we first took the norm of the  expression with the direct integral in order to  obtain
an ordinary integral, then  used  $|k_{i'}|$ appearing in the numerator to compensate $1/|k|$ resulting from the bound 
$\|\frac{1}{[H_{P,n-1}^{W+}]_k-z_{n+1}}\|_{\mcF_n} 
\leq \fr{c}{|k|}$ of Lemma~\ref{pull-through-resolvents},    and only in the end computed the integral.  (If we estimated instead $\fr{c}{|k|}\leq \fr{c}{\si_{n+1}}$, 
and computed the integral directly, incorporating $k_i'$, we could avoid the  factor $|\ln\, \veps|^{1/2}$).  On other occasions,
e.g. in the estimate (\ref{log-estimate-lemma}), we apply $|k_i|\leq \si_{n-1}$ to the $k_i$ in the numerator. Again, the logarithms could be avoided here
by streamlining the computation differently.}
\end{rem}
\bel\label{final-subsection-auxiliary-lemma-one} Under the assumptions of Theorem~\ref{main-technical-result} we have
\beqa
\frac{1}{H_{P,n-1}^{W+}-z_{n+1}}\mathcal{L}|_{n}^{n-1}\cdot \Gamma_{P,n-1} \phi_{P,n-1}=\mco(|\la|{\color{\red}|\ln\, \veps|^{1/2}}\si_{n-1}^{1-\de})
\eeqa
for $z_{n+1}\in \ga_{P,n+1}\cup \ti \ga_{P,n+1}$. 
\eel
\proof Making use of the direct integral representations from Subsection~\ref{direct-subsection}, we get
\beqa
\frac{1}{H_{P,n-1}^{W+}-z_{n+1}}\mathcal{L}|_{n}^{n-1}\cdot \Gamma_{P,n-1} \phi_{P,n-1}
\1 &\simeq&\1 -\la {\color{\red}\hint^{\oplus}_{\!\!\!\!\bmcA|^{n-1}_n}}  \fr{d^3k}{\sqrt{2}|k|^{3/2}} 
\frac{k_{i}}{\alpha_{P,n-1}(\hat{k})} \frac{1}{[H_{P,n-1}^{W+}]_{k}-z_{n+1}}\Gamma_{P,n-1} \phi_{P,n-1}\non\\
\1&=&\1 \mco(|\la|{\color{\red}|\ln\, \veps|^{1/2}}\si_{n-1}^{1-\de}), \label{log-estimate-lemma}
\eeqa
where we applied (\ref{estimate-Q}) and Theorem~\ref{thm:induction-convergence}. 
Here, the reader should note that (\ref{estimate-Q}) is used together with the identity
\begin{equation}
 \frac{1}{[H_{P,n-1}^{W+}]_k-z_{n+1}}  \Gamma_{P,n-1}\phi_{P,n-1}
  = \frac{1}{[H_{P,n-1}^{W+}]_k-z_{n+1}} Q_{P,n-1}^{\bot} \Gamma_{P,n-1}\phi_{P,n-1}\label{id}
\end{equation}
which follows from  the definition in (\ref{Ga-formula}).   \qed
\bel\label{middle-subsection-auxiliary-lemma} Under the assumptions of Theorem~\ref{main-technical-result} we have
\beqa
& &\frac{1}{H_{P,n-1}^{W+}-z_{n+1}}(\Gamma_{P,n-1})_{i} \frac{1}{H_{P,n-1}^{W+}-z'_{n+1}}
(  \mathcal{L}|_{n}^{n-1}  )_{i'}\phi_{P,n-1}=\mco(|\la|^{1/2}\si_{n-1}^{-\de}), \label{R-G-R-L-end-of-proof}\\
& &\frac{1}{H_{P,n-1}^{W+}-z_{n+1}}(\Gamma_{P,n-1})_{i} \frac{1}{H_{P,n-1}^{W+}-z'_{n+1}}
(  \mring{\mathcal{L}}|_{n}^{n-1}  )\phi_{P,n-1}=\mco(|\la|^{1/2}\si_{n-1}^{1-\de}), \label{R-G-R-L-dot-end-of-proof}
\eeqa
for $z_{n+1}, z'_{n+1}\in \ga_{P,n+1}\cup \ti \ga_{P,n+1}$.  Here we denoted
\beqa
(\mring{\mathcal{L}}|_{n}^{n-1}):= {\color{\red}\sum_{i=1}^{3}}[(\Ga_{P,n-1})_i,  (\mathcal{L}|_{n}^{n-1})_i]=\lambda \hint_{\!\!\sigma_{n}}^{\sigma_{n-1}}d^{3}k\frac{k^2(b(k)-b^{*}(k))}{\sqrt{2}|k|^{3/2}\alpha_{P,n-1}(\hat{k})}. \label{L-dot-formula-x}
\eeqa
\eel
\proof 
Making use of the direct integral representation from 
Subsection~\ref{direct-subsection}, we have
\beqa
& &\frac{1}{H_{P,n-1}^{W+}-z_{n+1}}(\Gamma_{P,n-1})_{i} \frac{1}{H_{P,n-1}^{W+}-z'_{n+1}}
(  \mathcal{L}|_{n}^{n-1}  )_{i'}\phi_{P,n-1}\non\\
& &\simeq-\la\hint^{\oplus}_{\!\!\!\! \bmcA|^{n-1}_n}  \fr{d^3k}{\sqrt{2}|k|^{3/2}} 
\frac{k_{i'}}{\alpha_{P,n-1}(\hat{k})}\frac{1}{[H_{P,n-1}^{W+}]_k-z_{n+1}}(\Gamma_{P,n-1}+k)_{i} \frac{1}{[H_{P,n-1}^{W+}]_k-z'_{n+1}}
\phi_{P,n-1}\non\\
& &=-\la\hint^{\oplus}_{\!\!\!\!\bmcA|^{n-1}_n}  \fr{d^3k}{\sqrt{2}|k|^{3/2}} 
\frac{k_{i'}}{\alpha_{P,n-1}(\hat{k})}\frac{1}{[H_{P,n-1}^{W+}]_k-z_{n+1}}(\Gamma_{P,n-1})_{i} \frac{1}{[H_{P,n-1}^{W+}]_k-z'_{n+1}}
\phi_{P,n-1} \label{needing-expansion}\\
& &\ph{4444444444444444444444444444444444444444444444444}+\mco(|\la||\ln\,\veps|^{1/2} ), \label{error-term-in-first-part}
\eeqa
where in the first step we made use of (\ref{second-direct}) and in the last step we applied (\ref{res-est}). Since $|\la|\leq \veps^8$, 
 the error term in (\ref{error-term-in-first-part}) contributes to the r.h.s. of (\ref{R-G-R-L-end-of-proof}). To treat 
(\ref{needing-expansion}), we apply the following resolvent expansion to the resolvent next to $\phi_{P,n-1}$:
\beqa
\frac{1}{[H_{P,n-1}^{W+}]_k-z'_{n+1}}
\1&=&\1\frac{1}{H_{P,n-1}^{W+}+\alpha_{P, n-1}(\hat k)|k|-z'_{n+1}}\label{res-exp-zero}\\
& &+\frac{1}{[H_{P,n-1}^{W+}]_k-z'_{n+1}}  [-k\cdot \Gamma_{P,n-1}-\frac{|k|^2}{2}]\frac{1}{H_{P,n-1}^{W+}+\alpha_{P, n-1}(\hat k)|k|-z'_{n+1}}.\quad\quad \label{res-exp}
\eeqa
{\color{\red}This is well defined because for} $\si_{n}\leq |k|\leq \si_{n-1}$,   making use of (\ref{conv-assumption-E}), (\ref{conv-assumption-c}) and
the definitions of $\ga_{P,n+1}, \ti\ga_{P,n+1}$, {\color{\red} we can state}
\beqa
|E_{P,n-1}+\De c_P|^{n-1}_n+\alpha_{P, n-1}(\hat k)|k|-\mrm{Re}\,z'_{n+1}|\geq c|k| \label{bound-denominator-k}
\eeqa
for some $c>0$. We first consider the contribution to (\ref{needing-expansion}) coming from (\ref{res-exp-zero}). It has the form
\beqa
& &-\la\hint^{\oplus}_{\!\!\!\!\bmcA|^{n-1}_n}  \fr{d^3k}{\sqrt{2}|k|^{3/2}} 
\frac{k_{i'}}{\alpha_{P,n-1}(\hat{k})}\frac{1}{[H_{P,n-1}^{W+}]_k-z_{n+1}}(\Gamma_{P,n-1})_{i} 
\frac{1}{H_{P,n-1}^{W+}+\alpha_{P, n-1}(\hat k)|k|-z'_{n+1}}\phi_{P,n-1} \non\\
& &\ph{44444444444444444444444444444444444444}=\mco( |\la|  |\ln\,\veps|^{1/2} \si_{n-1}^{-\de} )=\mco(|\la|^{1/2} \si_{n-1}^{-\de} ),
\eeqa
where we made use of (\ref{bound-denominator-k}), (\ref{estimate-Q}) and Theorem~\ref{thm:induction-convergence}.
The contribution to (\ref{needing-expansion}) coming from the term in  (\ref{res-exp}) involving $|k|^2$ is readily
shown to be $\mco(|\la|^{1/2})$ by (\ref{bound-denominator-k}), (\ref{res-est}), and (\ref{res-gamma-est}). Finally,
the contribution to (\ref{needing-expansion}) coming from the term in  (\ref{res-exp}) involving $k\cdot \Ga_{P,n-1}$
has the form
\beqa
\la\hint^{\oplus}_{\!\!\!\!\bmcA|^{n-1}_n}  \fr{d^3k}{\sqrt{2}|k|^{3/2}} 
\frac{k_{i'} \mco(|k|^{-1})}{\alpha_{P,n-1}(\hat{k})} \frac{1}{[H_{P,n-1}^{W+}]_k-z_{n+1}}
(\Gamma_{P,n-1})_{i}  \frac{1}{[H_{P,n-1}^{W+}]_k-z'_{n+1}}  (k\cdot \Gamma_{P,n-1})\phi_{P,n-1},
\eeqa
where the $\mco(|k|^{-1})$ insertion comes from (\ref{bound-denominator-k}). This expression is 
$\mco(|\la|^{1/2}\si_{n-1}^{-\de}|\ln\,\veps|^{1/2})$ by (\ref{res-gamma-est}), (\ref{estimate-Q}) and Theorem~\ref{thm:induction-convergence}. 
This concludes the proof of (\ref{R-G-R-L-end-of-proof}). Noting that $\mring{\mathcal{L}}|_{n}^{n-1}$ involves one additional
power of $k$ compared to  $\mathcal{L}|_{n}^{n-1}$, we obtain (\ref{R-G-R-L-dot-end-of-proof}) by a completely analogous argument. \qed
\subsection{Contribution to $I^q$ with $l\in \nat_0$, $l'\in \nat$, $l''=0$}\label{Main-result-second-subsection}

In the following lemma we merely recall several facts which appeared in earlier parts of the paper {\color{\red}and that will be used repeatedly in the present section.}
\bel\label{trivial-bounds-xx} Under the assumptions of Theorem~\ref{main-technical-result}
\beqa 
& &\|R (\Gamma_{P,n-1})_{i}\|_{\mcF_n}:=
\|\frac{1}{H_{P,n-1}^{W+}-z_{n+1}}(\Gamma_{P,n-1})_i\|_{\mcF_n}=\mco(\si_{n+1}^{-1}),\label{second-case-first-est}\\
& &\|RV\|_{\mcF_n}:=\|\frac{1}{H_{P,n-1}^{W+}-{\color{\red}z_{n+1}}}(-)H_{I}^{W}|_{n}^{n-1}\|_{\mcF_n}=\mco(|\la|^{1/2}), \label{R-prime-V}\\
& &\|R(\Delta\Gamma_{P}|_{n}^{n-1})_{i}\|_{\mcF_n}:=\|\frac{1}{H_{P,n-1}^{W+}-z_{n+1}} (\Delta\Gamma_{P}|_{n}^{n-1})_{i}\|_{\mcF_n}=\mco(\si_{n-1}^{-\de}), \label{second-case-second-estimate}  \\
& &\|R  (\Gamma_{P,n-1})_{i'}\phi_{P,n-1}\|:=\|\frac{1}{H_{P,n-1}^{W+}-z_{n+1}}(\Gamma_{P,n-1})_{i'}\phi_{P,n-1}\| =\mco(\si_{n-1}^{-\de}) \label{Q-R-V-second-est}
\eeqa
and the same estimates hold for $R$ replaced with $R', R'', R'''$ (cf. definitions (\ref{R-R'-def}), (\ref{R'''-V-def})).
\eel
\proof Estimates (\ref{second-case-first-est}) and (\ref{R-prime-V}) come from Lemma~\ref{lem:Basic-estimates}.
As for estimate~(\ref{second-case-second-estimate}), we recall that $\Delta\Gamma_{P}|_{n}^{n-1}:=
-\nabla E_{P,n-1}+\nabla E_{P,n}+\mathcal{I}|_{n}^{n-1}
+\mathcal{L}|_{n}^{n-1}$ and refer to (\ref{De-E-I}) and Lemma~\ref{lem:Basic-estimates}. The bound in (\ref{Q-R-V-second-est})
follows from Theorem~\ref{thm:induction-convergence}.  \qed\\
{\color{\red}The contribution studied in this subsection has the form}
\beqa
I^q_{\nat_0,\nat, 0}\1&:=&\1 (\tiQ_{P,n-1}^{\perp})^q\sum_{l=0}^{\infty}  \{ RV \}^{l} 
 R (\Gamma_{P,n-1}+\Delta\Gamma_{P}|_{n}^{n-1})_{i}  \sum_{l'=0}^{\infty}\{  R'V  \}^{l'} (R'V) R' 
 (\Gamma_{P,n-1}+\Delta\Gamma_{P}|_{n}^{n-1})_{i'}
 \phi_{P,n-1}\non\\
\1&=&\1 (\tiQ_{P,n-1}^{\perp})^q\sum_{l=0}^{\infty}  \{ RV \}^{l}  R (\Gamma_{P,n-1}+\Delta\Gamma_{P}|_{n}^{n-1})_{i}  
\sum_{l'=0}^{\infty}\{  R'V  \}^{l'} Q_{P,n-1}^{\perp}(R'V) R'  (\Gamma_{P,n-1})_{i'}\phi_{P,n-1}\label{second-sum-first-term} \quad\quad\quad \\
&&+ (\tiQ_{P,n-1}^{\perp})^q\sum_{l=0}^{\infty}  \{ RV \}^{l}  R (\Gamma_{P,n-1}+\Delta\Gamma_{P}|_{n}^{n-1})_{i}  
\sum_{l'=0}^{\infty}\{  R'V  \}^{l'} Q_{P,n-1}(R'V) R'  (\Gamma_{P,n-1})_{i'}\phi_{P,n-1}
\label{second-sum-second-term} \quad \\
&&+(\tiQ_{P,n-1}^{\perp})^q\sum_{l=0}^{\infty}  \{ RV \}^{l}  R  (\Gamma_{P,n-1}+\Delta\Gamma_{P}|_{n}^{n-1})_{i}  
\sum_{l'=0}^{\infty}\{  R'V  \}^{l'} (R'V) R'  (\Delta\Gamma_{P}|_{n}^{n-1})_{i'} \phi_{P,n-1} 
\label{second-sum-third-term} \\
\1&=&\1\mco(\si_{n-1}^{-2\de}).
\eeqa
To show this bound, let us first analyse (\ref{second-sum-first-term}). By (\ref{simple-sum-over-l}), 
$(\tiQ_{P,n-1}^{\perp})^q\sum_{l=0}^{\infty}  \{ RV \}^{l}=\mco(1)$ and  $\sum_{l'=0}^{\infty}\{  R'V  \}^{l'}=\mco(1)$, thus we can write
\beqa
(\ref{second-sum-first-term})\1 &=&\1\mco(1)R (\Gamma_{P,n-1}+\Delta\Gamma_{P}|_{n}^{n-1})_{i}  \sum_{l'=0}^{\infty}\{  R'V  \}^{l'}
\bigg(b^*\big( g_{i'}|^{n-1}_n\big)\phi_{P,n-1}+\mco(|\la|^{1/2}\si_{n-1}^{1-2\de})\bigg)\non\\
\1&=&\1\mco(1)R (\Gamma_{P,n-1}+\Delta\Gamma_{P}|_{n}^{n-1})_{i}  \sum_{l'=0}^{\infty}\{  R'V  \}^{l'}
b^*\big(g_{i'}|^{n-1}_n\big)\phi_{P,n-1}+\mco(\si_{n-1}^{-2\de}) \non\\
\1&=&\1\mco(1)R (\Gamma_{P,n-1}+\Delta\Gamma_{P}|_{n}^{n-1})_{i} \sum_{l'=0}^{\infty}\{  R'V  \}^{l'} 
R'Vb^*\big( g_{i'}|^{n-1}_n \big)\phi_{P,n-1} \label{above-xxx} \quad\quad\quad\\
& &+\mco(1)R (\Gamma_{P,n-1}+\Delta\Gamma_{P}|_{n}^{n-1})_{i}  
b^*\big(g_{i'}|^{n-1}_n\big)\phi_{P,n-1}+\mco(\si_{n-1}^{-2\de})\label{above-xxxx}\\
\1&=&\1\mco(\si_{n-1}^{-2\de}), \non
\eeqa
where  {\color{\red}$g_{i'}|^{n-1}_n$ is a function defined in (\ref{g_i}), cf. 
Lemma~\ref{b-star-lemma}. Here, }in the first step we used Lemma~\ref{b-star-lemma} and in the second step 
Lemma~\ref{trivial-bounds-xx}. The term in (\ref{above-xxx}) is estimated using 
Lemmas~\ref{interaction-x}, \ref{trivial-bounds-xx}. The term {\color{\red}in (\ref{above-xxxx})} is estimated  using 
Lemmas~\ref{R-Ga-g-phi},  \ref{R-De-Ga-g-phi}.
Now we analyse (\ref{second-sum-second-term}):
\beqa
& &(\ref{second-sum-second-term})
= \mco(1) R (\Gamma_{P,n-1}+\Delta\Gamma_{P}|_{n}^{n-1})_{i}  
 Q_{P,n-1}(R'V) R'  (\Gamma_{P,n-1})_{i'}\phi_{P,n-1} \label{second-sum-second-term-one}\\
& &\ph{444444}+\mco(1)  R (\Gamma_{P,n-1}+\Delta\Gamma_{P}|_{n}^{n-1})_{i}  
\sum_{l'=0}^{\infty}\{  R'V  \}^{l'} (R'V) Q_{P,n-1}(R'V) R'  (\Gamma_{P,n-1})_{i'}\phi_{P,n-1}\label{second-sum-second-term-two}\quad\quad\quad\\
& &\ph{444444}=\mco(\si_{n-1}^{-2\de}). \non
\eeqa
Expression~(\ref{second-sum-second-term-one}) is estimated using Lemma~\ref{trivial-bounds-xx}. Expression~(\ref{second-sum-second-term-two})
 is estimated using Lemmas~\ref{first-case-aux-lemma}, \ref{trivial-bounds-xx}.

\noindent
Finally~(\ref{second-sum-third-term})  is bounded using~ Lemma~\ref{last-lemma-second-case} and Lemma~\ref{trivial-bounds-xx}.

\bel\label{b-star-lemma} {\color{\red}(Control of (\ref{second-sum-first-term}))} Under the assumptions of Theorem~\ref{main-technical-result} and the inductive hypothesis
\beqa
Q_{P,n-1}^{\perp}(R'V) R'  (\Gamma_{P,n-1})_{i'}\phi_{P,n-1}
\1&:=&\1 Q_{P,n-1}^{\perp}\frac{1}{H_{P,n-1}^{W+}-z'_{n+1}} (-)H_{I}^{W}|_{n}^{n-1} \frac{1}{H_{P,n-1}^{W+}-z'_{n+1}} (\Gamma_{P,n-1})_{i'}\phi_{P,n-1}\non\\
\1&=&\1 {\color{\red}b^*\big(  g_{i'}|^{n-1}_n} \big)\phi_{P,n-1}+\mco(|\la|^{1/2}\si_{n-1}^{1-2\de}),
\eeqa
where {\color{\red} $ b^*(g_{i'}|^{n-1}_n):=\int d^3k g_{i'}(k)|^{n-1}_n b^*(k)$ with $g_{i'}|^{n-1}_n$ a function with support in  $\bar\mcA|^{n-1}_n$} which satisfies the pointwise bound
\beqa
 |g_{i'}(k)|^{n-1}_n|\leq |\la|\fr{c}{|k|^{3/2}\si_{n-1}^{\de}}. \label{g-bound}
\eeqa
\eel
\proof We recall that $H_I^{W}|_{n}^{n-1}:=(\mathcal{L}|_{n}^{n-1}+\mathcal{I}|_{n}^{n-1})\cdot\Gamma_{P,n-1}
+\Delta(H_{I}^{W}|_{n}^{n-1})_{\mr{mix}}+(H_{I}^{W}|_{n}^{n-1})_{\mr{quad}}$ and consider the respective terms.
First, by Lemma~\ref{lem:Basic-estimates} and Theorem~\ref{thm:induction-convergence}, 
we obtain that
\beqa
Q_{P,n-1}^{\perp}\frac{1}{H_{P,n-1}^{W+}-z'_{n+1}} (\Delta(H_{I}^{W}|_{n}^{n-1})_{\mr{mix}}+(H_{I}^{W}|_{n}^{n-1})_{\mr{quad}} ) \frac{1}{H_{P,n-1}^{W+}-z'_{n+1}} (\Gamma_{P,n-1})_{i'}\phi_{P,n-1}
\eeqa
is $\mco(|\la|^{1/2}\si_{n-1}\si_{n-1}^{-\de})=\mco(|\la|^{1/2}\si_{n-1}^{1-\de})$. Next, since $|\mathcal{I}|_{n}^{n-1}|\leq c_I |\lambda|^{2}\sigma_{n-1}$,
we have
\beqa
Q_{P,n-1}^{\perp}\frac{1}{H_{P,n-1}^{W+}-z'_{n+1}} (\mathcal{I}|_{n}^{n-1}\cdot\Gamma_{P,n-1}  ) \frac{1}{H_{P,n-1}^{W+}-z'_{n+1}} (\Gamma_{P,n-1})_{i'}\phi_{P,n-1}=\mco(|\la|^2\sigma_{n-1}^{1-2\de} ),
\eeqa
where we applied  the inductive hypothesis {\color{\red}of Theorem~\ref{main-technical-result}}. 
{\color{\red} Finally, we observe that in vector (\ref{R-LG-R-G-phi}) below we can omit  $Q_{P,n-1}^{\perp}$ since this vector is manifestly in its range (cf. Lemma~\ref{Q-shifting}). Afterwards, we use the direct integral representation  and then insert $Q_{P,n-1}+Q_{P,n-1}^{\bot}=1$ on $\mcF_{n-1}$ inside the direct integral:} 
\beqa
& &Q_{P,n-1}^{\perp}\frac{1}{H_{P,n-1}^{W+}-z'_{n+1}} ( \mathcal{L}|_{n}^{n-1}\cdot\Gamma_{P,n-1}  ) \frac{1}{H_{P,n-1}^{W+}-z'_{n+1}} (\Gamma_{P,n-1})_{i'}\phi_{P,n-1}\label{R-LG-R-G-phi}\\
&{\color{\red}\simeq }&-\la\hint^{\oplus}_{\!\!\!\bmcA|^{n-1}_n}  \fr{d^3k}{\sqrt{2}|k|^{3/2}} 
\frac{k_{i}}{\alpha_{P,n-1}(\hat{k})}\times \non\\
& &\ph{444444444}\times\frac{1}{[H_{P,n-1}^{W+}]_k-z'_{n+1}}Q_{P,n-1}^{\perp} (\Gamma_{P,n-1})_i \frac{1}{H_{P,n-1}^{W+}-z'_{n+1}} (\Gamma_{P,n-1})_{i'}\phi_{P,n-1}\label{second-case-second-est-0} \\
& & -\la\hint^{\oplus}_{\!\!\!\!\bmcA|^{n-1}_n}  \fr{d^3k}{\sqrt{2}|k|^{3/2}} \frac{k_{i}}{\alpha_{P,n-1}(\hat{k})}\times\non\\
& &\ph{444444444}\times \frac{1}{[H_{P,n-1}^{W+}]_k-z'_{n+1}} Q_{P,n-1}(\Gamma_{P,n-1})_i \frac{1}{H_{P,n-1}^{W+}-z'_{n+1}} (\Gamma_{P,n-1})_{i'}\phi_{P,n-1}. 
\label{second-case-second-est-1}
\eeqa
Now using  estimate~(\ref{estimate-Q}) we obtain
\beqa
\|(\ref{second-case-second-est-0})\|
\leq c|\la|\si_{n-1}|{\color{\red}\ln\, \veps|^{1/2}} \big\|Q_{P,n-1}^{\perp}\frac{1}{H_{P,n-1}^{W}-z_{n+1}} (\Gamma_{P,n-1})_i \frac{1}{H_{P,n-1}^{W+}-z_{n+1}} (\Gamma_{P,n-1})_{i'}\phi_{P,n-1}\big\|
\leq c|\la|^{1/2}\si_{n-1}^{1-2\de},\quad \label{unnumbered-formula-xxx}
\eeqa
where we made use of the  induction hypothesis. As for (\ref{second-case-second-est-1}), with the help of {\color{\red}(\ref{res-exp-zero})-}(\ref{res-exp}), (\ref{bound-denominator-k}), Lemma~\ref{pull-through-resolvents} and Theorem~\ref{thm:induction-convergence},  we can write
\beqa
\frac{1}{[H_{P,n-1}^{W+}]_k-z'_{n+1}} \phi_{P,n-1}=g_0(k)\phi_{P,n-1}+\mco(\si_{n-1}^{-\de}), \label{first-obs-xx}
\eeqa
where $g_0$ is a numerical function bounded by $c|k|^{-1}$. We also note that by Theorem~\ref{thm:induction-convergence}
\beqa
g_{i,i'}:=\fr{1}{\|\phi_{P,n-1}\|^2}\lan\phi_{P,n-1}, (\Gamma_{P,n-1})_i \frac{1}{H_{P,n-1}^{W+}-z'_{n+1}} (\Gamma_{P,n-1})_{i'}\phi_{P,n-1}\ran=\mco(\si_{n-1}^{-\de}).
\label{second-obs-xx}
\eeqa
Making use of (\ref{first-obs-xx}), (\ref{second-obs-xx}), (\ref{unnumbered-formula-xxx}) we get
\beqa
(\ref{second-case-second-est-1})=b^*\big( g_{i'}|^{n-1}_n\big)\phi_{P,n-1}+\mco({\color{\red}|\la| 
|\ln\, \veps|^{1/2}}\si_{n-1}^{1-2\de})
=b^*\big(g_{i'}|^{n-1}_n \big)\phi_{P,n-1}+\mco(|\la|^{1/2}\si_{n-1}^{1-2\de}),
\eeqa
where $g_{i'}|^{n-1}_n$ is the following function from $(\mcF|^{n-1}_n)^{(1)}$
\beqa
g_{i'}(k)|^{n-1}_n:=-\la\chi_{\ka}(k)|k|^{\alf}\fr{1}{\sqrt{2}|k|^{3/2}} \frac{k_{i}}{\alpha_{P,n-1}(\hat{k})}g_0(k)g_{i,i'}.\label{g_i}
\eeqa 
Clearly, it satisfies the pointwise bound $|g_{i'}(k)|^{n-1}_n|\leq |\la|\fr{c}{|k|^{3/2}\si_{n-1}^{\de}}$. \qed

\bel\label{interaction-x}  {\color{\red}(Control of (\ref{above-xxx}))} Under the assumptions of Theorem~\ref{main-technical-result} 
\beqa
R'Vb^*\big( g_{i'}|^{n-1}_n \big)\phi_{P,n-1}:=\frac{1}{H_{P,n-1}^{W+}-z'_{n+1}}(-)H_{I}^{W}|_{n}^{n-1}b^*\big( g_{i'}|^{n-1}_n \big)\phi_{P,n-1}=\mco(|\la|\si^{1-2\de}_{n-1}),
\eeqa
where $g_{i'}|^{n-1}_n$ was introduced in Lemma~\ref{b-star-lemma}. The same bound holds for $z'_{n+1}\in \ga_{P,n+1}$.
\eel
\proof  We recall that $H_I^{W}|_{n}^{n-1}:=(\mathcal{L}|_{n}^{n-1}+\mathcal{I}|_{n}^{n-1})\cdot\Gamma_{P,n-1}
+\Delta(H_{I}^{W}|_{n}^{n-1})_{\mr{mix}}+(H_{I}^{W}|_{n}^{n-1})_{\mr{quad}}$ and consider the respective terms.
First, by  {\color{\red} (\ref{3.53}) and (\ref{3.54}) in} Lemma~\ref{lem:Basic-estimates} and the fact that $\|g_{i'}|^{n-1}_n\|_2=\mco(|\la|^{1/2}\si_{n-1}^{-\de})$,
we obtain that
\beqa
\frac{1}{H_{P,n-1}^{W+}-z'_{n+1}} (\Delta(H_{I}^{W}|_{n}^{n-1})_{\mr{mix}}+(H_{I}^{W}|_{n}^{n-1})_{\mr{quad}} ) b^*\big( g_{i'}|^{n-1}_n \big)\phi_{P,n-1}
=\mco(|\la|\si_{n-1}^{1-\de}).
\eeqa
Next, making use of $|\mathcal{I}|_{n}^{n-1}|\leq c_I |\lambda|^{2}\sigma_{n-1}$, {\color{\red}the identity in  (\ref{id}) combined with (\ref{estimate-Q})} in Lemma~\ref{pull-through-resolvents}, 
Theorem~\ref{thm:induction-convergence} and bound (\ref{g-bound}), we get
\beqa
& &\frac{1}{H_{P,n-1}^{W+}-z'_{n+1}} (\mathcal{I}|_{n}^{n-1}\cdot\Gamma_{P,n-1}  ) b^*\big( g_{i'}|^{n-1}_n \big)\phi_{P,n-1}\non\\
& &\simeq (\mathcal{I}|_{n}^{n-1})_{j}\hint^{\oplus}_{\!\!\!\!\bmcA|^{n-1}_n} d^3k\,  g^{i'}(k)|^{n-1}_n \frac{1}{[H_{P,n-1}^{W+}]_k-z'_{n+1}} 
(\Gamma_{P,n-1}+k)_{j} \phi_{P,n-1}
=\mco(|\la|^2\sigma_{n-1}^{1-2\de} ),\quad\quad \label{upper-index-def}
\eeqa
where $ g_{i'}(k)|^{n-1}_n=:\chi_{\ka}(k)|k|^{\alf}  g^{i'}(k)|^{n-1}_n$, as we now indicate the factor $\chi_{\ka}(k)|k|^{\alf}$ by
`hat' over the integral. 
 Now we look at
\beqa
\frac{1}{H_{P,n-1}^{W+}-z'_{n+1}} (\mathcal{L}|_{n}^{n-1}\cdot\Gamma_{P,n-1}   ) b^*\big( g_{i'}|^{n-1}_n \big)\phi_{P,n-1}
\1&=&\1\frac{1}{H_{P,n-1}^{W+}-z'_{n+1}} ((\mathcal{L}|_{n}^{n-1})^{(+)}\cdot\Gamma_{P,n-1}   ) b^*\big( g_{i'}|^{n-1}_n \big)\phi_{P,n-1}\label{second-case-direct-int}\\
& &+\frac{1}{H_{P,n-1}^{W+}-z'_{n+1}} ((\mathcal{L}|_{n}^{n-1})^{(-)}\cdot\Gamma_{P,n-1}   ) b^*\big( g_{i'}|^{n-1}_n \big)\phi_{P,n-1}.\quad\quad\label{second-case-annihilation}
\eeqa
The direct integral representation, {\color{\red}an identity analogous to (\ref{id}) combined with (\ref{estimate-Q}) of Lemma~\ref{pull-through-resolvents},} and 
Theorem~\ref{thm:induction-convergence}  give
\beqa
& &(\ref{second-case-direct-int})\simeq -\fr{\la}{\sqrt{2}}\hint^{\oplus}_{\!\!\!\!(\bmcA|^{n-1}_n)^{\times 2} }d^3k_1d^3k_2\non\\ 
& &\ph{4444444}\bigg(\fr{1}{\sqrt{2}|k_1|^{3/2}} \frac{k_{1,j}}{\alpha_{P,n-1}(\hat{k}_1)}g^{i'}(k_2)|^{n-1}_n\frac{1}{[H_{P,n-1}^{W+}]_{k_1,k_2}-z'_{n+1}}
(\Gamma_{P,n-1}+k_2)_j\phi_{P,n-1}+\{1\leftrightarrow 2\}\bigg)\non\\
& &\ph{44444}=\mco(\la^2|\ln\, \veps|\si_{n-1}^{1-2\de})=\mco(|\la|\si_{n-1}^{1-2\de}).
\eeqa
As for (\ref{second-case-annihilation}), we have by Lemma~\ref{lem:Basic-estimates}, Theorem~\ref{thm:induction-convergence} and $|k|\leq \si_{n-1}$
\beqa
(\ref{second-case-annihilation})=-\la\bint_{\!\!\!\!\bmcA|^{n-1}_n}d^3k \fr{1}{\sqrt{2}|k|^{3/2}} \frac{k_{j}}{\alpha_{P,n-1}(\hat{k})}g^{i'}(k)|^{n-1}_n
\bigg(\frac{1}{H_{P,n-1}^{W+}-z'_{n+1}}\bigg)(\Gamma_{P,n-1}+k)_j \phi_{P,n-1},
\eeqa
which is $\mco( |\la|^2\si_{n-1}( \si_{n-1}^{-\de}+\veps^{-2})|\ln\,\veps|\si_{n-1}^{-\de})=\mco(|\la|\si_{n-1}^{1-2\de})$. \qed
\bel\label{R-Ga-g-phi}  {\color{\red}(Control of (\ref{above-xxxx}))} Under the assumptions of Theorem~\ref{main-technical-result} 
\beqa
R (\Gamma_{P,n-1})_ib^*\big(g_{i'}|^{n-1}_n\big)\phi_{P,n-1}:=\frac{1}{H_{P,n-1}^{W+}-z_{n+1}}\Gamma_{P,n-1}b^*\big(g_{i'}|^{n-1}_n\big)\phi_{P,n-1}=\mco(\si^{-2\de}_{n-1}),
\eeqa
where $g_{i'}|^{n-1}_n$ was introduced in Lemma~\ref{b-star-lemma}. 
\eel
\proof We write using the direct integral representation
\beqa
& &\frac{1}{H_{P,n-1}^{W+}-z_{n+1}}(\Gamma_{P,n-1})_ib^*\big(g_{i'}|^{n-1}_n\big)\phi_{P,n-1}\non\\
& &\ph{4}\simeq \hint^{\oplus}_{\!\!\!\!\bmcA|^{n-1}_n} d^3k\,  g_{i'}(k)|^{n-1}_n\frac{1}{[H_{P,n-1}^{W+}]_k-z_{n+1}}(\Gamma_{P,n-1}+k)_i\phi_{P,n-1}=\mco(|\la||\ln\,\veps|^{{\color{\red}\frac{1}{2}}}\si_{n-1}^{-2\de} ),\quad\quad 
\eeqa
where we made use of estimate~(\ref{g-bound}), {\color{\red}the identity in (\ref{id}),  (\ref{estimate-Q})} in Lemma~\ref{pull-through-resolvents}, and Theorem~\ref{thm:induction-convergence}. \qed
\bel\label{R-De-Ga-g-phi} {\color{\red}(Control of (\ref{above-xxxx}))} Under the assumptions of Theorem~\ref{main-technical-result} 
\beqa
& &R (\De\Gamma_{P}|^{n-1}_n)_i b^*\big(g_{i'}|^{n-1}_n\big)\phi_{P,n-1}:=\frac{1}{H_{P,n-1}^{W+}-z_{n+1}}(\De\Gamma_{P}|^{n-1}_n)_ib^*\big(g_{i'}|^{n-1}_n\big)\phi_{P,n-1}=\mco(\si^{-2\de}_{n-1}),\quad\quad\quad
\eeqa
where $g_{i'}|^{n-1}_n$ was introduced in Lemma~\ref{b-star-lemma}. 
\eel
\proof We recall that $\Delta\Gamma_{P}|_{n}^{n-1}=X|^{n-1}_n
+\mathcal{L}|_{n}^{n-1}$, where $X|^{n-1}_n:=-\nabla E_{P,n-1}+\nabla E_{P,n}+\mathcal{I}|_{n}^{n-1}$ 
is a vector in $\real^3$ satisfying
\beqa
|X|^{n-1}_n|\leq  
c|\la|^{1/4}\si_{n-1}^{1-\de}. \label{standard-bound-X-X}
\eeqa
Since $\|(H_{P,n-1}^{W+}-z_{n+1})^{-1}\|_{\mcF_n}=\mco(\si_{n+1}^{-1})$ 
and $\|g_{i'}|^{n-1}_n\|_2=\mco(|\la|^{1/2}\si_{n-1}^{-\de})$, we have
\beqa
\frac{1}{H_{P,n-1}^{W+}-z_{n+1}}( X|^{n-1}_n )_ib^*\big(g_{i'}|^{n-1}_n\big)\phi_{P,n-1}
=\mco(|\la|^{1/2} |\la|^{1/4}\veps^{-2}\si_{n-1}^{-2\de})=\mco(\si_{n-1}^{-2\de}).
\eeqa
Next, we consider the contribution with $\mathcal{L}|_{n}^{n-1}$
\beqa
& &\frac{1}{H_{P,n-1}^{W+}-z_{n+1}}(\mathcal{L}|_{n}^{n-1})_ib^*\big(g_{i'}|^{n-1}_n\big)\phi_{P,n-1}=
\frac{1}{H_{P,n-1}^{W+}-z_{n+1}}(\mathcal{L}|_{n}^{n-1})^{(+)}_ib^*\big(g_{i'}|^{n-1}_n\big)\phi_{P,n-1} \quad\quad\label{L+g}\\
& &\ph{444444444444}-\bigg(\la \int_{\bmcA|^{n-1}_n} d^3k\,\fr{1}{\sqrt{2}|k|^{3/2}} \frac{k_{i}}{\alpha_{P,n-1}(\hat{k})}  g_{i'}|^{n-1}_n(k) \bigg)\frac{1}{H_{P,n-1}^{W+}-z_{n+1}}\phi_{P,n-1}. \label{L-g} 
\eeqa
To the r.h.s. of (\ref{L+g}) we apply the direct integral representation
\beqa
& &\mathrm{r.h.s.}(\ref{L+g})\simeq -\fr{\la}{\sqrt{2}}{\color{\red}{\hint}^{\oplus}_{\1 (\bmcA|^{n-1}_n)^{\times 2} }}d^3k_1d^3k_2\non\\ 
& &\ph{444444}
\bigg(\fr{1}{\sqrt{2}|k_1|^{3/2}} \frac{k_{1,i}}{\alpha_{P,n-1}(\hat{k}_1)}g^{i'}(k_2)|^{n-1}_n\frac{1}{[H_{P,n-1}^{W+}]_{k_1,k_2}-z_{n+1}}
\phi_{P,n-1}+\{1\leftrightarrow 2\}\bigg),\quad
\eeqa
where   $g^{i'}|^{n-1}_n$ was defined below (\ref{upper-index-def}). 
{\bcb By Lemma~\ref{pull-through-resolvents}}, this gives the bound $\mco(|\la|^2|\ln\,\veps| \si_{n-1}^{-\de})=\mco(|\la| \si_{n-1}^{-\de})$.
Finally, the term in (\ref{L-g})  is clearly $\mco(|\la|^2 {\color{\red}|\ln\,\veps|}\veps^{-2}  \si_{n-1}^{-\de})=\mco(|\la|\si_{n-1}^{-\de})$.  \qed
\bel\label{last-lemma-second-case}  {\color{\red}(Control of (\ref{second-sum-third-term}))} Under the assumptions of Theorem~\ref{main-technical-result} 
\beqa
& &(R'V) R'  (\Delta\Gamma_{P}|_{n}^{n-1})_{i'} \phi_{P,n-1}\non\\
& &:=\frac{1}{H_{P,n-1}^{W+}-z'_{n+1}} (-)H_I^{W}|_{n}^{n-1} \frac{1}{H_{P,n-1}^{W+}-z'_{n+1}} (\Delta\Gamma_{P}|_{n}^{n-1})_{i'} \phi_{P,n-1}=
\mco(|\la|^{1/2}\si_{n-1}^{1-2\de}).\quad\quad
\eeqa
\eel
\proof We recall that $H_I^{W}|_{n}^{n-1}:=(\mathcal{L}|_{n}^{n-1}+\mathcal{I}|_{n}^{n-1})\cdot\Gamma_{P,n-1}
+\Delta(H_{I}^{W}|_{n}^{n-1})_{\mr{mix}}+(H_{I}^{W}|_{n}^{n-1})_{\mr{quad}}$ with $|\mathcal{I}|_{n}^{n-1}|\leq c_I |\lambda|^{2}\sigma_{n-1}$  and 
$\Delta\Gamma_{P}|_{n}^{n-1}=X|^{n-1}_n+\mathcal{L}|_{n}^{n-1}$ with $|X|^{n-1}_n|\leq c|\la|^{1/4}\si_{n-1}^{1-\de}$ and analyse the respective
terms.  First, we have by {\color{\red}(\ref{3.53}) and (\ref{3.54})} (in Lemma~\ref{lem:Basic-estimates}) and (\ref{second-case-second-estimate})
\beqa
& &\frac{1}{H_{P,n-1}^{W+}-z'_{n+1}} (\Delta(H_{I}^{W}|_{n}^{n-1})_{\mr{mix}}+(H_{I}^{W}|_{n}^{n-1})_{\mr{quad}} )\frac{1}{H_{P,n-1}^{W+}-z'_{n+1}} (\Delta\Gamma_{P}|_{n}^{n-1})_{i'} \phi_{P,n-1} =\mco(|\la|^{1/2}\si_{n-1}^{1-\de}). \ \ 
\eeqa
Next, we consider
\beqa
& &\frac{1}{H_{P,n-1}^{W+}-z'_{n+1}} (\mathcal{I}|_{n}^{n-1}\cdot\Gamma_{P,n-1} 
 )\frac{1}{H_{P,n-1}^{W+}-z'_{n+1}} ( X|^{n-1}_n)_{i'} \phi_{P,n-1}\non\\
& &=f(z'_{n+1})(\mathcal{I}|_{n}^{n-1})_j( X|^{n-1}_n)_{i'} \frac{1}{H_{P,n-1}^{W+}-z'_{n+1}} 
(\Gamma_{P,n-1} )_j\phi_{P,n-1}=\mco({\bc |\la|^2} \si_{n-1}^{1-2\de}),
\eeqa
where $f(z'_{n+1}):=(E_{P,n-1}+\De c_P|^{n-1}_n-z'_{n+1})^{-1}=\mco(\si_{n+1}^{-1})$ and we made
use of  Theorem~\ref{thm:induction-convergence}.  Next, we write,
making use of {\color{\red}(\ref{R-G-R-L-end-of-proof})} in Lemma~\ref{middle-subsection-auxiliary-lemma}
\beqa
& &\frac{1}{H_{P,n-1}^{W+}-z'_{n+1}} (\mathcal{I}|_{n}^{n-1}\cdot\Gamma_{P,n-1} 
 )\frac{1}{H_{P,n-1}^{W+}-z'_{n+1}} (\mathcal{L}|_{n}^{n-1})_{i'} \phi_{P,n-1}=\mco(|\la|^{5/2}\si_{n-1}^{1-\de}).
\eeqa
Now,  making use of the direct integral representation, we obtain
\beqa
& &\frac{1}{H_{P,n-1}^{W+}-z'_{n+1}} (\mathcal{L}|_{n}^{n-1}\cdot\Gamma_{P,n-1} 
 )\frac{1}{H_{P,n-1}^{W+}-z'_{n+1}} ( X|^{n-1}_n)_{i'}  \phi_{P,n-1}\non\\
& &\simeq f(z'_{n+1})( X|^{n-1}_n)_{i'}(-\la) \hint_{\!\!\!\!\bmcA|^{n-1}_n}^{\oplus} d^3k\,\fr{1}{\sqrt{2}|k|^{3/2}} \frac{k_{j}}{\alpha_{P,n-1}(\hat{k})}
\frac{1}{[H_{P,n-1}^{W+}]_k-z'_{n+1}} (\Gamma_{P,n-1})_j\phi_{P,n-1}\quad\non\\
& &=\mco(|\la|^{5/4}\veps^{-2}{\color{\red}|\ln\, \veps|^{1/2}}\si_{n-1}^{1-2\de} )=\mco(|\la|^{1/2}\si_{n-1}^{1-2\de}),
\eeqa
where we made use of {\color{\red}(\ref{id}) and (\ref{estimate-Q}) in} Lemma~\ref{pull-through-resolvents}, and Theorem~\ref{thm:induction-convergence}. 
Finally, we obtain from Lemma~\ref{R-L-R-L-whole-lemma}
\beqa
& &\frac{1}{H_{P,n-1}^{W+}-z'_{n+1}} (\mathcal{L}|_{n}^{n-1}\cdot\Gamma_{P,n-1} 
 )\frac{1}{H_{P,n-1}^{W+}-z'_{n+1}}  (\mathcal{L}|_{n}^{n-1})_{i'}  \phi_{P,n-1}=\mco(|\la| \si_{n-1}^{1-\de}). \label{R-L-R-L-whole--}
\eeqa
This concludes the proof. \qed
\bel\label{R-L-R-L-whole-lemma}  Under the assumptions of Theorem~\ref{main-technical-result}
\beqa
& &\frac{1}{H_{P,n-1}^{W+}-z_{n+1}} (\mathcal{L}|_{n}^{n-1}\cdot\Gamma_{P,n-1})
\frac{1}{H_{P,n-1}^{W+}-z'_{n+1}}  (\mathcal{L}|_{n}^{n-1})_{i'}  \phi_{P,n-1}=\mco(|\la| \si_{n-1}^{1-\de}), \label{R-L-R-L-whole-one}\\
& &\frac{1}{H_{P,n-1}^{W+}-z_{n+1}} (\mathcal{L}|_{n}^{n-1}\cdot\Gamma_{P,n-1})
\frac{1}{H_{P,n-1}^{W+}-z'_{n+1}}  \mring{\mathcal{L}}|_{n}^{n-1}  \phi_{P,n-1}=\mco(|\la| \si_{n-1}^{2-\de}) \label{R-L-R-L-whole-two}
\eeqa
for $z_{n+1}, z'_{n+1}\in \ga_{P,n+1}\cup\ti\ga_{P,n+1}$. Here we denoted
\beqa
(\mring{\mathcal{L}}|_{n}^{n-1}):={\color{\red}\sum_{i=1}^3}[(\Ga_{P,n-1})_i,  (\mathcal{L}|_{n}^{n-1})_i]=\lambda \hint_{\!\!\sigma_{n}}^{\sigma_{n-1}}d^{3}k\frac{k^2(b(k)-b^{*}(k))}{\sqrt{2}|k|^{3/2}\alpha_{P,n-1}(\hat{k})}. \label{L-dot-formula}
\eeqa
\eel
\proof We first consider~(\ref{R-L-R-L-whole-one}).  We write
\beqa
& &\frac{1}{H_{P,n-1}^{W+}-z_{n+1}} (\mathcal{L}|_{n}^{n-1}\cdot\Gamma_{P,n-1} 
 )\frac{1}{H_{P,n-1}^{W+}-z'_{n+1}}  (\mathcal{L}|_{n}^{n-1})_{i'}  \phi_{P,n-1}\label{R-L-R-L-whole}\\
& &\ph{44444444}=\frac{1}{H_{P,n-1}^{W+}-z_{n+1}} ((\mathcal{L}|_{n}^{n-1})^{(+)}\cdot\Gamma_{P,n-1} 
 )\frac{1}{H_{P,n-1}^{W+}-z'_{n+1}}  (\mathcal{L}|_{n}^{n-1})_{i'}  \phi_{P,n-1}\label{R-L+-R-L}\\
& &\ph{44444444}+\frac{1}{H_{P,n-1}^{W+}-z_{n+1}} ((\mathcal{L}|_{n}^{n-1})^{(-)}\cdot\Gamma_{P,n-1} 
 )\frac{1}{H_{P,n-1}^{W+}-z'_{n+1}}  (\mathcal{L}|_{n}^{n-1})_{i'}  \phi_{P,n-1}. \label{R-L--R-L}
\eeqa
Making use of the direct integral representation, we have
\beqa
& &(\ref{R-L+-R-L})\simeq \fr{\la^2}{\sqrt{2}}\hint^{\oplus}_{\!\!\!\!(\bmcA|^{n-1}_n)^{\times 2} }d^3k_1d^3k_2 
\bigg(\fr{1}{\sqrt{2}|k_1|^{3/2}} \frac{k_{1,j}}{\alpha_{P,n-1}(\hat{k}_1)}  \fr{1}{\sqrt{2}|k_2|^{3/2}} \frac{k_{2,i'}}{\alpha_{P,n-1}(\hat{k}_2)}\times\non\\
& &\ph{44444444}\times\frac{1}{[H_{P,n-1}^{W+}]_{k_1,k_2}-z_{n+1}}(\Ga_{P,n-1}+k_2)_j \frac{1}{[H_{P,n-1}^{W+}]_{k_2}-z'_{n+1}}\phi_{P,n-1}+\{1\leftrightarrow 2\}\bigg)\non\\
& &\ph{444444}=\fr{\la^2}{\sqrt{2}}\hint^{\oplus}_{\!\!\!\!(\bmcA|^{n-1}_n)^{\times 2}}d^3k_1d^3k_2 
\bigg(\fr{1}{\sqrt{2}|k_1|^{3/2}} \frac{k_{1,j}}{\alpha_{P,n-1}(\hat{k}_1)}  \fr{1}{\sqrt{2}|k_2|^{3/2}} \frac{k_{2,i'}}{\alpha_{P,n-1}(\hat{k}_2)}\times\non\\
& &\ph{44444444}\times\frac{1}{[H_{P,n-1}^{W+}]_{k_1,k_2}-z_{n+1}}(\Ga_{P,n-1})_j \frac{1}{[H_{P,n-1}^{W+}]_{k_2}-z'_{n+1}}\phi_{P,n-1}+\{1\leftrightarrow 2\}\bigg)+\mco(|\la|^2|\ln\, \veps|\si_{n-1}), \quad\quad 
\eeqa
where we made use of {\color{\red}(\ref{res-est})} of Lemma~\ref{pull-through-resolvents}. Now making use of the resolvent expansion~(\ref{res-exp-zero})-(\ref{res-exp})
we can rewrite a part of the integrand above as follows
\beqa
& &\frac{1}{[H_{P,n-1}^{W+}]_{k_1,k_2}-z_{n+1}}(\Ga_{P,n-1})_i \frac{1}{[H_{P,n-1}^{W+}]_{k_2}-z'_{n+1}}\phi_{P,n-1}\non\\
& &\ph{444}=f_1(k_2)\frac{1}{[H_{P,n-1}^{W+}]_{k_1,k_2}-z_{n+1}}(\Ga_{P,n-1})_i   \phi_{P,n-1}\non\\
& &\ph{444}+f_1(k_2)\frac{1}{[H_{P,n-1}^{W+}]_{k_1,k_2}-z_{n+1}}(\Ga_{P,n-1})_i \frac{1}{[H_{P,n-1}^{W+}]_{k_2}-z'_{n+1}}  [-k_2\cdot \Gamma_{P,n-1}-\frac{|k_2|^2}{2}]  \phi_{P,n-1}\non\\
& &\ph{444}=\mco(|k_2|^{-1}\si_{n-1}^{-\de}), \label{resolvent-expansion-step}
\eeqa
where we set $f_1(k_2):=(E_{P,n-1}+\De c_P|^{n-1}_n+\alpha_{P, n-1}(\hat k_2)|k_2|-z'_{n+1})^{-1}$, made use of the fact that  
$f_1(k_2)=\mco( |k_2|^{-1})$ by (\ref{bound-denominator-k}) and applied {\color{\red}(\ref{res-est}), (\ref{res-gamma-est}) and (\ref{id}) combined with (\ref{estimate-Q}) of } Lemma~\ref{pull-through-resolvents}, and Theorem~\ref{thm:induction-convergence}. Consequently,
\beqa
(\ref{R-L+-R-L})=\mco(|\la|^2|\ln\,\veps|\si_{n-1}^{1-\de})=\mco(|\la|\si_{n-1}^{1-\de}).
\eeqa
Next, we write 
\beqa
(\ref{R-L--R-L})\1&=&\1\la^2\bint_{\!\!\!\! \bmcA|^{n-1}_n} d^3k\,\fr{1}{2|k|^{3}} \frac{k_{i'} k_{j} }{\alpha_{P,n-1}(\hat{k})^2 } 
\bigg(\frac{1}{H_{P,n-1}^{W+}-z_{n+1}}(\Gamma_{P,n-1}+k)_j\frac{1}{[H_{P,n-1}^{W+}]_k-z'_{n+1}}  \phi_{P,n-1}\bigg)\non\\
\1&=&\1\la^2\bint_{\!\!\!\!\bmcA|^{n-1}_n} d^3k\,\fr{1}{2|k|^{3}} \frac{k_{i'} k_{j} }{\alpha_{P,n-1}(\hat{k})^2 } 
\bigg(\frac{1}{H_{P,n-1}^{W+}-z_{n+1}}(\Gamma_{P,n-1})_j\frac{1}{[H_{P,n-1}^{W+}]_k-z'_{n+1}}  \phi_{P,n-1}\bigg)\quad\quad\quad\label{R-L--R-L-one}\\
& &+\mco(|\la|^2{\color{\red}|\ln\,\veps|} \veps^{-2} \si_{n-1}),\non
\eeqa
where we made use of  {\color{\red}(\ref{res-est}) in Lemma~\ref{pull-through-resolvents} and (\ref{first}) in Lemma \ref{lem:Basic-estimates}}.  Now we rewrite a part of the integrand in (\ref{R-L--R-L-one}) using the resolvent expansion, analogously as in (\ref{resolvent-expansion-step}). We have
\beqa
& &\frac{1}{H_{P,n-1}^{W+}-z_{n+1}}(\Ga_{P,n-1})_i \frac{1}{[H_{P,n-1}^{W+}]_{k}-z'_{n+1}}\phi_{P,n-1}\non\\
& &\ph{4}=f_1(k)\frac{1}{H_{P,n-1}^{W+}-z_{n+1}}(\Ga_{P,n-1})_i   \phi_{P,n-1}\non\\
& &\ph{4}+f_1(k)\frac{1}{H_{P,n-1}^{W+}-z_{n+1}}(\Ga_{P,n-1})_i \frac{1}{[H_{P,n-1}^{W+}]_{k}-z'_{n+1}}  [-k\cdot \Gamma_{P,n-1}-\frac{|k|^2}{2}]  \phi_{P,n-1}
=\mco(|k|^{-1}\veps^{-2}\si_{n-1}^{-\de}),\quad  \label{resolvent-expansion-step-one}
\eeqa
where we used $|k|\leq \si_{n-1}$,  Lemmas~\ref{pull-through-resolvents}, \ref{lem:Basic-estimates} and Theorem~\ref{thm:induction-convergence}.
Substituting~(\ref{resolvent-expansion-step-one}) to (\ref{R-L--R-L-one}), we obtain
\beqa
(\ref{R-L--R-L})=\mco(|\la|^2|\ln\,\veps| \veps^{-2} \si_{n-1})+\mco( |\la|^2 |\ln\,\veps|\veps^{-2}\si_{n-1}^{1-\de})=\mco(|\la| \si_{n-1}^{1-\de}). 
\label{R-L--R-L-final}
\eeqa
This concludes the proof of (\ref{R-L-R-L-whole-one}).  As for (\ref{R-L-R-L-whole-two}), we see from (\ref{L-dot-formula})
that  $\mring{\mathcal{L}}|_{n}^{n-1}$ has one power of $k$ more than $(\mathcal{L}|_{n}^{n-1})_i$. Thus by the
same argument as above we obtain the bound by $\mco(|\la| \si_{n-1}^{2-\de})$. \qed

\subsection{Contribution to $I^q$ with $l\in \nat_0$, $l'\in \nat_0$, $l''\in\nat$}  \label{Main-result-third-subsection}
The contribution has the form  
\beqa
I^q_{\nat_0,\nat_0,\nat}\1&:=&\1 (\tiQ_{P,n-1}^{\perp})^q\sum_{l=0}^{\infty}   
\{ RV \}^{l}  R
 (\Gamma_{P,n-1}+\Delta\Gamma_{P}|_{n}^{n-1})_{i}  
\sum_{l'=0}^{\infty}\{  R'V  \}^{l'}  R' 
 (\Gamma_{P,n-1}+\Delta\Gamma_{P}|_{n}^{n-1})_{i'}\times\non\\
& &\ph{444444444444444444444444444444444}\times
\sum_{l''=0}^{\infty}\oint_{\gamma_{P,n+1}}dw_{n+1}\{ R''V \}^{l''} R''V  R''  \phi_{P,n-1}\non\\
\1&=&\1\mco(1) R
 (\Gamma_{P,n-1}+\Delta\Gamma_{P}|_{n}^{n-1})_{i}  
\sum_{l'=0}^{\infty}\{  R'V  \}^{l'}  R'  (\Gamma_{P,n-1}+\Delta\Gamma_{P}|_{n}^{n-1})_{i'}
\oint_{\gamma_{P,n+1}}dw_{n+1} R''V  R''  \phi_{P,n-1}\label{next-to-the-last-x-x-x} \\
& &\ph{44444}+\mco(\si_{n+1}^{-1})  R'  (\Gamma_{P,n-1}+\Delta\Gamma_{P}|_{n}^{n-1})_{i'}
\sum_{l''=0}^{\infty}\oint_{\gamma_{P,n+1}}dw_{n+1} (R''V)^{\ell''} R''VR''V  R''  \phi_{P,n-1},\quad\quad\quad 
\label{last-case-last-contribution}
\eeqa
where {\color{\red}  after splitting $ \sum_{l''=0}^{\infty}\{ R''V \}^{l''} =1  +\sum_{l''=0}^{\infty}\{ R''V \}^{l''} R''V  $ we have estimated 
\beqa
 (\tiQ_{P,n-1}^{\perp})^q\sum_{l=0}^{\infty}   
\{ RV \}^{l} =\mco(1)\quad 
\textrm{ and } 
\quad 
(\tiQ_{P,n-1}^{\perp})^q\sum_{l=0}^{\infty}   \{ RV \}^{l}  R
 (\Gamma_{P,n-1}+\Delta\Gamma_{P}|_{n}^{n-1})_{i}  \sum_{l'=0}^{\infty}\{  R'V  \}^{l'}=\mco(\si_{n+1}^{-1})  
\eeqa
in (\ref{next-to-the-last-x-x-x}) and (\ref{last-case-last-contribution}), respectively,  using Lemma~\ref{trivial-bounds-xx}.}  
To proceed, we set
$f(w_{n+1}):=(E_{P,n-1}+\De c_P|^{n-1}_n-w_{n+1})^{-1}=\mco(\si_{n+1}^{-1})$ so that $R''  \phi_{P,n-1}=f(w_{n+1})\phi_{P,n-1}$.
We first analyse~(\ref{next-to-the-last-x-x-x}). Since we have  
$\|R'  (\Delta\Gamma_{P}|_{n}^{n-1})_i\|_{\mcF_n}=\mco(\si_{n-1}^{-\de})$ , $\|R'(\Ga_{P,n-1})_i\|_{\mcF_n}=\mco(\si_{n+1}^{-1})$
by Lemma~\ref{trivial-bounds-xx}, and Lemma~\ref{first-case-aux-lemma} gives that 
$R''V\phi_{P,n-1}=\mco(|\la|^{1/2}\si_{n-1}^{1-\de})$, we 
obtain
\beqa
(\ref{next-to-the-last-x-x-x})
=\mco(1)\oint_{\gamma_{P,n+1}}dw_{n+1}f(w_{n+1})  R
 (\Gamma_{P,n-1})_{i}  
\sum_{l'=0}^{\infty}\{  R'V  \}^{l'}  R'  (\Gamma_{P,n-1})_{i'}
R''V \phi_{P,n-1}+\mco(\si_{n-1}^{-2\de}), \label{ingredients-stated-above}
\eeqa
where we also exploited that the length of the integration contour compensates $f(w_{n+1})=\mco(\si^{-1}_{n+1})$.
{\color{\red}Using the splitting $1=Q^{\bot}_{P,n-1}+Q_{P,n-1}$, we can write}
\beqa
(\ref{next-to-the-last-x-x-x})\1&=&\1 \mco(1)\oint_{\gamma_{P,n+1}}dw_{n+1}f(w_{n+1})  R
 (\Gamma_{P,n-1})_{i}  
\sum_{l'=0}^{\infty}\{  R'V  \}^{l'} Q^{\bot}_{P,n-1} R'  (\Gamma_{P,n-1})_{i'}
R''V \phi_{P,n-1} \quad \label{last-term-with-b-stars} \\
& &+ \mco(1)\oint_{\gamma_{P,n+1}}dw_{n+1}f(w_{n+1})  R
 (\Gamma_{P,n-1})_{i}  
\sum_{l'=0}^{\infty}\{  R'V  \}^{l'} R'VQ_{P,n-1} R'  (\Gamma_{P,n-1})_{i'}
R''V \phi_{P,n-1} \label{last-term-with-b-stars-x-one}\quad\quad\quad \\
& &+ \mco(1)\oint_{\gamma_{P,n+1}}dw_{n+1}f(w_{n+1})  R
 (\Gamma_{P,n-1})_{i}  Q_{P,n-1} R'  (\Gamma_{P,n-1})_{i'}
R''V \phi_{P,n-1}+\mco(\si_{n-1}^{-2\de}).   \label{last-term-with-b-stars-x-two}
\eeqa
We analyse (\ref{last-term-with-b-stars}). We have by Lemmas~\ref{R-G-R-V-lemma}, \ref{interaction-x}, \ref{R-Ga-g-phi}
\beqa
(\ref{last-term-with-b-stars})\1&=&\1 \mco(1)\oint_{\gamma_{P,n}}dw_{n+1}f(w_{n+1})  R
 (\Gamma_{P,n-1})_{i}  
\sum_{l'=0}^{\infty}\{  R'V  \}^{l'} R'Vb^*(g_{2,i}|^{n-1}_n)\phi_{P,n-1} \label{last-term-with-b-stars-one}\\
& &+ \mco(1)\oint_{\gamma_{P,n}}dw_{n+1}f(w_{n+1})  R
 (\Gamma_{P,n-1})_{i}  
b^*(g_{2,i}|^{n-1}_n)\phi_{P,n-1}
+\mco(\si_{n-1}^{-2\de}) \label{last-term-with-b-stars-two}\\
\1&=&\1\mco(\si_{n-1}^{-2\de}),\non
\eeqa
{\color{black}where $g_{2,i}|^{n-1}_n$ is as specified in Lemma~\ref{auxiliary-b-star-lemma} and analogous to the one introduced in Lemma~\ref{b-star-lemma} 
(the only difference being the inessential substitution $\ti\ga_{P,n+1}\ni z'_{n+1}\mapsto w_{n+1}\in \ga_{P,n+1}$).}

Now we consider (\ref{last-term-with-b-stars-x-one}) and (\ref{last-term-with-b-stars-x-two}). We have by 
the ingredients stated above (\ref{ingredients-stated-above}), namely  $R''V\phi_{P,n-1}=\mco(|\la|^{1/2}\si_{n-1}^{1-\de})$
and $\|R'(\Ga_{P,n-1})_i\|_{\mcF_n}=\mco(\si_{n+1}^{-1})$, together with 
$\|R'(\Ga_{P,n-1})_i\phi_{P,n-1}\|=\mco(\si_{n-1}^{-\de})$ of Lemma~\ref{trivial-bounds-xx}
\beqa
(\ref{last-term-with-b-stars-x-one})=\mco(\si_{n-1}^{-2\de}), \quad \textrm{and}\quad (\ref{last-term-with-b-stars-x-two})=\mco(\si_{n-1}^{-2\de}).
\eeqa
Thus altogether
\beqa
(\ref{next-to-the-last-x-x-x})=\mco(\si_{n-1}^{-2\de}).
\eeqa

Now we analyse (\ref{last-case-last-contribution}). {\color{\red}Using the splitting $1=Q^{\bot}_{P,n-1}+Q_{P,n-1}$ and $ \sum_{l''=0}^{\infty}\{ R''V \}^{l''} =1  +\sum_{l''=0}^{\infty}\{ R''V \}^{l''} R''V $, as well as the estimates
$\|R'(\Ga_{P,n-1})_i\|_{\mcF_n}=\mco(\si_{n+1}^{-1})$,  $\|R'  (\Delta\Gamma_{P}|_{n}^{n-1})_i\|_{\mcF_n}=\mco(\si_{n-1}^{-\de})$ of Lemma~\ref{trivial-bounds-xx}, the expression reads }
\beqa
(\ref{last-case-last-contribution})
\1&=&\1\mco(\si_{n+1}^{-1})R'  (\Gamma_{P,n-1}+\Delta\Gamma_{P}|_{n}^{n-1})_{i'}\sum_{l''=0}^{\infty}\oint_{\gamma_{P,n+1}}dw_{n+1}f(w_{n+1}) (R''V)^{\ell''} Q_{P,n-1}^{\bot}R''VR''V    \phi_{P,n-1} \label{last-case-last-contribution-one}\quad\quad\quad\\
& &+\mco(\si_{n+1}^{-1})R'  (\Gamma_{P,n-1}+\Delta\Gamma_{P}|_{n}^{n-1})_{i'}\oint_{\gamma_{P,n+1}}dw_{n+1}f(w_{n+1})  
Q_{P,n-1} R''VR''V   \phi_{P,n-1}\label{last-case-xx-x} \\
& &+\mco(\si_{n+1}^{-2})\sum_{l''=0}^{\infty}\oint_{\gamma_{P,n+1}}dw_{n+1} f(w_{n+1})(R''V)^{\ell''} R''VQ_{P,n-1}R''VR''V    \phi_{P,n-1}.
\label{last-case-xx-xx}
\eeqa
Let us consider (\ref{last-case-xx-x}). By Lemmas~\ref{trivial-bounds-xx} and \ref{first-case-aux-lemma}, we have that 
\beqa
R'  (\Gamma_{P,n-1}+\Delta\Gamma_{P}|_{n}^{n-1})_{i'}Q_{P,n-1}=\mco(\si_{n-1}^{-\de}), \quad Q_{P,n-1} R''VR''V   \phi_{P,n-1}=
\mco(|\la|\si_{n-1}^{1-\de}), \label{two-form-horisontal}
\eeqa
hence
\beqa
(\ref{last-case-xx-x})=\mco(\si_{n-1}^{-2\de}).
\eeqa
As for (\ref{last-case-xx-xx}), recalling from 
Lemma~\ref{first-case-aux-lemma} that $R''VQ_{P,n-1}=\mco(|\la|^{1/2}\si_{n-1}^{1-\de})$ and
making use of the second formula in (\ref{two-form-horisontal}) we get
\beqa
(\ref{last-case-xx-xx})=\mco(\si_{n-1}^{-2\de}).
\eeqa
Now making use of Lemmas~\ref{B-star-lemma} and \ref{trivial-bounds-xx} {\color{\red}combined with the splitting $ \sum_{l''=0}^{\infty}\{ R''V \}^{l''} =1  +\sum_{l''=0}^{\infty}\{ R''V \}^{l''} R''V $}, we have 
\beqa
(\ref{last-case-last-contribution-one})\1&=&\1\bigg\{\mco(\si_{n+1}^{-1})R'  (\Gamma_{P,n-1}+\Delta\Gamma_{P}|_{n}^{n-1})_{i'}\sum_{l''=0}^{\infty}\oint_{\gamma_{P,n+1}}dw_{n+1}f(w_{n+1}) (R''V)^{\ell''}\times\non\\
& &\ph{44444444}\times\big( B^*(G|^{n-1}_n)\phi_{P,n-1}+b^*\big(\mathcal{I}|_{n}^{n-1}\cdot g|^{n-1}_n \big)\phi_{P,n-1}\big)\bigg\}+\mco(\si_{n-1}^{-2\de})\non\\
\1&=&\1\mco(\si_{n+1}^{-1})R'  (\Gamma_{P,n-1}+\Delta\Gamma_{P}|_{n}^{n-1})_{i'} B^*(G|^{n-1}_n)\phi_{P,n-1}
\label{last-case-last-contribution-B-star-one}\\
& &+ \mco(\si_{n+1}^{-2})\sum_{l''=0}^{\infty}\oint_{\gamma_{P,n+1}}dw_{n+1}f(w_{n+1}) (R''V)^{\ell''}R''VB^*(G|^{n-1}_n)\phi_{P,n-1}
\label{last-case-last-contribution-B-star-two}\\
& &+\mco(\si_{n+1}^{-1})R'  (\Gamma_{P,n-1}+\Delta\Gamma_{P}|_{n}^{n-1})_{i'} b^*\big( \mathcal{I}|_{n}^{n-1}\cdot g|^{n-1}_n \big)\phi_{P,n-1}
\label{last-case-last-contribution-b-star-one}\\
& &+ \mco(\si_{n+1}^{-2})\sum_{l''=0}^{\infty}\oint_{\gamma_{P,n+1}}dw_{n+1}f(w_{n+1}) (R''V)^{\ell''}R''Vb^*\big( \mathcal{I}|_{n}^{n-1}\cdot g|^{n-1}_n \big)\phi_{P,n-1} \ \ \
\label{last-case-last-contribution-b-star-two}\\
& &+\mco(\si_{n-1}^{-2\de}). \non
\eeqa
We have by Lemma~\ref{interaction-x}, formula~(\ref{simple-sum-over-l}) and  $|\mathcal{I}|_{n}^{n-1}|\leq c_I |\lambda|^{2}\sigma_{n-1}$ that $(\ref{last-case-last-contribution-b-star-two})=\mco(\si_{n-1}^{-2\de})$. Similarly, by Lemmas~\ref{R-Ga-g-phi}, \ref{R-De-Ga-g-phi} and $|\mathcal{I}|_{n}^{n-1}|\leq c_I |\lambda|^{2}\sigma_{n-1}$ we have $(\ref{last-case-last-contribution-b-star-one})=\mco(\si_{n-1}^{-2\de})$. Furthermore, by Lemma~\ref{B-star-lemma-three} we have $(\ref{last-case-last-contribution-B-star-two})=\mco(\si_{n-1}^{-2\de})$ and 
by Lemmas~\ref{B-star-lemma-one}, \ref{B-star-lemma-two}, we have $(\ref{last-case-last-contribution-B-star-one})=\mco(\si_{n-1}^{-2\de})$. Thus
altogether
\beqa
(\ref{last-case-last-contribution-one})=\mco(\si_{n-1}^{-2\de}) \quad  \textrm{and} \quad (\ref{last-case-last-contribution})=\mco(\si_{n-1}^{-2\de}).
\eeqa
Thus the proof of Theorem~\ref{main-technical-result} is complete, given the auxiliary lemmas listed below.
\bel\label{R-G-R-V-lemma} {\color{\red}(Control of (\ref{last-term-with-b-stars}))} Under the assumptions of Theorem~\ref{main-technical-result} and the induction hypothesis
\beqa
& &Q^{\bot}_{P,n-1} R'  (\Gamma_{P,n-1})_{i'}R''V \phi_{P,n-1}
:=Q^{\bot}_{P,n-1}\frac{1}{H_{P,n-1}^{W+}-z'_{n+1}}(\Gamma_{P,n-1})_{i'}\frac{1}{H_{P,n-1}^{W+}-w_{n+1}} (-)H_{I}^{W}|_{n}^{n-1}\phi_{P,n-1}\non\\
& &\ph{44444444444444444444444}=b^*({\color{\red}g_{2,i}}|^{n-1}_n)\phi_{P,n-1}+\mco(|\la|^{1/4}\si_{n-1}^{1-2\de}),
\eeqa
where ${\color{\red}g_{2,i}}|^{n-1}_n$ is as specified in {\color{\red} Lemma~\ref{auxiliary-b-star-lemma} and analogous to the one introduced in Lemma~\ref{b-star-lemma} 
(the only difference being the inessential substitution $\ti\ga_{P,n+1}\ni z'_{n+1}\mapsto w_{n+1}\in \ga_{P,n+1}$).}
\eel
\proof We recall that $H_I^{W}|_{n}^{n-1}:=(\mathcal{L}|_{n}^{n-1}+\mathcal{I}|_{n}^{n-1})\cdot\Gamma_{P,n-1}
+\Delta(H_{I}^{W}|_{n}^{n-1})_{\mr{mix}}+(H_{I}^{W}|_{n}^{n-1})_{\mr{quad}}$ with $|\mathcal{I}|_{n}^{n-1}|\leq c_I |\lambda|^{2}\sigma_{n-1}$ and study the respective terms. We have
\beqa
Q^{\bot}_{P,n-1}\frac{1}{H_{P,n-1}^{W+}-z'_{n+1}}(\Gamma_{P,n-1})_{i'}\frac{1}{H_{P,n-1}^{W+}-w_{n+1}}\mathcal{I}|_{n}^{n-1}\cdot\Gamma_{P,n-1} \phi_{P,n-1}=\mco(|\la|^2\si_{n-1}^{1-2\de})
\eeqa
by the induction hypothesis {\color{\red}(see Remark \ref{max-mod-2})}. Furthermore, by Lemma~\ref{auxiliary-b-star-lemma},
\beqa
Q^{\bot}_{P,n-1}\frac{1}{H_{P,n-1}^{W+}-z'_{n+1}}(\Gamma_{P,n-1})_{i'}\frac{1}{H_{P,n-1}^{W+}-w_{n+1}} \mathcal{L}|_{n}^{n-1}\cdot\Gamma_{P,n-1}  \phi_{P,n-1}
=b^*({\color{\red}g_{2,i}}|^{n-1}_n)\phi_{P,n-1}+\mco(|\la|^{1/4}\si_{n-1}^{1-2\de}), 
\eeqa
where ${\color{\red}g_{2,i}}|^{n-1}_n$ {\color{\red}is provided} in this lemma.  Next, {\color{\red} in order to analyse the terms proportional to $\Delta(H_{I}^{W}|_{n}^{n-1})_{\mr{mix}}$ and $(H_{I}^{W}|_{n}^{n-1})_{\mr{quad}}$},  we recall that
\beqa
\Delta(H_{I}^{W}|_{n}^{n-1})_{\mr{mix}}\1 &:=&\1\h{\color{\red}\sum_{i=1}^{3}}[(\Ga_{P,n-1})_i,  (\mathcal{L}|_{n}^{n-1})_i]=
\fr{\lambda}{2}\hint_{\!\!\!\!\bmcA|^{n-1}_n } d^{3}k\frac{k^2(b(k)-b^{*}(k))}{\sqrt{2}|k|^{3/2}\alpha_{P,n-1}(\hat{k})},\\
(H_{I}^{W}|_{n}^{n-1})_{\mr{quad}} \1&:= &\1 \h(\mathcal{L}|_{n}^{n-1}+ \mathcal{I}|_{n}^{n-1} )^2,
\eeqa
and note that $\Delta(H_{I}^{W}|_{n}^{n-1})_{\mr{mix}}$ has the same structure as $\mathcal{L}|_{n}^{n-1}$ {\color{black} (see (\ref{Ln-definitions}))} but is by one power of $k$
more regular. To stress this fact we write $\mring{\mathcal{L}}|_{n}^{n-1}:=2\Delta(H_{I}^{W}|_{n}^{n-1})_{\mr{mix}} $ consistently with the
notation in Lemmas~\ref{RL-lemma}, \ref{R-L-R-L-whole-lemma}. We have by Lemma~\ref{middle-subsection-auxiliary-lemma}
\beqa
Q^{\bot}_{P,n-1}\frac{1}{H_{P,n-1}^{W+}-z'_{n+1}}(\Gamma_{P,n-1})_{i'}\frac{1}{H_{P,n-1}^{W+}-w_{n+1}} \mring{\mathcal{L}}|_{n}^{n-1}\phi_{P,n-1}
=\mco(|\la|^{1/2}\si_{n-1}^{1-\de}).
\eeqa
Now we study the three terms coming from $(H_{I}^{W}|_{n}^{n-1})_{\mr{quad}}$. We have by Lemma~\ref{auxiliary-double-l-lemma}
\beqa
Q^{\bot}_{P,n-1}\frac{1}{H_{P,n-1}^{W+}-z'_{n+1}}(\Gamma_{P,n-1})_{i'}\frac{1}{H_{P,n-1}^{W+}-w_{n+1}} \mathcal{L}|_{n}^{n-1}\cdot \mathcal{L}|_{n}^{n-1} \phi_{P,n-1}=\mco(|\la|\si_{n-1}^{1-\de}).
\eeqa
Next, making use of $|\mathcal{I}|_{n}^{n-1}|\leq c_I |\lambda|^{2}\sigma_{n-1}$ and 
Lemma~\ref{middle-subsection-auxiliary-lemma}, we get
\beqa
Q^{\bot}_{P,n-1}\frac{1}{H_{P,n-1}^{W+}-z'_{n+1}}(\Gamma_{P,n-1})_{i'}\frac{1}{H_{P,n-1}^{W+}-w_{n+1}} \mathcal{I}|_{n}^{n-1}\cdot \mathcal{L}|_{n}^{n-1} \phi_{P,n-1}=\mco(|\la|^{5/2}\si_{n-1}^{1-\de}).
\eeqa
Finally, by   $|\mathcal{I}|_{n}^{n-1}|\leq c_I |\lambda|^{2}\sigma_{n-1}$ and Theorem~\ref{thm:induction-convergence}
\beqa
Q^{\bot}_{P,n-1}\frac{1}{H_{P,n-1}^{W+}-z'_{n+1}}(\Gamma_{P,n-1})_{i'}\frac{1}{H_{P,n-1}^{W+}-w_{n+1}} (\mathcal{I}|_{n}^{n-1})^2 \phi_{P,n-1}=
\mco(|\la|^3\si_{n-1}^{1-\de}).
\eeqa
This completes the proof. \qed
\bel\label{B-star-lemma} {\color{\red}(Control of (\ref{last-case-last-contribution-one}) associated with (\ref{last-case-last-contribution}))} Under the assumptions of Theorem~\ref{main-technical-result} and the induction hypothesis
\beqa
& &Q_{P,n-1}^{\bot}R''VR''V\phi_{P,n-1}\non\\
& &:=Q_{P,n-1}^{\bot}\frac{1}{H_{P,n-1}^{W+}-w_{n+1}} (-)H_{I}^{W}|_{n}^{n-1} \frac{1}{H_{P,n-1}^{W+}-w_{n+1}}
(-)H_{I}^{W}|_{n}^{n-1}\phi_{P,n-1} \\ 
& &=B^*(G|^{n-1}_n)\phi_{P,n-1}+b^*\big(\mathcal{I}|_{n}^{n-1}\cdot g|^{n-1}_n \big)\phi_{P,n-1}+\mco(|\la|^{1/2}\si_{n-1}^{2-2\de}), \label{unexp}
\eeqa
where $g_i|^{n-1}_n$ is a 3-tuple of functions {\color{\red}with support in 
$\bar\mcA|^{n-1}_n$, $G|^{n-1}_n$ is a function with support in 
$\bar\mcA|^{n-1}_n \times  \bar\mcA|^{n-1}_n$ and  $B^*(G|^{n-1}_n):=\int d^3k_1d^3k_2\, G(k_1,k_2)|^{n-1}_n b^*(k_1)b^*(k_2)$.}
They satisfy the pointwise bounds:
\beqa
& &  |g(k)|^{n-1}_n|\leq c |\la| |k|^{-3/2}\si_{n-1}^{-\de},\\
& &|G(k_1,k_2)|^{n-1}_n|\leq c|\la|^2|k_1|^{-3/2} |k_2|^{-3/2} \si_{n-1}^{1-\de}.  \label{G-estimate}
\eeqa
\eel

\proof  We recall that $H_I^{W}|_{n}^{n-1}:=(\mathcal{L}|_{n}^{n-1}+\mathcal{I}|_{n}^{n-1})\cdot\Gamma_{P,n-1}
+\Delta(H_{I}^{W}|_{n}^{n-1})_{\mr{mix}}+(H_{I}^{W}|_{n}^{n-1})_{\mr{quad}}$ with 
$|\mathcal{I}|_{n}^{n-1}|\leq c_I |\lambda|^{2}\sigma_{n-1}$ and study the respective terms. \\

\nin\textbf{Step 1:} First, we consider
\beqa
& &Q_{P,n-1}^{\bot}\frac{1}{H_{P,n-1}^{W+}-w_{n+1}} (\mathcal{L}|_{n}^{n-1}+\mathcal{I}|_{n}^{n-1})\cdot\Gamma_{P,n-1} \frac{1}{H_{P,n-1}^{W+}-w_{n+1}}
(\mathcal{L}|_{n}^{n-1}+\mathcal{I}|_{n}^{n-1})\cdot\Gamma_{P,n-1}\phi_{P,n-1}\quad\quad\quad \label{(L+I)(L+I)}\\
& &\quad\quad\quad\quad  =Q_{P,n-1}^{\bot}\frac{1}{H_{P,n-1}^{W+}-w_{n+1}} (\mathcal{L}|_{n}^{n-1})^{(+)}\cdot\Gamma_{P,n-1} \frac{1}{H_{P,n-1}^{W+}-w_{n+1}}
\mathcal{L}|_{n}^{n-1}\cdot\Gamma_{P,n-1}\phi_{P,n-1}\label{L-G-L-G}\\
& &\quad\quad\quad\quad\ph{4} +Q_{P,n-1}^{\bot}\frac{1}{H_{P,n-1}^{W+}-w_{n+1}} (\mathcal{L}|_{n}^{n-1})^{(-)}\cdot\Gamma_{P,n-1} \frac{1}{H_{P,n-1}^{W+}-w_{n+1}}
\mathcal{L}|_{n}^{n-1}\cdot\Gamma_{P,n-1}\phi_{P,n-1}\label{Lm-G-L-G}\\
& &\quad\quad\quad\quad\ph{4}+Q_{P,n-1}^{\bot}\frac{1}{H_{P,n-1}^{W+}-w_{n+1}} \mathcal{L}|_{n}^{n-1}\cdot\Gamma_{P,n-1} \frac{1}{H_{P,n-1}^{W+}-w_{n+1}}
\mathcal{I}|_{n}^{n-1}\cdot\Gamma_{P,n-1}\phi_{P,n-1} \label{L-G-I-G}\\
& &\quad\quad\quad\quad\ph{4} +Q_{P,n-1}^{\bot}\frac{1}{H_{P,n-1}^{W+}-w_{n+1}} \mathcal{I}|_{n}^{n-1}\cdot\Gamma_{P,n-1} \frac{1}{H_{P,n-1}^{W+}-w_{n+1}}
\mathcal{L}|_{n}^{n-1}\cdot\Gamma_{P,n-1}\phi_{P,n-1} \label{I-G-L-G} \\
& &\quad\quad\quad\quad\ph{4} +Q_{P,n-1}^{\bot}\frac{1}{H_{P,n-1}^{W+}-w_{n+1}} \mathcal{I}|_{n}^{n-1}\cdot\Gamma_{P,n-1} \frac{1}{H_{P,n-1}^{W+}-w_{n+1}}
\mathcal{I}|_{n}^{n-1}\cdot\Gamma_{P,n-1}\phi_{P,n-1}. \label{I-G-I-G}
\eeqa
 We note that in (\ref{L-G-L-G}), (\ref{L-G-I-G}), (\ref{I-G-L-G}) we can skip the projection $Q_{P,n-1}^{\bot}$ as the respective vectors are manifestly in its range. 
Now we estimate the terms above: 
\beqa
(\ref{L-G-L-G})
\1&\simeq&\1 \fr{\la^2}{\sqrt{2}}\hint^{\oplus}_{\!\!\!\!(\bmcA|^{n-1}_n)^{\times 2}}d^3k_1d^3k_2 
\bigg(\fr{1}{\sqrt{2}|k_1|^{3/2}} \frac{k_{1,j}}{\alpha_{P,n-1}(\hat{k}_1)}  \fr{1}{\sqrt{2}|k_2|^{3/2}} \frac{k_{2,j'}}{\alpha_{P,n-1}(\hat{k}_2)}\times\non\\
& &\times\frac{1}{[H_{P,n-1}^{W+}]_{k_1,k_2}-w_{n+1}}(\Ga_{P,n-1}+k_2)_j \frac{1}{[H_{P,n-1}^{W+}]_{k_2}-w_{n+1}}(\Ga_{P,n-1})_{j'}\phi_{P,n-1}+\{1\leftrightarrow 2\}\bigg)\quad\quad\quad\label{L-G-L-G-1}\\
\1&=&\1 \fr{\la^2}{\sqrt{2}}\hint^{\oplus}_{\!\!\!\!(\bmcA|^{n-1}_n)^{\times 2} }d^3k_1d^3k_2 
\bigg(\fr{1}{\sqrt{2}|k_1|^{3/2}} \frac{k_{1,j}}{\alpha_{P,n-1}(\hat{k}_1)}  \fr{1}{\sqrt{2}|k_2|^{3/2}} \frac{k_{2,j'}}{\alpha_{P,n-1}(\hat{k}_2)}\times\non\\
& &\ph{4}\times\frac{1}{[H_{P,n-1}^{W+}]_{k_1,k_2}-w_{n+1}} Q_{P,n-1}^{\bot}(\Ga_{P,n-1})_j \frac{1}{[H_{P,n-1}^{W+}]_{k_2}-w_{n+1}}(\Ga_{P,n-1})_{j'}\phi_{P,n-1}+\{1\leftrightarrow 2\}\bigg)\quad\quad\quad\label{L-G-L-G-2}\\
& &+ \fr{\la^2}{\sqrt{2}}\hint^{\oplus}_{\!\!\!\!(\bmcA|^{n-1}_n)^{\times 2} }d^3k_1d^3k_2 
\bigg(\fr{1}{\sqrt{2}|k_1|^{3/2}} \frac{k_{1,j}}{\alpha_{P,n-1}(\hat{k}_1)}  \fr{1}{\sqrt{2}|k_2|^{3/2}} \frac{k_{2,j'}}{\alpha_{P,n-1}(\hat{k}_2)}\times\non\\
& &\ph{4}\times\frac{1}{[H_{P,n-1}^{W+}]_{k_1,k_2}-w_{n+1}} Q_{P,n-1}(\Ga_{P,n-1})_j \frac{1}{[H_{P,n-1}^{W+}]_{k_2}-w_{n+1}}(\Ga_{P,n-1})_{j'}\phi_{P,n-1}+\{1\leftrightarrow 2\}\bigg)\quad\quad\quad\label{term-which-complicates} \\
& &+\mco(|\la| \si_{n-1}^{2-\de}), \label{last-line-error-term}
\eeqa
where the error term in (\ref{last-line-error-term}) comes from the part of (\ref{L-G-L-G-1}) proportional to $k_{2,j}$ and is obtained using
Lemma~\ref{pull-through-resolvents} and Theorem~\ref{thm:induction-convergence}. As for the term in (\ref{L-G-L-G-2}),  using {\color{\red}(\ref{estimate-Q}) in} Lemma~\ref{pull-through-resolvents}   we can estimate a part of the integrand as follows
\beqa
& &\|\frac{1}{[H_{P,n-1}^{W+}]_{k_1,k_2}-w_{n+1}} Q_{P,n-1}^{\bot}(\Ga_{P,n-1})_j \frac{1}{[H_{P,n-1}^{W+}]_{k_2}-w_{n+1}}(\Ga_{P,n-1})_{j'}\phi_{P,n-1}\|\non\\
& &\leq c\| Q_{P,n-1}^{\bot} \frac{1}{H_{P,n-1}^{W+}-w_{n+1}} (\Ga_{P,n-1})_j \frac{1}{[H_{P,n-1}^{W+}]_{k_2}-w_{n+1}}(\Ga_{P,n-1})_{j'}\phi_{P,n-1}\|
=\mco(\veps^{-4}\si_{n-1}^{-2\de}),\quad\quad\quad
\eeqa
where in the last step we used Lemma~\ref{last-case-additional-lemma}.
Consequently,
\beqa
(\ref{L-G-L-G-2})=\mco(|\la|^2{\color{\red}|\ln\,\veps|}\veps^{-4}\si_{n-1}^{2-2\de})=\mco(|\la| \si_{n-1}^{2-2\de}).
\eeqa
Let us now consider the term in (\ref{term-which-complicates}). {\color{\red}Using the definition in (\ref{first-k-k-k-Hamiltonian}), in Lemma \ref{G-0-lemma} we show that}
\beqa
\frac{1}{[H_{P,n-1}^{W+}]_{k_1,k_2}-w_{n+1}}\phi_{P,n-1}=G_0(k_1,k_2)\phi_{P,n-1}+\mco(\si_{n-1}^{-\de}), \label{G-zero-intr}
\eeqa
where $G_0(k_1,k_2)=\mco((|k_1|+|k_2|)^{-1})$ is a  numerical function, symmetric under the exchange of variables. We also note that by 
Lemma~\ref{pull-through-resolvents}, {\color{\red}the identity in (\ref{id})}, Theorem~\ref{thm:induction-convergence},  and $\|(\Ga_{P,n-1})_j\phi_{P,n-1}\|\leq c$ (cf. Appendix~\ref{standard-resolvent-bounds-section})
we have
\beqa
G_{j,j'}(k):=\fr{1}{\|\phi_{P,n-1}\|^2}\lan \phi_{P,n-1},(\Ga_{P,n-1})_j \frac{1}{[H_{P,n-1}^{W+}]_{k}-w_{n+1}}(\Ga_{P,n-1})_{j'}\phi_{P,n-1}\ran=\mco(\si_{n-1}^{-\de}). \label{G-j-j'}
\eeqa
Making use of (\ref{G-zero-intr}) and (\ref{G-j-j'}), we have
\beqa
(\ref{term-which-complicates})=B^*(G|^{n-1}_n)\phi_{P,n-1}+\mco(|\la|^2{\color{\red}|\ln\,\veps|}\si_{n-1}^{2-2\de})
=B^*(G|^{n-1}_n)\phi_{P,n-1}+\mco(|\la|\si_{n-1}^{2-2\de}),
\eeqa
where $G|^{n-1}_n$ is a symmetric function {\color{\red}with support in 
$\bar\mcA|^{n-1}_n\times \bar\mcA|^{n-1}_n$}
given by
\beqa
{\color{\red}G(k_1,k_2)|^{n-1}_n}\1&:=&\1\la^2 \chi_{\ka}(k_1)\chi_{\ka}(k_2)|k_1|^{\alf} |k_2|^{\alf} \frac{k_{1,j}}{|k_1|^{3/2}\alpha_{P,n-1}(\hat{k}_1)}\frac{k_{2,j'}}{|k_2|^{3/2}\alpha_{P,n-1}(\hat{k}_2)} \times \non\\
& & \times G_0(k_1,k_2)\fr{1}{2}(G_{j,j'}(k_1)+G_{j,j'}(k_2))
\eeqa
and $B^*(G|^{n-1}_n)\phi_{P,n-1}:=\int d^3k_1 d^3k_2 \,G(k_1,k_2)|^{n-1}_nb^*(k_1)b^*(k_2)\phi_{P,n-1}$. 
Clearly, we have the pointwise bound
\beqa \label{bound-G}
|G(k_1,k_2)|^{n-1}_n|\leq |\la|^2\fr{c}{|k_1|^{3/2} |k_2|^{3/2}} \si_{n-1}^{1-\de},
\eeqa
where we used $|k_1||k_2|\leq |k_1|^{1/2}|k_2|^{1/2}(|k_1|+|k_2|)\leq \si_{n-1}(|k_1|+|k_2|)$.

Now we consider (\ref{Lm-G-L-G}). It has the form 
\beqa
(\ref{Lm-G-L-G})\1&=&\1Q_{P,n-1}^{\bot}\frac{1}{H_{P,n-1}^{W+}-w_{n+1}} (\mathcal{L}|_{n}^{n-1})^{(-)}\cdot\Gamma_{P,n-1} \frac{1}{H_{P,n-1}^{W+}-w_{n+1}}
\mathcal{L}|_{n}^{n-1}\cdot\Gamma_{P,n-1}\phi_{P,n-1}\non\\
\1&=&\1\la^2 \bint_{\!\!\!\! \bmcA|^{n-1}_n} d^3k  \fr{1}{2|k|^{3}} \frac{k_{i}k_{i'}  }{\alpha_{P,n-1}(\hat{k})^2}
  Q_{P,n-1}^{\bot}\frac{1}{H_{P,n-1}^{W+}-w_{n+1}}
(\Gamma_{P,n-1}+k)_i  \frac{1}{[H_{P,n-1}^{W+}]_k-w_{n+1}} (\Gamma_{P,n-1})_{i'}\phi_{P,n-1}.\non\\
\eeqa
The part of (\ref{Lm-G-L-G}) proportional to $k$ is $\mco(\la^2\veps^{-2}|\ln\,\veps|\si_{n-1}^{2-\de})=\mco(|\la| \si_{n-1}^{2-\de}  )$ due to Lemma~\ref{pull-through-resolvents}, {\color{\red}identity (\ref{id})}, Theorem~\ref{thm:induction-convergence} and {\color{\red}(\ref{first}) in Lemma~\ref{lem:Basic-estimates}}. The remaining part of
(\ref{Lm-G-L-G}) is $\mco( |\la|^2|\ln\,\veps| \veps^{-4} \si_{n-1}^{2-2\de})=\mco( |\la| \si_{n-1}^{2-2\de})$ by Lemma~\ref{last-case-additional-lemma}.
Thus altogether
\beqa
(\ref{Lm-G-L-G})=\mco( |\la| \si_{n-1}^{2-2\de}).
\eeqa

Next we estimate (\ref{L-G-I-G}). Proceeding as in the analysis of  (\ref{R-LG-R-G-phi}) and using  $|\mathcal{I}|_{n}^{n-1}|\leq c_I |\lambda|^{2}\sigma_{n-1}$
we have
\beqa
(\ref{L-G-I-G})=b^*\big({\color{\red}\sum_{i'=1}^{3}}(\mathcal{I}|_{n}^{n-1})_{i'}g_{1,i'}|^{n-1}_n \big)\phi_{P,n-1}+\mco(|\la|^{5/2}\si_{n-1}^{2-2\de}), \label{first-copy-4.90}
\eeqa
where $g_{1,i'}|^{n-1}_n$ is a 3-tuple of functions {\color{\red}with support in 
$ \bar\mcA|^{n-1}_n$ }  analogous to the one introduced in Lemma~\ref{b-star-lemma} 
(the only difference being the inessential substitution $\ti\ga_{P,n+1}\ni z'_{n+1}\mapsto w_{n+1}\in \ga_{P,n+1}$) which
satisfy the pointwise bound $|g_{1,i'}(k)|^{n-1}_n|\leq |\la|\fr{c}{|k|^{3/2}\si_{n-1}^{\de}}$.

Now we consider (\ref{I-G-L-G}). We have by Lemma~\ref{auxiliary-b-star-lemma} and $|\mathcal{I}|_{n}^{n-1}|\leq c_I |\lambda|^{2}\sigma_{n-1}$
\beqa
(\ref{I-G-L-G})=b^*\big({\color{\red}\sum_{i'=1}^{3}}(\mathcal{I}|_{n}^{n-1})_{i'} g_{2,i'}|^{n-1}_n \big)\phi_{P,n-1}+\mco(|\la|^{2}\si_{n-1}^{2-2\de}),
\eeqa
where $g_{2,i'}|^{n-1}_n$ is a 3-tuple of functions {\color{\red}with support in 
$\bar\mcA|^{n-1}_n$  } which satisfy the pointwise bound 
$|g_{2,i'}(k)|^{n-1}_n|\leq |\la|\fr{c}{|k|^{3/2}\si_{n-1}^{\de}}$.

Since it is readily seen that  $(\ref{I-G-I-G})=\mco(|\la|^4\si_{n-1}^{2-2\de})$ using  $|\mathcal{I}|_{n}^{n-1}|\leq c_I |\lambda|^{2}\sigma_{n-1}$ and the
inductive hypothesis, our analysis of (\ref{(L+I)(L+I)}) is complete. Summing up, we have shown that
\beqa
(\ref{(L+I)(L+I)})={\color{\red}\Big\{B^*(G|^{n-1}_n)+b^*\big(\sum_{i'=1}^{3}(\mathcal{I}|_{n}^{n-1})_{i'} g_{2,i'}|^{n-1}_n) +b^*\big(\sum_{i=1}^{3}(\mathcal{I}|_{n}^{n-1})_{i} g_{1,i}|^{n-1}_n \big)\Big\}\phi_{P,n-1}}+\mco(|\la|\si_{n-1}^{2-2\de}),\quad\quad
\eeqa
where  $|G(k_1,k_2)|^{n-1}_n|\leq c|\la|^2|k_1|^{-3/2} |k_2|^{-3/2} \si_{n-1}^{1-\de}$ and ${\color{\red} \{\sum_{i=1}^{3}|g_{\#,i}(k)|^{n-1}_n|^2\}^{\frac{1}{2}}}\leq c |\la| |k|^{-3/2}\si_{n-1}^{-\de}$
\vspace{0.2cm}
{\color{\red}with $\#=1,2$.}

\nin\textbf{Step 2:} Now we consider the following contribution to the expression from the statement of the lemma
\beqa
& &Q_{P,n-1}^{\bot}\frac{1}{H_{P,n-1}^{W+}-w_{n+1}} (\Delta(H_{I}^{W}|_{n}^{n-1})_{\mr{mix}}+(H_{I}^{W}|_{n}^{n-1})_{\mr{quad}})
\frac{1}{H_{P,n-1}^{W+}-w_{n+1}}
(\mathcal{L}|_{n}^{n-1}+\mathcal{I}|_{n}^{n-1})\cdot\Gamma_{P,n-1}\phi_{P,n-1}\quad\quad\quad\quad \label{(max+quad)(L+I)}\\
& &\ph{44444444444444444444444444444444444}=\mco(|\la|^{1/2}\si_{n-1})\frac{1}{H_{P,n-1}^{W+}-w_{n+1}} (\mathcal{L}|_{n}^{n-1}\cdot\Gamma_{P,n-1})\phi_{P,n-1} 
\label{(max+quad)(L+I)-one}\\
& &\ph{444444444444444444444444444444444444}+\mco(|\la|^{1/2}\si_{n-1})\frac{1}{H_{P,n-1}^{W+}-w_{n+1}} (\mathcal{I}|_{n}^{n-1}\cdot\Gamma_{P,n-1})\phi_{P,n-1}, 
\label{(max+quad)(L+I)-two}
\eeqa
where we made use of  {\color{\red}(\ref{3.53}) and (\ref{3.54}) in} Lemma~\ref{lem:Basic-estimates}.  Clearly $(\ref{(max+quad)(L+I)-two})=\mco(|\la|^{5/2}\si_{n-1}^{2-\de})$ due to
$|\mathcal{I}|_{n}^{n-1}|\leq c_I |\lambda|^{2}\sigma_{n-1}$ and Theorem~\ref{thm:induction-convergence}. 
Furthermore, we have by Lemma~\ref{final-subsection-auxiliary-lemma-one}   that $(\ref{(max+quad)(L+I)-one})=\mco(|\la|\si_{n-1}^{2-\de})$.
Thus altogether
\beqa
(\ref{(max+quad)(L+I)})=\mco(|\la|\si_{n-1}^{2-\de}).
\eeqa
\vspace{0.2cm}
\nin\textbf{Step 3:} Next we consider the following contribution to the expression from the statement of the lemma
\beqa
& &Q_{P,n-1}^{\bot}\frac{1}{H_{P,n-1}^{W+}-w_{n+1}} (\mathcal{L}|_{n}^{n-1}+\mathcal{I}|_{n}^{n-1})\cdot\Gamma_{P,n-1}\times\non\\ 
& &\ph{44444444444}\times  \frac{1}{H_{P,n-1}^{W+}-w_{n+1}}
  (\Delta(H_{I}^{W}|_{n}^{n-1})_{\mr{mix}}+(H_{I}^{W}|_{n}^{n-1})_{\mr{quad}})\phi_{P,n-1}\quad\quad\quad \label{(L+I)(mix+quad)}\\
 & &=Q_{P,n-1}^{\bot}\frac{1}{H_{P,n-1}^{W+}-w_{n+1}} \mathcal{I}|_{n}^{n-1}\cdot\Gamma_{P,n-1}
  \frac{1}{H_{P,n-1}^{W+}-w_{n+1}} \Delta(H_{I}^{W}|_{n}^{n-1})_{\mr{mix}}\phi_{P,n-1}  \label{I-G-mix}\\
& &\ph{44}+Q_{P,n-1}^{\bot}\frac{1}{H_{P,n-1}^{W+}-w_{n+1}} \mathcal{L}|_{n}^{n-1}\cdot\Gamma_{P,n-1}
  \frac{1}{H_{P,n-1}^{W+}-w_{n+1}} \Delta(H_{I}^{W}|_{n}^{n-1})_{\mr{mix}}\phi_{P,n-1}   \label{L-G-mix}\\
& &\ph{44}+Q_{P,n-1}^{\bot}\frac{1}{H_{P,n-1}^{W+}-w_{n+1}} \mathcal{I}|_{n}^{n-1}\cdot\Gamma_{P,n-1}
  \frac{1}{H_{P,n-1}^{W+}-w_{n+1}} (H_{I}^{W}|_{n}^{n-1})_{\mr{quad}}\phi_{P,n-1} \label{I-G-quad} \\
& &\ph{44}+Q_{P,n-1}^{\bot}\frac{1}{H_{P,n-1}^{W+}-w_{n+1}} \mathcal{L}|_{n}^{n-1}\cdot\Gamma_{P,n-1}
  \frac{1}{H_{P,n-1}^{W+}-w_{n+1}} (H_{I}^{W}|_{n}^{n-1})_{\mr{quad}}\phi_{P,n-1}. \label{L-G-quad-one}  
\eeqa
To analyse these expressions, norm-bounds involving  $\Delta(H_{I}^{W}|_{n}^{n-1})_{\mr{mix}}$, $(H_{I}^{W}|_{n}^{n-1})_{\mr{quad}}$
from Lemma~\ref{lem:Basic-estimates} are not sufficient and we have to recall the definitions:
\beqa
\Delta(H_{I}^{W}|_{n}^{n-1})_{\mr{mix}} \1&:=&\1{\color{\red}\sum_{i=1}^{3}}\h[(\Ga_{P,n-1})_i,  (\mathcal{L}|_{n}^{n-1})_i]=
\fr{\lambda}{2}\hint_{\!\!\!\!\bmcA|^{n-1}_n } d^{3}k\frac{k^2(b(k)-b^{*}(k))}{\sqrt{2}|k|^{3/2}\alpha_{P,n-1}(\hat{k})},\\
(H_{I}^{W}|_{n}^{n-1})_{\mr{quad}} \1&:=&\1 \h(\mathcal{L}|_{n}^{n-1}+ \mathcal{I}|_{n}^{n-1} )^2.
\eeqa
We note that $\Delta(H_{I}^{W}|_{n}^{n-1})_{\mr{mix}}$ has the same structure as $\mathcal{L}|_{n}^{n-1}$ but is by one power of $k$
more regular. To stress this fact we write $\mring{\mathcal{L}}|_{n}^{n-1}:=2\Delta(H_{I}^{W}|_{n}^{n-1})_{\mr{mix}} $ consistently with the
notation in Lemmas~\ref{RL-lemma}, \ref{middle-subsection-auxiliary-lemma}, \ref{R-L-R-L-whole-lemma}.

Thus we immediately get by  Lemma~\ref{middle-subsection-auxiliary-lemma} 
and  $|\mathcal{I}|_{n}^{n-1}|\leq c_I |\lambda|^{2}\sigma_{n-1}$,  that 
\beqa
(\ref{I-G-mix})=\mco(|\la|^{5/2}\si_{n-1}^{2-\de}).
\eeqa

Similarly, by Lemma~\ref{R-L-R-L-whole-lemma} 
we obtain
\beqa
(\ref{L-G-mix})=\mco(|\la|\si_{n-1}^{2-\de}).
\eeqa

Now proceeding to (\ref{I-G-quad}), we write
\beqa
 2\times (\ref{I-G-quad})
\1&=&\1Q_{P,n-1}^{\bot}\frac{1}{H_{P,n-1}^{W+}-w_{n+1}} \mathcal{I}|_{n}^{n-1}\cdot\Gamma_{P,n-1}
  \frac{1}{H_{P,n-1}^{W+}-w_{n+1}} (\mathcal{I}|_{n}^{n-1})^2  \phi_{P,n-1}  \label{I-G-quad-one}\\
& &+Q_{P,n-1}^{\bot}\frac{1}{H_{P,n-1}^{W+}-w_{n+1}} \mathcal{I}|_{n}^{n-1}\cdot\Gamma_{P,n-1}
  \frac{1}{H_{P,n-1}^{W+}-w_{n+1}} (2 \mathcal{I}|_{n}^{n-1}\cdot \mathcal{L}|_{n}^{n-1})   \phi_{P,n-1}  \label{I-G-quad-two} \\
& &+Q_{P,n-1}^{\bot}\frac{1}{H_{P,n-1}^{W+}-w_{n+1}} \mathcal{I}|_{n}^{n-1}\cdot\Gamma_{P,n-1}
  \frac{1}{H_{P,n-1}^{W+}-w_{n+1}} (\mathcal{L}|_{n}^{n-1}) \cdot (\mathcal{L}|_{n}^{n-1}) \phi_{P,n-1}.\quad\quad \label{I-G-quad-three} 
\eeqa
Using $|\mathcal{I}|_{n}^{n-1}|\leq c_I |\lambda|^{2}\sigma_{n-1}$, the fact that $(H_{P,n-1}^{W+}-w_{n+1})^{-1}\phi_{P,n-1}=\mco(\si_{n+1}^{-1})$
and Theorem~\ref{thm:induction-convergence}, we have 
\beqa
(\ref{I-G-quad-one})=\mco(|\la|^6 \veps^{-2}\si^{2-\de}_{n-1})=
\mco(|\la|^5 \si^{2-\de}_{n-1}). 
\eeqa
Next, by Lemma~\ref{middle-subsection-auxiliary-lemma} and $|\mathcal{I}|_{n}^{n-1}|\leq c_I |\lambda|^{2}\sigma_{n-1}$
\beqa
(\ref{I-G-quad-two})=\mco(|\la|^{9/2}\si_{n-1}^{2-\de}).
\eeqa
Now we consider (\ref{I-G-quad-three}). We have by Lemma~\ref{auxiliary-double-l-lemma} and  $|\mathcal{I}|_{n}^{n-1}|\leq c_I |\lambda|^{2}\sigma_{n-1}$
\beqa
(\ref{I-G-quad-three})=\mco(|\la|^3\si_{n-1}^{2-\de}).
\eeqa
Thus, altogether, we have
\beqa
(\ref{I-G-quad})=\mco(|\la|^3\si_{n-1}^{2-\de}).
\eeqa

Finally, we consider (\ref{L-G-quad-one}).  We have
\beqa
2\times (\ref{L-G-quad-one})\1&=&\1Q_{P,n-1}^{\bot}\frac{1}{H_{P,n-1}^{W+}-w_{n+1}} (\mathcal{L}|_{n}^{n-1})^{(+)}\cdot\Gamma_{P,n-1}
  \frac{1}{H_{P,n-1}^{W+}-w_{n+1}}( \mathcal{I}|_{n}^{n-1} )^2  \phi_{P,n-1}\label{L-G-quad-I-I}  \\ 
& &+Q_{P,n-1}^{\bot}\frac{1}{H_{P,n-1}^{W+}-w_{n+1}} \mathcal{L}|_{n}^{n-1}\cdot\Gamma_{P,n-1}
  \frac{1}{H_{P,n-1}^{W+}-w_{n+1}}(2 \mathcal{I}|_{n}^{n-1}\cdot (\mathcal{L}|_{n}^{n-1})^{(+)} ) \phi_{P,n-1}\label{Lp-G-quad-I-L}  \\ 
& &+Q_{P,n-1}^{\bot}\frac{1}{H_{P,n-1}^{W+}-w_{n+1}} (\mathcal{L}|_{n}^{n-1})^{(+)}\cdot\Gamma_{P,n-1}
  \frac{1}{H_{P,n-1}^{W+}-w_{n+1}}(  (\mathcal{L}|_{n}^{n-1})^{(+)}\cdot (\mathcal{L}|_{n}^{n-1})^{(+)} )  \phi_{P,n-1}\label{Lp-G-quad-Lp-L}  \\ 
& &+Q_{P,n-1}^{\bot}\frac{1}{H_{P,n-1}^{W+}-w_{n+1}} (\mathcal{L}|_{n}^{n-1})^{(-)}\cdot\Gamma_{P,n-1}
  \frac{1}{H_{P,n-1}^{W+}-w_{n+1}}(  (\mathcal{L}|_{n}^{n-1})^{(+)}\cdot (\mathcal{L}|_{n}^{n-1})^{(+)} )  \phi_{P,n-1}\label{Lm-G-quad-Lp-L}  \\ 
& &+Q_{P,n-1}^{\bot}\frac{1}{H_{P,n-1}^{W+}-w_{n+1}} (\mathcal{L}|_{n}^{n-1})^{(+)}\cdot\Gamma_{P,n-1}
  \frac{1}{H_{P,n-1}^{W+}-w_{n+1}}(  (\mathcal{L}|_{n}^{n-1})^{(-)}\cdot (\mathcal{L}|_{n}^{n-1})^{(+)} )  \phi_{P,n-1}.\label{Lp-G-quad-Lm-L} \quad\quad\quad
\eeqa
 {\color{\red}Making use of  $|\mathcal{I}|_{n}^{n-1}|\leq c_I |\lambda|^{2}\sigma_{n-1}$ and applying Lemma~\ref{final-subsection-auxiliary-lemma-one}, we easily see}
\beqa
(\ref{L-G-quad-I-I})=\mco(\la^{{\color{\red}4}} \si_{n-1}^{2-\de}).
\eeqa
Next, applying Lemma~\ref{R-L-R-L-whole-lemma}  and using $|\mathcal{I}|_{n}^{n-1}|\leq c_I |\lambda|^{2}\sigma_{n-1}$, we have
\beqa
(\ref{Lp-G-quad-I-L})=\mco(|\la|^3\si_{n-1}^{2-\de}).
\eeqa 
Now we consider (\ref{Lp-G-quad-Lp-L}) {\color{\red}where we can omit $Q_{P,n-1}^{\bot}$}. We can write
\beqa
(\ref{Lp-G-quad-Lp-L})
\1&\simeq&\1 \fr{(-\la)^3}{\sqrt{3!}}\hint^{\oplus}_{\!\!\!\! (\bmcA|^{n-1}_n)^{\times 3}}d^3k_1d^3k_2 d^3k_3\non\\
& &\bigg\{\fr{1}{\sqrt{2}|k_1|^{3/2}} \frac{k_{1,j}}{\alpha_{P,n-1}(\hat{k}_1)}  \fr{1}{\sqrt{2}|k_2|^{3/2}} \frac{k_{2,j}}{\alpha_{P,n-1}(\hat{k}_2)}
\fr{1}{\sqrt{2}|k_3|^{3/2}} \frac{k_{3,j'}}{\alpha_{P,n-1}(\hat{k}_3)}\times\non\\
& & \times\frac{1}{[H_{P,n-1}^{W+}]_{k_1,k_2,k_3}-w_{n+1}}
 (\Gamma_{P,n-1}+k_1+k_2)_{j'}
  \frac{1}{[H_{P,n-1}^{W+}]_{k_1,k_2}-w_{n+1}}  \phi_{P,n-1}+\{\textrm{perm} \}\bigg\},\label{three-floating-k}
\eeqa
where $\{\textrm{perm} \}$ denotes all the remaining permutations of the variables $k_1,k_2,k_3$ and summation over $j,j'$ is understood. The part
of (\ref{three-floating-k}) which is proportional to $k_1+k_2$ is $\mco(|\la|^3|\ln\,\veps |^{3/2}\si_{n-1}^2)$ by  {\color{\red}(\ref{res-est})}, (\ref{res-gamma-est}) in Lemma~\ref{pull-through-resolvents}.
Furthermore, we have by Lemmas~\ref{G-0-lemma}, \ref{pull-through-resolvents} and Theorem~\ref{thm:induction-convergence}
\beqa
& &\frac{1}{[H_{P,n-1}^{W+}]_{k_1,k_2,k_3}-w_{n+1}}
 (\Gamma_{P,n-1})_{j'}
  \frac{1}{[H_{P,n-1}^{W+}]_{k_1,k_2}-w_{n+1}}  \phi_{P,n-1}\non\\
  & &\ph{44444}=G_0(k_1,k_2)\frac{1}{[H_{P,n-1}^{W+}]_{k_1,k_2,k_3}-w_{n+1}}
 (\Gamma_{P,n-1})_{j'}  \phi_{P,n-1}
 +\frac{1}{[H_{P,n-1}^{W+}]_{k_1,k_2,k_3}-w_{n+1}}  (\Gamma_{P,n-1})_{j'}\mco(\si_{n-1}^{-\de})\non\\
& &\ph{44444}=\mco( |k_1|^{-1}\si_{n-1}^{-\de}).\quad \label{three-floating-k-contribution}
\eeqa
By substituting (\ref{three-floating-k-contribution}) to (\ref{three-floating-k}), we have 
\beqa
(\ref{Lp-G-quad-Lp-L})=\mco(|\la|^2\si_{n-1}^{2-\de}).
\eeqa
Next, we estimate (\ref{Lm-G-quad-Lp-L}). We have 
\beqa
(\ref{Lm-G-quad-Lp-L})
\1&\simeq&\1 2(-\la)^3\hint^{\oplus}_{\!\!\!\! \bmcA|^{n-1}_n}  d^3k'\,  \fr{1}{\sqrt{2}|k'|^{3/2}} \frac{k'_{j'}}{\alpha_{P,n-1}(\hat{k'})}\bint_{\!\!\!\!\bmcA|^{n-1}_n} d^3k \fr{1}{2|k|^{3} } \frac{ k_{j'}k_{j} }{\alpha_{P,n-1}(\hat{k})^2}     
     \times \non\\
& &\times \frac{1}{[H_{P,n-1}^{W+}]_{k'}-w_{n+1}} (\Gamma_{P,n-1}+k+k'  )_j 
  \frac{1}{[H_{P,n-1}^{W+}]_{k,k'}-w_{n+1}} \phi_{P,n-1}. \label{L-minus-term}
\eeqa
We note that by Lemma~ \ref{pull-through-resolvents} the term proportional to $k+k'$ in this expression is $\mco(|\la|^3 |\ln\,\veps|^{3/2} \si_{n-1}^2)$. Furthermore, we have by Lemmas~\ref{G-0-lemma}, \ref{pull-through-resolvents} and Theorem~\ref{thm:induction-convergence},
\beqa
& &\frac{1}{[H_{P,n-1}^{W+}]_{k'}-w_{n+1}}
 (\Gamma_{P,n-1})_{j'}
  \frac{1}{[H_{P,n-1}^{W+}]_{k,k'}-w_{n+1}}  \phi_{P,n-1}\non\\
& &\ph{}=G_0(k,k')\frac{1}{[H_{P,n-1}^{W+}]_{k'}-w_{n+1}}
 (\Gamma_{P,n-1})_{j'}  \phi_{P,n-1}+\frac{1}{[H_{P,n-1}^{W+}]_{k'}-w_{n+1}}  (\Gamma_{P,n-1})_{j'}\mco(\si_{n-1}^{-\de})
 =\mco( |k'|^{-1}\si_{n-1}^{-\de}). \quad\quad\quad \label{three-floating-k-contribution-x}
\eeqa
By substituting~(\ref{three-floating-k-contribution-x}) to (\ref{L-minus-term}), we have altogether
\beqa
(\ref{Lm-G-quad-Lp-L})=\mco(|\la|^2 \si_{n-1}^{2-\de}).
\eeqa 
Finally, we consider (\ref{Lp-G-quad-Lm-L}). {\color{black}Making use of (\ref{L+L-formula-x}) and Lemma~\ref{final-subsection-auxiliary-lemma-one}, we have}
\beqa
 (\ref{Lp-G-quad-Lm-L})=\mco({\color{\red}|\la|^2 |\ln\veps|} \veps^{-2}\si_{n-1})\frac{1}{H_{P,n-1}^{W+}-w_{n+1}} (\mathcal{L}|_{n}^{n-1})^{(+)}\cdot\Gamma_{P,n-1}  \phi_{P,n-1}=\mco(|\la|\si^{2-\de}_{n-1} ).
\eeqa
Thus altogether
\beqa
(\ref{L-G-quad-one})=\mco(|\la|\si^{2-\de}_{n-1} ).
\eeqa

Summing up,
\beqa
(\ref{(L+I)(mix+quad)})=\mco(|\la|\si^{2-\de}_{n-1} ).
\eeqa
\vspace{0.2cm}
\nin\textbf{Step 4:} Finally, we consider the following contribution to the expression from the statement of the lemma
\beqa
& &Q_{P,n-1}^{\bot}\frac{1}{H_{P,n-1}^{W+}-w_{n+1}} (\Delta(H_{I}^{W}|_{n}^{n-1})_{\mr{mix}}+(H_{I}^{W}|_{n}^{n-1})_{\mr{quad}})\times\non\\ 
& &\ph{444}\times\frac{1}{H_{P,n-1}^{W+}-w_{n+1}}(\Delta(H_{I}^{W}|_{n}^{n-1})_{\mr{mix}}+(H_{I}^{W}|_{n}^{n-1})_{\mr{quad}})\phi_{P,n-1}
=\mco(|\la|^{1/2}\si_{n-1}^2), \ \
\eeqa
where we made use of {\color{\red}(\ref{3.53}) and (\ref{3.54}) in} Lemma~\ref{lem:Basic-estimates}. This concludes the proof. \qed
\bel\label{B-star-lemma-three} (Control of (\ref{last-case-last-contribution-B-star-two}) associated with (\ref{last-case-last-contribution-one})) Under the assumptions of Theorem~\ref{main-technical-result}
\beqa
R''VB^*(G|^{n-1}_n)\phi_{P,n-1}:=\frac{1}{H_{P,n-1}^{W+}-w_{n+1}} (-)H_{I}^{W}|_{n}^{n-1} B^*(G|^{n-1}_n)\phi_{P,n-1}=\mco(|\la|^{3/2}\si_{n-1}^{2-2\de}),
\eeqa
where $B^*(G|^{n-1}_n)$ appeared in Lemma~\ref{B-star-lemma}.
\eel
\proof   We recall that $H_I^{W}|_{n}^{n-1}:=(\mathcal{L}|_{n}^{n-1}+\mathcal{I}|_{n}^{n-1})\cdot\Gamma_{P,n-1}
+\Delta(H_{I}^{W}|_{n}^{n-1})_{\mr{mix}}+(H_{I}^{W}|_{n}^{n-1})_{\mr{quad}}$ and consider the respective terms.
First, by  {\color{\red}(\ref{3.53})-(\ref{3.54})} in Lemma~\ref{lem:Basic-estimates} {\color{\red} combined with the fact that $\|G|^{n-1}_n\|_2=\mco(|\la|\si_{n-1}^{1-\de})$ and $B\big( G|^{n-1}_n \big)\phi_{P,n-1}=0$,}
we obtain that
\beqa
\frac{1}{H_{P,n-1}^{W+}-w_{n+1}} (\Delta(H_{I}^{W}|_{n}^{n-1})_{\mr{mix}}+(H_{I}^{W}|_{n}^{n-1})_{\mr{quad}} ) B^*\big( G|^{n-1}_n \big)\phi_{P,n-1}
=\mco(|\la|^{3/2}\si_{n-1}^{2-\de}).
\eeqa
Next, making use of $|\mathcal{I}|_{n}^{n-1}|\leq c_I |\lambda|^{2}\sigma_{n-1}$,   {\color{\red}(\ref{estimate-Q})} 
in Lemma~\ref{pull-through-resolvents} {\color{\red}combined with an identity analogous to (\ref{id})}, 
Theorem~\ref{thm:induction-convergence}, (\ref{res-est}) in Lemma~\ref{pull-through-resolvents} (for the $k_1+k_2$ term below), and the bound in (\ref{G-estimate}), we get
\beqa
& &\frac{1}{H_{P,n-1}^{W+}-w_{n+1}} (\mathcal{I}|_{n}^{n-1}\cdot\Gamma_{P,n-1}  ) B^*\big( G|^{n-1}_n \big)\phi_{P,n-1}\non\\
& &\simeq (\mathcal{I}|_{n}^{n-1})_{j}\hint^{\oplus}_{\!\!\!\!(\bmcA|^{n-1}_n)^{\times 2}} d^3k_1d^3k_2\,  G'(k_1,k_2)|^{n-1}_n 
\frac{1}{[H_{P,n-1}^{W+}]_{k_1,k_2}-w_{n+1}} 
(\Gamma_{P,n-1}+k_1+k_2)_{j} \phi_{P,n-1}\non\\
& &=\mco(|\la|^3\sigma_{n-1}^{2-2\de} ),\quad\quad
\eeqa
where we defined  $G'|^{n-1}_n$ by {\color{\red}$G(k_1,k_2)|^{n-1}_n=: \chi_{\ka}(k_1)\chi_{\ka}(k_2)|k_1|^{\alf} |k_2|^{\alf} G'(k_1,k_2)|^{n-1}_n$} and
accounted for  the factor $\chi_{\ka}(k_1)\chi_{\ka}(k_2)|k_1|^{\alf} |k_2|^{\alf}$ by the `hat' over the integral.  
 Now we look at
\beqa
& &\frac{1}{H_{P,n-1}^{W+}-w_{n+1}} (\mathcal{L}|_{n}^{n-1}\cdot\Gamma_{P,n-1}   ) B^*\big(G|^{n-1}_n \big)\phi_{P,n-1}\non\\
& &\ph{4444444444444}=\frac{1}{H_{P,n-1}^{W+}-w_{n+1}} ((\mathcal{L}|_{n}^{n-1})^{(+)}\cdot\Gamma_{P,n-1}   ) B^*\big( G|^{n-1}_n \big)\phi_{P,n-1}\label{second-case-direct-int-x}\\
& &\ph{444444444444444}+\frac{1}{H_{P,n-1}^{W+}-w_{n+1}} ((\mathcal{L}|_{n}^{n-1})^{(-)}\cdot\Gamma_{P,n-1}   ) B^*\big( G|^{n-1}_n \big)\phi_{P,n-1}.\label{second-case-annihilation-x}
\eeqa
The direct integral representation, {\color{\red}(\ref{3.53}), (\ref{estimate-Q})} in Lemma~\ref{pull-through-resolvents} {\color{\red}combined with an identity analogous to (\ref{id})},
Theorem~\ref{thm:induction-convergence}, (\ref{res-est}) in Lemma~\ref{pull-through-resolvents} (for the $k_2+k_3$ term below) and the bound in~(\ref{G-estimate}),  give
\beqa
& &(\ref{second-case-direct-int-x})\simeq -\fr{\la}{\sqrt{3!}}\hint^{\oplus}_{\!\!\!\!(\bmcA|^{n-1}_n)^{\times 3} }d^3k_1d^3k_2 d^3k_3\,
\bigg(\fr{1}{\sqrt{2}|k_1|^{3/2}} \frac{k_{1,j}}{\alpha_{P,n-1}(\hat{k}_1)}G'(k_2,k_3)|^{n-1}_n\times\non\\
& &\ph{44444444444444444}\times\frac{1}{[H_{P,n-1}^{W+}]_{k_1,k_2,k_3}-w_{n+1}}
(\Gamma_{P,n-1}+k_2+k_3)_j\phi_{P,n-1}+ \{\textrm{perm}\}  \bigg)\non\\
& &\ph{444444}=\mco(|\la|^3|\ln\, \veps|^{3/2}\si_{n-1}^{2-2\de})=\mco(|\la|^2\si_{n-1}^{2-2\de}),
\eeqa
where summation over $j$ is understood.
As for (\ref{second-case-annihilation-x}), we have by (\ref{res-est}) in Lemma~\ref{pull-through-resolvents} 
{\color{\red}(used only for the term proportional to $k+k'$ below)},  {\color{\red}(\ref{estimate-Q})} in Lemma~\ref{pull-through-resolvents} {\color{\red}combined with the identity in (\ref{id})}, Theorem~\ref{thm:induction-convergence}, the bound in~(\ref{G-estimate}), and {\color{\red}the constraint} $\si_n\leq |k|, |k'| \leq \si_{n-1}$
\beqa
(\ref{second-case-annihilation-x})\1&\simeq&\1-2\la\hint^{\oplus}_{\!\!\!\!\bmcA|^{n-1}_n}d^3k' \bint_{\!\!\!\!\bmcA|^{n-1}_n}d^3k \fr{1}{\sqrt{2}|k|^{3/2}} \frac{k_{j}}{\alpha_{P,n-1}(\hat{k})}G'(k,k')|^{n-1}_n\times\\
& &\times \frac{1}{[H_{P,n-1}^{W+}]_{k'}-w_{n+1}}(\Gamma_{P,n-1}+k+k')_j \phi_{P,n-1}\non\\
\1&=&\1\mco\big(|\la|^3\si_{n-1}^{2-\de}|\ln\,\veps|^{3/2}(\si_{n-1}^{-\de}+1+\veps^{-1})\big)=\mco(\la^2\si_{n-1}^{2-2\de}),
\eeqa
which concludes the proof. \qed

\bel\label{B-star-lemma-two} (Control of (\ref{last-case-last-contribution-B-star-one}) 
associated with (\ref{last-case-last-contribution-one})) Under the assumptions of Theorem~\ref{main-technical-result}
\beqa
R' ( \Delta\Gamma_{P}|_{n}^{n-1})_{i'}B^*(G|^{n-1}_n)\phi_{P,n-1}
:=\frac{1}{H_{P,n-1}^{W+}-z'_{n+1}}( \Delta\Gamma_{P}|_{n}^{n-1})_{i'} B^*(G|^{n-1}_n)\phi_{P,n-1}=\mco(|\la| \si_{n-1}^{1-2\de} ),\quad \quad
\eeqa
where $B^*(G|^{n-1}_n)$ appeared in Lemma~\ref{B-star-lemma}.
\eel
\proof We recall that $\Delta\Gamma_{P}|_{n}^{n-1}=X|^{n-1}_n
+\mathcal{L}|_{n}^{n-1}$, where $X|^{n-1}_n:=-\nabla E_{P,n-1}+\nabla E_{P,n}+\mathcal{I}|_{n}^{n-1}$ 
is a vector in $\real^3$ satisfying
\beqa
|X|^{n-1}_n|\leq  
c|\la|^{1/4}\si_{n-1}^{1-\de}. \label{standard-bound-X-X-X}
\eeqa
Since $\|(H_{P,n-1}^{W+}-z_{n+1})^{-1}\|_{\mcF_n}=\mco(\si_{n+1}^{-1})$ 
and $\|G|^{n-1}_n\|_2={\color{\red}\mco(|\la|^2|\ln \epsilon|\si_{n-1}^{1-\de})}=\mco(|\la|\si_{n-1}^{1-\de})$ {\color{\red}(see (\ref{bound-G}))}, we have
\beqa
\frac{1}{H_{P,n-1}^{W+}-z'_{n+1}}( X|^{n-1}_n )_iB^*\big({\color{\red} G|^{n-1}_n}\big)\phi_{P,n-1}
=\mco( |\la|^{5/4}\veps^{-2}\si_{n-1}^{1-2\de})=\mco(|\la|\si_{n-1}^{1-2\de}).
\eeqa
Next, we consider the contribution with $\mathcal{L}|_{n}^{n-1}$. We have
\beqa
\frac{1}{H_{P,n-1}^{W+}-z'_{n+1}}(\mathcal{L}|_{n}^{n-1} )_{i'} B^*(G|^{n-1}_n)\phi_{P,n-1}\1&=&\1
\frac{1}{H_{P,n-1}^{W+}-z'_{n+1}}
(\mathcal{L}|_{n}^{n-1} )^{(+)}_{i'} B^*(G|^{n-1}_n)\phi_{P,n-1}\quad\quad\label{B-star-lemma-two-plus}\\
& &+\frac{1}{H_{P,n-1}^{W+}-z'_{n+1}}
(\mathcal{L}|_{n}^{n-1} )^{(-)}_{i'} B^*(G|^{n-1}_n)\phi_{P,n-1}.\quad\quad\label{B-star-lemma-two-minus}
\eeqa
{\color{\red} Making use of (\ref{res-est}) in Lemma~\ref{pull-through-resolvents} and of (\ref{G-estimate}), similarly as in (\ref{three-floating-k}), we have}
\beqa
(\ref{B-star-lemma-two-plus})\1&\simeq&\1 \fr{(-\la)}{\sqrt{3!}}\hint^{\oplus}_{\!\!\!\! (\bmcA|^{n-1}_n)^{\times 3}}d^3k_1d^3k_2 d^3k_3\non\\
& &\bigg\{\fr{1}{\sqrt{2}|k_1|^{3/2}} \frac{k_{1,i'}}{\alpha_{P,n-1}(\hat{k}_1)}  
G'(k_2,k_3)|^{n-1}_n
\frac{1}{[H_{P,n-1}^{W+}]_{k_1,k_2,k_3}-z'_{n+1}}  \phi_{P,n-1}+\{\textrm{perm}\} \bigg\},\non\\ \label{three-floating-k-B-star}
\1&=&\1{\color{\red}\mco(|\la|^3|\ln \epsilon|^{3/2}\si_{n-1}^{1-\de})}=\mco(|\la|^{2}\si_{n-1}^{1-\de} ), \label{above-G-prime}
\eeqa
where we {\color{\red} recall that} $G(k_1,k_2)|^{n-1}_n=: \chi_{\ka}(k_1)\chi_{\ka}(k_2)|k_1|^{\alf} |k_2|^{\alf} G'(k_1,k_2)|^{n-1}_n$.
Next we write, similarly as in (\ref{L-minus-term}),  applying Lemma~\ref{pull-through-resolvents},  (\ref{G-estimate}) and {\color{\red}the constraint} $\si_n\leq |k|, |k'| \leq \si_{n-1}$
\beqa
& &(\ref{B-star-lemma-two-minus})\simeq 2(-\la)\hint^{\oplus}_{\!\!\!\! \bmcA|^{n-1}_n}  d^3k'\, \bint_{\!\!\!\!\bmcA|^{n-1}_n} d^3k \fr{1}{\sqrt{2}|k|^{3/2} } \frac{ k_{i'} }{\alpha_{P,n-1}(\hat{k})}   G'(k,k')|^{n-1}_n 
 \frac{1}{[H_{P,n-1}^{W+}]_{k'}-z'_{n+1}} \phi_{P,n-1}\non\\
& &\ph{444444}=\mco(|\la|^{3}|\ln\,\veps|^{3/2} \veps^{-1}\si_{n-1}^{1-\de})=\mco(|\la|^{2} \si_{n-1}^{1-\de}).
\eeqa
This concludes the proof. \qed

\bel\label{B-star-lemma-one} (Control of (\ref{last-case-last-contribution-B-star-one})
associated with (\ref{last-case-last-contribution-one})) Under the assumptions of Theorem~\ref{main-technical-result}
\beqa
R' (\Gamma_{P,n-1})_{i'}B^*(G|^{n-1}_n)\phi_{P,n-1}:=\frac{1}{H_{P,n-1}^{W+}-z'_{n+1}}(\Gamma_{P,n-1})_{i'}B^*(G|^{n-1}_n)\phi_{P,n-1}=\mco(|\la| \si_{n-1}^{1-2\de} ),
\eeqa
where $B^*(G|^{n-1}_n)$ appeared in Lemma~\ref{B-star-lemma}.
\eel
\proof Making use of the direct integral representation, of {\color{\red}(\ref{estimate-Q})} in Lemma~\ref{pull-through-resolvents} {\color{\red}combined with an identity analogous to (\ref{id})} and of (\ref{res-est}) in Lemma~\ref{pull-through-resolvents} (for the $k_1+k_2$ term below), we get
\beqa
& &\frac{1}{H_{P,n-1}^{W+}-z'_{n+1}}(\Gamma_{P,n-1})_{i'}B^*(G|^{n-1}_n)\phi_{P,n-1}\non\\
& &\simeq\sqrt{2}\hint^{\oplus}_{\!\!\!\!(\bmcA|^{n-1}_n)^{\times 2}}d^3k_1d^3k_2\, G'(k_1,k_2)|^{n-1}_n  \frac{1}{[H_{P,n-1}^{W+}]_{k_1,k_2}-z'_{n+1}}
(\Gamma_{P,n-1}+k_1+k_2)_{i'}\phi_{P,n-1}\non\\
& &=\mco(|\la|^2|\ln\,\veps|\si_{n-1}^{1-2\de} )=\mco(|\la| \si_{n-1}^{1-2\de} ),
\eeqa
where we exploited (\ref{G-estimate}) and $G'$ is defined below~(\ref{above-G-prime}). This concludes the proof. \qed

\bel\label{G-0-lemma}  Under the assumptions of Theorem~\ref{main-technical-result}
\beqa
\frac{1}{[H_{P,n-1}^{W+}]_{k_1,k_2}-z_{n+1}}\phi_{P,n-1}=G_0(k_1,k_2)\phi_{P,n-1}+\mco(\si_{n-1}^{-\de}), \label{G-0-lemma-formula}
\eeqa
where $\si_n\leq |k_1|,|k_2|\leq \si_{n-1}$, $z_{n+1}\in\ga_{P,n+1}\cup \ti\ga_{P,n+1}$ and $G_0(k_1,k_2)=\mco((|k_1|+|k_2|)^{-1})$ is a symmetric numerical function {\color{\red}with support in  $\bar\mcA|^{n-1}_n\times \bar\mcA|^{n-1}_n$.}    
\eel
\proof We recall that
\beqa
 [H_{P,n-1}^{W+}]_{k_1,k_2}:=H_{P,n-1}^{W+}+(k_1+k_2)\cdot \Gamma_{P,n-1}+\frac{(k_1+k_2)^2}{2}
+\alpha_{P, n-1}(\hat k_1)|k_1| +\alpha_{P, n-1}(\hat k_2)|k_2|.\ \ \ 
\eeqa
{\color{\red}Set} $a_{P,n-1}(k_1,k_2):=\alpha_{P, n-1}(\hat k_1)|k_1| +\alpha_{P, n-1}(\hat k_2)|k_2|$ and write the following
resolvent expansion
\begin{eqnarray}
& &\frac{1}{[H_{P,n-1}^{W+}]_{k_1,k_2}-z_{n+1}}=\frac{1}{H_{P,n-1}^{W+}+a_{P,n-1}(k_1,k_2)-z_{n+1}} \label{second-trunc-fin-one}\\
& &\ph{44444444444444444}+\frac{1}{[H_{P,n-1}^{W+}]_{k_1,k_2}-z_{n+1}} 
[-(k_1+k_2)\cdot \Gamma_{P,n-1}-\frac{(k_1+k_2)^2}{2}]
\times\label{second-trunc-fin-intermediate} \\
& &\ph{444444444444444444444444444444444444}\times\frac{1}{H_{P,n-1}^{W+}+a_{P,n-1}(k_1,k_2)-z_{n+1}}.
\label{second-trunc-fin-two}
\end{eqnarray}
The term $G_0(k_1,k_2)\phi_{P,n-1}$ in (\ref{G-0-lemma-formula}) originates from the first term on the r.h.s. of (\ref{second-trunc-fin-one}), i.e.
\beqa
G_0(k_1,k_2):=(E_{P,n-1}+\De c_P|^{n-1}_n+a_{P,n-1}(k_1,k_2)-z_{n+1})^{-1}.
\eeqa
We have
\beqa
& &|E_{P,n-1}+\De c_P|^{n-1}_n+a_{P,n-1}(k_1,k_2)-\mrm{Re}\,z_{n+1}|\non\\
& &\ph{44444444444}\geq a_{P,n-1}(k_1,k_2)-|E_{P,n-1}-E_{P,n}|-|\De c_P|^{n-1}_n|-\fr{\si_{n+1}}{3}\non\\
& &\ph{44444444444}\geq \fr{1}{3}(|k_1|+|k_2|)-(1/6+1/20)(|k_1|+|k_2|)\geq c'(|k_1|+|k_2|), \quad \label{k-one+k-two}
\eeqa
where $c'>0$. Here we made use of the definition $\alpha_{P, n-1}(\hat k):=(1-\nabla E_{P,n-1}\cdot \hat k)$, the bounds
$|\De c_P|^{n-1}_n|, |E_{P,n-1}-E_{P,n}|\leq (1/20)\si_{n+1}{\color{\red}\leq (1/40)(|k_1|+|k_2|)}$, $\nabla E_{P,n}\leq (1/3)+c|\la|$ and 
$|\mrm{Re}\, z_{n+1}-E_{P,n}|\leq \si_{n+1}/3{\color{\red}\leq (1/6)(|k_1|+|k_2|)}$. This proves that $G_0(k_1,k_2)=\mco((|k_1|+|k_2|)^{-1})$.

Let us now {\color{\red}explain} the error term in (\ref{G-0-lemma-formula}). We get from (\ref{second-trunc-fin-intermediate})-(\ref{second-trunc-fin-two}), {\color{\red}(\ref{estimate-Q})} in Lemma~\ref{pull-through-resolvents} {\color{\red}combined with an identity analogous to (\ref{id})}, Theorem~\ref{thm:induction-convergence} and (\ref{k-one+k-two})
that  the part of the  error term proportional to $(k_1+k_2)\cdot \Gamma_{P,n-1}$ is $\mco(\si_{n-1}^{-\de})$. The part proportional to $(k_1+k_2)^2$
is $\mco(1)$ by (\ref{res-est}) in Lemma~\ref{pull-through-resolvents}. \qed

\bel\label{last-case-additional-lemma} Under the assumptions of Theorem~\ref{main-technical-result} and the inductive hypothesis
\beqa
\|Q_{P,n-1}^{\bot}\frac{1}{H_{P,n-1}^{W+}-w_{n+1}} (\Ga_{P,n-1})_j \frac{1}{[H_{P,n-1}^{W+}]_{k}-w_{n+1}}(\Ga_{P,n-1})_{j'}\phi_{P,n-1}\|
=\mco(\veps^{-4}\si_{n-1}^{-2\de})  \label{L-G-L-G-2-1-1}
\eeqa
for $\si_n\leq |k|\leq \si_{n-1}$.
\eel
\proof 
To be able to use the induction hypothesis we expand the resolvent of $[H_{P,n-1}^{W+}]_{k}$ in 
(\ref{L-G-L-G-2-1-1}) to the second order:
\beqa
& &\frac{1}{[H_{P,n-1}^{W+}]_{k}-w_{n+1}}
:=\frac{1}{H_{P,n-1}^{W+}+k\cdot \Gamma_{P,n-1}+\frac{|k|^2}{2}+\alpha_{P, n-1}(\hat k)|k|-w_{n+1}}\non\\
&=&\frac{1}{H_{P,n-1}^{W+}+\alpha_{P, n-1}(\hat k)|k|-w_{n+1}}\label{0}\\
& &+\frac{1}{H_{P,n-1}^{W+}+\alpha_{P, n-1}(\hat k)|k|-w_{n+1}}[-k\cdot \Gamma_{P,n-1}-\frac{|k|^2}{2}]\frac{1}{H_{P,n-1}^{W+}+\alpha_{P, n-1}(\hat k)|k|-w_{n+1}}\quad\quad \label{1}\\
& &+\frac{1}{[H_{P,n-1}^{W+}]_{k}-w_{n+1}}\bigg\{[-k\cdot \Gamma_{P,n-1}-\frac{|k|^2}{2}]\frac{1}{H_{P,n-1}^{W+}+\alpha_{P, n-1}(\hat k)|k|-w_{n+1}}\bigg\}^2.\quad\label{3}
\eeqa
We set $\ti w_{n+1}:=w_{n+1}-\alpha_{P, n-1}(\hat k)|k|$ and first consider the contribution of (\ref{0}) to 
(\ref{L-G-L-G-2-1-1}). We have
\beqa
& &\|Q_{P,n-1}^{\bot}\frac{1}{H_{P,n-1}^{W+}-w_{n+1}} (\Ga_{P,n-1})_j \frac{1}{H_{P,n-1}^{W+}-\ti{w}_{n+1}}(\Ga_{P,n-1})_{j'}\phi_{P,n-1}\|\non\\
& &\leq \sup_{z'_{n+1}\in \ti\ga_{P,n+1} }\|Q_{P,n-1}^{\bot}\frac{1}{H_{P,n-1}^{W+}-w_{n+1}} (\Ga_{P,n-1})_j \frac{1}{H_{P,n-1}^{W+}-z'_{n+1}}(\Ga_{P,n-1})_{j'}\phi_{P,n-1}\|\leq \si_{n-1}^{-2\de},\quad\quad
\eeqa
where we used that $\alpha_{P, n-1}(\hat k)|k|\geq 0$,  applied the maximal modulus principle and the inductive hypothesis via 
relation~(\ref{alternative-inductive-hypothesis}).

Now we consider the contribution of the term involving $\Ga_{P,n-1}$ in (\ref{1}) to (\ref{L-G-L-G-2-1-1}):
\beqa
& &\|Q_{P,n-1}^{\bot}\frac{1}{H_{P,n-1}^{W+}-w_{n+1} } (\Ga_{P,n-1})_j 
\frac{1}{H_{P,n-1}^{W+}-\ti w_{n+1}}(k\cdot \Gamma_{P,n-1}) \frac{1}{H_{P,n-1}^{W+}-\ti w_{n+1}}   (\Ga_{P,n-1})_{j'}\phi_{P,n-1}  \|\non\\
& &=\|Q_{P,n-1}^{\bot}\frac{1}{H_{P,n-1}^{W+}-w_{n+1} } (\Ga_{P,n-1})_j Q_{P,n-1}^{\bot}
\frac{1}{H_{P,n-1}^{W+}-\ti w_{n+1}}(k\cdot \Gamma_{P,n-1}) \frac{1}{H_{P,n-1}^{W+}-\ti w_{n+1}}   (\Ga_{P,n-1})_{j'}\phi_{P,n-1}  \|\quad\quad\quad \label{Q-bot-part-ind-hyp}\\
& &\ph{4}+\|Q_{P,n-1}^{\bot}\frac{1}{H_{P,n-1}^{W+}-w_{n+1} } (\Ga_{P,n-1})_j Q_{P,n-1}
\frac{1}{H_{P,n-1}^{W+}-\ti w_{n+1}}(k\cdot \Gamma_{P,n-1}) \frac{1}{H_{P,n-1}^{W+}-\ti w_{n+1}}   (\Ga_{P,n-1})_{j'}\phi_{P,n-1}  \|.\label{Q-part-main-technical-result}\quad\quad\quad 
\eeqa
With the help of {\color{\red}(\ref{3.11-Gamma})}   in Lemma~\ref{lem:Basic-estimates}, {\color{\red}the constraint} $|k|\leq \si_{n-1}$, {\color{\red}the} maximal modulus principle, and the inductive hypothesis, we have
\beqa
(\ref{Q-bot-part-ind-hyp})\leq c\veps^{-2}\si_{n-1}^{-2\de}.
\eeqa
Next, applying Theorem~\ref{thm:induction-convergence} (twice)  in combination with the maximal modulus principle,  Lemma~\ref{tilda-w-lemma} 
and $|k|\leq \si_{n-1}$ we obtain
\beqa
(\ref{Q-part-main-technical-result})\leq c\veps^{-2}\si_{n-1}^{-2\de}.
\eeqa

Next, we consider the contribution of the term involving $|k|^2$ in (\ref{1}) to (\ref{L-G-L-G-2-1-1}). By Theorem~\ref{thm:induction-convergence} combined
with the maximal modulus principle, Lemma~\ref{tilda-w-lemma} and {\color{\red}(\ref{first}) in} Lemma~\ref{lem:Basic-estimates}, we obtain
\beqa
 \|Q_{P,n-1}^{\bot}\frac{1}{H_{P,n-1}^{W}-w_{n+1}} (\Ga_{P,n-1})_j 
\frac{1}{H_{P,n-1}^{W+}-\ti w_{n+1}}\fr{|k|^2}{2} \frac{1}{H_{P,n-1}^{W+}-\ti w_{n+1}}   (\Ga_{P,n-1})_{j'}\phi_{P,n-1}  \|=\mco(\veps^{-4}\si_{n-1}^{-\de}).  
\eeqa

Finally, we study the contribution of (\ref{3}) to (\ref{L-G-L-G-2-1-1}). Expression~(\ref{3}) consist of four
terms which we denote $(k\Ga, k\Ga), (k\Ga,k^2), (k^2,k\Ga), (k^2,k^2)$ in the obvious order.
The contribution of $(k\Ga, k\Ga)$ to (\ref{L-G-L-G-2-1-1}) can be bounded by
\beqa
& &\mco(\si_{n+1}^{-1})\| \frac{1}{[H_{P,n-1}^{W+}]_{k}-w_{n+1}} (k\cdot \Gamma_{P,n-1})
\times\non\\& &\ph{444444} \times\frac{1}{H_{P,n-1}^{W+}-\ti w_{n+1}}    (k\cdot \Gamma_{P,n-1})\frac{1}{H_{P,n-1}^{W+}-\ti w_{n+1}}    (\Ga_{P,n-1})_{j'}\phi_{P,n-1}  \| \label{kGa-kGa-zero}\\
\1&=&\1\mco(\si_{n+1}^{-1})\| \frac{1}{[H_{P,n-1}^{W+}]_{k}-w_{n+1}} (k\cdot \Gamma_{P,n-1})\times\non\\& &\ph{444444} \times Q_{P,n-1}^{\bot}\frac{1}{H_{P,n-1}^{W+}-\ti w_{n+1}}    (k\cdot \Gamma_{P,n-1})\frac{1}{H_{P,n-1}^{W+}-\ti w_{n+1}}    (\Ga_{P,n-1})_{j'}\phi_{P,n-1}  \| \label{kGa-kGa-one}\\
& &+\mco(\si_{n+1}^{-1})\| \frac{1}{[H_{P,n-1}^{W+}]_{k}-w_{n+1}} (k\cdot \Gamma_{P,n-1})Q_{P,n-1} \times\non\\& &\ph{444444} \times Q_{P,n-1}\frac{1}{H_{P,n-1}^{W+}-\ti w_{n+1}}    (k\cdot \Gamma_{P,n-1})\frac{1}{H_{P,n-1}^{W+}-\ti w_{n+1}}    (\Ga_{P,n-1})_{j'}\phi_{P,n-1}  \|. \label{kGa-kGa-two}
\eeqa
Thus we have by {\color{\red}(\ref{res-gamma-est})} in Lemma~\ref{pull-through-resolvents}, $|k|\leq \si_{n-1}$, the maximal modulus principle, and the inductive hypothesis
\beqa
(\ref{kGa-kGa-one})\leq c\veps^{-2}\si_{n-1}^{-2\de}.
\eeqa
Now by Lemma~\ref{pull-through-resolvents}, Theorem~\ref{thm:induction-convergence} combined with the maximal modulus principle, $|k|\leq \si_{n-1}$ and Lemma~\ref{tilda-w-lemma},
\beqa
(\ref{kGa-kGa-two})\leq c\veps^{-4}\si_{n-1}^{-2\de}.
\eeqa
Thus altogether
\beqa
(\ref{kGa-kGa-zero})\leq c\veps^{-{\color{\red}4}}\si_{n-1}^{-2\de}.
\eeqa

Now the contribution of $(k\Ga,k^2)$ to (\ref{L-G-L-G-2-1-1}) has the form
\beqa
& &\mco(\si_{n+1}^{-1})\| \frac{1}{[H_{P,n-1}^{W+}]_{k}-w_{n+1}} (k\cdot \Gamma_{P,n-1})
\times\non\\& &\ph{444444} \times\frac{1}{H_{P,n-1}^{W+}-\ti w_{n+1}}    \fr{|k|^2}{2}\frac{1}{H_{P,n-1}^{W+}-\ti w_{n+1}}    (\Ga_{P,n-1})_{j'}\phi_{P,n-1}  \| \leq c{\bc \veps^{-4} }\si_{n-1}^{-\de},
\eeqa
 where we made use of {\color{black}$\ti w_{n+1}=w_{n+1}-\alpha_{P, n-1}(\hat k)|k|$}, Theorem~\ref{thm:induction-convergence} in combination with the maximal modulus principle,
{\color{\red}(\ref{res-gamma-est})} in Lemma~\ref{pull-through-resolvents}, Lemma~\ref{tilda-w-lemma} and $|k|\leq \si_{n-1}$.  

Next, the contribution of $(k^2, k\Ga)$ to (\ref{L-G-L-G-2-1-1}) has the form
\beqa
& &\mco(\si_{n+1}^{-1})\| \frac{1}{[H_{P,n-1}^{W+}]_{k}-w_{n+1}} \fr{|k|^2}{2}
\times\non\\& &\ph{444} \times\frac{1}{H_{P,n-1}^{W+}-\ti w_{n+1}}    (k\cdot \Gamma_{P,n-1})\frac{1}{H_{P,n-1}^{W+}-\ti w_{n+1}}    (\Ga_{P,n-1})_{j'}\phi_{P,n-1}  \|\leq c\veps^{-4}\si_{n-1}^{-\de},  \label{k-2-k-Ga-bound}
\eeqa
where we made use of Theorem~\ref{thm:induction-convergence} in combination with the maximal modulus principle, {\color{\red}(\ref{res-est}) in} Lemma~\ref{pull-through-resolvents}, $|k|\leq \si_{n-1}$ and Lemma~\ref{tilda-w-lemma}.

Finally, the contribution of $(k^2, k^2)$ to (\ref{L-G-L-G-2-1-1}) has the form
\beqa
& &\mco(\si_{n+1}^{-1})\| \frac{1}{[H_{P,n-1}^{W+}]_{k}-w_{n+1}} \fr{|k|^2}{2}
\times\non\\& &\ph{444444} \times\frac{1}{H_{P,n-1}^{W+}-\ti w_{n+1}}    \fr{|k|^2}{2}\frac{1}{H_{P,n-1}^{W+}-\ti w_{n+1}}    (\Ga_{P,n-1})_{j'}\phi_{P,n-1}  \| \leq c\veps^{-4}\si_{n-1}^{1-\de},
\eeqa
where we made use of the same ingredients as in (\ref{k-2-k-Ga-bound}). Thus the proof is complete. \qed

\bel\label{tilda-w-lemma} Under the assumptions of Theorem~\ref{main-technical-result}
\beqa
& &\|\frac{1}{H_{P,n-1}^{W+}-\ti w_{n+1}}\|_{\mcF_n}\leq \fr{c}{\si_{n+1}}, \label{tilda-w-lemma-first}\\
& &\|\frac{1}{H_{P,n-1}^{W+}-\ti w_{n+1}}    (\Gamma_{P,n-1})_i\|_{\mcF_n}\leq \fr{c}{\si_{n+1}},\label{tilda-w-lemma-second}
\eeqa
where $\ti w_{n+1}=w_{n+1}-\alpha_{P, n-1}(\hat k)|k|$, $\si_{n}\leq |k|\leq \si_{n-1}$, $w_{n+1}\in\ga_{P,n+1}$.
\eel
\proof We show only (\ref{tilda-w-lemma-second}) as the proof of (\ref{tilda-w-lemma-first}) is analogous and simpler. We write
\beqa
\frac{1}{H_{P,n-1}^{W+}-\ti w_{n+1}}    (\Gamma_{P,n-1})_i=\frac{1}{H_{P,n-1}^{W+}-\ti w_{n+1}}  Q_{P,n-1}  (\Gamma_{P,n-1})_i
+\frac{1}{H_{P,n-1}^{W+}-\ti w_{n+1}}  Q_{P,n-1}^{\bot} (\Gamma_{P,n-1})_i. \quad
\eeqa
The term involving $Q_{P,n-1} $ is estimated by a computation using (\ref{bound-denominator-k}), (\ref{bound-on-QA}), (\ref{res-gamm-bound})
and $ |k|^{-1}\leq \si_n^{-1}\leq \si_{n+1}^{-1}$. As for the term with $ Q_{P,n-1}^{\bot}$, making use of the maximal modulus principle and {\color{\red}(\ref{3.11-Gamma}) in}
Lemma~\ref{lem:Basic-estimates}, {\bcb we can write for any $\phi\in \mcF_n$, $\|\phi\|\leq 1$, }
\beqa
\|\frac{1}{H_{P,n-1}^{W+}-\ti w_{n+1}}  Q_{P,n-1}^{\bot} (\Gamma_{P,n-1})_i\phi\| \leq\sup_{z_{n+1}\in \ti\ga_{P,n+1}}
\|\frac{1}{H_{P,n-1}^{W+}- z_{n+1}}  Q_{P,n-1}^{\bot} (\Gamma_{P,n-1})_i\phi\|\leq \fr{c}{\si_{n+1}},\quad
\eeqa
uniformly in $\phi$. This concludes the proof. \qed
\bel\label{auxiliary-double-l-lemma} Under the assumptions of Theorem~\ref{main-technical-result} 
\beqa
& &\frac{1}{H_{P,n-1}^{W+}-z'_{n+1}}(\Gamma_{P,n-1})_{i}\frac{1}{H_{P,n-1}^{W+}-z_{n+1}} \mathcal{L}|_{n}^{n-1}\cdot \mathcal{L}|_{n}^{n-1} \phi_{P,n-1}=\mco(|\la|\si_{n-1}^{1-\de}),
\eeqa
where $z'_{n+1}, z_{n+1}\in \ga_{P,n+1}\cup \ti\ga_{P,n+1}$.
\eel
\proof First, we note that
\beqa
(\mathcal{L}|_{n}^{n-1})^{(-)} \cdot (\mathcal{L}|_{n}^{n-1})^{(+)} \phi_{P,n-1}\1&=&\1\la^2\bint_{\!\!\!\! \bmcA|^{n-1}_n } d^3k\fr{k^2}{2|k|^{3}\alpha_{P,n-1}(\hat{k})^2 }\phi_{P,n-1}\non\\
\1&=&\1\phi_{P,n-1}\,\mco(\la^2 |\ln\,\veps| \si_{n-1}^2). \label{L+L-formula-x} 
\eeqa
The first contribution to the expression from the statement of the lemma is therefore
\beqa
& &\frac{1}{H_{P,n-1}^{W+}-z'_{n+1}}(\Gamma_{P,n-1})_{i}\frac{1}{H_{P,n-1}^{W+}-z_{n+1}} ((\mathcal{L}|_{n}^{n-1})^{(-)}\cdot (\mathcal{L}|_{n}^{n-1})^{(+)})  \phi_{P,n-1}\non\\
& &\ph{44444444444444444444444}=\mco(\la^2 |\ln\,\veps|\veps^{-2} \si_{n-1}^{1-\de})=\mco(|\la|\si_{n-1}^{1-\de}),
\eeqa
where we made use of Theorem~\ref{thm:induction-convergence}. 
The remaining contribution is
\beqa
& &\frac{1}{H_{P,n-1}^{W+}-z'_{n+1}}(\Gamma_{P,n-1})_{i}\frac{1}{H_{P,n-1}^{W+}-z_{n+1}} (\mathcal{L}|_{n}^{n-1})^{(+)}\cdot (\mathcal{L}|_{n}^{n-1})^{(+)} \phi_{P,n-1}\label{I-G-quad-three-xxx}\\
& &\ph{4444}\simeq \la^2 \sqrt{2}   \hint^{\oplus}_{\!\!\!\!(\bmcA|^{n-1}_n)^{\times 2}}d^3k_1d^3k_2 \fr{1}{\sqrt{2}|k_1|^{3/2}} \frac{k_{1,i}}{\alpha_{P,n-1}(\hat{k}_1)}
\fr{1}{\sqrt{2}|k_2|^{3/2}} \frac{k_{2,i}}{\alpha_{P,n-1}(\hat{k}_2)}\times\non\\
& &\ph{444444} \times \frac{1}{[H_{P,n-1}^{W+}]_{k_1,k_2}-z'_{n+1}} (\Gamma_{P,n-1}+k_1+k_2)_i
  \frac{1}{[H_{P,n-1}^{W+}]_{k_1,k_2}-z_{n+1}}  \phi_{P,n-1}. \ \ \ \  \label{I-G-quad-three-repeated-x}
\eeqa
The part of (\ref{I-G-quad-three-repeated-x}) proportional to $k_1+k_2$ is $\mco(|\la|^2|\ln\,\veps| \si_{n-1})=
\mco( |\la| \si_{n-1})$ by {\color{\red}(\ref{res-est})} in
Lemma~\ref{pull-through-resolvents}. To control the remaining part of (\ref{I-G-quad-three-repeated-x}),
we write, making use of Lemma~\ref{G-0-lemma}
\beqa
& &\frac{1}{[H_{P,n-1}^{W+}]_{k_1,k_2}-z'_{n+1}}  (\Gamma_{P,n-1})_i
  \frac{1}{[H_{P,n-1}^{W+}]_{k_1,k_2}-z_{n+1}}  \phi_{P,n-1}\non\\
& &=G_0(k_1,k_2)\frac{1}{[H_{P,n-1}^{W+}]_{k_1,k_2}-z'_{n+1}}  (\Gamma_{P,n-1})_i  \phi_{P,n-1}\non\\
& &\ph{44444444444444}+\frac{1}{[H_{P,n-1}^{W+}]_{k_1,k_2}-z'_{n+1}}  (\Gamma_{P,n-1})_i\mco(\si_{n-1}^{-\de})
=\mco((|k_1|+|k_2|)^{-1}\si_{n-1}^{-\de}), \quad\quad \label{R-k-Ga-R-k-xx}
\eeqa
where we applied {\color{\red}(\ref{estimate-Q})} in Lemma~\ref{pull-through-resolvents} {\color{\red}combined with an identity analogous to (\ref{id})}, Theorem~\ref{thm:induction-convergence} and the fact that $G_0(k_1,k_2)=\mco((|k_1|+|k_2|)^{-1})$.
Substituting  (\ref{R-k-Ga-R-k-xx}) to (\ref{I-G-quad-three-repeated-x}), we obtain altogether
\beqa
(\ref{I-G-quad-three-xxx})=\mco(\la^2 |\ln\,\veps| \si_{n-1}^{1-\de})= \mco(|\la| \si_{n-1}^{1-\de}),
\eeqa
which completes the proof. \qed

\bel\label{auxiliary-b-star-lemma} Under the assumptions of Theorem~\ref{main-technical-result} 
\beqa
& &\frac{1}{H_{P,n-1}^{W+}-z'_{n+1}}(\Gamma_{P,n-1})_{i}\frac{1}{H_{P,n-1}^{W+}-z_{n+1}} (\mathcal{L}|_{n}^{n-1}\cdot \Gamma_{P,n-1})  \phi_{P,n-1}\non\\
& &\ph{4444444444444444444444}=b^*({\color{\red}g_{2,i}}|^{n-1}_n)\phi_{P,n-1}+\mco(|\la|^{1/4}\si_{n-1}^{1-2\de}),
\eeqa
where $z'_{n+1}, z_{n+1}\in \ga_{P,n+1}\cup \ti\ga_{P,n+1}$ and {\color{\red}$g_{2,i}|^{n-1}_n$} is a 3-tuple of functions {\color{\red}with support in $\bar\mcA|^{n-1}_n$} which satisfy the pointwise bound 
\beqa
|{\color{\red}g_{2,i}(k)|^{n-1}_n}|\leq |\la|\fr{c}{|k|^{3/2}\si_{n-1}^{\de}}.
\eeqa
\eel
\proof We use the direct integral representation
\beqa
& &\frac{1}{H_{P,n-1}^{W+}-z'_{n+1}} (\Gamma_{P,n-1})_i \frac{1}{H_{P,n-1}^{W+}-z_{n+1}}
\mathcal{L}|_{n}^{n-1}\cdot\Gamma_{P,n-1}\phi_{P,n-1}\non\\
&\simeq&(-\la) \hint^{\oplus}_{\!\!\!\bmcA|^{n-1}_n}  \fr{d^3k}{\sqrt{2}|k|^{3/2}} 
\frac{k_{i'}}{\alpha_{P,n-1}(\hat{k})} \times \non\\
& &\times \frac{1}{[H_{P,n-1}^{W+}]_k-z'_{n+1}} (\Gamma_{P,n-1}+k)_i \frac{1}{[H_{P,n-1}^{W+}]_k-z_{n+1}}
(\Gamma_{P,n-1})_{i'}\phi_{P,n-1}.\label{four-four-four-x} 
\eeqa
By {\color{\red}(\ref{estimate-Q})} {\color{\red}combined with the identity in (\ref{id})} Lemma~\ref{pull-through-resolvents} and Theorem~\ref{thm:induction-convergence}, the part of (\ref{four-four-four-x}) proportional to $k$ is $\mco(|\la||\ln\,\veps|^{1/2} \si_{n-1}^{1-\de})=\mco(|\la|^{1/2} \si_{n-1}^{1-\de})$.
We write the remaining part as follows
\beqa
& &(-\la) \hint^{\oplus}_{\!\!\!\bmcA|^{n-1}_n}  \fr{d^3k}{\sqrt{2}|k|^{3/2}} 
\frac{k_{i'}}{\alpha_{P,n-1}(\hat{k})} \times \non\\
& &\ph{44444}\times \frac{1}{[H_{P,n-1}^{W+}]_k-z'_{n+1}} Q^{\bot}_{P,n-1}(\Gamma_{P,n-1})_i \frac{1}{[H_{P,n-1}^{W+}]_k-z_{n+1}}
(\Gamma_{P,n-1})_{i'}\phi_{P,n-1}   \label{I-G-L-G-one-x} \\
& &+(-\la) \hint^{\oplus}_{\!\!\!\bmcA|^{n-1}_n}  \fr{d^3k}{\sqrt{2}|k|^{3/2}} 
\frac{k_{i'}}{\alpha_{P,n-1}(\hat{k})} \times \non\\
& &\ph{44444}\times \frac{1}{[H_{P,n-1}^{W+}]_k-z'_{n+1}} Q_{P,n-1}(\Gamma_{P,n-1})_i \frac{1}{[H_{P,n-1}^{W+}]_k-z_{n+1}}
(\Gamma_{P,n-1})_{i'}\phi_{P,n-1} \label{I-G-L-G-two-x}.
\eeqa
We have $(\ref{I-G-L-G-one-x})=\mco(|\la||\ln\,\veps|^{1/2}\veps^{-4}\si_{n-1}^{1-2\de})=\mco(|\la|^{1/4} \si_{n-1}^{1-2\de})$, where we made use {\color{black} of (\ref{estimate-Q}) in}
Lemma~\ref{pull-through-resolvents},  and of Lemma \ref{last-case-additional-lemma}.
The analysis of (\ref{I-G-L-G-two-x}) is  very similar to the discussion of (\ref{second-case-second-est-1}), thus we can write 
\beqa
(\ref{I-G-L-G-two-x})=b^*\big({\color{\red}g_{2,i}}|^{n-1}_n \big)\phi_{P,n-1}+\mco(|\la|^{1/2}\si_{n-1}^{1-2\de}),
\eeqa
where ${\color{\red}g_{2,i}}|^{n-1}_n$ is a 3-tuple of functions {\color{\red}with support in 
$\bar\mcA|^{n-1}_n$}  which satisfy the pointwise bound 
$|g_{2,i'}(k)|^{n-1}_n|\leq |\la|\fr{c}{|k|^{3/2}\si_{n-1}^{\de}}$. This completes the proof. \qed

\section{Proof of Theorem~\ref{preliminaries-on-spectrum}} \label{derivatives-section}
\setcounter{equation}{0}

\subsection{ Proof of Theorem~\ref{preliminaries-on-spectrum}  (c) } \label{gs}

\nin An immediate consequence of Theorem~\ref{thm:induction-convergence} (and its proof) is the convergence of the sequence $\{\phi_{P,n}\}_{n\in\mathbb{N}_0}$ to the limiting non-zero vector $\phi_P$, which is  the ground state of the transformed Hamiltonian with removed infrared cut-off $H^W_{P,n=\infty}$.
To study the properties of this sequence, we  define an antiunitary $J$ on $\cal F$ as follows: for $\varphi \in \cal F$ with momentum wave functions 
$\{ \varphi^q(k_1,\dots, k_q)\}_{q\in \mathbb{N}_0}$ we set 
\begin{equation}
 J\{ \varphi^q(k_1,\dots, k_q)\}_{q\in \mathbb{N}_0}:=\{ \overline{\varphi^q(k_1,\dots, k_q)}\}_{q\in \mathbb{N}_0},
\end{equation}
where $J\Om=\Om$ is understood. Now we prove the following corollary of Theorem~\ref{thm:induction-convergence}.
\begin{cor}\label{vector-existence} Let  $\{\phi_{P,n}\}_{n\in\mathbb{N}_0}$ be the sequence of vectors appearing in  Theorem~\ref{thm:induction-convergence} 
and denote  $\cphi_{P,n}:=\phi_{P,n}/\|\phi_{P,n}\|$. Then, under the assumptions of Theorem~\ref{thm:induction-convergence}:  
\begin{enumerate}
\item[(a)] $\|\phi_{P,n+1}-\phi_{P,n}\|\leq 2|\la|^{1/4}\si_n^{1/2}$ and $\|\cphi_{P,n+1}-\cphi_{P,n}\|\leq  8|\la|^{1/4}\si_n^{1/2} $. Therefore, the sequence  
$\{\phi_{P,n}\}_{n\in\mathbb{N}_0}$ (resp.  $\{\cphi_{P,n}\}_{n\in\mathbb{N}_0}$) is convergent as $n\to\infty$ to the non-zero limiting vector $\phi_{P}$ (resp. $\cphi_{P}$).
\item[(b)]  $\lan  \cphi_{P,n}, \Om\ran>  1/2$. 
\item[(c)] Let $Q_{P,n}$ be the spectral projection onto the ground state energy of $H^W_{P,n}${\color{black}; see (\ref{3.37})}. Then $\cphi_{P,n}= Q_{P,n} \Om/\|  Q_{P,n}\Om\|$.
\end{enumerate}
\end{cor}
\proof  The estimates in part~(a) follow from Theorem~\ref{thm:induction-convergence} ii), iii), and formula~(\ref{n+1-vector-bound}).  The convergence of the sequences to non-zero vectors  then follows by a telescopic argument and  Theorem~\ref{thm:induction-convergence}~iii).  

To prove part (b), we recall that $\phi_{P,0}:=\Om$ and note that by part~(a)
\beqa
\|\Om-\cphi_{P,n}\|\leq 8|\la|^{1/4}\sum_{n=0}^{\infty}{\bcb \si_n^{1/2}}<1 
\eeqa
for $|\la_0|$ sufficiently small,  {\color{black}so that} we obtain $\mathrm{Re}\lan \cphi_{P,n}, \Om\ran>1/2$. Noting that, by definition~(\ref{phi-definitions}),  $J\cphi_{P,n}=\cphi_{P,n}${\color{black}, we deduce that $\lan \cphi_{P,n}, \Om\ran$ is real} and complete the proof of  part (b) of the corollary.

To prove part (c), we note that by $\lan \cphi_{P,n}, \Om\ran>1/2$, the uniqueness of the ground state of $H^W_{P,n}$, and 
$JQ_{P,n}\Om=Q_{P,n}\Om$ we have that $\cphi_{P,n}=s(n)Q_{P,n}\Om/\|   Q_{P,n}\Om\|$, where $s(n)\in \{\pm 1\}$. Clearly, $s(0)=1$. {\bcb To show that $s(n)=1$ for all $n\geq 0$, suppose the opposite, i.e.} that
$s(n_*)=-1$ for some $n_*\in \nat$. Then
\beqa
1>\|\Om-\cphi_{P,n_*}\|\geq {\bcb \lan \Om, \Om- \cphi_{P,n_*}\ran}=1-\lan \Om,  \cphi_{P,n_*}\ran=1+\|Q_{P,n_*}\Om\|,
\eeqa
which is a contradiction.   \qed

Now we describe the passage from the discrete  to the continuous infrared cut-off, following  \cite{Pi03}. 
For any $\de\in (0,1/4)$  we fix a sequence of discrete cut-offs  $\rho_n:=(\veps^*(\de))^n$  
as specified in Theorem~\ref{main-technical-result}. Now for any given (continuous) infrared cut-off $\si\in(0,{\bcb \veps^*(\de)}]$ there is exactly one 
$n_{\si}\in \nat$ s.t. $\rho_{n_{\si}+1}<\si\leq \rho_{n_{\si}}$. Defining $\veps_{\si}:=\si^{1/n_{\si}}$,  we obtain a new sequence of discrete cut-offs  $\si_n:=\veps_{\si}^n$ with the property $\si_{n_{\si}}=\si$. Clearly, $\veps^*(\de)^{1+1/n_{\si}}<\veps_{\si}\leq  \veps^*(\de)$, hence $\veps_{\si}\in (\veps^*(\de)^2, \veps^*(\de)]$ and   $\veps_{\si}\to  \veps^*(\de) $ for $\si\to 0$. 
\begin{defn}\label{rule}
{\color{black}For any sequence $\{A_n^{(\veps)}\}_{n\in\nat}$ of quantities depending on the discrete cut-off (where we write the dependence on $\veps$ explicitly), 
we can set $A_{\si}:=A_{n_{\si}}^{(\veps_{\si})}$.  By applying  this rule we recover $H_{P,\si}$, introduced  in (\ref{infrared-cut-off-Hamiltonian}), and
also define  $W_{P,\si}$, $H^W_{P,\si}$, $E_{P,\si}$, $\Ga_{P,\si}$, $\phi_{P,\si}$,  $\hat{\phi}_{P,\si}$,  $Q_{P,\si}$, $\hat{Q}_{P,\si}$, $\bar{Q}_{P,\si}$ which will be used below.  }
\end{defn} 
We cannot apply this rule to $\ga_{P,n}$, since it depends both
on $n$ and $n-1$. Instead, we declare
\beqa
\gamma_{P,\sigma}:=\left\{w\in\mathbb{C}\,\, \big| \,\, |w-E_{P,\sigma}|=\frac{\sigma}{6}\right\}. \label{continuous-contour}
\eeqa
\newcommand{\bsi}{\bar{\si}}
As for the ground state vector of $H_{P,\si}$, it will be convenient to set its phase as follows: 
\begin{defn}\label{abuse-definition}  We define
$\check{\psi}_{P,\si}:= W^*_{P,\si}\cphi_{P,\si}$, where $\cphi_{P,\si}:=\phi_{P,\si}/\|\phi_{P,\si}\|$,
and $\phi_{P,\si}:=\phi_{P,n_{\si}}^{(\veps_{\si})}$ was constructed in Corollary~\ref{phi-vectors-corollary}. 
In view of Corollary~\ref{vector-existence} (c),  $\cphi_{P,\si}$ can equivalently be defined as
\beqa
\cphi_{P,\si}:=\fr{  \oint_{\ga_{P,\si}}dw\, \fr{1}{H^W_{P,\si}-w}\Om    }{ \| \oint_{\ga_{P,\si}}dw\, \fr{1}{H^W_{P,\si}-w}\Om\| }.
\eeqa
\end{defn}
After this preparation, we proceed to the proof of Theorem~\ref{preliminaries-on-spectrum}  (c). 
To this end, we establish a counterpart  of Corollary~\ref{vector-existence} (a) for the continuous cut-off, that is {\color{black}showing} that
$\|\cphi_{P,\bar{\si}}-\cphi_{P,\si}\|\leq 8 |\la|^{1/4}\si^{1/2}$ for $0<\bar{\si}<\si\leq  {\bc \veps^*(\de)}  $. We cannot apply the arguments from 
the proof of the corollary and Theorem~\ref{thm:induction-convergence} directly, because $\veps_{\si}\neq \veps_{\bsi}$ in general.  
Instead, we proceed as follows: we 
find $n\in \nat_0$ s.t.  $\bsi\in [\si_{n+1}, \si_n)$ and we define 
$\hat{H}^W_{P,\bsi}$, $\hat{Q}_{P, \bsi}$ as indicated {\color{black}in Definition \ref{rule} using the sequence of cut-offs  $\{\veps_{\bar{\si}}^n\}_{n=0}^{\infty}$.}
By adapting our considerations from
Subsection~\ref{shift-of-the-cut-off-subsection}, we can write
\beqa
\hat{H}_{P,\bar{\si}}=H_{P,n}^W+\Delta c_P|_{\bar{\si}}^{n}+H_I^{W}|_{\bar{\si}}^{n}, \label{n+1-eta-Hamiltonian}
\eeqa
where
\beqa
H_{I}^{W}|_{\bar{\si}}^{n}:=(\mathcal{L}|_{\bar{\si}}^{n}+\mathcal{I}|^n_{\bar{\si}})\cdot\Gamma_{P,n}
+\Delta(H_{I}^{W}|_{\bar{\si}}^{n})_{\mr{mix}}+(H_{I}^{W}|_{\bar{\si}}^{n})_{\mr{quad}}
\eeqa
and the quantities $\mathcal{L}|^n_{\bar{\si}}, \mathcal{I}|^n_{\bar{\si}}, \Delta c_{P}|^n_{\bar{\si}}, \Delta(H_{I}^{W}|^n_{\bar{\si}})_{\mr{mix}},
\Delta(H_{I}^{W}|_{\bar{\si}}^{n})_{\mr{mix}}, (H_{I}^{W}|_{\bar{\si}}^{n})_{\mr{quad}}$ are obtained by replacing {\color{black}$\si_{n+1}=\epsilon_{\sigma}^{n+1}$} with $\bar{\si}$
in their counterparts from Section~\ref{Prel-sec}.
Using that $\bar{\si}\geq \si_{n+1}$, it is easy to check that these expressions satisfy the estimates from Lemma~\ref{lem:Basic-estimates}
and that considerations about direct integral representations from Subsection~\ref{direct-subsection} can be  adapted to the present situation.
Furthermore, the discussion from Appendix~\ref{pull-through-appendix}  can be repeated for the modified  cut-off as follows. We set
\beqa
& &W_{P,\bar{\si}}:=\exp(-\lambda\hint^{\kappa}_{\!\!\!\bar{\si}}\frac{d^3k}
{\sqrt{2}|k|^{\frac{3}{2}}\alpha_{P,\bar{\si}}(\hat{k} ) }\{  b(k)-b^{*}(k) \}), \\
& &\widetilde{W}'_{P,\bar{\si}}:=\exp(-\lambda\hint^{\kappa}_{\!\!\!\bar{\si}}\frac{d^3k}{\sqrt{2}|k|^{\frac{3}{2}}\alpha_{P,n}(\hat{k})}\{b(k)-b^{*}(k)\}),
\eeqa
{\color{black} where $\alpha_{P,\bar{\si}}(\hat{k} ):=(1-\hat{k}\cdot\nabla E_{P, \bar{\si} })$},  and define  by analogy with (\ref{phi-definitions})
\beqa
\hat{\phi}'_{P,\bar{\si}}:=\hat{Q}_{P,\bar{\si}}\phi_{P,n}, \quad {\phi}'_{P,\bar{\si}}:=W_{P,\bar{\si}}(\widetilde{W}'_{P,\bar{\si}})^*\hat{\phi}'_{P,\bar{\si}},
\eeqa
where we use the prime to distinguish these vectors from $\hat{\phi}_{P,\bsi}$ and $\phi_{P,\bsi}$ defined 
{\color{black} using the rule in Definition \ref{rule}}. Now  by repeating the arguments from Lemmas~\ref{vector-shift-Lemma}, \ref{gradient-shift-lemma}  we  obtain:
\bel\label{vector-shift-Lemma-xx}  Let $|P|\leq P_{\max}=1/3$ and $|\la|\in (0,\la_0]$. Then 
\beqa
|\nabla E_{P,\bar{\si}}-\nabla E_{P,n}|\1&\leq&\1 c_{1}[\lambda^{2}\sigma_{n}+\|\frac{\hat{\phi}'_{P,\bar{\si}}}{\|\hat{\phi}'_{P,\bar{\si}}\|}-\frac{\phi_{P,n}}{\|\phi_{P,n}\|}\|], \\
\|\big(W_{P,\bar{\si}}(\wt{W}'_{P,\bar{\si}})^*-1\big)\hat{\phi}'_{P,\bar{\si}}\|\1&\leq&\1 c|\lambda|\,  |\nabla E_{P,\bar{\si}}-\nabla E_{P,n}|\,
|\ln \bar{\si}| \,\|\hat{\phi}'_{P,\bar{\si}}\|, 
\label{vector-difference-xx}\\
\|\big(W_{P,\bar{\si}} W_{P,n}^*-1\big)\hat{\phi}'_{P,\bar{\si}}\|\1&\leq&\1 c|\lambda|\,  |\nabla E_{P,\bar{\si}}-\nabla E_{P,n}|\,|\ln \bar{\si}| \,\|\hat{\phi}'_{P,\bar{\si}}\|\non\\
& &+c|\la|  {\bcb (\alf)^{-1}}\si_n^{\alf}\,\|\hat{\phi}'_{P,\bar{\si}}\|, 
\label{new-vector-difference-xx}
\eeqa
where (only) for (\ref{new-vector-difference-xx}) we assumed $\alf>0$ and (\ref{vector-difference-xx}), (\ref{new-vector-difference-xx}) also hold after replacing $\hat{\phi}'_{P,\bar{\si}}$ with $\phi'_{P,\bar{\si}}$.
\eel
\nin Now we are ready to prove the following proposition:
\begin{prop}\label{eta-lemma-cont-cut} Let $|P|\leq P_{\max}=1/3$ and $|\la|\in (0,\la_0]$. Then   
\begin{enumerate}
\item[(a)] For $\alf\geq 0$, $\|\cphi_{P, \bar{\si}}-\cphi_{P,n}\|\leq c |\la|^{1/4}\si_n^{1/2}$, 
\item[(b)] For $\alf>0$,  $\|\cpsi_{P, \bar{\si}}-\cpsi_{P,n}\|\leq c |\la|^{1/8} {\bcb (\alf)^{-1} }\si_n^{\alf}$,
\end{enumerate}
where {\color{black}$\cphi_{P,  \bar{\si} }:=Q_{P, \bar{\si}}\Om/\| Q_{P, \bar{\si}}\Om\|$} and $\cpsi_{P, \bar{\si}}:=W^*_{P,\bar{\si}}\cphi_{P, \bar{\si}}$ according to Definition~\ref{abuse-definition}. 
\end{prop}
\proof Using the information in (\ref{n+1-eta-Hamiltonian})-(\ref{vector-difference-xx}) and repeating the steps from the 
proof of Claim ii) and Claim iii) of Theorem~\ref{thm:induction-convergence}, we obtain for some $0<\de<1/4$
\beqa
\|\hat{\phi}'_{P, \bar{\si}}-\phi_{P,n}\|\leq  |\lambda|^{\frac{1}{4}}\sigma_{n}^{1-\delta}, \quad \|\phi'_{P, \bar{\si}}-\hat{\phi}'_{P,\bar{\si}} \|\leq   |\la|^{1/4} \bar{\si}^{1/2}, \quad \|\phi'_{P,\bar{\si}}\|\geq (1/2).
\eeqa
(We note that the application of Claim i) of Theorem~\ref{thm:induction-convergence} within the above discussion is unproblematic,
since it relies on $\bar{\si}$-independent\footnote{{\color{black}To be more precise they are not \emph{explicitly} $\bar{\si}$-dependent, indeed the value  $n$ depends  on $\sigma$ and $\bar{\sigma}$.}} ingredients $H_{P,n}^W$ and $\Ga_{P,n}$ in (\ref{n+1-eta-Hamiltonian})). 
Now setting $\cphi'_{P,\bar{\si}}:=\phi'_{P, \bar{\si}} /\|\phi'_{P, \bar{\si}}\|$, we immediately get
\beqa
\|\cphi'_{P, \bar{\si}}-\cphi_{P,n}\|\leq  8|\la|^{1/4}\si_n^{1/2}.  \label{primed-unprimed-estimate}
\eeqa
Let us now prove that $\cphi'_{P, \bar{\si} }=\cphi_{P, \bar{\si}}$: We recall from Corollary~\ref{vector-existence} that  
$\cphi_{P, n }=Q_{P,n }\Om/\| Q_{P, n }\Om\|$,
where $Q_{P, n}$ is the  ground state projection of $H^W_{P,n}$. Since, by construction,
$J \cphi'_{P, \bar{\si}  }=\cphi'_{P, \bar{\si}}$, we have  $\cphi'_{P,\bar{\si} }=(\pm)Q_{P, \bar{\si}} \Om/\|  Q_{P, \bar{\si}  } \Om\|$.
Suppose there is a $(-)$ sign in the last formula. Then, by (\ref{primed-unprimed-estimate}) and Corollary~\ref{vector-existence}~(b)
\beqa
8|\la|^{1/4}\si_{n}^{1/2}\geq \bigg\| \fr{Q_{P, \bar{\si} } \Om}{\| Q_{P, \bar{\si}} \Om\| }
+\fr{Q_{P,  n} \Om}{\|  Q_{P,  n   } \Om\| }\bigg\|\geq {\color{black} \lan \Om, \cphi_{P, \bar{\si} }\ran+ \lan \Om, \cphi_{P, n}\ran}>1\,,
\eeqa
{\color{black} where we have used that both $\lan \Om, \cphi_{P, \bar{\si}  }\ran$ and $\lan \Om, \cphi_{P, n}\ran$ are larger than $ \frac{1}{2}$.}
As this is a contradiction (possibly after reducing $|\la_0|$) we obtain that  $\cphi'_{P, \bar{\si}}=Q_{P, \bar{\si} } \Om/\|  Q_{P, \bar{\si}  } \Om\|\equiv \cphi_{P,  \bar{\si}}$.
Together with (\ref{primed-unprimed-estimate}), this concludes the proof of part (a) of the proposition.

Proceeding  to part (b), we write,  making use of (\ref{primed-unprimed-estimate}) and Lemma~\ref{vector-shift-Lemma-xx}, 
\beqa
\|\cpsi_{P,\bar{\si}}-\cpsi_{P,n}\|\1&\leq&\1 \|(W_{P, \bar{\si}} W^*_{P,n}-1)\cphi_{P, \bar{\si}}\|+\|\cphi_{P, \bar{\si}}-\cphi_{P,n}\|\non\\
\1&\leq&\1  c[\la^2\si_n+4|\la|^{1/4}\si_n^{1-\de}] |\ln \bar{\si}| +c|\la| {\bcb (\alf)^{-1}}\si_n^{\alf}+   8|\la|^{1/4}\si_n^{1/2} \non\\
\1&\leq&\1 c|\la|^{1/8} {\bcb (\alf)^{-1} }\si_n^{\alf},
\eeqa
where we used that $0<\alf\leq 1/2$. This concludes the proof. \qed\\
\nin {\color{black} We recall that,  according to Definition \ref{rule},  $\cpsi_{P,n_{\sigma}}=\cpsi^{(\epsilon_{\sigma})}_{P,n_{\sigma}}=:\cpsi_{P,\sigma}$ and $\sigma_{n_{\sigma}}=\sigma$.} Then, by the telescopic argument (to step from $n_{\sigma}$ to $n$) Proposition~\ref{eta-lemma-cont-cut} yields immediately the following corollary.
Its second part  gives  Theorem~\ref{preliminaries-on-spectrum}  (c).
\begin{cor} For $|\la|\in (0,\la_0]$  and  $0<\bar{\si}<\si\leq {\bcb \epsilon^*(\de)}$ we have
\begin{enumerate}
\item[(a)] For $\alf\geq 0$, $\|\cphi_{P,\bar{\si}}-\cphi_{P,\si}\|\leq c |\la|^{1/4}{\bcb \si^{1/2}}$, 
\item[(b)] For $\alf>0$,  $\|\cpsi_{P,\bar{\si}}-\cpsi_{P,\si}\|\leq c |\la|^{1/8} {\bcb (\alf)^{-1} } \si^{\alf}$.
\end{enumerate}
Consequently, the limits $\cphi_{P}:=\lim_{\si\to 0}\cphi_{P,\si}$ and $\cpsi_{P}:=\lim_{\si\to 0}\cpsi_{P,\si}$ exist. (The latter only for $\alf>0$). 
\end{cor}

\subsection{Proof of Theorem~\ref{preliminaries-on-spectrum}  (b)}

By a straightforward modification of the discussion from Lemmas~\ref{upper-bound-lemma}-\ref{energy-shift-one} 
we obtain $|E_{P,\bsi}-E_{P,n}|\leq c|\la|^2\si_n$, which by the telescopic argument gives $|E_{P,\bsi}-E_{P,\si}|\leq c|\la|^2\si$
for $0<\bsi<\si\leq 1$.  The statement about convexity
of $S\ni P\mapsto E_{P}$ is proven in Appendix~\ref{convexity}.


\subsection{Proof of Theorem~\ref{preliminaries-on-spectrum}  (a) }

We start with the following corollary of Theorems~\ref{thm:induction-convergence}, \ref{main-technical-result}.
 \begin{cor}\label{technical-corollary}
 There is $\la_0>0$ and $\de_{\la_0}\in (0,1/4)$  s.t. for all $\la\in (0,\la_0]$, ${\bcb \si\in (0,  \veps^*(\de)}]$ and $P\leq P_{\max}=1/3$ the following estimates hold true
\beqa
& &\|\frac{1}{H_{P,\sigma}^{W}-E_{P,\sigma}}(\Gamma_{P,\sigma})_{i}\check{\phi}_{P,\sigma}\|\leq \fr{1}{\si^{\de_{\la_0}} },
\label{eq:inequality-cont-0}\\
& &\|\frac{Q_{P,\sigma}^{\perp}}{H_{P,\sigma}^{W}-E_{P,\sigma}}(\Gamma_{P,\sigma})_{i}\frac{1}{H_{P,\sigma}^{W}-E_{P,\sigma}}(\Gamma_{P,\sigma})_{i'}\check{\phi}_{P,\sigma}\|\leq \fr{1}{\si^{2\de_{\la_0}} }.\label{eq:inequality-cont-1}
\eeqa
Clearly, (\ref{eq:inequality-cont-0}), (\ref{eq:inequality-cont-1}) remain valid after replacing $\la_0$ by some $\ti{\la}_0\in (0,\la_0]$. The resulting
function $\ti{\la}_0\to\de_{\ti{\la}_0} $ can be chosen s.t. $\lim_{\ti{\la}_0\to 0}\de_{\ti{\la}_0}=0$.
\end{cor}
\proof We discuss only (\ref{eq:inequality-cont-1}), since the treatment of (\ref{eq:inequality-cont-0}) is analogous and simpler. Fix $\de\in (0,1/4)$, 
$\veps^*(\de)$ and $\la_0^*(\veps^*(\de))$ as in Theorem~\ref{main-technical-result}.
Now, as described {\bcb above} Definition \ref{rule},   for any ${\bcb \si\in (0,  \veps^*(\de)}]$  
we obtain $\veps_{\si}\in (\veps^*(\de)^2  ,\veps^*(\de)]$ and $n_{\si}\in\nat$ s.t. $\si=\veps_{\si}^{n_{\si}}$. Then, using also the maximal modulus principle, we obtain from Theorem~\ref{main-technical-result}
\beqa
\|\frac{Q_{P,\sigma}^{\perp}}{H_{P,\sigma}^{W}-E_{P,\sigma}}(\Gamma_{P,\sigma})_{i}\frac{1}{H_{P,\sigma}^{W}-E_{P,\sigma}}(\Gamma_{P,\sigma})_{i'}\check{\phi}_{P,\sigma}\|\leq \fr{1}{\si^{2\de} }
\eeqa
for all $|\la|\leq \veps_{\si}^8$, hence for all $|\la|\leq   \veps^*(\de)^{16}$. Now we define  the function of  $0< \tilde{\lambda}_0 \leq \lambda^*_0(\epsilon^*(\delta))$ 
\beqa
\de(\ti{\la}_0):=\inf\{\,\de\in (0,1/4)\,|\, \ti{\la}_0\leq \veps^*(\de)^{16}\,\}.
\eeqa
Since for any $\de\in (0,1/4)$ we can find a non-zero  $\veps^*(\de)$ we have $\lim_{\ti\la_0\to 0}\de(\ti{\la}_0)=0$. Thus setting
e.g. $\de_{\ti{\la}_0}:=\de(\ti{\la}_0)+\ti{\la}_0$ (to account for the fact that the infimum may be outside of the set) we conclude the proof. \qed
\newcommand{\nss}{\si}

Now we compute  the first and the second derivative with respect to $P$ of the normalized ground state vector $\cpsi_{P,\si}$ of the Hamiltonian $H_{P,\si}$
introduced in Definition~\ref{abuse-definition}.  Since $\cpsi_{P,\si}=\bar{Q}_{P,\si}W_{P,\si}^*\cphi_{P,\si}/\| \bar{Q}_{P,\si}W_{P,\si}^*\cphi_{P,\si}\|$, 
the families of operators $\{H_{P,\si}\}_{P\in \mathbb{R}^3}$, $\{H^W_{P,\si}\}_{P\in \mathbb{R}^3}$ are analytic of type A, and the eigenvalue  $E_{P,\si}$ is isolated on the relevant subspaces (see Proposition \ref{prop:Hamiltonian-results}), we conclude that the derivatives of $\cpsi_{P,\si}$ w.r.t. $P$ exist. We write
\beqa
\cpsi_{P,\si}= \oint_{\ga_{P,\si}}dw\, \fr{1}{H_{P,\si}-w}\cpsi_{P,\si}
\eeqa
and note that the r.h.s is invariant under small changes of the integration contour. Thus we can write
\beqa
\pa_{i}\check\psi_{P,\si}=\oint_{\gamma_{P,\si}}dw\frac{1}{H_{P,\si}-w}(-)(P-P_{\pho})_{i}\frac{1}{H_{P,\si}-w}\check{\psi}_{P,\si}
+\check\psi_{P,\si}\lan\check\psi_{P,\si},\pa_i\check\psi_{P,\si}\ran,
\eeqa
where we set $\pa_i:=\pa_{P_i}$.
By normalization of $\check\psi_{P,\si}$ we have $\pa_i\|\check \psi_{P,\si}\|^2=0$,
i.e.,  $\mathrm{Re}\lan\check\psi_{P,\si},\pa_i\check\psi_{P,\si}\ran=0$.  Moreover, by Definition~\ref{abuse-definition}
\beqa
\lan \check\psi_{P,\si},\pa_i\check\psi_{P,\si}\ran=
\lan J\check\psi_{P,\si},J\pa_i\check\psi_{P,\si}\ran=
\ov{\lan \check\psi_{P,\si},\pa_i\check\psi_{P,\si}\ran},
\eeqa
hence also $\mathrm{Im}\lan \check\psi_{P,\si},\pa_i\check\psi_{P,\si}\ran=0$.
Thus we obtain
\begin{equation}
\partial_{i}\check{\psi}_{P,\si}=\oint_{\gamma_{P,\si}}dw\frac{1}{H_{P,\si}-w}(-)(P-P_{\pho})_{i}\frac{1}{H_{P,\si}-w}\check{\psi}_{P,\si}. \label{first-deriv-section5}
\end{equation}
Therefore
\begin{eqnarray}
\partial_{i'}\partial_{i}\check{\psi}_{P,\si}\1&= &\1 \bigg( \oint_{\gamma_{P,\si}}dw\frac{1}{H_{P,\si}-w}(-)(P-P_{\pho})_{i'}\frac{1}{H_{P,\si}-w}(-)(P-P_{\pho})_{i}\frac{1}{H_{P,\si}-w}\check{\psi}_{P,\si}+{\bcc \{i\leftrightarrow i'\} }   \bigg) \\
& &+\bigg( \oint_{\gamma_{P,\si}}dw\frac{1}{H_{P,\si}-w}(-)(P-P_{\pho})_{i}\frac{1}{H_{P,\si}-w}  \bigg)\ 
\bigg(\oint_{\gamma_{P,\si}}d\mu\frac{1}{H_{P,\si}-\mu}(-)(P-P_{\pho})_{i'}\frac{1}{H_{P,\si}-\mu}\bigg)\check{\psi}_{P,\si}\quad\quad\quad
\label{derivative-computation-two}\\
\1&= &\1 \bigg( \oint_{\gamma_{P,\si}}dw\frac{1}{H_{P,\si}-w}(-)(P-P_{\pho})_{i'}\frac{1}{H_{P,\si}-w}(-)(P-P_{\pho})_{i}\frac{1}{H_{P,\si}-w}\check{\psi}_{P,\si}+{\bcc \{i\leftrightarrow i'\} }   \bigg) \label{second-deriv-first-term}\\
& & +\check{\psi}_{P,\si}\langle\check{\psi}_{P,\si},(-)(P-P_{\pho})_{i}\frac{\bar{Q}_{P,\si}^{\perp}}{(H_{P,\si}-E_{P,\si})^{2}}(-)(P-P_{\pho})_{i'}\check{\psi}_{P,\si}\rangle. \label{second-deriv-last-line}
\end{eqnarray}
{\bcc In the first step of this computation we noted that when $\pa_{i'}$ acts on the factor $(P-P_{\pho})_i$ in (\ref{first-deriv-section5}), the resulting
integrand has a (trivial) pole of the second order with vanishing residuum.  
 In the second step  we inserted twice the decomposition $1=\bar{Q}_{P,\si}+\bar{Q}_{P,\si}^{\bot}$, namely  next to the two integral signs in 
  (\ref{derivative-computation-two}). The $(\bar{Q}_{P,\si}^{\bot}, \bar{Q}_{P,\si}^{\bot})$--term vanishes since the corresponding $w$-integrand is holomorphic inside $\ga_{P,\si}$. The $(\bar{Q}_{P,\si}^{\bot}, \bar{Q}_{P,\si})$--term vanishes since the corresponding  $\mu$-integrand has a (trivial) pole of second order with vanishing residuum. 
The $(\bar{Q}_{P,\si}, \bar{Q}_{P,\si})$--term vanishes, since both the $w$- and $\mu$-integrands have (trivial) poles of the second order with vanishing residuum. The remaining 
$(\bar{Q}_{P,\si}, \bar{Q}_{P,\si}^{\bot})$--term was evaluated using the Cauchy integral formula and is stated in (\ref{second-deriv-last-line}).   }

Next, using the identity
\beqa
W_{P,\si}(-)(P-P_{\pho})W_{P,\si}^{*}=\Gamma_{P,\si}-\nabla E_{P,\si} \label{key-identity-W}
\eeqa
{\bcc we rearrange  formulas (\ref{first-deriv-section5}) and (\ref{second-deriv-first-term})-(\ref{second-deriv-last-line}). First, we write }
\begin{eqnarray}
W_{P,\si}\partial_{i}\check{\psi}_{P,\si}\1&= &\1 W_{P,\si}\oint_{\gamma_{P,\si}}dw\frac{1}{H_{P,\si}-w}(-)(P-P_{\pho})_{i}\frac{1}{H_{P,\si}-w}\check{\psi}_{P,\si}\\
\1&= &\1 \oint_{\gamma_{P,\si}}dw\frac{1}{H_{P,\si}^{W}-w}(\Gamma_{P,\si}-\nabla E_{P})_{i}\frac{1}{H_{P,\si}^{W}-w}\check{\phi}_{P,\si}\\
\1&= &\1 \frac{1}{H_{P,\si}^{W}-E_{P,\si}}(\Gamma_{P,\si})_{i}\check{\phi}_{P,\si}, \label{first-derivative-state}
\end{eqnarray} 
{\bcc where the term proportional to $\nabla E_{P,\si}$ gave a pole of second order with vanishing residuum, and to the term proportional to $\Gamma_{P,\si}$ we applied the
Cauchy integral formula. It was important here that $\Ga_{P,\si}\cphi_{P,\si}$ is in the range of $Q_{P,\si}^{\bot}$, to ensure the required holomorphy. 
Second, we compute  }
\beqa
W_{P,\si}\partial_{i'}\partial_{i}\check{\psi}_{P,\si}
\1&= &\1 \bigg(\oint_{\gamma_{P,\si}}dw\frac{1}{H_{P,\si}^{W}-w}(\Gamma_{P,\si}-\nabla E_{P,\si})_{i'}\frac{1}{H_{P,\si}^{W}-w}(\Gamma_{P,\si}-\nabla E_{P,\si})_{i}\frac{1}{H_{P,\si}^{W}-w}\check{\phi}_{P,\si}+ \{i\leftrightarrow i'\}  \bigg)\quad\quad\label{eq:second-derivative/1}\\
& &+\check{\phi}_{P,\si}\langle\check{\phi}_{P,\si},(\Gamma_{P,\si}-\nabla E_{P,\si})_{i}\frac{\bar{Q}_{P,\si}^{\perp}}{(H_{P,\si}-E_{P,\si})^{2}} (\Gamma_{P,\si}-\nabla E_{P,\si})_{i'}   \check{\phi}_{P,\si}\rangle\label{first-computation-zero} \\
\1&= &\1\bigg( \frac{Q_{P,\si}^{\perp}}{H_{P,\si}^{W}-E_{P,\si}}(\Gamma_{P,\si})_{i'}\frac{1}{H_{P,\si}^{W}-E_{P,\si}}(\Gamma_{P,\si})_{i}\check{\phi}_{P,\si} + \{i\leftrightarrow i'\}  \bigg) \label{second-d-v} \\
& & -\check{\phi}_{P,\si}\langle \check{\phi}_{P,\si},(\Gamma_{P,\si})_{i'}\frac{1}{(H_{P,\si}^{W}-E_{P,\si})^{2}}(\Gamma_{P,\si})_{i}\check{\phi}_{P,\si}\rangle. 
\label{last-term-second-derivative-section-5-x}
\eeqa
{\bcc Let us justify this computation:  Making use of the fact that $\lan \cphi_{P,\si},  (\Ga_{P,\si})_i \cphi_{P,\si}\ran=0$, we can write    }
\beqa
(\ref{first-computation-zero})=\check{\phi}_{P,\si}\langle\check{\phi}_{P,\si},(\Gamma_{P,\si})_{i}\frac{1}{(H_{P,\si}-E_{P,\si})^{2}} 
(\Gamma_{P,\si})_{i'}   \check{\phi}_{P,\si}\rangle, \label{first-computation-zero-section-5}
\eeqa
 and we note  the sign and $(i\leftrightarrow i')$ difference w.r.t. to (\ref{last-term-second-derivative-section-5-x}). Let us now discuss 
(\ref{eq:second-derivative/1}): The $(\nabla E_{P,\si}, \nabla E_{P,\si})$--terms vanish as  (trivial) poles of the third order with vanishing residua. Thus we can write
\beqa
(\ref{eq:second-derivative/1}) \1&=&\1 \bigg(\oint_{\gamma_{P,\si}}dw\frac{1}{H_{P,\si}^{W}-w}(\Gamma_{P,\si})_{i'}\frac{1}{H_{P,\si}^{W}-w}
(\Gamma_{P,\si})_{i}\frac{1}{H_{P,\si}^{W}-w}\check{\phi}_{P,\si}+ \{i\leftrightarrow i'\}  \bigg)  \label{surviving-term}\\
& &+\bigg(\oint_{\gamma_{P,\si}}dw\frac{1}{H_{P,\si}^{W}-w}(\Gamma_{P,\si})_{i'}\frac{1}{H_{P,\si}^{W}-w}(-\nabla E_{P,\si})_{i}\frac{1}{H_{P,\si}^{W}-w}\check{\phi}_{P,\si}+ \{i\leftrightarrow i'\}  \bigg) \label{Delta-E-one}\\
& & +\bigg(\oint_{\gamma_{P,\si}}dw\frac{1}{H_{P,\si}^{W}-w}(-\nabla E_{P,\si})_{i'}\frac{1}{H_{P,\si}^{W}-w}(\Gamma_{P,\si})_{i}\frac{1}{H_{P,\si}^{W}-w}\check{\phi}_{P,\si}+ \{i\leftrightarrow i'\}  \bigg). \label{Delta-E-two}
\eeqa
{\bcc Now we note that $(\ref{Delta-E-one})+(\ref{Delta-E-two})=0$. Indeed, by combining the explicitly stated term in (\ref{Delta-E-one}) with
the $\{i\leftrightarrow i'\}$ term in (\ref{Delta-E-two}) (and vice versa) we obtain}
\beqa
(\ref{Delta-E-one})+(\ref{Delta-E-two})=(-\nabla E_{P,\si})_{i}\oint_{\gamma_{P,\si}}dw\, \fr{d}{dw}\bigg(\frac{1}{H_{P,\si}^{W}-w}(\Gamma_{P,\si})_{i'}\frac{1}{H_{P,\si}^{W}-w}\bigg)
\check{\phi}_{P,\si}+\{i\leftrightarrow i'\}=0, 
\eeqa  
{\bcc where in the last step we used the fact that the derivative of a meromorphic function w.r.t. $w$ can only have a vanishing residuum, which follows by
inspection of the Laurent series. Given this, we can come back to (\ref{surviving-term}) and compute}
\beqa
(\ref{eq:second-derivative/1}) \1&=&\1 \bigg(\oint_{\gamma_{P,\si}}dw \,  \frac{1}{E_{P,\si}-w}    \frac{Q_{P,\si}^{\perp}}{H_{P,\si}^{W}-w}(\Gamma_{P,\si})_{i'}\frac{1}{H_{P,\si}^{W}-w} 
(\Gamma_{P,\si})_{i}\check{\phi}_{P,\si}+ \{i\leftrightarrow i'\}  \bigg) \label{eq:second-derivative-section5}\\
& &+\bigg(\oint_{\gamma_{P,\si}}dw \,  \frac{1}{(E_{P,\si}-w)^2} \cphi_{P,\si}\lan \cphi_{P,\si}, (\Gamma_{P,\si})_{i'}\frac{1}{H_{P,\si}^{W}-w}
(\Gamma_{P,\si})_{i}\check{\phi}_{P,\si}\ran+ \{i\leftrightarrow i'\}  \bigg) \label{non-trivial-pole-of-second-order} \\
\1&=&\1  \bigg(  \frac{Q_{P,\si}^{\perp}}{H_{P,\si}^{W}-E_{P,\si}}(\Gamma_{P,\si})_{i'}\frac{1}{H_{P,\si}^{W}-E_{P,\si}}
(\Gamma_{P,\si})_{i}\check{\phi}_{P,\si}+ \{i\leftrightarrow i'\}  \bigg) \label{last-term-second-derivative-section-5-zero}\\
& &-\bigg(\cphi_{P,\si}\lan \cphi_{P,\si}, (\Gamma_{P,\si})_{i'}\frac{1}{(H_{P,\si}^{W}-E_{P,\si})^2}
(\Gamma_{P,\si})_{i}\check{\phi}_{P,\si}\ran+ \{i\leftrightarrow i'\}  \bigg), \label{last-term-second-derivative-section-5}
\eeqa
where we applied the Cauchy integral formula to (\ref{eq:second-derivative-section5}) and computed the residuum of a meromorphic
function with a (non-trivial) pole of second order in (\ref{non-trivial-pole-of-second-order}). We remark that (\ref{last-term-second-derivative-section-5})
can alternatively be obtained from the Cauchy formula and
\beqa
\oint_{\gamma_{P,\si}}dw \, \fr{d}{dw} \bigg(\frac{1}{(E_{P,\si}-w)} \cphi_{P,\si}\lan \cphi_{P,\si}, (\Gamma_{P,\si})_{i'}\frac{1}{H_{P,\si}^{W}-w}
(\Gamma_{P,\si})_{i}\check{\phi}_{P,\si}\ran\bigg)=0.
\eeqa
Making use of (\ref{last-term-second-derivative-section-5-zero}), (\ref{last-term-second-derivative-section-5}), (\ref{first-computation-zero-section-5})
we justify the computation (\ref{eq:second-derivative/1})-(\ref{last-term-second-derivative-section-5-x}). \vspace{0.2cm}

Now we proceed to  the derivatives of the ground state energy up to third degree,  starting from the identity
\begin{equation}
\pa_i E_{P,\si}=\lan \check{\psi}_{P,\si}, (P-P_{\pho})_i\check{\psi}_{P,\si}\ran. \label{first-deriv}
\end{equation}
We first compute
\beqa
\pa_{i'}\pa_i E_{P,\si}\1&=&\1\de_{i,i'}+2\lan \pa_{i'}\check{\psi}_{P,\si}, (P-P_{\pho})_i\check{\psi}_{P,\si}\ran \label{main-formula-derivative-E}\\
\1&=&\1 \de_{i,i'}- 2\lan W_{P,\si} \pa_{i'}\check{\psi}_{P,\si}, (\Ga_{P,\si}-\nabla E_{P,\si})_i\check{\phi}_{P,\si}\ran \\
\1&=&\1 \de_{i,i'}- 2\lan \check{\phi}_{P,\si}, (\Gamma_{P,\si})_{i'} \frac{1}{H_{P,\si}^{W}-E_{P,\si}}  (\Ga_{P,\si})_i\check{\phi}_{P,\si}\ran, \label{E-sec-der}
\eeqa
where in the first step we used that $J\cpsi_{P,\si}=\cpsi_{P,\si}$, in the second step we applied (\ref{key-identity-W}) and in the last 
step we substituted (\ref{first-derivative-state}) and used that $\lan \cphi_{P,\si}, \Ga_{P,\si}\cphi_{P,\si}\ran=0$.

Next, coming back to (\ref{main-formula-derivative-E}) and using  $\lan \pa_{i'}\check{\psi}_{P,\si}, \check{\psi}_{P,\si}\ran=0$ and (\ref{key-identity-W}), 
we obtain
\beqa
\pa_{i''}\pa_{i'}\pa_i E_{P,\si}\1&=&\1 2\lan \pa_{i''}\pa_{i'}\check{\psi}_{P,\si}, (P-P_{\pho})_i\check{\psi}_{P,\si}\ran\non\\
& &+2\lan \pa_{i'}\check{\psi}_{P,\si}, (P-P_{\pho})_i\pa_{i''}\check{\psi}_{P,\si}\ran\non\\
\1&=&\1 -2\lan W_{P,\si} \pa_{i''}\pa_{i'}\check{\psi}_{P,\si}, (\Ga_{P,\si}-\nabla E_{P,\si})_i\check{\phi}_{P,\si}\ran \label{third-derivative-energy-one}\\
& &-2\lan W_{P,\si}\pa_{i'}\check{\psi}_{P,\si}, (\Ga_{P,\si}-\nabla E_{P,\si})_i W_{P,\si}\pa_{i''}\check{\psi}_{P,\si}\ran. \label{third-derivative-energy-two}
\eeqa
Now making use of (\ref{first-derivative-state}), we can write
\beqa
(\ref{third-derivative-energy-two})=-2\lan \frac{1}{H_{P,\si}^{W}-E_{P,\si}}(\Gamma_{P,\si})_{i'}\check{\phi}_{P,\si}, (\Ga_{P,\si}-\nabla E_{P,\si})_i \frac{1}{H_{P,\si}^{W}-E_{P,\si}}(\Gamma_{P,\si})_{i''}\check{\phi}_{P,\si}\ran. \label{third-derivative-energy-two-repeated}
\eeqa
Furthermore, exploiting (\ref{second-d-v})-(\ref{last-term-second-derivative-section-5-x}) we obtain
\beqa
(\ref{third-derivative-energy-one})\1&=&\1-2\bigg(\lan \frac{Q_{P,\si}^{\perp}}{H_{P,\si}^{W}-E_{P,\si}}(\Gamma_{P,\si})_{i''}\frac{1}{H_{P,\si}^{W}-E_{P,\si}}(\Gamma_{P,\si})_{i'}\check{\phi}_{P,\si}, (\Ga_{P,\si})_i\check{\phi}_{P,\si}\ran+\{i'\leftrightarrow i''\}\bigg)\non\\
& &-2(\nabla E_{P,\si})_i \langle \check{\phi}_{P,\si},(\Gamma_{P,\si})_{i''}\frac{1}{(H_{P,\si}^{W}-E_{P,\si})^{2}}(\Gamma_{P,\si})_{i'}\check{\phi}_{P,\si}\rangle.
\label{third-derivative-last-term}
\eeqa
Noting that $(\Gamma_{P,\si})_{i'}\check{\phi}_{P,\si}=J(\Gamma_{P,\si})_{i'}\check{\phi}_{P,\si}$, we can swop the indices $i', i''$ in (\ref{third-derivative-last-term}),
which then cancels with the part of (\ref{third-derivative-energy-two-repeated}) proportional to $(\nabla E_{P,\si})_i$. Thus, altogether,
\beqa
\pa_{i''}\pa_{i'}\pa_i E_{P,\si}\1&=&\1-2\bigg(\lan \frac{Q_{P,\si}^{\perp}}{H_{P,\si}^{W}-E_{P,\si}}(\Gamma_{P,\si})_{i''}\frac{1}{H_{P,\si}^{W}-E_{P,\si}}(\Gamma_{P,\si})_{i'}\check{\phi}_{P,\si}, (\Ga_{P,\si})_i\check{\phi}_{P,\si}\ran+\{i'\leftrightarrow i''\}\bigg)\non\\
& &-2\lan \frac{1}{H_{P,\si}^{W}-E_{P,\si}}(\Gamma_{P,\si})_{i'}\check{\phi}_{P,\si}, (\Ga_{P,\si})_i \frac{1}{H_{P,\si}^{W}-E_{P,\si}}(\Gamma_{P,\si})_{i''}\check{\phi}_{P,\si}
\ran\non\\
\1&=&\1- \bigg(\lan \check{\phi}_{P,\si},(\Gamma_{P,\si})_{i} \frac{1}{H_{P,\si}^{W}-E_{P,\si}}   (\Ga_{P,\si})_{i'} \frac{1}{H_{P,\si}^{W}-E_{P,\si}}(\Gamma_{P,\si})_{i''}\check{\phi}_{P,\si}\ran+\{\mathrm{perm}\}\bigg),
\eeqa
where $\{\mathrm{perm}\}$ denotes all the remaining permutations of the $i,i',i''$ indices and in the last step we made use again of $(\Gamma_{P,\si})_{i'}\check{\phi}_{P,\si}=J(\Gamma_{P,\si})_{i'}\check{\phi}_{P,\si}$ and of the fact that $\lan \cphi_{P,\si}, \Ga_{P,\si}\cphi_{P,\si}\ran=0$.

Making use of the formulas above and Corollary~\ref{technical-corollary}, we immediately obtain
\beqa
& &|\pa_{i}E_{P,\si}|\leq c, \quad | \pa_{i''}\pa_{i'}\pa_i  E_{P,\si}|, \   \|\pa_i\cpsi_{P,\si}\|, \  \|\pa_{i'}\pa_i\cpsi_{P,\si}\|   \leq c/\si^{\de_{\la_0}}, 
\eeqa
The remaining estimate $|\pa_{i'}\pa_{i}E_{P,\si}|\leq c$ from Theorem~\ref{preliminaries-on-spectrum}  (a) and  the convexity statement
are proven in Appendix~\ref{convexity}.

\appendix

\section{Proof of Proposition~\ref{prop:Hamiltonian-results} } \label{proof-of-basic-lemma}
\setcounter{equation}{0}
In the proof of Proposition~\ref{prop:Hamiltonian-results}, which is contained in Lemmas~\ref{energy-shift}, \ref{resolvent}, \ref{inductive-theorem}, \ref{upper-bound-lemma}, \ref{upper-bound-lemma-one}, and formula~(\ref{Gap-bound}) below, we follow mostly \cite{BDP12}, 
but most ideas date back to \cite{Pi03}. 
First, we define
\beqa
& &\Phi|_{n}^{m}:=\hint_{\!\!\!\si_{n}}^{\si_{m}}d^3k\frac{1}{\sqrt{2|k|} }\{b(k)+b^{*}(k)\}, \label{Phi}\\
& &H_{\pho}|_n^m:=\int_{\si_n\leq |k|\leq \si_m} d^3k\,|k|\,b^*(k)b(k), \\
& &P_{\pho}|_n^m:=\int_{\si_n\leq |k|\leq \si_m} d^3k\,k\,b^*(k)b(k), \label{P-pho}
\eeqa
for $m\leq n$ as operators on (a domain in) $\mcF$. To analyse these quantities we will often use the energy bounds {\color{\red}reported in Lemmata \ref{energy-bounds}-\ref{simple-bound-one} below}:
\bel\cite[formula (22)]{BDP12}\label{energy-bounds} Let $f\in L^2(\real^3)$ be s.t. the integrals on the r.h.s. below are well defined and $\psi\in D(H^{1/2}_{P,\free})$. Then, 
for $m\geq n$,
\beqa
& &\|\int_{\si_n\leq |k|\leq \si_m} d^3k\, f(k)b(k)\psi\|\leq \bigg(\int_{\si_n\leq |k|\leq \si_m} d^3k\, 
\bigg|\fr{f(k)}{\sqrt{|k|}}\bigg|^2\bigg)^{1/2}
\|(H_{\pho}|_n^m)^{1/2}\psi\|,  \label{first-energy-bound-xx}\\
& & \|\int_{\si_n\leq |k|\leq \si_m} d^3k\, f(k)b^*(k)\psi\|\leq \bigg(\int_{\si_n\leq |k|\leq \si_m} d^3k\, 
\bigg|\fr{f(k)}{\sqrt{|k|}}\bigg|^2\bigg)^{1/2}
\|(H_{\pho}|_n^m)^{1/2}\psi\|\non\\ 
& &\ph{444444444444444444444444}+\bigg(\int_{\si_n\leq |k|\leq \si_m} d^3k\, |f(k)|^2\bigg)^{1/2}\|\psi\|.
\eeqa 
\eel
\nin Next, we fix values of various parameters:
\beqa
& & 0<\epsilon\leq 1/2, \quad |\la|\leq \epsilon^2, 
\label{first-constraints} \\
& & \theta=1/12, \ \ \quad 0<\xi<1/3,  \quad C^{\la}_{\nabla E}=1/3+\mathcal{O}(\la), \quad 
\eeqa
where $|O(\la)|\leq c|\la|$. We demand  that
\beqa
1-\theta-C^{\la}_{\nabla E}\geq 2\xi. \label{parameter-restriction}
\eeqa
We  list some  preparatory facts.
\bel\cite[Theorem 8]{Gro72}\label{Gross} $E_{0,n}\leq E_{P,n}, \quad P\in \real^3$.
\eel
\bel\label{energy-bound-zero}\cite[Lemma 3.2]{BDP12} Let $H_{P,\free}$ be the full free Hamiltonian {\color{\red}defined in (\ref{free-Hamiltonian})}. Then  there exists a finite 
constant $c_a>0$ s.t. for all $P\in \real^3$
\beqa
\lan \psi, H_{P,\free}\psi\ran\leq \fr{1}{1-|\la|c_a}\big[\lan \psi, H_{P,n}\psi\ran+|\la|c_{a}\lan \psi,\psi\ran\big]
\eeqa 
for $|\la|\leq 1, 1/c_a$ and $\psi\in D(H_{P,\free}^{1/2})$.
\eel
\proof {\color{\red}From the identity $H_{P,n}=H_{P,\free}+\la \Phi|^0_n$, where the last term denotes the interaction with cut-off $\si_n$, and from Lemma~\ref{energy-bounds}, we can write for $\psi\in D(H_{P,\free}^{1/2})$}
\beqa
\lan \psi, H_{P,\free}\psi\ran=\lan \psi, H_{P,n}\psi\ran-\lan \psi, \la \Phi|^0_n\psi\ran\,.
\eeqa
{\color{\red}Furthermore, the} energy bounds of Lemma~\ref{energy-bounds} give
\beqa
|\lan \psi, \la \Phi|^0_n\psi\ran|\leq |\la| c_a(\lan \psi, H_{P,\free}\psi\ran+\lan \psi, \psi\ran)
\eeqa
{\color{\red} for some $c_a>0$.}
Altogether $\lan \psi, H_{P,\free}\psi\ran{\color{\red}\leq}\lan \psi, H_{P,n}\psi\ran+|\la| c_a(\lan \psi, H_{P,\free}\psi\ran+\lan \psi, \psi\ran)$,
which gives the claim. \qed\\
\nin In the following we will assume that $|\la|\in (0, \la_0]$ for some $0<\la_0\leq 1/(2c_a)$. The value of $\la_0$ will be
tacitly reduced in the course of our discussion but it will remain non-zero.   
\bel\cite[Corollary 5.4]{BDP12}\label{simple-bound-one} The following inequalities hold true:
$-|\la|c_a\leq E_{P,n}\leq \h P^2, \quad  P\in \real^3, \quad |\la|\in (0,\la_0]$.
\eel
\proof We note that $E_{P,n}\leq \lan\Om,H_{P,n}\Om\ran\leq  P^2/2$.
Moreover, Lemma~\ref{energy-bound-zero} gives for {\color{\red} $\psi\in D(H_{P,n}) \subset D(H_{P,\free}^{1/2})$}
\beqa
0\leq (1-|\la|c_a)\inf_{\|\psi\|=1}\lan \psi, H_{P,\free}\psi\ran\leq \inf_{\|\psi\|=1}   \big[\lan \psi, H_{P,n}\psi\ran+|\la|c_{a}\lan \psi,\psi\ran\big]= E_{P,n}+|\la|c_{a}
\textrm{ \qed}.
\eeqa 

\nin Now we are ready to prove estimate (\ref{eq:difference-bound}) of Lemma~\ref{prop:Hamiltonian-results}.
\bel\label{energy-shift} The following inequality holds true:  $E_{P-k,n}-E_{P,n}\geq-C^{\la}_{\nabla E}|k|,\quad k\in\mathbb{R}^{3}, \,\, \text{for}\,\,  |P|\leq P_{\max}=1/3, \quad |\la|\in (0,\la_0]$.
\eel
\begin{rem} The bound $|\nabla E_{P,n}|\leq C^{\la}_{\nabla E}$ easily follows.
\end{rem}
\proof The argument follows closely \cite[Section 6]{CFP09} and \cite[Corollary 5.4]{BDP12}. {\color{\red}By definition of $E_{P,n}:=\inf \pmb{\sigma}(H_{P,n})$}, 
\beqa
E_{P-k,n}-E_{P,n} \1&=& \1\inf_{\|\psi\|=1}[\lan \psi, (H_{P-k,n}-H_{P,n})\psi\ran+\lan \psi, H_{P,n}\psi\ran-E_{P,n} ]\non\\
 \1&\geq & \1\inf_{\|\psi\|=1}\bigg[ \fr{|k|^2}{2}-|k|  |\lan \psi, (P-P_{\pho})   \psi\ran|+\lan \psi, H_{P,n}\psi\ran-E_{P,n} \bigg],
\label{two-estim}
\eeqa
where $ {\color{\red}\psi\in D(H^{1/2}_{P,\free})}$.
By Lemma~\ref{energy-bound-zero} and $\|\psi\|=1$
\beqa
|\lan \psi, (P-P_{\pho})   \psi\ran|^2 \leq \lan \psi, (P-P_{\pho})^2  \psi\ran\leq 2\lan\psi,H_{P,\free}\psi\ran\leq \fr{2}{1-|\la|c_a}\big[\lan \psi, H_{P,n}\psi\ran
+|\la|c_{a}\lan\psi,\psi\ran\big].\label{one-estim} 
\eeqa
From  (\ref{one-estim}) and (\ref{two-estim}) we get
\beqa
E_{P-k,n}-E_{P,n} \1&\geq& \1\inf_{\|\psi\|=1}\bigg[\fr{|k|^2}{2}-|k|\fr{\sqrt{2}}{\sqrt{1-|\la|c_a}} \sqrt{\lan \psi, H_{P,n}\psi\ran +|\la|c_{a}}+\lan \psi, H_{P,n}\psi\ran-E_{P,n}\bigg]\non\\
 \1&\geq& \1\inf_{\La\geq 0 }\bigg[\fr{|k|^2}{2}-|k|\fr{\sqrt{2}}{\sqrt{1-|\la|c_a}} \sqrt{\La+E_{P,n} +|\la|c_{a}}+
\La \bigg]=:\inf_{\La\geq 0} f(\La).
\eeqa
The infimum is either achieved at $\La^*=0$ or at the point where the derivative of $f$ vanishes, i.e.,
\beqa\label{a16}
\La^*=\fr{|k|^2}{4}\fr{2}{1-|\la|c_a}-(E_{P,n}+|\la|c_a).
\eeqa
$\Lambda$ is by construction a nonnegative quantity. First we consider the case $\La^*=0$. We have
\beqa
f(0)\geq -|k|\fr{\sqrt{2}}{\sqrt{1-|\la|c_a}} \sqrt{E_{P,n} +|\la|c_{a}}.
\eeqa
By Lemma~\ref{simple-bound-one}, we get
\beqa\label{a18}
E_{P,n} +|\la|c_{a}\leq \fr{P^2}{2}+|\la|{\color{\red}c_a}\leq \fr{P^2_{\max}}{2}+|\la|{\color{\red}c_a}={\color{\red}\fr{1}{2}\bigg(\frac{1}{3}\bigg)^2+|\la|c_a}.
\eeqa
Therefore
\beqa
f(0)\geq-|k|P_{\max}(1+\mco(\la)). \label{first-case}
\eeqa
Now we look at the case of $\La^*>0$ . We have
\beqa
f(\La^*)=\fr{k^2}{2}\bigg(1- \fr{1}{1-|\la|c_a}\bigg)-(E_{P,n}+|\la|c_a). \label{function-Lambda}
\eeqa
We proceed as in \cite{CFP09}: It is clear from Lemma~\ref{simple-bound-one} that $E_{P,n}+|\la|c_a\geq 0$.
The constraint $\La^*> 0$ can be fulfilled only if {\color{\red}(see (\ref{a16})) }
\beqa \label{a21}
\sqrt{E_{P,n}+|\la|c_a} < \fr{|k|}{\sqrt{2}}(1+c|\la|).
\eeqa
{\color{\red}By combining (\ref{a21}) with (\ref{a18})}, we have
\beqa
E_{P,n}+|\la|c_a < \fr{|k|}{\sqrt{2}}(1+c|\la|)\sqrt{E_{P,n}+|\la|c_a}\leq \fr{|k|}{\sqrt{2}}(1+c|\la|)\sqrt{\h\bigg(\fr{1}{3}\bigg)^2+ |\la|c_a}\leq \fr{|k|}{3}(1+c'|\la|).\quad
\eeqa
Thus we get from (\ref{function-Lambda}) that for $|k|\leq 1$,
\beqa
f(\La^*)\geq -k^2c|\la|-\fr{|k|}{3}(1+c'|\la|)\geq-\fr{|k|}{3}(1+c''|\la|).
\eeqa
Recall that $P_{\max}=1/3$. Therefore, using (\ref{first-case}), we have altogether for $|k|\leq 1$
\beqa
E_{P-k,n}-E_{P,n}\geq -C_{\nabla E}^{\la} |k| \label{step}
\eeqa
 for $C^{\la}_{\nabla E}=1/3+\mco(\la)$. Now we consider the case $|k|\geq 1$. We have by Lemma~\ref{Gross}
\beqa
& &E_{P-k,n}-E_{P,n}=(E_{P-k,n}-E_{0,n})+(E_{0,n}-E_{P,n})\geq (E_{0,n}-E_{P,n})\non\\
& &\ph{444444444444444444444444444444}\geq -{\color{\red}C^{\lambda}_{\nabla E}} |P_{\max}|\geq -{\color{\red}C^{\lambda}_{\nabla E}}|k|,
\eeqa
where in the {\bcb second line } we used (\ref{step}) in the case $k=P$. \qed\\
The following statement is similar to \cite[Lemma 5.5]{BDP12}, except that  we introduce an additional parameter $\eta$ 
in order to suitably adjust our constants. 

\bel\label{previous} Let $|\la|\in (0, \la_0]$ and $|P|\leq P_{\max}=1/3$. Assume that for some 
$7/4> \eta\geq 1$:
\begin{enumerate}
\item $E_{P,n-1}$ is the non-degenerate eigenvalue of 
$H_{P,n-1}\res\mcF_{n-1}$ with eigenvector $\psi_{P,n-1}$.
\item $\mathrm{Gap}(H_{P,n-1}\res\mcF_{n-1})\geq \eta\xi\si_{n-1}$.
\end{enumerate}
This implies that $E_{P,n-1}$ is also the non-degenerate ground state
energy of $H_{P,n-1}\res \mcF_{n}$ with eigenvector $\psi_{P,n-1}\otimes \Om$. Furthermore,
\beqa
\mathrm{Gap}(H_{P,n-1}\res\mcF_{n})\geq 
\inf_{\mcF_n\ni \psi\bot \psi_{P,n-1}\otimes \Om} \lan H_{P,n-1}-1\otimes\theta H_{\pho}|^{n-1}_n-E_{P,n-1}\ran_{\psi}\geq 2\xi\si_n, 
\label{Gap-bound}
\eeqa
where $\lan\,\cdot\, \ran_{\psi}:=\fr{\lan \psi, \,\cdot\, \psi\ran}{\lan \psi, \psi\ran}$ and the infimum is taken over non-zero $\psi\in D(H_{P,\free})$. 
\eel
\begin{rem} {\color{\red}From (\ref{infimum-bounds})  below and from Lemma \ref{energy-shift}},  it can be concluded  that for $(1-\theta-C_{\nabla E}^{\la})\geq 0$
\beqa
\inf_{\mcF_n\ni \psi\bot \psi_{P,n-1}\otimes \Om} \lan H_{P,n-1}-1\otimes \theta H_{\pho}|^{n-1}_n-E_{P,n-1}\ran_{\psi}\geq 0 
\label{infimum-remark}
\eeqa
without assuming $1.,2.$.
\end{rem}
\proof 
Using 1. a computation gives  $H_{P,n-1}(\psi_{P,n-1}\otimes \Om)=E_{P,n-1}(\psi_{P,n-1}\otimes \Om)$,
since the interaction (but not $H_{\pho}$, $P_{\pho}$ inside of $H_{P,n-1}$)  is cut-off at $\si_{n-1}$. To prove the  simplicity of the eigenvalue,  we write
\beqa
\inf_{\mcF_n\ni {\color{black}\varphi\bot \psi_{P,n-1}}\otimes \Om} 
\lan H_{P,n-1}-E_{P,n-1}\ran_{{\color{black}\varphi}}\geq 
\inf_{\mcF_n\ni {\color{black}\varphi}\bot \psi_{P,n-1}\otimes \Om} 
\lan H_{P,n-1}-1\otimes \theta H_{\pho}|^{n-1}_n-E_{P,n-1}\ran_{{\color{black}\varphi}} \,\,\,\,\,\, \label{pre-minimum}
\eeqa
which is the  first inequality in (\ref{Gap-bound}). Set $P_{\psi}:=(|\psi_{P,n-1}\ran\lan \psi_{P,n-1}|)/\|\psi_{P,n-1}\|^2$  
and $P_0:=|\Om\ran \lan \Om|$. Note that $X:=(P_{\psi}\otimes P_0)^{\bot}\mcF_n$ is an invariant
subspace of the operator $A:= H_{P,n-1}-1\otimes \theta H_{\pho}|^{n-1}_n-E_{P,n-1}$, thus we can
consider this operator on this subspace. Furthermore, $(1\otimes P_0)$ commutes with $(P_{\psi}\otimes P_0)$
and $A$, so we can write
\beqa
\mathrm{r.h.s.}(\ref{pre-minimum})= \inf\sib(A\res X)=\min\big( \inf\sib( (1\otimes P_0)A\res X), \ \inf\sib( (1\otimes P_0)^{\bot}A\res X)\big).
\eeqa
By considering $\lan (1\otimes P_0)A\ran_{{\color{black}\varphi}}$ for ${\color{black}\varphi}=\sum_{i,i'}c_{i,i'}e^{(1)}_i\otimes e^{(2)}_{i'}$, where the orthonormal bases $\{e^{(j)}_i\}_{i\in\nat_0}$ are chosen so that $e^{(1)}_0=\psi_{P,n}/ \|\psi_{P,n}\|$ and $e^{(2)}_0=\Om$,  it is easy to see that $\inf\sib( (1\otimes P_0)A\res X)=\mathrm{Gap}(H_{P,n-1}\res\mcF_{n-1})$.  Let us now  analyse the
second term: Let $N$ be the number operator on $\mcF|^{n-1}_{n}$ and $P_{m}$  its spectral projections corresponding to eigenvalues $m\in \nat_0$.
Since $1\otimes P_{m}$ commute with $A$ and  $(P_{\psi}\otimes P_0)$, we can write 
\beqa
\inf\sib( (1\otimes P_0)^{\bot}A\res X)=\inf_{m\geq 1}\inf\sib((1\otimes P_{m})(1\otimes P_0)^{\bot}A\res X)\geq 
\inf_{m\geq 1}\inf\sib( (1\otimes P_{m})A). \label{sigmas}
\eeqa 
Now we can  write
\beqa
& &(1\otimes P_{m})A\simeq \int^{\oplus}_{K_m} d^{3m}k\,(P_{m}A)(k), \\
& &(P_{m}A)(k):= \h\big(P-P_{\pho}-\sum_{j=1}^mk_j \big)^2+H_{\pho}+
\la \Phi|_{n-1}^0+(1-\theta)\sum_{j=1}^m|k_j|-E_{P,n-1},
\eeqa
where $K_m:=\{(k_1, \ldots, k_m)\in \real^{3m}\,|\, \si_n\leq|k_i|\le \si_{n-1}\,\}$.
Consequently,
\beqa
\inf_{m\geq 1}\inf\sib((1\otimes  P_{m})A)=\inf_{m\geq 1}\inf_{k\in K_m} \sib\big((P_{{\color{\red}m}}A)(k)\big)=
\inf_{m\geq 1}\inf_{k\in K_m}\inf_{\vp\in \mcF_{n-1}}\lan (P_{{\color{\red}m}}A)(k)\ran_{\vp}.
\eeqa
Thus we have
\beqa
& &\inf_{k\in K_m}\inf_{\vp\in \mcF_{n-1}}\lan (P_{m}A)(k)\ran_{\vp}   \non\\
& &\geq {\color{black}\inf_{k\in K_m}\inf_{\vp\in \mcF_{n-1}}} \big\lan \h\big(P-P_{\pho}-\sum_{j=1}^mk_j \big)^2+H_{\pho}+
\la \Phi|^0_{n-1}+(1-\theta)\sum_{j=1}^m
|k_j| -E_{P,n-1}\big\ran_{\vp}\non\\
& &\geq \inf_{ k_j \in [\si_n,\si_{n-1})    }\big( (1-\theta)\sum_{j=1}^m |k_j|+E_{P-\sum_{j=1}^m k_j, n-1} -E_{P,n-1}\big). \label{infimum-bounds}
\eeqa
Now we note that by Lemma~\ref{energy-shift}
\beqa
E_{P-\sum_{j=1}^m k_j, n-1}-E_{P,n-1}\geq -
C^{\la}_{\nabla E}\sum_{i=1}^m|k_i|.
\eeqa
Hence, by (\ref{parameter-restriction}),
\beqa
(\ref{infimum-bounds})\geq \inf_{ k_j \in [\si_n,\si_{n-1})    }\bigg( (1-\theta-C^{\la}_{\nabla E})\sum_{j=1}^m |k_j|     \bigg)\geq  
(1-\theta-C^{\la}_{\nabla E})\si_{n}\geq 2\xi \si_n.
\eeqa
Summing up,
\beqa
{\color{\red}\inf_{\mcF_n\ni \psi\bot \psi_{P,n-1}\otimes \Om} 
\lan H_{P,n-1}-E_{P,n-1}\ran_{\psi}}\geq \min\{ \mathrm{Gap}(H_{P,n-1}\res\mcF_{n-1}), 2\xi \si_n\}=\min(\eta\xi\si_{n-1},2\xi\si_n)\geq 2\xi\si_n.\quad 
\eeqa
Here we made use of the fact that $\epsilon \leq 1/2$ and $\eta\geq 1$. This also
implies that: a) $\psi_{P,n-1}\otimes \Om$ is the non-degenerate ground state of $H_{P,n-1}\res\mcF_{n}$ since we looked at all the vectors
orthogonal to it; b) {\color{\red}$\mathrm{Gap}(H_{P,n-1}\res\mcF_{n})=\inf_{\mcF_n\ni \psi\bot \psi_{P,n-1}\otimes \Om} 
\lan H_{P,n-1}-E_{P,n-1}\ran_{\psi}\geq 2\xi\si_n$} . \qed

\bel\label{resolvent}\cite[Lemma 5.6]{BDP12} For $n\geq 1$ and $|\la|\in (0,\la_0]$ 
the assumptions of Lemma~\ref{previous}
imply that the resolvent
\beqa\label{res-a38}
\fr{1}{H_{P,n}-z}
\eeqa
restricted to $\mcF_n$ is well-defined in the domain
\beqa
\fr{1}{4}\xi\si_n\leq |E_{P,n-1}-z|\leq \eta\xi\si_n. \label{contour-bounds}
\eeqa
\eel
\proof We recall that $H_{P,n}\res \mcF_n=H_{P,n-1}\res \mcF_{n}
+\la\Phi|^{n-1}_n$. We note that by Lemma~\ref{previous}  $z$ is in the resolvent set of $H_{P,n-1}\res \mcF_{n}$. 
 Thus for $A(z):=H_{P,n-1}\res \mcF_{n}-z$ we can define $A(z)^{-1}$ and $A(z)^{-1/2}$ (for some choice of the square root). Setting
 $B:=\la\Phi|^{n-1}_n$ we can thus define the resolvent in {\color{\red}(\ref{res-a38})} by
\beqa
(A(z)+B)^{-1}
:=A(z)^{-1/2}\sum_{j=0}^{\infty}[A(z)^{-1/2}BA(z)^{-1/2}]^j A(z)^{-1/2},
\eeqa
provided that
\beqa
\|A(z)^{-1/2}BA(z)^{-1/2}\|_{\mcF_n}<1 \label{ABA}
\eeqa
within the specified restrictions. We denote by $B^{(\pm)}$ the creation and annihilation part of $B$. We have by the energy bounds from Lemma~\ref{energy-bounds}
\beqa
& &\| B^{(-)}A(z)^{-1/2}\|_{\mcF_n}\leq |\la| c((1-\epsilon)\si_{n-1})^{1/2}\|(H_{\pho}|^{n-1}_{n})^{1/2} A(z)^{-1/2}\|_{\mcF_n}, \\
& &\| A(z)^{-1/2} B^{(+)} \|_{\mcF_n}=\|B^{(-)} (A(z)^{-1/2})^*\|_{\mcF_n}\leq {\color{black}|\la|}c((1-\epsilon)\si_{n-1})^{1/2}\|(H_{\pho}|^{n-1}_{n})^{1/2} (A(z)^{-1/2})^*\|_{\mcF_n},
\eeqa
and we have $\|(H_{\pho}|^{n-1}_{n})^{1/2} (A(z)^{-1/2})^*\|_{\mcF_n}=\|(H_{\pho}|^{n-1}_{n})^{1/2} (A(z)^{-1/2})\|_{\mcF_n}$,
since $H_{\pho}|^{n-1}_{n}$ and $H_{P,n-1}\res \mcF_n$ commute. By the same token
\beqa
& &\|(H_{\pho}|^{n-1}_{n})^{1/2} (H_{P,n-1}-z)^{-1/2}\|_{\mcF_n}\non\\
& &=\theta^{-1/2}\|(\theta H_{f}|^{n-1}_{n})^{1/2} 
(H_{P,n-1}-\theta H_{\pho}|^{n-1}_{n}-E_{P,n-1}+\theta H_{\pho}|^{n-1}_{n}+E_{P,n-1}-z)^{-1/2}\|_{\mcF_n}\non\\
& &\leq \theta^{-1/2} \|(\theta H_{\pho}|^{n-1}_{n})^{1/2} 
(2\xi\si_n-\eta\xi\si_n+\theta H_{\pho}|^{n-1}_{n})^{-1/2}\|_{\mcF_n}\leq \theta^{-1/2},
\label{A-eighty-four}
\eeqa
where we made use of {\color{\red}(\ref{Gap-bound})} of Lemma~\ref{previous} {\color{\red}combined with} the fact that 
$\theta H_{\pho}|^{n-1}_{n}$ effectively projects onto the orthogonal
complement of the ground state of $H_{P,n-1}$ on $\mcF_n$. Now using 
Lemma~\ref{previous}, we get
\beqa
\|A(z)^{-1/2}\|_{\mcF_n}^2 \1&=& \1\|(H_{P,n-1}-E_{P,n-1}+E_{P,n-1} -z)^{-1/2}\|_{\mcF_n}^2\non\\
 \1&\leq& \1 \max\bigg\{ \fr{1}{(1/4)\xi\si_n}, \fr{1}{(2-\eta)(\xi\si_n)}  \bigg\}=\fr{4}{\xi\si_n}, \label{bound-A-app}
\eeqa
where $(2-\eta)\xi\si_n=2\xi\si_n-\eta\xi\si_n$ comes from  (\ref{Gap-bound})
and (\ref{contour-bounds}) and we note that $(2-\eta)>1/4$. 
Thus altogether, 
\beqa\label{bound-a46}
(\ref{ABA})\leq c|\la|\bigg( \fr{\si_{n-1}}{\si_n} \bigg)^{1/2}\leq c|\la|^{3/4}<1,
\eeqa
since $|\la|\leq \epsilon^2$ and $\theta=1/12$ and the last inequality may require to reduce $\la_0$.
\qed\\
\nin\textbf{ Now we choose $\eta=4/3$  and $\xi=1/4$ s.t. $\eta\xi=1/3$ and define the contour}
\beqa
\ga_{P,n}:=\bigg\{\,z\in\complex\,\bigg|\, |E_{P,n-1}-z|=\fr{1}{3}\si_n\bigg\}.
\eeqa
\bel\label{inductive-theorem}\cite[Theorem 4.8]{BDP12} For all $n\geq 0$ and $|\la|\in (0,\la_0]$ the following holds true:
\begin{enumerate}
\item[i)] $E_{P,n}:=\inf \sib(H_{P,n}\res \mcF)=\inf \sib(H_{P,n}\res \mcF_n)$ is the non-degenerate ground state energy of $H_{P,n}\res \mcF_n$. 
\item[ii)] $\mathrm{Gap}(H_{P,n}\res \mcF_n)\geq \eta\xi\si_n=(1/3)\si_n$.
\item[iii)] The vectors $\psi_{P,0}=\Om$ and $\psi_{P,n}:=\bar Q_{P,n}\psi_{P,n-1}$, where {\bcb $\bar Q_{P,n}: \mcF_n\to \mcF_n$},
\beqa
\bar Q_{P,n}:=\oint_{\ga_{P,n}}\fr{d{\bcb w_n} }{H_{P,n}-{\bcb w_n}}, \quad n\geq 1\,, \label{Q-def-app-A}
\eeqa
are well defined and non-zero. The vector $\psi_{P,n}$ is the unique (up to phase and normalization) ground state of $H_{P,n}\res \mcF_n$. 
\end{enumerate}
\eel
\proof The proof is by induction. Statements i), ii), iii) for $(n-1)$ will be referred to
as assumptions a-i), a-ii), a-iii) while the same statements for $n$ as claims c-i), c-ii), c-iii).

For $n=0$ the claims can be verified as follows:  We show that $P^2/2$ is an isolated eigenvalue of $H_{P,0} \res  \mcF_0$  with a gap larger than $(1/3)\si_0=1/3$. We have
\beqa
H_{P,0}=\h (P-P_{\pho})^2+H_{\pho} 
\eeqa
and we consider this operator on $\mcF_0:=\Ga(L^2(\real^3\backslash B_{\si_0}), d^3k)$.
We note that  $\Om$ is an eigenvector and the corresponding eigenvalue is $P^2/2$. 
The gap on the $m$-particle subspace is determined  by the following function of $|k_i|\geq 1$:
\beqa
f(k_1,\ldots, k_m) \1&=& \1\h (P-\sum_{i=1}^mk_i)^2+\sum_{i=1}^m|k_i|-\fr{1}{2}P^2\geq -P\cdot \sum_{i=1}^m k_i+\sum_{i=1}^m|k_i|\non\\
 \1&\geq& \1(1-|P_{\max}|)\sum_{i=1}^m|k_i|\geq 2/3=(2/3)\si_0\geq (1/3) \si_0.
\eeqa
Hence $P^2/2=\inf\sib(H_{P,0} \res \mcF_0)=\inf\sib(H_{P,0} \res \mcF)=:E_{P,0}$ and the gap on $\mcF_{0}$ is bounded as required.

Now we proceed with the inductive argument:
\begin{enumerate}
\item We have by a-i), a-ii) and Lemma~\ref{previous}
\beqa
\mathrm{Gap}(H_{P,n-1}\res \mcF_n)\geq \fr{1}{2} \si_n. \label{gap-bound}
\eeqa
By Lemma~\ref{resolvent} we get that the resolvent $(H_{P,n}-z)^{-1}\res \mcF_n$  is well defined for $(1/16)\si_n\leq  |E_{P,n-1}-z|\leq  (1/3)\si_n$. (Since the resolvent set is open, it actually must be well defined in a slightly larger set). 

\item Now we use Kato perturbation theory: We note that
$H_{P,n}\res \mcF_n=H_{P,n-1}\res\mcF_n+\la^{*}\Phi|^{n-1}_n$ for $\la^*=\la$. We introduce auxiliary Hamiltonians $H_{P,n}^{\la^*}$ defined as above with $|\la^*|\leq |\la|$. We note that the corresponding expression for the ground state projection $\bar Q_{P,n}^{\la^*}$   is well-defined for small $\la^*$ by the analytic perturbation theory and the bound (\ref{gap-bound}) on the gap of the `unperturbed' Hamiltonian. But by Lemma~\ref{resolvent} the formula can be extended  to $\la^*=\la$.
(Indeed, if some piece of the spectrum would hit the contour of integration as $|\la^*|$
increases, the norm of the resolvent would blow up. But this cannot happen by Lemma~\ref{resolvent}). 
Thus Kato theory gives that $E_{P,n}$ is an isolated, simple eigenvalue of $H_{P,n}$. Since $E_{P,n}\leq E_{P,n-1}$ 
by Lemma~\ref{upper-bound-lemma} below, we also have
that 
\beqa
\mathrm{Gap}(H_{P,n}\res \mcF_n)\geq (1/3) \si_n,
\eeqa  
 due to the upper bound in (\ref{contour-bounds}).

\item It suffices to show that the vector  $\psi_{P,n}=\bar Q_{P,n}\psi_{P,n-1}$ is  non-zero. 
We consider the difference
\beqa
\|\psi_{P,n}-\psi_{P,n-1}\|=\|\bar Q_{P,n}\psi_{P,n-1}-\psi_{P,n-1}\|\,.
\eeqa
Similarly as in the proof of Lemma~\ref{resolvent} we set $A({\bcb w_n}):=(H_{P,n-1}\res\mcF_n-{\bcb w_n})$, $B:= \la\Phi|^{n-1}_n$ and write
\beqa
 \bar Q_{P,n}\psi_{P,n-1}-\psi_{P,n-1}
=\oint_{\ga_{P,n}} d{\bcb w_n} \,
 A({\bcb w_n})^{-1/2}\sum_{j=1}^{\infty} \bigg( A({\bcb w_n})^{-1/2} B A({\bcb w_n})^{-1/2}\bigg)^j A({\bcb w_n})^{-1/2}\psi_{P,n-1}.
\eeqa
Thus taking into account  (\ref{bound-A-app}), (\ref{bound-a46})  from the proof of Lemma~\ref{resolvent} and the length
of the integration contour, we obtain
\beqa
{\bcb \|\psi_{P,n}-\psi_{P,n-1}\|}\leq c|\la|^{3/4}\|\psi_{P,n-1}\|,
\eeqa
from which we conclude  $\|\psi_{P,n}\|\geq (1-c|\la|^{3/4})\|\psi_{P,n-1}\|$. \qed
\end{enumerate}
\begin{rem} \label{energy-consistency} {\bcb We recall that $E_{P,n}:=\inf\sib(H_{P,n}\res\mcF)$ according to the original definition in 
 (\ref{original-definition}).  Since, by construction, the corresponding eigenvector of $H_{P,n}\res\mcF$ has the form $\psi_{P,n}\otimes \Om $, where
 $\psi_{P,n}\in \mcF_{n}$,  we have $E_{P,n}=\inf\pmb{\si}( H_{P,n} \res \mcF_{n} )$.  This explains the second equality in  Lemma~\ref{inductive-theorem} (i). }
\end{rem}

{\color{black} In Lemmata~\ref{upper-bound-lemma} and \ref{upper-bound-lemma-one} we prove the estimates stated in (\ref{eq:monotonicity}) and (\ref{eq:energy-shift}) of Proposition~\ref{prop:Hamiltonian-results}. Lemma~\ref{completing-square} is needed for the proof of Lemma \ref{upper-bound-lemma-one}.}
\bel \cite[Lemma 1.2]{Pi03}\label{upper-bound-lemma} Recall that $E_{P,n-1}:=\inf  \sib(H_{P,{\color{black}n-1}}\res \mcF)$, $E_{P,n}:=\inf  \sib(H_{P,n}\res \mcF)$. Suppose that $H_{P,n-1}\res \mcF_{n-1}$  
has an eigenvector $\psi_{P,n-1}$ and the corresponding eigenvalue coincides with $E_{P,n-1}$.
Then $E_{P,n-1}\geq E_{P,n}$
\eel
\proof  The statement follows from
\beqa
E_{P,n-1}=\fr{\lan \psi_{P,n-1}\otimes \Om, H_{P,n} \psi_{P,n-1}\otimes \Om\ran}{ \|\psi_{P,n-1}\otimes\Om\|^2 }\geq E_{P,n}. 
\textrm{\qed}
\eeqa
\bel\label{completing-square}\cite[Lemma 5.9]{BDP12} {\color{black}For some $c>0$ the following  inequality holds true:} $\theta H_{\pho}|_{n}^{n-1}+\la \Phi|_{n}^{n-1}\geq -\la^2\fr{c}{\theta}\si_{n-1}.$
\eel
\proof  The statement follows from the computation
\beqa
& &\int_{\si_{n}}^{\si_{n-1}}d^3k\, |k| \theta\big(b^*(k)b(k)+\fr{\la {\bcb  \chi_{\ka}(k)} |k|^{\alf}}{\sqrt{2|k|} |k|\theta}b(k)+ \fr{\la {\bcb \chi_{\ka}(k) }   |k|^{\alf}}{\sqrt{2|k|}|k|\theta}b^*(k) \big)\non\\
&=&\int_{\si_{n}}^{\si_{n-1}}d^3k\, |k|\theta\big(b^*(k)+\fr{\la{\bcb \chi_{\ka}(k)} |k|^{\alf}  }{\sqrt{2|k|} |k|\theta}\big)\big(b(k)+\fr{\la {\bcb \chi_{\ka}(k)}|k|^{\alf} }{\sqrt{2|k|} |k|\theta}\big)-\int_{\si_{n}}^{\si_{n-1}}d^3k\, |k| \theta \bigg( \fr{\la  {\bcb \chi_{\ka}(k)} |k|^{\alf}}{\sqrt{2|k|} |k|\theta}\bigg)^2\non\\
&\geq&-\fr{1}{\theta}\la^2\int_{\si_{n}}^{\si_{n-1}}d^3k\, |k| \bigg( \fr{ {\bcb \chi_{\ka}(k)}|k|^{\alf}}{\sqrt{2|k|} |k|}\bigg)^2. \textrm{ \qed }
\eeqa
\bel\label{energy-shift-one}\cite[Lemma 1.2]{Pi03}\label{upper-bound-lemma-one} Let $|P|\leq P_{\max}=1/3$ and $|\la|\in (0, \la_0]$. Then $|E_{P,n-1}-E_{P,n}|\leq c_{\Delta E}\lambda^{2}\sigma_{n-1}.$
\eel
\proof  We have: 
\beqa
E_{P,n}=\inf \sib\bigg( H_{P,n-1} \res \F_{n}+1\otimes \la \Phi|_{n}^{n-1}   \bigg).
\eeqa 
Since by Lemma~\ref{completing-square} $\theta H_{\pho}|_{n}^{n-1}+\la \Phi|_{n}^{n-1}\geq -\la^2\fr{c}{\theta}\si_{n-1}$, 
we get
\beqa
E_{P,n}\geq \inf \sib\big\{ H_{P,n-1} \res \F_{n}-1\otimes \theta H_{\pho}|_{n}^{n-1}- \la^2\fr{c}{\theta}\si_{n-1} \big\}.
\eeqa
Setting $A:=H_{P,n-1} \res \F_{n}-1\otimes \theta H_{\pho}|_{n}^{n-1}- \la^2\fr{c}{\theta}\si_{n-1}$ and noting that $\psi_{P,n-1}\otimes \Om$ is an eigenvector of $A$, we can  write
\beqa
E_{P,n}\geq \min\{ \, \lan A \ran_{\psi_{P,n-1}\otimes \Om},  \inf_{\mcF_n\ni \psi\bot \psi_{P,n-1}\otimes \Om  }\lan A\ran_{\psi}\}.
\eeqa
Clearly, $\lan A \ran_{\psi_{P,n-1}\otimes \Om}=E_{P,n-1}-\la^2\fr{c}{\theta}\si_{n-1}$. As for the second quantity under the minimum
\beqa
\inf_{\mcF_n\ni \psi\bot \psi_{P,n-1}\otimes \Om  }\lan A\ran_{\psi} 
=\inf_{\mcF_n\ni \psi\bot \psi_{P,n-1}\otimes \Om  }\lan H_{P,n-1} \res \F_{n}-1\otimes\theta H_{\pho}|_{n}^{n-1}-E_{P,n-1}\ran_{\psi}+E_{P,n-1}-\la^2\fr{c}{\theta}\si_{n-1}. \,\,\,\,
\eeqa
The first term above is positive by (\ref{infimum-remark}). Thus we get 
\beqa
E_{P,n}\geq E_{P,n-1}-\la^2\fr{c}{\theta}\si_{n-1}. \textrm{\qed}
\eeqa
\section{  Proof of Lemma~\ref{lem:Basic-estimates} } \label{standard-resolvent-bounds-section}
\setcounter{equation}{0}

We consider only the case $z_{n+1}\in \tigam_{P,n+1}$ as the case
of contours  $\ga_{P,n+1}$ is analogous and simpler. Let  us start with some general considerations. Let $q\in \{1/2,1\}$ and $X$ be an operator on $\mathcal{F}$ which maps $\mcF_n\to \mcF_m$ for some finite $m$, s.t.
\beqa
 \|(i+H^W_{P,n-1})^{-q}X\|_{\mcF_n}<\infty.
\eeqa
We note for future reference that
\beqa
\|\tiQ_{P,n-1}X\|_{\mcF_n}=|i+E_{P,n-1}|^q \, \|\tiQ_{P,n-1}(i+H_{P,n-1}^{W})^{-q}X\|_{\mcF_n}\leq 
c\|(i+H_{P,n-1}^W)^{-q}X\|_{\mcF_n}<\infty,
\label{bound-on-QA}
\eeqa
where $\tiQ_{P,n-1}$ was defined in (\ref{3.37}).
For $z$ outside of the spectrum of $H_{P,n-1}^{W+}$ we consider the following expression on $\mcF_n$
\beqa
r(z):=\bigg(\frac{1}{H_{P,n-1}^{W+}-z}\bigg)^q X
= \bigg(\frac{1}{E_{P,n-1}^{+}-z}\bigg)^q \tiQ_{P,n-1}X+ \tiQ_{P,n-1}^{\bot}\bigg( \frac{1}{H_{P,n-1}^{W+}-z}\bigg)^q (i+ H_{P,n-1}^W)^q (i+ H_{P,n-1}^W)^{-q}X, \label{r(z)}
\eeqa
where $E_{P,n-1}^{+}:=E_{P,n-1}+\De c_{P}|^{n-1}_n$.
We {\color{\red}call $r_1(z)$ and $r_2(z)$ the two terms on the r.h.s. of (\ref{r(z)})}. Clearly,
\beqa
\|r_1(z_{n+1})\|_{\mcF_n}\leq \fr{c}{\si^q_{n+1}} 
\|(i+H_{P,n-1}^W)^{-q}X\|_{\mcF_n}, 
\eeqa
since  for sufficiently small $\la_0$ we can ensure {\bcb (cf. Remark~\ref{remark-intro}) }
\beqa
|\De c_{P}|^{n-1}_n|, \ |E_{P,n-1}-E_{P,n}|\leq (1/20)\si_{n+1}.
\eeqa
Furthermore, we estimate
\beqa
\|r_2(z_{n+1})\|_{\mcF_n}\1&\leq& \1 \big\| \tiQ_{P,n-1}^{\bot}  \bigg(\frac{1}{H_{P,n-1}^{W+}-z_{n+1}}\bigg)^q (i+H_{P,n-1}^W)^q\big\|_{\mcF_n}\, 
\|(i+H_{P,n-1}^W)^{-q}X\|_{\mcF_n}\leq \fr{c}{\si_{{\color{black}n+1}}^q} \|(i+H_{P,n-1}^W)^{-q}X\|_{\mcF_n},\quad \quad
\eeqa
where we made use of the gap estimate~(\ref{last-gap-estimate}). 
Summarizing the above discussion, we have
\beqa
& &\big\|\frac{1}{H_{P,n-1}^{W+}-z_{n+1}} X\big\|_{\mcF_n}\leq \fr{c}{\si_{n+1}}\|(i+H^W_{P,n-1})^{-1}X\|_{\mcF_n}, \label{n+1-estimate-X}\\
& &\big\| \bigg(\frac{1}{H_{P,n-1}^{W+}-z_{n+1}}\bigg)^{1/2} X\big\|_{\mcF_n}\leq \fr{c}{\si^{1/2}_{n+1}}\|(i+H^W_{P,n-1})^{-1/2}X\|_{\mcF_n}.
 \label{n+1-estimate-X-new}
\eeqa
These simple-minded estimates will yield some (but not all)  bounds from Lemma~\ref{lem:Basic-estimates}.   
\bel\label{simple-resolvent-bounds} Under the assumptions of Lemma~\ref{lem:Basic-estimates}
\beqa
& &\|\frac{1}{H_{P,n-1}^{W+}-z_{n+1}}\|_{\mathcal{F}_{n}}\leq\frac{c_{0}}{\sigma_{n+1}}, 
\label{resolvent-estimate-xxx}\\
& &\|\frac{1}{H_{P,n-1}^{W+}-z_{n+1}}(\Gamma_{P,n-1})_{i}\|_{\mathcal{F}_{n}}\leq \frac{c_{0}}{\sigma_{n+1}}, \label{eq:resolvent-estimate-x}\\
& &\|\frac{1}{H_{P,n-1}^{W+}-z_{n+1}}(\mathcal{I}|_{n}^{n-1})_{i}\|_{\mathcal{F}_{n}}\leq c_2|\la|^{3/2}.
\label{I-estimate-x}
\eeqa
\eel
\proof The estimate in (\ref{resolvent-estimate-xxx}) follows
from (\ref{n+1-estimate-X}) {\color{\red}with $X=1$}.
Given (\ref{n+1-estimate-X}),   to verify (\ref{eq:resolvent-estimate-x}), it suffices to check that
\beqa
\|(i+H^{W}_{P,n-1})^{-1}\Ga_{P,n-1}\|_{\mcF_n}\leq c. \label{res-gamm-bound}
\eeqa
Since $\Ga_{P,n-1}:=\nabla E_{P,n-1}- W_{P,n-1}(P-P_{\pho})W_{P,n-1}^*$, we have by~(\ref{eq:gradient-bound})
\beqa
\|(i+H^{W}_{P,n-1})^{-1}\Ga_{P,n-1}\|_{\mcF_n}\leq 
\|(i+H_{P,n-1})^{-1}(P-P_{\pho})\|_{\mcF_n}+c.
\eeqa
Furthermore, by Lemma~\ref{energy-bound-zero}
\beqa
\|(i+H_{P,n-1})^{-1/2}(P-P_{\pho})\|^2_{\mcF_n}\leq 
c\|(i+H_{P,n-1})^{-1/2}H_{P, \free}((i+H_{P,n-1})^{-1/2})^*\|_{\mcF_n}\leq c'. \label{free-bound}
\eeqa
As for (\ref{I-estimate-x}), we have by definition~(\ref{I-definitions}) that 
$|(\mathcal{I}|_{n}^{n-1})_{i}|\leq c|\la|^2\si_{n-1}$, which gives together with (\ref{resolvent-estimate-xxx}) and $|\la|\leq \veps^4$
the required bound. \qed
\bel\label{RL-lemma} Under the assumptions of Lemma~\ref{lem:Basic-estimates}
\beqa
& &\|  \frac{1}{H_{P,n-1}^{W+}-z_{n+1}} \mathcal{L}|^{n-1}_{n}    \|_{\mcF_n}\leq c_2|\la|^{1/2}. 
\label{L-first}
\eeqa
If $\mcL|^{n-1}_n$ is replaced with  $\mathcal{L}'|_{n}^{n-1}:=[ H_{\pho}, \mathcal{L}|_{n}^{n-1}]$
or $\mring{\mathcal{L}}|^{n-1}_{n}=\sum_{j=1}^3[(P_{\pho})_j, (\mathcal{L}|^{n-1}_{n})_j]$ the bounds
can be improved to $c_2|\la|^{1/2}\si_{n-1}$.
\eel
\proof We set ${\bcb A:=H_{P,n-1}+\Delta c_{P}|_{n}^{n-1}-\bar z_{n+1}}$. {\color{\red}Since $[ W_{P,n-1}, \mathcal{L}|^{n-1}_{n}]=0$, we have}
\beqa
& &{\color{\red}\|  \frac{1}{H_{P,n-1}^{W+}-z_{n+1}} \mathcal{L}|^{n-1}_{n}    \|_{\mcF_n}=\| {\bcb  (A^{-1})^*} \mathcal{L}|^{n-1}_{n}    \|_{\mcF_n}. }
\eeqa
By the energy bounds (Lemma~\ref{energy-bounds}) and the fact that $A^{-1/2}$ and $(H_{\pho}|^{n-1}_{n})^{1/2}$ commute we have
\beqa
\| {\bcb (A^{-1/2})^*}\mathcal{L}|_{n}^{n-1} \|_{\mcF_n}=\|\mathcal{L}|_{n}^{n-1} A^{-1/2}\|_{\mcF_n}\1 &\leq& \1 |\la| c \si^{1/2}_{n-1}
\|(H_{\pho}|^{n-1}_{n})^{1/2} A^{-1/2}\|_{\mcF_n}+|\la|c\si_{n-1}\|A^{-1/2}\|_{\mcF_n} \non\\
\1 &\leq& \1 |\la|c\si_{n-1}^{1/2}+|\la|c\fr{\si_{n-1}}{(\si_{n+1})^{1/2}}.
\eeqa
In the last step we used (\ref{resolvent-estimate-xxx}) and {\color{black}the} spectral theorem which give 
$\|A^{-1/2}\|_{\mcF_n}\leq \|A^{-1}\|_{\mcF_n}^{1/2}\leq c(\si_{n+1})^{-1/2}$,  and the following computation similar to (\ref{A-eighty-four}) 
{\color{black}\beqa
\|(H_{\pho}|^{n-1}_{n})^{1/2} A^{-1/2}\|_{\mcF_n}
\1&=&\1\|(H_{\pho}|^{n-1}_{n})^{1/2}  \big(H_{P,n-1}+\Delta c_{P}|_{n}^{n-1}-z_{n+1}\big)^{-1/2}  \|_{\mcF_n},\non\\
\1&\leq& \1\theta^{-1/2}\|(\theta H_{\pho}|^{n-1}_{n})^{1/2} \big( H_{P,n-1}-\theta H_{\pho}|^{n-1}_{n}-E_{P,n-1}+
\mco(\si_{n+1})+\theta H_{\pho}|^{n-1}_{n}-{\bcb i\mrm{Im}\, z_{n+1}} \big)^{-1/2}  \|_{\mcF_n}\non\\
\1&\leq&\1 \theta^{-1/2}\|(\theta H_{\pho}|^{n-1}_{n})^{1/2} \big( \si_n/2+
\mco(\si_{n+1})+\theta H_{\pho}|^{n-1}_{n}-{\bcb i\mrm{Im}\, z_{n+1}} \big)^{-1/2}  \|_{\mcF_n}\leq c, \label{last-estimate-XXX}
\eeqa }
where we made use of the fact that  $H_{P,n-1}-\theta H_{\pho}|^{n-1}_{n}-E_{P,n-1}\geq \si_n/2$
on the subspace orthogonal to $\psi_{P,n-1}\otimes \Om$ (cf. (\ref{Gap-bound})). Since
{\color{black}$|\mco(\si_{n+1})|:=\sup_{z_{n+1}\in \gamma_{n+1}\cup  \ti{\gamma}_{n+1} }|E_{P,n-1}-E_{P,n}+\De c_{P}|^{n-1}_n+{\bcb E_{P,n} -\mrm{Re}\, z_{n+1}}|\leq \si_{n+1}/2$}, we obtain
the last estimate in (\ref{last-estimate-XXX}). Alluding again to the fact that $\|A^{-1/2}\|_{\mcF_n}\leq c(\si_{n+1})^{-1/2}$ and using $|\la|\leq \veps^4$ we conclude the proof of (\ref{L-first}). The remaining statements in the lemma follow
from obvious modifications of the above discussion. \qed
\bel\label{LG-lemma} Under the assumptions of Lemma~\ref{lem:Basic-estimates}
\beqa
\|  \frac{1}{H_{P,n-1}^{W+}-z_{n+1}} \mathcal{L}|^{n-1}_{n}\cdot \Ga_{P,n-1} \|_{\mcF_n}\leq c_2|\la|^{1/2}. \label{L-Ga-est-n+1} 
\label{L-Ga-est-n}
\eeqa
\eel
\proof {\color{\red}Since $\Ga_{P,n-1}:=\nabla E_{P,n-1}- W_{P,n-1}(P-P_{\pho})W_{P,n-1}^*$ and $[ W_{P,n-1}, \mathcal{L}|^{n-1}_{n}]=0$,} we note the following inequality
\beqa
\|  \frac{1}{H_{P,n-1}^{W+}-z_{n+1}} \mathcal{L}|^{n-1}_{n}\cdot \Ga_{P,n-1} \|_{\mcF_n}
\1 &\leq & \1 \|  \frac{1}{H_{P,n-1}+\De c_P|^{n-1}_n-z_{n+1}} \mathcal{L}|^{n-1}_{n}\cdot (P-P_{\pho}) \|_{\mcF_n} \label{LG-firststep}\\
& &+ \, \|  \frac{1}{H_{P,n-1}+\De c_P|^{n-1}_n-z_{n+1}} \mathcal{L}|^{n-1}_{n}\|_{\mcF_n}
|\nabla E_{P,n-1}|.\label{LG-secondstep}
\eeqa
By Lemma~\ref{RL-lemma}, {\color{black}using $[ W_{P,n-1}, \mathcal{L}|^{n-1}_{n}]=0$ one more time,} we have $(\ref{LG-secondstep})\leq c|\la|^{1/2}$. As for (\ref{LG-firststep}),
we write
\beqa
(\ref{LG-firststep})\1 &=&\1 \|  \frac{1}{H_{P,n-1}+\De c_P|^{n-1}_n-z_{n+1}} \mring{\mathcal{L}}|^{n-1}_{n}  \|_{\mcF_n}\label{LG-first-estimate}\\
& &+\| \frac{1}{H_{P,n-1}+\De c_P|^{n-1}_n-z_{n+1}}(P-P_{\pho}) \cdot\mathcal{L}|^{n-1}_{n}  \|_{\mcF_n},
\label{LG-second-estimate}
\eeqa
where $\mring{\mathcal{L}}|^{n-1}_{n}=-\sum_{j=1}^3[(\mathcal{L}|^{n-1}_{n})_j,(P_{\pho})_j]$.  
By the last statement in Lemma~\ref{RL-lemma} {\color{black}and $[ W_{P,n-1}, \mathcal{L}|^{n-1}_{n}]=0$,} we have that
$(\ref{LG-first-estimate})\leq c|\la|^{1/2}\si_{n-1}$. To (\ref{LG-second-estimate}) we apply the energy bounds
(Lemma~\ref{energy-bounds}). Setting ${\bcb A:=H_{P,n-1}+\Delta c_{P}|_{n}^{n-1}-\bar z_{n+1}}$, we have
\beqa
\|\mathcal{L}|_{n}^{n-1}{\color{\red}\cdot} (P-P_{\pho})A^{-1}\|_{\mcF_n} \1 &=& \1
 c|\la| \si_{n-1}^{1/2}\|(H_{\pho}|^{n-1}_n)^{1/2}(P-P_{\pho})A^{-1}\|_{\mcF_n}
+c|\la| \si_{n-1}\|(P-P_{\pho})A^{-1}\|_{\mcF_n}\non\\
\1 &\leq& \1 c|\la|\bigg( \fr{\si_{n-1}^{1/2}}{\si_{n+1}^{1/2}}+\fr{\si_{n-1}}{\si_{n+1}} \bigg)=c|\la|^{1/2}.
\label{A-P-P}
\eeqa
Here we made use of $\|(P-P_{\pho})A^{-1}\|_{\mcF_n}\leq c/\si_{n+1}$, which is shown using (\ref{n+1-estimate-X})  and (\ref{free-bound}). Furthermore, {\color{\red}using that $[H_{\pho}|^{n-1}_n, P-P_{\pho}]=[H_{\pho}|^{n-1}_n, A]=0$},  
\begin{eqnarray}
\|(H_{\pho}|^{n-1}_n)^{1/2}(P-P_{\pho})A^{-1}\|_{\mcF_n}
\1 & =& \1 {\color{\red}\|(P-P_{\pho})A^{-1/2}(H_{\pho}|^{n-1}_n)^{1/2}A^{-1/2}\|_{\mcF_n}}\nonumber\\
\1 &\leq& \1 \|(P-P_{\pho})A^{-1/2}\|_{\mcF_n}
\|(H_{\pho}|^{n-1}_n)^{1/2}A^{-1/2}\|_{\mcF_n}\leq \fr{c}{(\si_{n+1})^{1/2}},
\end{eqnarray}
{\color{\red}where in the last inequality we exploit (\ref{last-estimate-XXX}) and (\ref{free-bound}) combined with   (\ref{n+1-estimate-X-new}).}
This completes the proof. \qed
\bel\label{RLL-lemma} Under the assumptions of Lemma~\ref{lem:Basic-estimates}
\beqa
& &\|  \frac{1}{H_{P,n-1}^{W+}-z_{n+1}} (\mathcal{L}|^{n-1}_{n})^2 \|_{\mcF_n}\leq  c|\la|^{3/2}\si_{n-1}. \label{L-L-est-n+1}
\label{L-L-est-n}
\eeqa
\eel
\proof  
 We set as before ${\bcb A:=H_{P,n-1}+\Delta c_{P}|_{n}^{n-1}-\bar{z}_{n+1}}$, {\color{black}we then use $[ W_{P,n-1}, \mathcal{L}|^{n-1}_{n}]=0$} and apply the energy bounds (Lemma~\ref{energy-bounds}):
\beqa
\|(\mathcal{L}|_{n}^{n-1})^2A^{-1}\|_{\mcF_n}\leq c|\la|\si_{n-1}^{1/2}
\|(H_{\pho}|_{n}^{n-1})^{1/2}\mathcal{L}|_{n}^{n-1}A^{-1}\|_{\mcF_n}
+c|\la|\si_{n-1}\|\mathcal{L}|_{n}^{n-1}A^{-1}\|_{\mcF_n}. \label{iterated-energy-bounds}
\eeqa
We first estimate the second term on the r.h.s. above applying again the energy bounds
\beqa
c|\la|\si_{n-1}\|\mathcal{L}|_{n}^{n-1}A^{-1}\|_{\mcF_n}\1 &\leq& \1 
c|\la|^2(\si_{n-1})^{3/2}\| (H_{\pho}|_{n}^{n-1})^{1/2}A^{-1}\|_{\mcF_n}+c|\la|^2(\si_{n-1})^2\|A^{-1}\|_{\mcF_n}\non\\
\1 &\leq& \1 c|\la|^2\bigg((\si_{n-1})^{3/2}\si_{n+1}^{-1/2}+(\si_{n-1})^2\si_{n+1}^{-1}\bigg)\non\\
\1 &\leq& \1 c|\la|^{3/2} \si_{n-1},
\eeqa
where we made use  of (\ref{last-estimate-XXX}), $\|A^{-1/2}\|_{\mcF_n}\leq c(\si_{n+1})^{-1/2}$ and    $|\la|^{1/2}\leq \epsilon^2$. Now we consider the first term on the r.h.s. of (\ref{iterated-energy-bounds}):
\beqa
\|(H_{\pho}|_{n}^{n-1})^{1/2}\mathcal{L}|_{n}^{n-1}A^{-1}\|_{\mcF_n}^2
\1 &=&\1 \|{\color{\red}(A^{-1})^*}\mathcal{L}|_{n}^{n-1}    H_{\pho}|_{n}^{n-1} \mathcal{L}|_{n}^{n-1}A^{-1}\|_{\mcF_n}\non\\
\1 &\leq& \1 \|{\color{\red}(A^{-1})^*}\mathcal{L}|_{n}^{n-1}   \mathcal{L}'|_{n}^{n-1}A^{-1}\|_{\mcF_n}+
\|{\color{\red}(A^{-1})^*}\mathcal{L}|_{n}^{n-1} \mathcal{L}|_{n}^{n-1} H_{\pho}|_{n}^{n-1}A^{-1}\|_{\mcF_n}\non\\
\1 &\leq& \1 c|\la|\si_{n-1}+c\|{\color{\red}(A^{-1})^*}(\mathcal{L}|_{n}^{n-1})^2\|_{\mcF_n},
\eeqa
where {\color{\red}$\mathcal{L}'|_{n}^{n-1}:=[H_{\pho}, \mathcal{L}|_{n}^{n-1}]$} and we applied Lemma~\ref{RL-lemma} and (\ref{last-estimate-XXX}) in the last step.
Thus altogether we get
\beqa
\|(\mathcal{L}|_{n}^{n-1})^2A^{-1}\|_{\mcF_n}\leq c|\la|\si_{n-1}^{1/2}
\big(|\la|\si_{n-1}+\|{\color{\red}(A^{-1})^*}(\mathcal{L}|_{n}^{n-1})^2\|_{\mcF_n}\big)^{1/2}+c|\la|^{3/2} \si_{n-1}.\label{b.28}
\eeqa
Set $x:=\|(\mathcal{L}|_{n}^{n-1})^2A^{-1}\|_{\mcF_n}$ and shift the term $c|\la|^{3/2} \si_{n-1}$ on the l.h.s. of (\ref{b.28}). If $x-c|\la|^{3/2} \si_{n-1}\leq 0$
then the proof is complete. Otherwise, we can square both sides of the resulting inequality which implies the following relation
\beqa
x^2-c_1|\la|^{3/2}\si_{n-1}x-c_2|\la|^{3}\si_{n-1}^2\leq 0,
\eeqa
with $c_1,c_2\geq 0$.  This  leads to  $x\leq c|\la|^{3/2}\si_{n-1}$. \qed
\bel\label{quad-lemma} Under the assumptions of Lemma~\ref{lem:Basic-estimates}
\beqa
& &\|  \frac{1}{H_{P,n-1}^{W+}-z_{n+1}} (H_{I}^W|^{n-1}_n )_{\mrm{quad}}\|_{\mcF_n}\leq  c|\la|^{3/2} \si_{n-1}.\label{H-I-quad}
\label{H-I-quad-I}
\eeqa
\eel
\proof We recall that $(H_{I}^W|^{n-1}_n )_{\mrm{quad}}=\h(\mathcal{L}|_{n}^{n-1}+ \mathcal{I}|_{n}^{n-1} )^2 $. Thus the statement follows from Lemmas~\ref{simple-resolvent-bounds}, \ref{RL-lemma}, \ref{RLL-lemma}, 
and from $|(\mathcal{I}|_{n}^{n-1})_{i}|\leq c|\la|^2\si_{n-1}$. \qed
\bel\label{mix-lemma} Under the assumptions of Lemma~\ref{lem:Basic-estimates}
\beqa
& &\|  \frac{1}{H_{P,n-1}^{W+}-z_{n+1}} \De(H_{I}^W|^{n-1}_n )_{\mrm{mix}}\|_{\mcF_n}\leq  c|\la|^{1/2} \si_{n-1}. \label{H-I-mix} 
\label{H-I-mix-I-x}
\eeqa
\eel
\proof We recall that $\Delta(H_{I}^{W}|_{n}^{n-1})_{\mr{mix}}=\h\sum_{j=1}^3[(\Ga_{P,n-1})_j,  (\mathcal{L}|_{n}^{n-1})_j]$. Thus (\ref{H-I-mix}) corresponds to
the {\color{\red}second result} of Lemma~\ref{RL-lemma}.  \qed
\bel Under the assumptions of Lemma~\ref{lem:Basic-estimates}
\beqa
& &\|  \frac{1}{H_{P,n-1}^{W+}-z_{n+1}} (H_{I}^W|^{n-1}_n )\|_{\mcF_n}\leq  c|\la|^{1/2}. \label{H-I-mix-1} 
\label{H-I-mix-I}
\eeqa
\eel
\proof Given that $H_{I}^W|^{n-1}_n:=(\mathcal{L}|_{n}^{n-1}+\mathcal{I}|_{n}^{n-1})\cdot\Gamma_{P,n-1}+ (H_{I}^W|^{n-1}_n )_{\mrm{quad}}+\De(H_{I}^W|^{n-1}_n )_{\mrm{mix}}$, the statement follows from 
Lemmas~\ref{simple-resolvent-bounds}, \ref{LG-lemma}, \ref{quad-lemma}, \ref{mix-lemma}
 and $|(\mathcal{I}|_{n}^{n-1})_{i}|\leq c|\la|^2\si_{n-1}$. \qed

\section{Proof of Theorem~\ref{thm:induction-convergence}} \label{app-induction-convergence}
\setcounter{equation}{0}

 {\bcc First, we check that the hypotheses holds for $n=0$. Statement i) is obvious since 
$\Ga_{P,0}\Om=0$.
As for ii), we have
\beqa
\hat\phi_{P,1}-\Om=\oint_{\ga_{P,2}} \fr{dw_2}{\hat H^W_{P,1}-w_2}\Om-\Om. \label{ii-claim}
\eeqa
We use the resolvent expansion (\ref{eq:resolvent-expansion}):
\beqa
\frac{1}{\hat{H}_{P,1}^{W}-w_{2}}\Om=
\frac{1}{H_{P,0}^{W}+\Delta c_{P}|_{1}^{0}-w_{2}}
\sum_{j=0}^{\infty}\{-H_I^{W}|_{1}^{0}\frac{1}{H_{P,0}^{W}+\Delta c_{P}|_{1}^{0}-w_{2}}\}^{j}\Om,\label{eq:resolvent-expansion-zero-x}
\eeqa
for $w_2\in \ga_{P,2}$. The $j=0$ term gives zero contribution to (\ref{ii-claim}) due to $H_{P,0}^{W}\Om=E_{P,0}\Om$, $|E_{P,0}-E_{P,1}|\leq (1/20)\si_2$, $ \Delta c_{P}|_{1}^{0}=(1/20)\si_2$ and $|w_2-E_{P,1}|=\si_2/3$.  Thus we have 
\beqa
\big\|\hat\phi_{P,1}-\Om\big\|&=&\big\|\oint_{\ga_{P,2}}dw_2\frac{1}{H_{P,0}^{W}+\Delta c_{P}|_{1}^{0}-w_{2}}
\sum_{j=1}^{\infty}\{-H_I^{W}|_{1}^{0}\frac{1}{H_{P,0}^{W}+\Delta c_{P}|_{1}^{0}-w_{2}}\}^{j}\Om\big\|\non\\
&\leq& \si_2\fr{c_0}{\si_2} c_2|\la|^{1/2}\fr{1}{1-c_2|\la|^{1/2}}\leq |\la|^{1/4}\si_0^{1-\de},
\label{n=0-verification}
\eeqa
where we made use of estimates (\ref{first}) and (\ref{H-I-standard-estimates}) from Lemma~\ref{lem:Basic-estimates}, (\ref{new-la-restrictions}) and $\si_0=1$.
Now iii) follows from ii) via the inverse triangle inequality. 

Now we proceed to the inductive argument.} The claims in i), ii), and iii) are at the $n+1-$th step by definition, and they will be denoted as c-i), c-ii), and c-iii),  respectively. At the $n-$th step they will be denoted as assumptions a-i), a-ii), and a-iii), respectively. {\bcc All operator norms in the proof are meant on {\color{\red}$B(\mcF_n)$ (the bounded operators on $\mcF_n$)} unless stated otherwise}.\\

\noun{Claim-i)} \emph{(Here we show the implication: a-i), a-ii),
a-iii) $\rightarrow$} \emph{c-i)})\\
We use the unitary operator $\widetilde{W}_{P,n}W_{P,n}^{*}$ to switch
from the given expression to the expression with `hats'
\begin{equation}
\|\,\frac{1}{H_{P,n}^{W}-z_{n+1}}(\Gamma_{P,n})_{i}\phi_{P,n}\|=\|\widetilde{W}_{P,n}W_{P,n}^{*}\,\frac{1}{H_{P,n}^{W}-z_{n+1}}(\Gamma_{P,n})_{i}\phi_{P,n}\|=\|\,\frac{1}{\hat{H}_{P,n}^{W}-z_{n+1}}(\hat{\Gamma}_{P,n})_{i}\hat{\phi}_{P,n}\|. \label{claim-i)}
\end{equation}
In (\ref{claim-i)}) we proceed with the full expansion of $\hat{\phi}_{P,n}$, $\frac{1}{\hat{H}_{P,n}^{W}-z_{n+1}}$,
and $(\hat{\Gamma}_{P,n})_{i}$, i.e., we study
\begin{eqnarray}
\sum_{l=0}^{\infty}\sum_{l'=0}^{\infty}\oint_{\gamma_{P,n+1}}dw_{n+1}\{\frac{1}{H_{P,n-1}^{W+}-z_{n+1}}(-)H_{I}^{W}|_{n}^{n-1}\}^{l}\frac{1}{H_{P,n-1}^{W+}-z_{n+1}}(\Gamma_{P,n-1}+\Delta\Gamma_{P}|_{n}^{n-1})_{i}\times\non\\
\times\{\frac{1}{H_{P,n-1}^{W+}-w_{n+1}}(-)H_{I}^{W}|_{n}^{n-1}\}^{l'}\frac{1}{H_{P,n-1}^{W+}-w_{n+1}}\phi_{P,n-1}. \label{eq:expanded-object-1-1}
\end{eqnarray}
By splitting $\Delta\Gamma_{P}|_{n}^{n-1}$ into $\Delta'\Gamma_{P}|_{n}^{n-1}+\mathcal{I}|_{n}^{n-1}$   {\color{\red}(see (\ref{eq:delta'-gamma}) and (\ref{def-deltagamma}))}
we can write 
\begin{eqnarray}
(\ref{eq:expanded-object-1-1}) & = & \sum_{l=0}^{\infty}\sum_{l'=0}^{\infty}\oint_{\gamma_{P,n+1}}dw_{n+1}\{\frac{1}{H_{P,n-1}^{W+}-z{}_{n+1}}(-)H_{I}^{W}|_{n}^{n-1}\}^{l}\frac{1}{H_{P,n-1}^{W+}-z{}_{n+1}}\times \nonumber\\
 &  & \quad\quad\quad\quad\times(\Gamma_{P,n-1}+\Delta'\Gamma_{P}|_{n}^{n-1})_i\{\frac{1}{H_{P,n-1}^{W+}-w_{n+1}}(-)H_{I}^{W}|_{n}^{n-1}\}^{l'}\frac{1}{H_{P,n-1}^{W+}-w_{n+1}}\phi_{P,n-1}  \label{eq:expanded-object-1-step-1}  \\
 &  & +\mathcal{R}_{P,n-1},
\end{eqnarray}
where $\mathcal{R}_{P,n-1}$ is {\color{\red}the term proportional to $\mathcal{I}|_{n}^{n-1}$. Using Lemma~\ref{lem:Basic-estimates} 
we can estimate} 
\begin{equation}
\|\mathcal{R}_{P,n-1}\|\leq\sigma_{n+1}\cdot\frac{1}{1-c_{2}|\lambda|^{\frac{1}{2}}}c_{2}|\lambda|^{\frac{1}{2}}\cdot\frac{1}{1-c_{2}|\lambda|^{\frac{1}{2}}}\cdot\frac{c_{0}}{\sigma_{n+1}}.\label{eq:first-summand}
\end{equation}
(We recall that according to our notation  the numerical factor $-\frac{1}{2\pi i}$ is hidden in the symbol $\oint_{\gamma_{P,n}}$).
Next we split the term in (\ref{eq:expanded-object-1-step-1}) into

\begin{eqnarray}
& &\sum_{l=0}^{\infty} \{ \frac{1}{H_{P,n-1}^{W+}-z{}_{n+1}}(-)H_{I}^{W}|_{n}^{n-1} \}^{l} \frac{1}{H_{P,n-1}^{W+}-z{}_{n+1}}(\Gamma_{P,n-1}+\Delta'\Gamma{}_{P}|_{n}^{n-1})_i\phi_{P,n-1} \label{eq:zero-order-expansion}\\
& &\ph{444444444}+\sum_{l=0}^{\infty}\sum_{l'=1}^{\infty}\oint_{\gamma_{P,n+1}}dw_{n+1} \{ \frac{1}{H_{P,n-1}^{W+}-z{}_{n+1}}(-)H_{I}^{W}|_{n}^{n-1}\}^{l} \frac{1}{H_{P,n-1}^{W+}-z_{n+1}}\times\quad\non\\
& &\ph{444444444444}\times(\Gamma_{P,n-1}+\Delta'\Gamma_{P}|_{n}^{n-1})_i\{\frac{1}{H_{P,n-1}^{W+}-w_{n+1}}(-)H_{I}^{W}|_{n}^{n-1}\}^{l'}\frac{1}{H_{P,n-1}^{W+}-w_{n+1}}\phi_{P,n-1}. \label{eq:geq-first-order-expansion}
\end{eqnarray}

\nin\emph{Term (\ref{eq:zero-order-expansion})}\\
We re-state this term:
\begin{equation}
\sum_{l=0}^{\infty}\{\frac{1}{H_{P,n-1}^{W+}-z{}_{n+1}}(-)H_{I}^{W}|_{n}^{n-1}\}^{l}\frac{1}{H_{P,n-1}^{W+}-z{}_{n+1}}(\Gamma_{P,n-1}+\Delta'\Gamma_{P}|_{n}^{n-1})_{i}\phi_{P,n-1}.\label{eq:zero-order-expansion-bis}
\end{equation}
As for the part of (\ref{eq:zero-order-expansion-bis}) proportional
to $(\Delta'\Gamma_{P}|_{n}^{n-1})_{i}{\bcc:=(-\nabla E_{P,n-1}+\nabla E_{P,n}+\mcL|_n^{n-1})_i}$,
notice that:
\begin{enumerate}
\item In $(\mathcal{L}|_{n}^{n-1})_{i}$ only the term proportional to the
creation operator, i.e., $(\mathcal{L}|_{n}^{n-1})_{i}^{(+)}$, gives
a nonzero contribution. This contribution  can be bounded by (see Lemma~\ref{lem:Basic-estimates})
\begin{eqnarray}
  \sup_{z_{n+1}\in\ti\gamma_{P,n+1}}\Big\{\sum_{l=0}^{\infty}||\{\frac{1}{H_{P,n-1}^{W+}-z{}_{n+1}}(-)H_{I}^{W}|_{n}^{n-1}\}\|^{l}\|\frac{1}{H_{P,n-1}^{W+}-z{}_{n+1}}(\mathcal{L}|_{n}^{n-1})_{i}^{(+)}\phi_{P,n-1}\|\Big\}
  \leq\frac{1}{1-c{}_{2}|\lambda|^{\frac{1}{2}}}\cdot c_{2}|\lambda|^{\frac{1}{2}}.
\end{eqnarray}

\item The contribution coming from $-\nabla E_{P,n-1}+\nabla E_{P,n}$
can be bounded by 
\begin{eqnarray}
  \sup_{z_{n+1}\in\ti\gamma_{P,n+1}}\Big\{\sum_{l=0}^{\infty}||\{\frac{1}{H_{P,n-1}^{W+}-z{}_{n+1}}(-)H_{I}^{W}|_{n}^{n-1}\}\|^{l}|\frac{1}{E_{P,n-1}+{\bcc \Delta c_P|_n^{n-1} } -z{}_{n+1}}|\Big\}
 c_{1}[\lambda^{2}\sigma_{n-1}+\|\frac{\hat{\phi}_{P,n}}{\|\hat{\phi}_{P,n}\|}-\frac{\phi_{P,n-1}}{\|\phi_{P,n-1}\|}\|]\non\\
\leq  \frac{1}{1-c{}_{2}|\lambda|^{\frac{1}{2}}}\cdot\frac{ c_{0} }{\sigma_{n+1}}\cdot c_{1}\cdot[\lambda^{2}\sigma_{n-1}
+4|\lambda|^{\frac{1}{4}}\sigma_{n-1}^{1-\delta}].\quad \label{reference-for-later}
\end{eqnarray}
Here we have used (\ref{grad-difference}), {\color{\red}Lemma~\ref{lem:Basic-estimates}},  and
a-ii), a-iii). (Notice that $\|\hat{\phi}_{P,n}\|=\|\phi_{P,n}\|$ and $\|\phi_{P,n-1}\|\geq \|\phi_{P,n}\|$
because of the definitions in (\ref{phi-definitions})).  \\

\end{enumerate}
As for the term in (\ref{eq:zero-order-expansion-bis}) proportional
to \foreignlanguage{english}{$\Gamma_{P,n-1}$,  its norm is bounded
by }
\begin{eqnarray}
 & & \sum_{l=0}^{\infty}||\frac{1}{H_{P,n-1}^{W+}-z{}_{n+1}}(-)H_{I}^{W}|_{n}^{n-1}\|^{l}\|\frac{1}{H_{P,n-1}^{W+}-z{}_{n+1}}(\Gamma_{P,n-1})_{i}\phi_{P,n-1}\|\non\\
& &\leq  \sum_{l=0}^{\infty}||\frac{1}{H_{P,n-1}^{W+}-z{}_{n+1}}(-)H_{I}^{W}|_{n}^{n-1}\|^{l}{\bcc \sup_{z'_{n}\in\ti\gamma_{P,n}}
 \|\frac{1}{H^{W}_{P,n-1}-z'_{n}}(\Gamma_{P,n-1})_{i}\phi_{P,n-1}\| } \leq {\bcc \frac{1}{1-c{}_{2}|\lambda|^{\frac{1}{2}}}\frac{1}{\sigma_{n-1}^{\delta}}, } \label{Cauchy-theorem}
\end{eqnarray}
{\bcc where in the first step we applied the maximal modulus principle (Lemma~\ref{maximal-modulus-appl-one}) after using 
the gap estimate (\ref{eq:gap-estimate}) and the fact that $\Ga_{P,n-1}\phi_{P,n-1}$ is orthogonal to $\phi_{P,n-1}$ (as well as (\ref{conv-assumption-E}), (\ref{conv-assumption-c})) to ensure the holomorphy of the expression. It was essential here that $\mrm{Re}\,z_{n+1}-\De c_P|^{n-1}_n\leq \mrm{Re}\,z'_{n}$. In the last step we used the induction hypothesis a-i). (Cf.  Remark \ref{max-mod-1} for a similar argument).}

Summing up, we have
\begin{equation}
\|(\ref{eq:zero-order-expansion})\|\leq\frac{1}{1-c{}_{2}|\lambda|^{\frac{1}{2}}}\cdot c_{2}|\lambda|^{\frac{1}{2}}+\frac{1}{1-c{}_{2}|\lambda|^{\frac{1}{2}}}\cdot\frac{ c_{0} }{\sigma_{n+1}}\cdot c_{1}\cdot[\lambda^{2}\sigma_{n-1}+4|\lambda|^{\frac{1}{4}}\sigma_{n-1}^{1-\delta}]+\bigg(\frac{1}{1-c{}_{2}|\lambda|^{\frac{1}{2}}}\bigg)\frac{1}{\sigma_{n-1}^{\delta}}.\label{eq:second-summand}
\end{equation}
\emph{Term (\ref{eq:geq-first-order-expansion})}\\
Firstly, we notice that the norm of the term proportional to $\Delta'\Gamma_{P}|_{n}^{n-1}$,
i.e., 
\begin{eqnarray}
\sum_{l=0}^{\infty}\sum_{l'=1}^{\infty}\oint_{\gamma_{P,n+1}}dw_{n+1}\{\frac{1}{H_{P,n-1}^{W+}-z{}_{n+1}}(-)H_{I}^{W}|_{n}^{n-1}\}^{l}\frac{1}{H_{P,n-1}^{W+}-z{}_{n+1}}(\Delta'\Gamma_{P}|_{n}^{n-1})_i\times\non\\
\quad\quad\quad\quad\quad\quad\quad\times\{\frac{1}{H_{P,n-1}^{W+}-w_{n+1}}(-)H_{I}^{W}|_{n}^{n-1}\}^{l'-1}\frac{1}{H_{P,n-1}^{W+}-w_{n+1}}(-)H_{I}^{W}|_{n}^{n-1}\frac{1}{H_{P,n-1}^{W+}-w_{n+1}}\phi_{P,n-1}\,,
\end{eqnarray}
is bounded by
\begin{eqnarray}
 & &\sigma_{n+1}\frac{1}{1-c{}_{2}|\lambda|^{\frac{1}{2}}}\{\frac{c_{0}}{\sigma_{n+1}}c_{1}[\lambda^{2}\sigma_{n-1}+\|\frac{\hat{\phi}_{P,n}}{\|\hat{\phi}_{P,n}\|}-\frac{\phi_{P,n-1}}{\|\phi_{P,n-1}\|}\|]+c_{2}|\lambda|^{\frac{1}{2}}\}\bigg(\frac{1}{1-c_{2}|\lambda|^{\frac{1}{2}}}\bigg)c_{2}|\lambda|^{\frac{1}{2}}\frac{c_{0}}{\sigma_{n+1}}\nonumber \\
& &\leq  c_{0}c_{2}|\lambda|^{\frac{1}{2}}\bigg(\frac{1}{1-c_{2}|\lambda|^{\frac{1}{2}}}\bigg)^{2}\{\frac{c_{0}}{\sigma_{n+1}}c_{1}[\lambda^{2}\sigma_{n-1}+4|\lambda|^{\frac{1}{4}}\sigma_{n-1}^{1-\delta}]+c_{2}|\lambda|^{\frac{1}{2}}\},\label{eq:third-summand}
\end{eqnarray}
where we have used (\ref{grad-difference}), a-ii), a-iii)  {\bcc and  Lemma~\ref{lem:Basic-estimates}}.

\noindent
As for the term proportional to $\Gamma_{P,n-1}$, i.e., 
\begin{eqnarray}
\sum_{l=0}^{\infty}\sum_{l'=1}^{\infty}\oint_{\gamma_{P,n+1}}dw_{n+1}\{\frac{1}{H_{P,n-1}^{W+}-z{}_{n+1}}(-)H_{I}^{W}|_{n}^{n-1}\}^{l}\frac{1}{H_{P,n-1}^{W+}-z{}_{n+1}}(\Gamma_{P,n-1})_i\times\non\\
\quad\quad\quad\quad\quad\times\{\frac{1}{H_{P,n-1}^{W+}-w_{n+1}}(-)H_{I}^{W}|_{n}^{n-1}\}^{l'-1}\frac{1}{H_{P,n-1}^{W+}-w_{n+1}}
(-)H_{I}^{W}|_{n}^{n-1}\frac{1}{H_{P,n-1}^{W+}-w_{n+1}}\phi_{P,n-1} \,,\quad \label{new-number-added}
\end{eqnarray}
for each summand in the series
\beqa
\sum_{l'=1}^{\infty}\{\frac{1}{H_{P,n-1}^{W+}-w_{n+1}}(-)H_{I}^{W}|_{n}^{n-1}\}^{l'-1}\frac{1}{H_{P,n-1}^{W+}-w_{n+1}}(-)H_{I}^{W}|_{n}^{n-1}\frac{1}{H_{P,n-1}^{W+}-w_{n+1}}\phi_{P,n-1}
\eeqa
consider the first operator $H_{I}^{W}|_{n}^{n-1}$ from the right
\beqa
H_{I}^{W}|_{n}^{n-1}=\h(\Gamma_{P,n-1}\cdot(\mathcal{L}|_{n}^{n-1}+\mathcal{I}|_{n}^{n-1})+\mr{h.c.})+(H_{I}^{W}|_{n}^{n-1})_{\mr{quad}}.
\eeqa
 It can be re-written as 
\beqa
H_{I}^{W}|_{n}^{n-1}=(\mathcal{L}|_{n}^{n-1}+\mathcal{I}|_{n}^{n-1})\cdot\Gamma_{P,n-1}+\Delta(H_{I}^{W}|_{n}^{n-1})_{\mr{mix}}+(H_{I}^{W}|_{n}^{n-1})_{\mr{quad}}.
\eeqa
The contribution proportional to $\Delta(H_{I}^{W}|_{n}^{n-1})_{\mr{mix}}+(H_{I}^{W}|_{n}^{n-1})_{ \mr{quad} }$
in (\ref{new-number-added}), i.e., 
\begin{eqnarray}
& &\sum_{l=0}^{\infty}\sum_{l'=1}^{\infty}\oint_{\gamma_{P,n+1}}dw_{n+1}\{\frac{1}{H_{P,n-1}^{W+}-z{}_{n+1}}(-)H_{I}^{W}|_{n}^{n-1}\}^{l}\frac{1}{H_{P,n-1}^{W+}-z{}_{n+1}}( \Gamma_{P,n-1} )_i\times\non\\
 & &\quad\quad\quad\times\{\frac{1}{H_{P,n-1}^{W+}-w_{n+1}}(-)(H_{I}^{W}|_{n}^{n-1})\}^{l'-1}\frac{1}{H_{P,n-1}^{W+}-w_{n+1}} \times
\non\\
& &\ph{444444444444444}\times(-)\{\Delta(H_{I}^{W}|_{n}^{n-1})_{\mr{mix}}+(H_{I}^{W}|_{n}^{n-1})_{\mr{quad}}\}\frac{1}{H_{P,n-1}^{W+}-w_{n+1}}\phi_{P,n-1} 
\end{eqnarray}
is bounded in norm by
\begin{eqnarray}
 \sigma_{n+1}\frac{1}{1-c_{2}|\lambda|^{\frac{1}{2}}} \frac{c_{0}}{\sigma_{n+1}}\frac{1}{1-c{}_{2}|\lambda|^{\frac{1}{2}}}(c_{4}|\lambda|^{\frac{1}{2}}+c_{3}|\lambda|)\sigma_{n-1}\frac{c_{0}}{\sigma_{n+1}}
\leq  c_{0}^{2}(c_{4}+c_{3})|\lambda|^{\frac{1}{2}}\frac{\sigma_{n-1}}{\sigma_{n+1}}\bigg(\frac{1}{1-c{}_{2}|\lambda|^{\frac{1}{2}}} \bigg)^{2},\label{eq:fourth-summand}
\end{eqnarray}
where we have used  {\bcc   Lemma~\ref{lem:Basic-estimates}}.

As for the contribution proportional to $(\mathcal{L}|_{n}^{n-1}+\mathcal{I}|_{n}^{n-1})\cdot\Gamma_{P,n-1}$,
i.e., 
\begin{eqnarray}
& &\sum_{l=0}^{\infty}\sum_{l'=1}^{\infty}\oint_{\gamma_{P,n+1}}dw_{n+1}\{\frac{1}{H_{P,n-1}^{W+}-z{}_{n+1}}(-)H_{I}^{W}|_{n}^{n-1}\}^{l}\frac{1}{H_{P,n-1}^{W+}-z{}_{n+1}}(\Gamma_{P,n-1})_i\times\non\\
& &\quad\quad \times\{\frac{1}{H_{P,n-1}^{W+}-w_{n+1}}(-)H_{I}^{W}|_{n}^{n-1}\}^{l'-1}\frac{1}{H_{P,n-1}^{W+}-w_{n+1}}(-)(\mathcal{L}|_{n}^{n-1}+\mathcal{I}|_{n}^{n-1})\cdot\Gamma_{P,n-1}\times \non\\
& &\ph{4444444444444444444444444444444444444444444444}\times\frac{1}{H_{P,n-1}^{W+}-w_{n+1}}\phi_{P,n-1}, \label{splitting-label}
\end{eqnarray}
we split it into two pieces. The norm of the  summand proportional to $\mathcal{I}|_{n}^{n-1}\cdot\Gamma_{P,n-1}$ can be easily estimated in terms of 
\begin{eqnarray}
& &  \sigma_{n+1}\frac{1}{1-c{}_{2}|\lambda|^{\frac{1}{2}}}\cdot\frac{1}{1-c{}_{2}|\lambda|^{\frac{1}{2}}}\cdot\frac{c_{0}}{\sigma_{n+1}}c_I|\lambda|^{2}\sigma_{n-1}\sum_{j=1}^{3}\|\frac{1}{H_{P,n-1}^{W+}-w_{n+1}}(\Gamma_{P,n-1})_{j}\phi_{P,n-1}\|\frac{c_{0}}{\sigma_{n+1}}\non\\
& &\ph{444444444444444444444444444444444444} \leq  c^2_{0}\bigg(\frac{1}{1-c{}_{2}|\lambda|^{\frac{1}{2}}}\bigg)^{2}\cdot\bigg(\frac{\sigma_{n-1}}{\sigma_{n+1}}\bigg)\cdot  c_I |\lambda|^{2}\cdot\frac{3}{\sigma_{n-1}^{\delta}},\label{fourth-bis summand}
\end{eqnarray}
where we have used that $|\mathcal{I}|_{n}^{n-1}|\leq c_I |\lambda|^{2}\sigma_{n-1}$, $c_I$ being a universal constant, {\bcc   Lemma~\ref{lem:Basic-estimates}, the maximal modulus principle (Lemma~\ref{maximal-modulus-appl-one} and  {\color{\red}Remark \ref{max-mod-1}}) and 
the induction hypothesis a-i)}.
 {\bcc As for the other summand, making use of the direct integral representations from Subsection~\ref{direct-subsection} we can write
\beqa
& &\frac{1}{H_{P,n-1}^{W+}-w_{n+1}}(\mathcal{L}|_{n}^{n-1}\cdot\Gamma_{P,n-1})\frac{1}{H_{P,n-1}^{W+}-w_{n+1}}\phi_{P,n-1}\non\\
& &\ph{444} \simeq\mco(\si_{n+1}^{-1})(-\la)\hint^{\oplus}_{\!\!\!\!\bmcA|^{n-1}_n}  \fr{d^3k}{\sqrt{2}|k|^{3/2}} 
\frac{k_{i}}{\alpha_{P,n-1}(\hat{k})} \frac{1}{[H_{P,n-1}^{W+}]_k-w_{n+1}} (\Gamma_{P,n-1})_i\phi_{P,n-1}, \label{first-proof-direct-int}
\eeqa
where $\mco(\si_{n-1}^{-1}):=(E_{P,n-1}+\De c_P|^{n-1}_n-w_{n+1})^{-1}$. Now by (\ref{estimate-Q}) and  the induction hypothesis {\color{\red}combined with Remark \ref{max-mod-1}} we get 
\beqa
\|(\ref{first-proof-direct-int})\|\leq c |\la|\bigg(\fr{\si_{n-1}}{\si_{n+1}}\bigg) |\ln\, \veps|^{1/2}\fr{1}{\si_{n-1}^{\de}}.
\label{first-proof-direct-int-x}
\eeqa
Hence the norm of the summand in (\ref{splitting-label}) proportional to $(\mathcal{L}|_{n}^{n-1}\cdot\Gamma_{P,n-1})$ is bounded by
\beqa
\sigma_{n+1}\frac{1}{1-c{}_{2}|\lambda|^{\frac{1}{2}}}\cdot\frac{1}{1-c{}_{2}|\lambda|^{\frac{1}{2}}}\cdot\frac{c_{0}}{\sigma_{n+1}} 
c |\la|\bigg(\fr{\si_{n-1}}{\si_{n+1}}\bigg) |\ln\, \veps|^{1/2}\fr{1}{\si_{n-1}^{\de}}=c_0 \bigg(\frac{1}{1-c{}_{2}|\lambda|^{\frac{1}{2}}}\bigg)^2
c |\la|\bigg(\fr{\si_{n-1}}{\si_{n+1}}\bigg) |\ln\, \veps|^{1/2}\fr{1}{\si_{n-1}^{\de}}. \label{eq:fifth-summand-new}
\eeqa
}

The sum of all contributions (\ref{eq:first-summand}), (\ref{eq:second-summand}),
(\ref{eq:third-summand}), (\ref{eq:fourth-summand}),  (\ref{fourth-bis summand}), and (\ref{eq:fifth-summand-new}), 
yields
\begin{eqnarray}
 & & \sigma_{n+1}\cdot\frac{1}{1-c_{2}|\lambda|^{\frac{1}{2}}}c_{2}|\lambda|^{\frac{1}{2}}\cdot\frac{1}{1-c_{2}|\lambda|^{\frac{1}{2}}}\cdot\frac{c_{0}}{\sigma_{n+1}}\non\\
& &+  \frac{1}{1-c{}_{2}|\lambda|^{\frac{1}{2}}}\cdot|\lambda|^{\frac{1}{2}}c_{2}+\frac{1}{1-c{}_{2}|\lambda|^{\frac{1}{2}}}\cdot
\frac{ c_{0} }{\sigma_{n+1}}\cdot c_{1}\cdot[\lambda^{2}\sigma_{n-1}+4|\lambda|^{\frac{1}{4}}\sigma_{n-1}^{1-\delta}] +\bigg(\frac{1}{1-c{}_{2}|\lambda|^{\frac{1}{2}}}\bigg)\frac{ 1 }{\sigma_{n-1}^{\delta}}\non\\
& &+  c_{0}c_{2}|\lambda|^{\frac{1}{2}}\bigg(\frac{1}{1-c_{2}|\lambda|^{\frac{1}{2}}}\bigg)^{2}\{\frac{c_{0}}{\sigma_{n+1}}c_{1}[\lambda^{2}\sigma_{n-1}+4|\lambda|^{\frac{1}{4}}\sigma_{n-1}^{1-\delta}]+c_{2}|\lambda|^{\frac{1}{2}}\}\non\\
 & &+  c_{0}^{2}(c_{4}+c_{3})|\lambda|^{\frac{1}{2}}\frac{\sigma_{n-1}}{\sigma_{n+1}}\bigg(\frac{1}{1-c{}_{2}|\lambda|^{\frac{1}{2}}}\bigg)^{2}+c^2_{0}\bigg(\frac{1}{1-c{}_{2}|\lambda|^{\frac{1}{2}}}\bigg)^{2}\cdot\bigg(\frac{\sigma_{n-1}}{\sigma_{n+1}}\bigg)\cdot c_{I}|\lambda|^{2}\cdot\frac{3 }{\sigma_{n-1}^{\delta}}\non\\
& &{\bcc +c_0 \bigg(\frac{1}{1-c{}_{2}|\lambda|^{\frac{1}{2}}}\bigg)^2
c |\la|\bigg(\fr{\si_{n-1}}{\si_{n+1}}\bigg) |\ln\, \veps|^{1/2}\fr{1}{\si_{n-1}^{\de}} }
\leq  \frac{\mathcal{\mathcal{C}}^{(i)}}{\sigma_{n-1}^{\delta}}
\end{eqnarray}
for some universal constant $\mathcal{C}^{(i)}$, {\bcc where we made use of $|\la|\leq \veps^8$. (The power eight is needed in the second line above in the term involving $|\la|^{1/4}$).
Thus, by setting $\epsilon(\delta):=\min\{\frac{1}{2};(1/\mathcal{C}^{(i)})^{\frac{1}{\delta}}\}$
and choosing $\la_0^{(i)}\equiv\la^{(i)}_0(\veps(\de))>0$ (within all the earlier restrictions) such that
\beqa
\lambda_{0}^{(i)} \leq \epsilon(\de)^8, \label{first-restriction-la-eps} 
\eeqa
we have proven that c-i) is implied by a-i), a-ii), a-iii) for all $|\la|\in (0,\la_0^{(i)}]$ and $\veps\in (0, \veps(\de)]$ s.t.
$|\la|\leq \veps^8$.} \\

\emph{\noun{Claim-ii)}}\emph{ (Here we show the implication: c-i)
$\rightarrow$} \emph{c-ii))}\\
{\bcc We choose $\veps$ and $\la$ as specified above.} 
We set $\m:=n+1$ and start from the difference
\begin{eqnarray}
   \hat{\phi}_{P,n+1}-\phi_{P,n}
\1 &=& \1\sum_{j=1}^{\infty}\oint_{\gamma_{P,\m+1}}dw_{\m+1}\{\frac{1}{H_{P,\m-1}^{W+}-w_{\m+1}}(-)H_{I}^{W}|_{\m}^{\m-1}\}^{j-1}\times\non\\
 & &\times\frac{1}{H_{P,\m-1}^{W+}-w_{\m+1}}(-)[(\mathcal{L}|_{\m}^{\m-1}+\mathcal{I}|_{\m}^{\m-1})\cdot
\Gamma_{P,\m-1}+\Delta(H_{I}^{W+}|_{\m}^{\m-1})_{\rm{mix}}+(H_{I}^{W}|_{\m}^{\m-1})_{\rm{quad}}]\times\non\\
& &\ph{44444444444444444444444444444444444444444444}\times\frac{1}{H_{P,\m-1}^{W+}-w_{\m+1}}\phi_{P, \m-1}. \label{needed-for-smooth-cut-off}
\end{eqnarray}
 By   Lemma \ref{lem:Basic-estimates},
the contribution proportional to $\Delta(H_{I}^{W}|_{\m}^{\m-1})_{\mr{mix}}+(H_{I}^{W}|_{\m}^{\m-1})_{\mr{quad}}$
can be bounded in norm by
\beqa
\sigma_{\m+1}\frac{1}{1-c_{2}|\lambda|^{\frac{1}{2}}}\cdot(c_{4}|\lambda|^{\frac{1}{2}}+c_{3}|\lambda|)\sigma_{\m-1}\frac{c_{0}}{\sigma_{\m+1}}.
\eeqa
{\bcc (Clearly, the operator norms involved in these estimates are on $B(\mcF_m)$ rather than on $B(\mcF_n)$).}
We can rewrite the rest as follows 
\begin{eqnarray}
& &   -\sum_{\q=1}^{3}(\mathcal{I}|_{\m}^{\m-1})_{\q}\sum_{j=1}^{\infty}\oint_{\gamma_{P,\m+1}}dw_{\m+1}\bigg(\frac{1}{E_{P,\m-1}^+-w_{\m+1}}\bigg)
\{\frac{1}{H_{P,\m-1}^{W+}-w_{\m+1}}(-)H_{I}^{W}|_{\m}^{\m-1}\}^{j-1}\times \non\\
& &\ph{44444444444444444444444444444444444444}\times\frac{1}{H_{P,\m-1}^{W+}-w_{\m+1}}(\Gamma_{P,\m-1})_{\q}\phi_{P,\m-1}
\label{I-app-B}\\
& &-\sum_{\q=1}^{3}\sum_{j=1}^{\infty}\oint_{\gamma_{P,\m+1}}dw_{\m+1}\bigg(\frac{1}{E_{P,\m-1}^+-w_{\m+1}}\bigg)\{\frac{1}{H_{P,\m-1}^{W+}-w_{\m+1}}(-)H_{I}^{W}|_{\m}^{\m-1}\}^{j-1}\times\non\\
& &\ph{44444444444444444444444444444444444444}\times\frac{1}{H_{P,\m-1}^{W+}-w_{\m+1}}(\mathcal{L}|_{\m}^{\m-1})_{\q}^{(+)} (\Gamma_{P,\m-1})_{\q}\phi_{P,\m-1},  \label{C-29} 
\end{eqnarray} 
where $E_{P,\m-1}^+:=E_{P,\m-1}+\Delta c_P|^{m-1}_{m}$.
{\bcc  Using that $|(\mathcal{I}|_{\m}^{\m-1})_{\q}|\leq c_{I}|\lambda|^{2}\sigma_{\m-1}$ and applying the maximal modulus principle (Lemma~\ref{maximal-modulus-appl-one}  and  {\color{\red}Remark \ref{max-mod-1}}) and c-i) to (\ref{I-app-B}), we obtain
\beqa
\|(\ref{I-app-B})\|\leq \frac{3c_0\cdot c_{I}|\lambda|^{2}\sigma_{\m-1}}{1-c_{2}|\lambda|^{\frac{1}{2}}}
 \frac{1}{\sigma_{\m-1}^{\delta}}.
\eeqa
Next, arguing as in (\ref{first-proof-direct-int}), (\ref{first-proof-direct-int-x}) above and making use of c-i) we obtain
\beqa
\big\|\frac{1}{H_{P,\m-1}^{W+}-w_{\m+1}}(\mathcal{L}|_{\m}^{\m-1})_{\q}^{(+)} (\Gamma_{P,\m-1})_{\q}\phi_{P,\m-1}\big\|\leq 
c |\la|  \si_{m-1} |\ln\, \veps|^{1/2}\fr{1}{\si_{m-1}^{\de}},
\eeqa
and therefore
\beqa
\|(\ref{C-29})\|\leq \frac{3c_{0}}{1-c_{2}|\lambda|^{\frac{1}{2}} }c |\la|  \si_{m-1} |\ln\, \veps|^{1/2}\fr{1}{\si_{m-1}^{\de}}.
\eeqa
}
Summing up, we have
\begin{eqnarray}
\|\hat{\phi}_{P,n+1}-\phi_{P,n}\| \1 & \leq & \1 \frac{c_{0}}{1-c_{2}|\lambda|^{\frac{1}{2}}}\cdot(c_{4}|\lambda|^{\frac{1}{2}}+c_{3}|\lambda|)\sigma_{\m-1}\non\\
 &  & +\frac{3c_0\cdot c_{I}|\lambda|^{2}\sigma_{\m-1}}{1-c_{2}|\lambda|^{\frac{1}{2}}}
 \frac{1}{\sigma_{\m-1}^{\delta}}+{\bcc \frac{3c_{0}}{1-c_{2}|\lambda|^{\frac{1}{2}}} c |\la|  \si_{m-1} |\ln\, \veps|^{1/2}\fr{1}{\si_{m-1}^{\de}}  }\non\\
 \1& \leq & \1 \mathcal{C}^{(ii)} |\lambda|^{\frac{1}{2}}\sigma_{\m-1}^{1-\delta}= \mathcal{C}^{(ii)} |\lambda|^{\frac{1}{2}}\sigma_{n}^{1-\delta}
\end{eqnarray}
for some universal constant $\mathcal{C}^{(ii)}$, and finally
\beqa
\|\hat{\phi}_{P,n+1}-\phi_{P,n}\|\leq|\lambda|^{\frac{1}{4}}\sigma_{n}^{1-\delta} \label{second-restriction-la-eps}
\eeqa
for 
${\bcc |\lambda|\in (0,  \lambda_{0}^{(ii)}], \textrm{ where }  \lambda_{0}^{(ii)}\equiv\min\{\lambda_{0}^{(i)},[\frac{1}{\mathcal{C}^{(ii)}   }]^{4}\}. }$ \\

\noun{Claim-iii)} \emph{(Here we show the implication: c-ii) and a-iii)
$\rightarrow$} \emph{c-iii))}\\
{\bcc We choose $\veps$ and $\la$ within the restrictions specified below formulas (\ref{first-restriction-la-eps}) and (\ref{second-restriction-la-eps}).}
{\bcc By relation~(\ref{phi-definitions})
\begin{equation}
\|\phi_{P,n+1}-\hat{\phi}_{P,n+1}\|=\|\big(W_{P,n+1}\wt{W}_{P,n+1}^*-1\big)\hat{\phi}_{P,n+1}\|. \label{starting-claim-three}
\end{equation}
Using Lemma~\ref{vector-shift-Lemma} and relation~(\ref{grad-difference}),  {\bcb the r.h.s of} (\ref{starting-claim-three}) 
can be estimated less than}
\begin{eqnarray}
& &c|\lambda||\nabla E_{P,n+1}-\nabla E_{P,n}||\ln\sigma_{n+1}| \, \|\hat\phi_{P,n+1}\|  \non\\
& &\ph{4444444444}\leq  c|\lambda|c_{1}|\ln\sigma_{n+1}|[\lambda^{2}\sigma_{n}+\|\frac{\hat{\phi}_{P,n+1}}{\|\hat{\phi}_{P,n+1}\|}-\frac{\phi_{P,n}}{\|\phi_{P,n}\|}\|] \, \|\hat\phi_{P,n+1}\|  \nonumber\\
& &\ph{4444444444} \leq  c|\lambda|c_{1}|\ln\sigma_{n+1}|[\lambda^{2}\sigma_{n}\|\hat\phi_{P,n+1}\|   +\|\hat{\phi}_{P,n+1}-\phi_{P,n}\|+\big|\|\phi_{P,n}\|-\|\hat{\phi}_{P,n+1}\|\big| ]\nonumber\\
& &\ph{4444444444} \leq  c|\lambda|c_{1}|\ln\sigma_{n+1}|[\lambda^{2}\sigma_{n}+2|\lambda|^{\frac{1}{4}}\sigma_{n}^{1-\delta} ] 
  \leq  \mathcal{C}^{'(iii)} |\la|^{\fr{5}{4}} \sigma_{n}^{1-\delta}|\ln\sigma_{n+1}|,\label{vector-difference-x-bis}
\end{eqnarray}
 where {\bcb we made use of c-ii)}, the fact that $\|\phi_{P,n+1}\|\leq 1$, and
 $\mathcal{C}^{'(iii)}$ is some universal constant. Furthermore, we can write
\beqa
\mathcal{C}^{'(iii)} |\lambda|^{\frac{5}{4}}\sigma_{n}^{1-\delta}|\ln\sigma_{n+1}| 
\1 & \leq& \1 \mathcal{C}^{'(iii)} |\lambda|^{\frac{5}{4}}   \fr{\sigma_{n+1}^{1-\delta}}{\epsilon^{1-\delta}}   |\ln\sigma_{n+1}| 
= \mathcal{C}^{'(iii)} (|\la|^{1/4} \sigma_{n+1}^{1/4} )   |\lambda|   \fr{\sigma_{n+1}^{3/4-\delta}}{\epsilon^{1-\delta} }   
|\ln\sigma_{n+1}| \non\\
\1 &\leq& \1 \mathcal{C}^{'(iii)} |\lambda|^{1/2}   (|\la|^{1/4} \sigma_{n+1}^{1/2} )   \fr{|\lambda|^{1/2} }{\epsilon}   
\sigma_{n+1}^{1/4} |\ln\sigma_{n+1}| 
\leq \mathcal{C}^{(iii)} |\lambda|^{1/2}   ( |\la|^{1/4} \sigma_{n+1}^{1/2} ), \label{shifted-trivial-computation}
\eeqa
{\color{black}for some universal constant  $\mathcal{C}^{(iii)}(=c\,\mathcal{C}^{'(iii)})$}, where we  used  that   $\fr{|\lambda|^{\frac{1}{4}}}{\epsilon}\leq 1$ and $\sigma_{n+1}^{{\color{black}1/4}} |\ln\sigma_{n+1}|\leq c$.
Altogether, we have by (\ref{starting-claim-three}), (\ref{vector-difference-x-bis}), (\ref{shifted-trivial-computation})
\beqa
\|\phi_{P,n+1}-\hat{\phi}_{P,n+1}\|\leq  \mathcal{C}^{(iii)} |\lambda|^{1/2}   ( |\la|^{1/4} \sigma_{n+1}^{1/2} ). \label{n+1-vector-bound}
\eeqa
Thus we obtain by the inverse triangle inequality, (\ref{n+1-vector-bound}) and c-ii)
\begin{eqnarray}
\|\phi_{P,n+1}\| \1 & \geq & \1 \|\phi_{P,n}\|-\|\phi_{P,n+1} -\hat{\phi}_{P,n+1}\|- \|\hat\phi_{P,n+1}-\phi_{P,n}\|\geq \|\phi_{P,n}\|-  (\mathcal{C}^{(iii)} |\lambda|^{1/2} +1) 
|\lambda|^{\frac{1}{4}}\sigma_{n}^{1/2}. \label{n+1-estimate}   
\end{eqnarray}
{\bcc
 {\bcb We choose $\la_0$ s.t. $\mathcal{C}^{(iii)} |\lambda_0|^{1/2}\leq 1$ and obtain from (\ref{n+1-estimate})  and a-iii)}
\beqa
\|\phi_{P,n+1}\| \geq  1-  {\bcb 2}\sum_{j=0}^{n}\{|\lambda|^{\frac{1}{4}}\sigma_{j}^{1/2}
\} 
\geq  1- {\bcb 2}\fr{|\la|^{1/4}  }{1-\veps^{1/2} }.
\label{sum-over-j}
\eeqa
We obtain from the last formula that c-ii) and a-iii) imply c-iii) provided that
\beqa
|\la|\in (0, \la_0], \quad \la_0\equiv \min\bigg( \la_0^{(ii)},   \fr{1}{(\mathcal{C}^{(iii)})^2  }, 
\bigg( \fr{ 1-\veps(\de)^{1/2}}{4}     \bigg)^4 \bigg). \label{epsilon-restrictions}
\eeqa
This concludes the proof. \qed
}

\section{A pull through identity and its consequences}\label{pull-through-appendix}
\setcounter{equation}{0}

We refer to (\ref{Froehlich-form}) below as a pull through identity. It has several useful consequences { (see e.g. (\ref{vector-difference-x-bis}) above)} which will be studied in this appendix. The analysis of pull-through estimates dates back to \cite{Fr73}, but our  treatment of technical aspects also profited from \cite{KM12}.
We will use the domain
\beqa
C^{\infty}(H_{P,\free}):=\bigcap_{\ell\geq 0}D(H_{P,\free}^{\ell}).  
\eeqa
 We also recall the standard bounds, valid for any $P\in S$ and $\ell \in \nat$ 
 \beqa
\|H_{P, \free}^{\ell}(H_{P,n}+i)^{-\ell} \|\leq c,     \label{ell-energy-bounds}      
\eeqa
For $\ell=1$ these estimates follow from Lemmas~\ref{energy-bounds}, \ref{energy-bound-zero}. For $\ell\geq 1$ we refer to 
\cite{FGS01, Ma10, DM15} for the standard arguments. 

By Lemma~\ref{energy-bounds}, for  $\psi\in C^{\infty}(H_{P,\free}) $ we can define $\real^{3m}\ni (k_1, \ldots, k_m) \mapsto b(k_1)\ldots b(k_m)\psi$ as a vector-valued distribution on $S(\real^{3m})$  which gives rise
to a continuous map on $L^2_{\om}(\real^3)^{\times m}$, where $L^2_{\om}(\real^3):=\{\,  f\in L^2(\real^3)\,|\,  \| |k|^{-1/2}f \|_2<\infty \, \}$. For some vectors $\psi\in  C^{\infty}(H_{P,\free})$ we can also define $b(k_1)\ldots b(k_m)\psi$ pointwise in $(k_1, \ldots, k_m)$ as follows: 
\begin{defn}\label{defeta}
Let $\eta\in C_0^{\infty}(\real^3)$ be a spherically symmetric function s.t. $\eta\geq 0$, $\int \eta(k') d^3k'=1$,  so that $\eta_k^{\eps}(k'):=\eps^{-3}\eta((k'-k)\eps^{-1})$ is an approximating sequence of $k'\mapsto \de(k'-k)$ as $\eps\to 0$. Then, for any $(k_1,\ldots, k_m)\in \real^{3m}$,   we say that 
$\psi\in  C^{\infty}(H_{P,\free})$ is in $D(b(k_1)\ldots b(k_m) )$  if the following limit exists in norm 
\beqa
{\bcb [b(k_1)\ldots  b(k_m)\psi]:=\lim_{\eps\to 0} b(\eta_{k_1}^{\eps}) \ldots b(\eta_{k_m}^{\eps})\psi.} \label{domain-b(k)}
\eeqa
For $m=0$ we set by
convention $D(b(k_1)\ldots b(k_m) ):=C^{\infty}(H_{P,\free})$. 
\end{defn}
 
 Lemma \ref{lemma-distr} below shows that this definition is consistent with  the definition as a distribution and, consequently, that the dependence on  $\eta$   is inessential after smearing. 
In the following lemma we will  use the notation $\un k_m:=k_1+\cdots+k_m$ and $|\un k|_m:=|k_1|+\cdots+|k_m|$.

{\color{black}

\bel\cite{Fr73}\label{pull-through-lemma} Let $|\la|\in (0,\la_0]$,  $|P|\leq P_{\max}=1/3$ and $m\in \nat_0$. Then:
\begin{enumerate}
\item[(a)]  For $k_1, \ldots, k_m\in \real^3\backslash \{0\}$, $m\geq 1$,  
the following bound holds true
\beqa
\big\|\frac{1}{E_{P,n}-|\un k|_m -H_{P-\un{k}_m,n}}\big\| \leq \fr{c}{|\un k|_m}. \label{resolvent-k-bound}
\eeqa
\item[(b)]  $\psi_{P,n}\in D(b(k_1)\ldots b(k_m) )$ for  $k_1, \ldots, k_m\in \real^3\backslash \{0\}$ and satisfies
\beqa
[b(k_1) \ldots b(k_m)\psi_{P,n}]= \frac{1}{E_{P,n}-|\un k|_m -H_{P-\un{k}_m,n}}\sum_{i=1}^m \vv^{\si_n}(k_i)[b(k_1)\ldots\check{i}\ldots b(k_m)\psi_{P,n}].
\label{Froehlich-form}
\eeqa
{\bcb If $|k_i|=\si_n$ for some $i=1, \ldots, m$, the corresponding $\vv^{\si_n}(k_i)$ must be multiplied by $\h$.}

\item[(c)]  For any $f_1, \ldots f_m\in L^2(\real^3)\cap L^1(\real^3)$, supported away from zero\footnote{For consistency with the discussion above 
Definition~\ref{defeta}, we note that if $f\in L^2(\real^3)$ is supported away from zero, then it is automatically in $L^2_{\om}(\real^3)$. },
\beqa
\int d^3k_1\ldots d^3k_m\, \bar f_1(k_1)\ldots \bar f_m(k_m)[b(k_1) \ldots b(k_m)\psi_{P,n}]=b(f_1)\ldots b(f_m)\psi_{P,n},
\eeqa
where the l.h.s. is a weak integral and on the r.h.s. the usual definition of the annihilation operators is understood. 

 \end{enumerate}

\eel
\proof  To prove (a)  we note that,  by definition of $E_{P-\un{k}_m,n}$, we have $H_{P-\un{k}_m,n}-E_{P-\un{k}_m,n}\geq 0$  and by (\ref{eq:difference-bound}) $E_{P-\un{k}_m,n}-E_{P,n}+|\un k|_m\geq \big(2/3-c|\lambda|\big)|\un k|_m> 0$ for $|\un k|_m\neq 0$ (cf. also Remark~\ref{energy-consistency}). 
Therefore  $H_{P-\un{k}_m,n}+|\un k|_m-E_{P,n} \geq c'|\un k|_{m}>0$  and (\ref{resolvent-k-bound}) follows. 

{\bcb To prove (b) and (c) we first assume that $|k_i|\neq \si_n$, $i=1, \ldots, m$,  and only in the last part of the proof we will explain how to drop this assumption.}    We proceed by induction, namely 
we suppose that  for  $m'=0,1,\ldots, m-1$ the following inductive assumptions hold
\begin{enumerate}
\item[a-(b)] $\psi_{P,n}\in D(b(k_1)\ldots b(k_{m'}))$ for all $ k_1, \ldots k_{m'}\in \real^3\backslash \{0\}$ and  
\beqa
[b(k_1) \ldots b(k_{m'})\psi_{P,n}]= \frac{1}{E_{P,n}-|\un k|_{m'} -H_{P-\un{k}_{m'},n}}\sum_{i=1}^{m'} \vv^{\si}(k_i)[b(k_1)\ldots\check{i}\ldots b(k_{m'})\psi_{P,n}]. \label{Froehlich-before-iteration}
\eeqa

\item[a-(c)] $\int d^3k_1\ldots d^3k_{m'}\, \bar f_1(k_1)\ldots \bar f_{m'}(k_{m'})[b(k_1) \ldots b(k_{m'})\psi_{P,n}]=b(f_1)\ldots b(f_{m'})\psi_{P,n}$ for
any $f_1, \ldots f_{m'}\in L^2(\real^3)\cap L^1(\real^3)$ supported away from zero.
\end{enumerate}
{\bcb For $m=1$ the only content of our inductive assumption is that $\psi_{P,n}\in C^{\infty}(H_{P,\free})$, which  holds by (\ref{ell-energy-bounds})}.
We will conclude from these assumptions statements (b) and (c) of the lemma for $m$, thereby closing the inductive argument. As a preparation, we note that by iteration
of (\ref{Froehlich-before-iteration}) we get (cf. \cite{Fr73})
\beqa
[ b(q_{1})\ldots b(q_{m-1}) \psi_{P,n}]=\sum_{\pi\in S_{m-1}}\{(-1)^{m-1}\prod_{i=m-1}^{1}\frac{1}{H_{P-\un{q}_{\pi,i},n}-E_{P,n}+|\un{q}|_{\pi,i}}\vv^{\sigma_n}(q_{\pi(i)}) \psi_{P,n}\} \label{product-of-b},
\eeqa
where $\pi\in S_{m-1}$ is any permutation of $(1,\dots,m-1)$ and $\un{q}_{\pi,i}=\sum_{i'=1}^{i}q_{\pi_{i'}}$, $|\un{q}|_{\pi,i}=\sum_{i'=1}^{i}|q_{\pi_{i'}}|$. (We renamed here the variables from $k_i$ to $q_i$ to facilitate applications in the later part of the proof, where $q_i$ will denote smearing variables). By the bound from part (a) of the lemma and a combinatorial argument from \cite{Fr73} we obtain a standard bound
\beqa
\| [ b(q_{1})\ldots b(q_{m-1}) \psi_{P,n}]\|\leq \prod_{i=1}^{m-1}\fr{c \vv^{\si_n}(q_i)  }{|q_i|}. \label{Froehlich-bound-AD}
\eeqa 
Making use of a-(c) and (\ref{Froehlich-bound-AD}) we also obtain for $k_1, \ldots, k_{m-1}$ in  any compact set $K\subset \real^3\backslash \{0\}$ and any continuous functions $g_1, \ldots, g_{m-1}$ 
\beqa
\| b(g_1\eta_{k_1}^{\eps}) \ldots b(g_{m-1}\eta_{k_{m-1}}^{\eps})\psi_{P,n}\| \1  &\leq& \1 \prod_{i=1}^{m-1}\int d^3q_i\, \fr{c \vv^{\si_n}(q_i)  }{|q_i|} |g_i(q_i) |
\eta_{k_i}^{\eps}(q_i)\non\\
\1&\leq& \1 \prod_{i=1}^{m-1}  \int d^3q_i\, \eta(q_i)|g_i(k_i+\eps q_i) |\fr{\chi_{[\si_n,\ka)}(k_i+\eps q_i) | k_i+\eps q_i|^{\alf}  } 
{ |k_i+\eps q_i |^{3/2} } \non\\
\1 & \leq &\1  C_{K,m-1}\prod_{i=1}^{m-1} \sup_{q\in \supp\,\eta} |g_i(k_i+\eps q) |\leq C_{K,m-1}' . \label{g-v-expression-x}
\eeqa
 The last bound holds uniformly in $k_1, \ldots, k_{m-1}\in K$  and in  $\eps$ s.t. $|k_i+\eps q|>\h |k_i|$ for all $q\in \supp\, \eta$. (The constants $C_{K,m-1}, C_{K,m-1}'$ do not depend on $n, \alf$. But the latter constant depends on $g_i$). The estimate (\ref{g-v-expression-x}) will be important for our
 proof of part (c) of the lemma for $m$ via Lemma~\ref{lemma-distr}. 

Let us now proceed to the proof of (\ref{Froehlich-form}). In the following, the expressions $b(q_1)\ldots b(q_m)\psi$, for $\psi\in C^{\infty}(H_{P,\free})$,
are understood in the sense of distributions unless stated otherwise. More precisely, the following smearing is understood 
\begin{equation}
b(f_m)\dots b(f_1)\psi = \int d^3q_1\dots d^3q_m\, \bar{f}_m(q_m)  \dots \bar{f}_1(q_1)b(q_m)\ldots b(q_1)\psi \label{App-D-smearing}
\end{equation}
with functions $f_1, \ldots, f_m\in L^2(\real^3)\cap L^1(\real^3)$ supported away from zero.
 
 By canonical commutation relations, for any $\psi\in C^{\infty}(H_{P,\free})$
\beqa
b(\ti k)H_{P-k,\si}\psi=(H_{P-k-\ti k,\si}+|\ti k|) b(\ti k)\psi+\vv^{\si}(\ti k)\psi  \label{Froehlich-pull-through-formula}
\eeqa
holds in the sense of distributions. Next,   by iteration of (\ref{Froehlich-pull-through-formula}), we obtain that
\beqa
b(q_m)\ldots b(q_1)H_{P,n}\psi=(H_{P-\un{q}_m,n}+|\un q|_m)   b(q_m)\ldots b(q_1)\psi+
\sum_{i=1}^m\vv^{\si}(q_i)b(q_m)\ldots {\color{\red}\check{i}} \ldots b(q_1)\psi, \label{n-part-pull-x}
\eeqa
where analogous smearing as in (\ref{App-D-smearing}) is understood. 
To move on, we set $\psi\equiv \psi_{P,n}$ in (\ref{n-part-pull-x})
and introduce the following notation:
\beqa
\un{q}\1&:=&\1(q_{m-1}, \ldots, q_1), \\  
\un{q}^i\1&:=&\1(q_{m-1}, \ldots \check{i} \ldots q_1), \\
\un{q}^{i, i'}\1&:=&\1(q_{m-1}, \ldots \check{i} \ldots \check{i'}\ldots q_1), \textrm{ for }  i\neq i', \\
A_{\un{q}}(q_m)\1&:=&\1(E_{P,n}-|\un q|_{m-1}- |q_m|- H_{P-\un q_{m-1}- q_m ,n}), \label{A-defin-x}\\
B_{m-1}(  \un{q})\1&:=&\1b(q_{m-1})\ldots b(q_1)\,,\,\,\,\, \quad   {\color{black}B_{m-1}=1\quad \text{for}\,\, m=1}, \label{B-first-def}\\ 
B_{m-1}^i(  \un{q}^i  )\1&:=&\1 b(q_{m-1})\ldots \check{i} \ldots b(q_1) \,,\,\, {\color{black}B^i_{m-1}=0\quad \text{for}\,\, m=1,} \\
B_{m-1}^i(  \un{q}^{i,i'}  )\1&:=&\1 b(q_{m-1})\ldots \check{i} \ldots \check{i'} \ldots b(q_1) \,,\,\, {\color{black}B^{i,i'}_{m-1}=0\quad \text{for}\,\, m=1,2.}\label{B-second-def}
\eeqa
We use $H_{P,n}\psi_{P,n}=E_{P,n}\psi_{P,n}$,  and combine the term on the l.h.s. of (\ref{n-part-pull-x})
with the first term on the r.h.s. so as to obtain
\beqa
A_{\un{q}}(q_m)b(q_m) B_{m-1}(\un q) \psi_{P,n}
=\vv^{\si}(q_m) B_{m-1}(\un q)\psi_{P,n} +    \sum_{i=1}^{m-1} \vv^{\si_n}(q_i)b(q_m)  B_{m-1}^i(\un q^i) \psi_{P,n}, 
\label{aux-compt}
\eeqa
where we also separated the $i=m$ contribution to the sum on the r.h.s. of  (\ref{n-part-pull-x}). Now we pick some 
$k_1, \ldots, k_m\in K$,  where $K$ is compact and $0\notin K$, write
$A_{\un{q}}(q_m)= A_{\un{k}}(k_m)+ (A_{\un{q}}(q_m)- A_{\un{k}}(k_m) )$ and obtain from (\ref{aux-compt}):
\beqa
A_{\un{k}}(k_m)  b(q_m) B_{m-1}(\un q) \psi_{P,n} \1 &=& \1 \vv^{\si}(q_m) B_{m-1}(\un q)\psi_{P,n} +    \sum_{i=1}^{m-1} \vv^{\si_n}(q_i)b(q_m)  B_{m-1}^i(\un q^i) \psi_{P,n}\non\\
\1 & & \1 +( A_{\un{k}}(k_m)  -  A_{\un{q}}(q_m) ) b(q_m) B_{m-1}(\un q) \psi_{P,n}. \label{dividing-A}
\eeqa 
Next we divide both sides of (\ref{dividing-A}) by $A_{\un{k}}(k_m) $, which is legitimate as it does not involve any smearing variables $q_i$, $i=1, \ldots, m$ and, by part (a)
of the lemma, $\| A_{\un{k}}(k_m)^{-1}\|\leq  c|k_m|^{-1}$.  
 This gives:
\beqa
b(q_m) B_{m-1}(\un q) \psi_{P,n} \1 &=& \1 A_{\un{k}}(k_m)^{-1}\vv^{\si}(q_m) B_{m-1}(\un q)\psi_{P,n} +  
 A_{\un{k}}(k_m)^{-1} \sum_{i=1}^{m-1} \vv^{\si_n}(q_i)b(q_m)  B_{m-1}^i(\un q^i) \psi_{P,n} \label{bB-formula-one-x}\\
& &\ph{4444444444444444}+A_{\un{k}}(k_m)^{-1}(  A_{\un{k}}(k_m) -  A_{\un{q}}(q_m)    ) b(q_m) B_{m-1}(\un q) \psi_{P,n}. \label{bB-formula-two-x} 
\eeqa
To proceed, we introduce the smearing function $\eta_{\un k, k_m}^{\eps}(q_{m}, \ldots, q_1):=  \eta_{k_m}^{\eps}(q_{m})
 \ldots \eta_{k_1}^{\eps}(q_1)$ and set
\beqa
& &B^{\eps}_{m-1}(\un k):= b( \eta_{k_{m-1}}^{\eps}   ) \ldots b( \eta_{k_1}^{\eps}  ), \\
& &B^{\eps,i}_{m-1}(\un k^i):= b( \eta_{k_{m-1}}^{\eps}   ) \ldots \check{i}\ldots b( \eta_{k_1}^{\eps}  ), \label{B-eps-not}\\
& &B^{\eps,i,i'}_{m-1}(\un k^{i,i'}):= b( \eta_{k_{m-1}}^{\eps}   ) \ldots \check{i}\ldots  \check{i'} \ldots b( \eta_{k_1}^{\eps}  ).
\eeqa
By smearing with the respective test-functions we obtain from (\ref{bB-formula-one-x})--(\ref{bB-formula-two-x}) 
\beqa
b( \eta_{k_m}^{\eps} ) B^{\eps}_{m-1}(\un k) \psi_{P,n} \1 &=& \1 
A_{\un{k}}(k_m)^{-1}\lan  \eta_{k_m}^{\eps}, \vv^{\si}\ran B^{\eps}_{m-1}(\un k)\psi_{P,n} +  
 A_{\un{k}}(k_m)^{-1} \sum_{i=1}^{m-1}   \lan \eta_{k_i}^{\eps}, \vv^{\si_n}\ran \,    b( \eta_{k_m}^{\eps} )  B_{m-1}^{\eps,i}(\un k^i) \psi_{P,n}\quad\quad 
 \label{bB-formula-one}\\
& &+\int d^{3(m-1)}\un{q} d^3q_m\, \eta_{\un k, k_m}^{\eps}(\un{q}, q_m ) \, A_{\un{k}}(k_m)^{-1}(  A_{\un{k}}(k_m) -  A_{\un{q}}(q_m)    ) 
b(q_m) B_{m-1}(\un q) \psi_{P,n}.\quad\quad\quad \label{bB-formula-two} 
\eeqa
\textbf{Analysis of (\ref{bB-formula-one})}. This is the leading term and we will show that it converges as $\eps\to 0$ to the expression
on the r.h.s. of (\ref{Froehlich-form}). To close the inductive argument we also need to show a uniform bound analogous
to (\ref{g-v-expression-x}).

Since  $k_m\neq 0$, the first part of the lemma gives  $\|A_{\un{k}}(k_m)^{-1}\|\leq c|k_m|^{-1}<\infty$. Furthermore, introducing for future reference some continuous functions $g_i$, we have
\beqa
\lan  g_i\eta_{k_i}^{\eps}, \vv^{\si}\ran= \int d^3q\, \eta(q) g_i(k_i+\eps q)  \fr{\chi_{[\si_n,\ka)}(k_i+\eps q) | k_i+\eps q|^{\alf}  } 
{ \sqrt{2|k_i+\eps q |} },  \label{leading-term-one}
\eeqa
hence 
\beqa
\lim_{\eps\to 0} \lan g_i\eta_{k_i}^{\eps}, \vv^{\si}\ran= g_{i}(k_i)\vv^{\si}(k_i)\quad \textrm{and}\quad  |\lan g_i \eta_{k_i}^{\eps}, \vv^{\si}\ran| \leq  c_K\sup_{q\in \supp\eta} |g_i(k_i+\eps q) |\leq   c'_K. \label{eta-v-estimate}
\eeqa
The first relation above is the only place in the proof where the additional assumption $|k_i|\neq \si_n$ enters to ensure the continuity of the integrand in $\eps$ needed for a dominated convergence argument. (See the last paragraph of the proof for the case $|k_i|=\si_n$). 
The last bound in (\ref{eta-v-estimate}) holds uniformly in $k_i \in K$  and in  $\eps$ s.t. $|k_i+\eps q|>\h |k_i|$ for all $q\in \supp\, \eta$. (The constant $c_{K}$ does not depend on $n, \alf$. The constant $c_{K}'$  does depend on $g_i$). Next, by a-(b) we have $\psi_{P,n}\in D(b(k_1)\ldots b(k_{m'}))$. This  gives the existence of the following
limits
\beqa
\lim_{\eps\to 0}B^{\eps}_{m-1}(\un k)\psi_{P,n}\1 &=& \1 [b(k_1)\ldots b(k_{m-1}) \psi_{P,n}], \\
\lim_{\eps\to 0}  b(\eta_{k_m}^{\eps} )  B_{m-1}^{\eps,i}(\un k^i) \psi_{P,n} \1 &=& \1   [b(k_1)\ldots \check{i} \ldots b(k_{m}) \psi_{P,n}].
\eeqa
Furthermore, estimate (\ref{g-v-expression-x}) gives
\beqa
\|B^{\eps}_{m-1}(\un k)\psi_{P,n}\|\leq C_{K,m-1 }, \quad  \|b(\eta_{k_m}^{\eps} )  B_{m-1}^{\eps,i}(\un k^i) \psi_{P,n}\|\leq C_{K,m-1}, \label{leading-terms-bounds}
\eeqa
uniformly in $k_1\ldots, k_m\in K$ and $\eps\in (0, \eps_{K})$ for some $\eps_K>0$. Thus (\ref{leading-term-one})-(\ref{leading-terms-bounds})  give
\beqa
& &\lim_{\eps\to 0}(\mrm{r.h.s.}(\ref{bB-formula-one}))= \frac{1}{E_{P,n}-|\un k|_m -H_{P-\un{k}_m,n}}\sum_{i=1}^m \vv^{\si_n}(k_i)[b(k_1)\ldots\check{i}\ldots b(k_m)\psi_{P,n}]=(\mrm{r.h.s.}(\ref{Froehlich-form})), \\
& &\|(\mrm{r.h.s.}(\ref{bB-formula-one}))\|\leq C'_{K,m},
\eeqa
uniformly in $k_1\ldots, k_m\in K$ and $\eps\in (0, \eps'_{K})$ for some $\eps'_K>0$. This concludes our analysis of (\ref{bB-formula-one}).

\vspace{0.2cm}

\nin\textbf{Analysis of (\ref{bB-formula-two})}. This is an error term and we are aiming at an estimate
\beqa
\|(\ref{bB-formula-two})\|\leq \eps^{1/2} C_{K,m}, \label{error-term-bound}
\eeqa
uniformly in $k_1,\ldots, k_m\in K$ and $\eps\in (0, \eps_{K})$ for some $\eps_K>0$. This will give $\lim_{\eps\to 0}  \|(\ref{bB-formula-two})\|=0$ together
with the uniformity in $k_i$ and $\eps$ needed to close the inductive argument. 

To this end, we  note that
\beqa
& &{\bcb (  A_{\un{k}}(k_m) -  A_{\un{q}}(q_m))=L_{\un k, k_m}(\un q, q_m)-(P-P_{\pho})\cdot M_{\un k, k_m}(\un q, q_m)}, \quad \textrm{where} \label{A-difference}\\
 & &L_{\un k, k_m}(\un q, q_m):=\sum_{j=1}^{m}\bigg\{ (|k_j|-|q_j|)+(k_j-q_j)\cdot \sum_{j'=1}^m(k_{j'}+q_{j'}) \bigg\}, \quad  M_{\un k, k_m}(\un q, q_m):=\sum_{j=1}^m (k_j-q_j).
  \eeqa
The key property of the expressions $L_{\un k, k_m}(\un q, q_m)$ and $M_{\un k, k_m}(\un q, q_m)$  is that they vanish when all $q_i\to k_i$ and therefore  
\beqa
\eta_{\un k, k_m}^{\eps}(\un{q}, q_m )L_{\un k, k_m}(\un q, q_m)\to 0, \quad  \eta_{\un k, k_m}^{\eps}(\un{q}, q_m )M_{\un k, k_m}(\un q, q_m)\to 0 \quad \textrm{for} \quad \eps\to 0,
\eeqa
in a sense to be specified below. This mechanism will be used to establish (\ref{error-term-bound}). 
 
Before exploiting the above observation,  we  rewrite (\ref{bB-formula-two}) making use of (\ref{bB-formula-one-x})-(\ref{bB-formula-two-x}):
\beqa
& &\1\int d^{3(m-1)}\un{q} d^3q_m\, \eta_{\un k, k_m}^{\eps}(\un{q}, q_m ) \, A_{\un{k}}(k_m)^{-1}(  A_{\un{k}}(k_m) -  A_{\un{q}}(q_m)) b(q_m) B_{m-1}(\un q) \psi_{P,n} \non\\
\!&=&\!  \int d^{3(m-1)}\un{q} d^3q_m\, \eta_{\un k, k_m}^{\eps}(\un{q}, q_m ) \, A_{\un{k}}(k_m)^{-1}(  A_{\un{k}}(k_m) -  A_{\un{q}}(q_m)) A_{\un{k}}(k_m)^{-1}\vv^{\si}(q_m) B_{m-1}(\un q)\psi_{P,n} \label{iterate-one}\\
\!& +&\!  \int d^{3(m-1)}\un{q} d^3q_m\, \eta_{\un k, k_m}^{\eps}(\un{q}, q_m ) \, A_{\un{k}}(k_m)^{-1}(  A_{\un{k}}(k_m) -  A_{\un{q}}(q_m)) A_{\un{k}}(k_m)^{-1} \sum_{i=1}^{m-1} \vv^{\si_n}(q_i)b(q_m)  B_{m-1}^i(\un q^i) \psi_{P,n} \label{iterate-two} \quad\quad\quad \\
\!&+&\! \int d^{3(m-1)}\un{q} d^3q_m\, \eta_{\un k, k_m}^{\eps}(\un{q}, q_m ) \, \{ A_{\un{k}}(k_m)^{-1}(  A_{\un{k}}(k_m) -  A_{\un{q}}(q_m)) \}^2  b(q_m) B_{m-1}(\un q) \psi_{P,n},\label{iterate-three}
\eeqa
and we will show that (\ref{iterate-one}), (\ref{iterate-two}), (\ref{iterate-three}) satisfy  bounds of the form (\ref{error-term-bound}). 

 Each contribution (\ref{iterate-one}) and (\ref{iterate-two}) can be divided into two parts coming from the  two terms
on the r.h.s. of (\ref{A-difference}), which we denote $(\ldots)_L$ and $(\ldots)_M$. We will consider only the $M$-parts as the analysis of $L$-parts is analogous. Since 
\beqa
\|A_{\un{k}}(k_m)^{-1}\|\leq C_K,\, \quad  \sum_{\ell=1,2,3}\|A_{\un{k}}(k_m)^{-1}(P-P_{\pho})^{\ell}\|\leq C_K, \label{standard-en-bound-D}
\eeqa
where $\ell$ is a vector index, we can write
\beqa
\|(\ref{iterate-one})_{M}\| \1&\leq&\1 C_K\sup_{\ell=1,2,3}\| \int d^{3(m-1)}\un{q} d^3q_m\, \eta_{\un k, k_m}^{\eps}(\un{q}, q_m )  M^{\ell}_{\un k, k_m}(\un q, q_m)\vv^{\si}(q_m) B_{m-1}(\un q)\psi_{P,n}\|\non\\
\1&\leq &\1  C_K \sup_{\ell=1,2,3} |\lan (k_m-q_m)^{\ell}\eta^{\eps}_{k_m}, \vv^{\si}  \ran| \, \|B^{\eps}_{m-1}(\un k)\psi_{P,n}\| \non\\
\1&  &\1 +C_K   |\lan \eta^{\eps}_{k_m}, \vv^{\si}  \ran | \sup_{\ell=1,2,3}\sum_{j=1}^{m-1}\| b( (k_j-q_j)^{\ell} \eta^{\eps}_{k_j}  )  B^{\eps,j}_{m-1}(\un k^j)\psi_{P,n}\|.%
\eeqa
By the second relation in (\ref{eta-v-estimate}) with $i=m$ and $g_m(q):=(k_m-q)$, we have
\beqa
|\lan (k_m-q_m)^{\ell}\eta^{\eps}_{k_m}, \vv^{\si}  \ran|\leq c_K\eps, \quad   |\lan \eta^{\eps}_{k_m}, \vv^{\si}  \ran|\leq c_K,
\eeqa
where the latter relation corresponds to $g_m\equiv 1$. Furthermore, by the first estimate in  (\ref{g-v-expression-x}) we can write
\beqa
\| b( (k_j-q_j)^{\ell} \eta^{\eps}_{k_j}  )  B^{\eps,j}_{m-1}(\un k^j)\psi_{P,n}\|\leq \eps C_K, \quad  \|B^{\eps}_{m-1}(\un k)\psi_{P,n}\|\leq C_K.
\eeqa
Thus altogether
\beqa
\|(\ref{iterate-one})_{M}\|\leq  \eps C'_K.
\eeqa
Next, we note that each term in the sum in (\ref{iterate-two}) differs from (\ref{iterate-one}) only by renaming the indices $i\leftrightarrow m$. Thus by analogous arguments
we have
\beqa
\|(\ref{iterate-two})_{M}\|\leq  \eps C'_K \quad \textrm{ and altogether } \quad \|(\ref{iterate-one}) + (\ref{iterate-two})\|\leq \eps C''_K. \label{intermediate-estimate-to-the-expression}
\eeqa

Now we consider (\ref{iterate-three}). This contribution can be divided into four parts coming from the  two terms
on the r.h.s. of (\ref{A-difference}). We will denote them $(\ref{iterate-three})_{LL}$,  $(\ref{iterate-three})_{LM}$, $(\ref{iterate-three})_{ML}$, $(\ref{iterate-three})_{MM}$.
As they all have very similar structure, it suffices to consider $(\ref{iterate-three})_{MM}$. Making use again of  (\ref{standard-en-bound-D}), we obtain
\beqa
& &\!\!\!\!\! \|(\ref{iterate-three})_{MM}\|\quad \non\\
\1&\leq&\1 C_K \sup_{\ell, \ell'=1,2,3}\|A_{\un{k}}(k_m)^{-1}(P-P_{\pho})^{\ell}  \int d^{3(m-1)}\un{q} d^3q_m\, \eta_{\un k, k_m}^{\eps}(\un{q}, q_m )    M^{\ell}_{\un k, k_m}(\un q, q_m) M^{\ell'}_{\un k, k_m}(\un q, q_m)  b(q_m) B_{m-1}(\un q) \psi_{P,n}\|\non\\
\1&\leq &\1 C_K  \sup_{\ell, \ell'=1,2,3} \|  A_{\un{k}}(k_m)^{-1}(P-P_{\pho})^{\ell}  b((k_m-q_m)^{\ell}(k_m-q_m)^{\ell'} \eta^{\eps}_{k_m} ) B_{m-1}^{\eps}(\un k)\psi_{P,n}\|  \label{last-iterative-term-x}\\
 \1&   &\1+C_K    \sup_{\ell, \ell'=1,2,3}      \sum_{j=1}^{m-1}\| A_{\un{k}}(k_m)^{-1}(P-P_{\pho})^{\ell}   b( (k_m-q_m)^{\ell}  \eta^{\eps}_{k_m} ) b( (k_j-q_j)^{\ell'}  \eta^{\eps}_{k_j} )   B_{m-1}^{\eps,j}(\un k^j)\psi_{P,n}\| \label{last-iterative-term-xx} \\
\1&   &\1+C_K    \sup_{\ell, \ell'=1,2,3}      \sum_{j=1}^{m-1}\| A_{\un{k}}(k_m)^{-1}(P-P_{\pho})^{\ell}   b(  \eta^{\eps}_{k_m} ) b( (k_j-q_j)^{\ell}  (k_j-q_j)^{\ell'} \eta^{\eps}_{k_j} )   B_{m-1}^{\eps,j}(\un k^j)\psi_{P,n}\| \label{last-iterative-term-xxx} \\
  \1&   &\1+C_K    \sup_{\ell, \ell'=1,2,3}      \sum_{\substack{j,j'=1 \\ j\neq j' }}^{m-1}\| A_{\un{k}}(k_m)^{-1}(P-P_{\pho})^{\ell}   b(  \eta^{\eps}_{k_m} ) b( (k_j-q_j)^{\ell}   \eta^{\eps}_{k_j} )  b( (k_{j'}-q_{j'})^{\ell'} \eta^{\eps}_{k_{j'}} ) B_{m-1}^{\eps,j,j'}(\un k^{j,j'})\psi_{P,n}\|. \label{last-iterative-term-xxxx}\quad\quad\quad
 \eeqa
To analyze the above expressions we make the following observations: It is an easy application of Lemmas~\ref{energy-bounds} and \ref{energy-bound-zero} that 
\beqa
\|A_{\un{k}}(k_m)^{-1}(P-P_{\pho})^{\ell} H_{\pho}^{1/2}\|\leq c_K.
\eeqa
Consequently, using again Lemma~\ref{energy-bounds}, we obtain
\beqa
& &\|A_{\un{k}}(k_m)^{-1}(P-P_{\pho})^{\ell} \,\,  b((k_m-q_m)^{\ell}(k_m-q_m)^{\ell'} \eta^{\eps}_{k_m} )\|\leq c\bigg(\int d^3q_m\, |q_m|^{-1}    (k_m-q_m)^4   \eta^{\eps}_{k_m}(q_m)^2   \bigg)^{1/2}\non\\
\1& &\1 \ph{4444444444444444444444} \leq c'\eps^{1/2}\bigg(\int \fr{d^3q_m}{\eps^3}\, |q_m|^{-1}    \fr{|q_m-k_m|^4}{\eps^4} \eta\bigg(\fr{q_m-k_m}{\eps}\bigg)^2   \bigg)^{1/2}\non\\
\1& &\1 \ph{4444444444444444444444}=c\eps^{1/2}\big(\int  d^3q_m'\, |\eps q_m'+k_m|^{-1} |   (q_m')^2\eta( q_m')|^2   \big)^{1/2}\leq c'\eps^{1/2}|k_m|^{-1/2}. \label{first-iterative-est}
\eeqa
An analogous consideration gives
\beqa
& &\|A_{\un{k}}(k_m)^{-1}(P-P_{\pho})^{\ell}  b((k_m-q_m)^{\ell} \eta^{\eps}_{k_m} )\|\leq c' \eps^{-1/2}|k_m|^{-1/2}.
\eeqa
Furthermore, by estimate (\ref{g-v-expression-x}), 
\beqa
\|B_{m-1}^{\eps,j}(\un k^j)\psi_{P,n}\| \leq C_{K,m-1}, \quad   \|b( (k_j-q_j)^{\ell'}  \eta^{\eps}_{k_j} )   B_{m-1}^{\eps,j}(\un k^j)\psi_{P,n}\|\leq \eps C_{K,m-1}. \label{last-iterative-est}
\eeqa
Taking (\ref{first-iterative-est})--(\ref{last-iterative-est}) into account, we get
\beqa
\| \ref{last-iterative-term-x}\|\leq \eps^{1/2}C_K, \quad \| \ref{last-iterative-term-xx}\|\leq \eps^{1/2}C_K.
\eeqa
Now we note that (\ref{last-iterative-term-xxx}), (\ref{last-iterative-term-xxxx}) have the same structure as (\ref{last-iterative-term-x}), (\ref{last-iterative-term-xx}), respectively, and
therefore satisfy analogous bounds. Consequently $\|(\ref{iterate-three})_{MM}\|\leq \eps^{1/2}C_K$ and such bound holds also for the remaining contributions to (\ref{iterate-three}). 
Together  with (\ref{intermediate-estimate-to-the-expression}), this concludes the proof of (\ref{error-term-bound}).

\vspace{0.2cm}

We have therefore proven property (b) from the statement of the lemma for $m$. Furthermore, we have verified that
\beqa
\| b(\eta_{k_1}^{\eps}) \ldots b(\eta_{k_{m}}^{\eps})\psi_{P,n}\|\leq C_{K,m}
\eeqa
uniformly in $k_1, \ldots, k_m\in K$ and $\eps\in (0, \eps_{K})$ for some $\eps_K>0$. Therefore, by Lemma~\ref{lemma-distr} we obtain
statement (c) from the lemma for $m$, which concludes the inductive argument under the additional assumption that $|k_i|\neq \si_n$, $i=1,\ldots, m$.

{\bcb Let us now explain how to drop this assumption at a cost of providing factors $\h$ in formula~(\ref{Froehlich-form}): It suffices to reestablish the first relation in (\ref{eta-v-estimate}) in the case $|k_i|=\si_n$. We have, making use of (\ref{leading-term-one}), and assuming that $\eps$ is sufficiently small so that $0<|k_i+\eps q| \ll \ka$ for $q\in \supp\,\eta$, 
 \beqa
 \lan g_i\eta_{k_i}^{\eps}, \vv^{\si}\ran\1& = &\1 \int d^3q\, \eta(q) g_i(k_i+\eps q)  \fr{\chi_{[\si_n,\ka)}(k_i+\eps q) | k_i+\eps q|^{\alf}  } 
{ \sqrt{2|k_i+\eps q |} } 
 =   g(k_i)\fr{|k_i|^{\alf}}{\sqrt{2|k_i|  } }  \int_{\si_n\leq |k_i+\eps q|} d^3q\, \eta(q) +O(\eps), \quad\quad
\eeqa
where  $\mco(\eps)$ denotes an error term which tends to zero as $\eps\to 0$ and we made use of continuity of $k\mapsto g(k)\fr{|k|^{\alf}}{\sqrt{2|k|  } }$ near $k_i\neq 0$.
Now by squaring both sides of the relation $\si_n\leq |k_i+\eps q|$ and using $|k_i|^2=\si_n^2$ we get
\beqa
\cos\,\theta\geq -\fr{\eps |q|}{2|k_i|},
\eeqa
where $\theta$ is the angle between $k_i$ and $q$. Exploiting the spherical symmetry of $\eta$ to choose the third axis of the reference frame 
in the direction of $k_i$ and passing to spherical coordinates, we have
\beqa
 \int_{\si_n\leq |k_i+\eps q|} d^3q\, \eta(q) \1 &=& \1 2\pi \int |q|^2 d|q|\, \eta(|q|)\int_{\cos\,\theta\geq -\fr{\eps |q|}{2|k_i|}} d\cos\theta\, \non\\
 \1 &=& \1 2\pi \int |q|^2 d|q|\, \eta(|q|)  (1+  \fr{\eps |q|}{2|k_i|} )  \non\\  
 \1 &=& \1 \h \int d^3q\, \eta(q)+O(\eps).
 \eeqa
Since $\int d^3q\,  \eta(q)=1$, this concludes the argument.} \qed 

\bel \label{lemma-distr} Let  $f_i\in L^1(\real^3)\cap L^2(\real^3)$, $i=1,\ldots. m$,  be compactly supported outside of zero and  let $K$ be a  compact set,  not containing zero, 
whose interior contains all $\supp\, f_i$.  Suppose that $\psi\in D(b(k_1)\ldots b(k_m))$ for all $k_i\in K$. Suppose furthermore that
\beqa
 \|b(\eta_{k_1}^{\eps}) \ldots  b(\eta_{k_m}^{\eps})  \psi\|\leq C_{K,m}, \label{boundedness-assumption-D}
\eeqa
uniformly in $k_1, \ldots, k_m\in K$ and $\eps\in (0, \eps_K)$ for some $\eps_K>0$. 
Then  $K^{\times m}\ni (k_1,\ldots, k_m) \to  [b(k_1)\ldots b(k_m)\psi]$ is weakly measurable and
\beqa
\int d^3k_1\ldots d^3k_m\, \bar{f}_1(k_1) \ldots \bar{f}_m(k_m) [b(k_1)\ldots b(k_m)\psi]=b(f_1)\ldots b(f_m)\psi, \label{lemma-D1-estimate}
\eeqa
where the l.h.s. is a weak integral and the r.h.s. is the usual definition.
\eel
}
\begin{rem}  We prove this lemma for functions from  $L^1(\real^3)\cap L^2(\real^3)$ in order to accommodate $h_{n+1}$  from the proof of Lemma~\ref{vector-shift-Lemma}. 
For the assumption  (\ref{boundedness-assumption-D}) see the proof
 of Lemma~\ref{pull-through-lemma}. 
\end{rem}
\proof  First we note that by iterating (\ref{first-energy-bound-xx}), we obtain for any functions $f_1, \ldots, f_m$ as in the statement of the lemma
and $\psi\in C^{\infty}(H_{P,\free})$
\beqa
\|b(f_1)\ldots b(f_m)\psi\|\leq c_m(\psi) \|f_1\|_2\ldots \|f_m\|_2,  \label{iterated-energy-bounds-x}
\eeqa  
where the constant $c_m(\psi)$ depends on $\psi$ and we used that $f_i$ are supported away from zero. 
Now let us set $\un k=(k_1, \ldots, k_m)$ and write $F(\un k):=f_1(k_1)\ldots f_m(k_m)$.
By standard density arguments, for any $\theta>0$ we can find a function $F_{\theta}(\un k):=f_{1, \theta}(k_1)\ldots f_{m, \theta}(k_m)$, with $f_{i, \theta} \in C_0^{\infty}(\real^{3m})$, s.t.
\beqa
\|f_i-f_{i,\theta}\|_{2}\leq \theta. \label{L-1-property}
\eeqa
Furthermore,  we can ensure that all $f_{i,\theta}$ are supported in the compact set $K\subset \real^3\backslash \{0\}$ from the statement of the lemma. Then we also have 
\beqa
\|f_i-f_{i,\theta}\|_{1}\leq |K|^{1/2}\theta^{1/2},  \label{two-estimates} 
\eeqa
where $|K|$ is the volume of $K$. Using (\ref{iterated-energy-bounds-x}), (\ref{L-1-property}),  we obtain 
\beqa
\|(b(f_1)\ldots b(f_m)) \psi-(b(f_{\theta,1}) \ldots b(f_{\theta,m} ))\psi\|\leq C_1\theta. \label{b-b-energy-bounds}
\eeqa
 Next, we note that for any $\psi'\in \mcF$ and any fixed $\eps\in (0,\eps_{K}]$ the function given by  $K^{\times m}\ni (k_1,\ldots, k_m) \mapsto  
 \lan \psi', b(\eta_{k_1}^{\eps})\ldots b(\eta_{k_m}^{\eps})  \psi\ran$ is measurable. 
This is easily seen by expressing $\psi$, $\psi'$ in terms of their Fock space components and noting that the resulting sum of measurable functions converges for any 
$(k_1,\ldots, k_m)\in  K^{\times m}$.
Similarly,  since $\psi\in D(b(k_1)\ldots b(k_m))$, we obtain that  $\lan\psi', [b(k_1)\ldots b(k_m)\psi]\ran=\lim_{\eps\to 0} \lan  \psi', 
b(\eta_{k_1}^{\eps})\ldots b(\eta_{k_m}^{\eps})  \psi\ran$ is a measurable function.   
Moreover, using the   boundedness assumption~(\ref{boundedness-assumption-D}),  the dominated convergence theorem and estimate  in (\ref{two-estimates}),
we can write
\beqa
& &\big| \int d^{3m}\un{k}\, (\bar F(\un k)-\bar F_{\theta}(\un k))\lan \psi', [b(k_1)\ldots b(k_m)  \psi] \ran \big|  =  \lim_{\eps\to 0}\big| \int d^{3m}\un{k}\, (\bar F(\un k)-\bar F_{\theta}(\un k))\lan \psi', b(\eta_{k_1}^{\eps})\ldots b(\eta_{k_m}^{\eps})  \psi \ran \big|\non\\
\1 & & \1 \ph{44444444444444444444444444444}\leq C_{K,m}\|\psi'\| \sum_{i=1}^m (\|f_1\|_1\ldots \|f_{i-1}\|_1  K^{1/2}\theta^{1/2} \|f_{i+1,\theta}\|_1\ldots \|f_{m, \theta}\|_1). \label{integral-of-bracket}
\eeqa
In view of the estimates (\ref{b-b-energy-bounds}) and (\ref{integral-of-bracket}) it suffices to prove (\ref{lemma-D1-estimate}) for $f_{i, \theta}$.
 For this purpose, by  (\ref{iterated-energy-bounds-x})
 and the fact that $f_{i, \theta}$ are supported away from zero, we get for sufficiently small $\eps$
\beqa
& &\|\int d^{3m}\un{k}\,  \bar{f}_{1, \theta}(k_1)\ldots \bar{f}_{m, \theta}(k_m)(b(\eta_{k_1}^{\eps}) \ldots  b(\eta_{k_m}^{\eps}))\psi-(b(f_{1, \theta}) \ldots b(f_{m, \theta}))\psi\|\1\non\\   
\1& &\1 \leq c_m\sum_{i=1}^m ( \| (\eta^{\eps}_{(\cdot)}*f_{1, \theta})\|_2\ldots  \| (\eta^{\eps}_{(\cdot)}*f_{i-1, \theta})\|_2 ) \,\, \|(\eta^{\eps}_{(\cdot)}*f_{i, \theta}) -f_{i, \theta}\|_2\,\, ( \|f_{i+1, \theta} \|_2\ldots \|f_{m, \theta}\|_2). \label{first-identity-D(b(k))}
\eeqa
Now we analyse the relevant factor
\beqa
(\eta^{\eps}_{(\cdot)} *f_{i, \theta})  (k') -f(k')\1&=&\1\int d^3k \, \eps^{-3}\eta({\color{black}(k'-k)\eps^{-1}})f_{i,\theta}(k)-f_{i, \theta}(k')\non\\
\1&=&\1 \int d^3k\, \eta(k)\big(f_{i, \theta}(k'{\color{black}-}\eps k)-f_{i, \theta}(k')\big)\non\\
\1&=&\1 -\eps\int_0^1d\la\,  \int d^3k\, \eta(k)\nabla f_{i, \theta}(k'{\color{black}-}\eps\la k)\cdot k \label{difference-D(b(k))}
\eeqa
By substituting~(\ref{difference-D(b(k))}) to (\ref{first-identity-D(b(k))}), making use of
\beqa
\bigg|\int_0^1d\la\,  \int d^3k\, \eta(k)\nabla f_{i, \theta}(k'{\color{black}-}\eps\la k)\cdot k\bigg|^2\leq c \int_0^1d\la\,  \int d^3k\, \big|\eta(k)\nabla f_{i, \theta}(k'{\color{black}-}\eps\la k)\cdot k\big|^2,
\eeqa
(which follows from the compactness of the region of integration in $k$ and $\la$),  the r.h.s of (\ref{first-identity-D(b(k))}) tends to zero as $\epsilon\to 0$ for any fixed $\theta$.
\qed\\
\nin In view of the above lemma we will simply write $b(k)\psi$ for $[b(k)\psi]$.  For  the proof of Theorem~\ref{preliminaries-on-spectrum} we only need 
Lemma~\ref{pull-through-lemma} for $m=1$, but the case $m\geq 1$   is needed in Theorem~\ref{main-theorem-spectral} which we prove in \cite{DP16}.
\bel\cite{Pi03}\label{vector-shift-Lemma}  Let $|P|\leq P_{\max}=1/3$ and $|\la|\in (0,\la_0]$. Then 
\beqa
\|\big(W_{P,n+1}\wt{W}_{P,n+1}^*-1\big)\hat{\phi}_{P,n+1}\|\1&\leq&\1 c|\lambda|\,  |\nabla E_{P,n+1}-\nabla E_{P,n}|\,|\ln\sigma_{n+1}| \,\|\hat{\phi}_{P,n+1}\|, 
\label{vector-difference}\\
\|\big(W_{P,n+1} W_{P,n}^*-1\big)\hat{\phi}_{P,n+1}\|\1&\leq&\1 c|\lambda|\,  |\nabla E_{P,n+1}-\nabla E_{P,n}|\,|\ln\sigma_{n+1}| \,\|\hat{\phi}_{P,n+1}\|
+c|\la|(\alf)^{{\bcb -1}}
\si_n^{\alf}\,\|\hat{\phi}_{P,n+1}\|, 
\label{new-vector-difference}
\eeqa
where (only) for (\ref{new-vector-difference}) we assumed $\alf>0$ and both estimates also hold after replacing $\hat{\phi}_{P,n+1}$ with 
$\phi_{P,n+1}$.
\eel
\proof We will prove the two estimates in parallel. Let us set
\beqa
& &\ti f_{n+1}^{\eps}(k):=\la \chi_{[\sigma_{n+1},\ka ) }(k)|k|^{\alf}  \frac{1}{\sqrt{2}|k|^{\frac{3}{2}}\alpha_{P,n}(\hk)}-
\eps \la \chi_{[\sigma_{n+1}, \sigma_n] }(k)|k|^{\alf}  \frac{1}{\sqrt{2}|k|^{\frac{3}{2}}\alpha_{P,n}(\hk)}, \\
& &f_{n+1}(k):=\la\chi_{[\sigma_{n+1}, \ka)}(k)|k|^{\alf} \frac{1}{\sqrt{2}|k|^{\frac{3}{2}}\alpha_{P,n+1}(\hk)}, 
\eeqa
where $\eps\in \{0,1\}$ and define $h^{\eps}_{n+1}:=f_{n+1}-\ti f^{\eps}_{n+1}$.  We  estimate using spectral calculus 
\beqa
\|\big(\e^{(b^*(f_{n+1})-b(f_{n+1}))} \e^{-(b^*(\ti f^{\eps}_{n+1})-b(\ti f^{\eps}_{n+1}))}-1\big)\hat\phi_{P,n+1}\|
\1 &\leq& \1 c\|(b^*(h^{\eps}_{n+1})-b(h^{\eps}_{n+1}))\hat\phi_{P,n+1}\|\non\\
\1 &\leq& \1 c\|h^{\eps}_{n+1}\|\,\|\hat\phi_{P,n+1}\|+c'\|b(h^{\eps}_{n+1})\hat\phi_{P,n+1}\|, \quad\quad\quad \label{auxiliary-b-formula}
\eeqa
where the case $\eps=0$ (resp. $\eps=1$) corresponds to estimate~(\ref{vector-difference}) (resp. (\ref{new-vector-difference})).  
We recall the definitions $\cphi_{P,n}:=\phi_{P,n}/\|\phi_{P,n}\|$, 
 $\phi_{P,n}:= W_{P,n} \wt{W}_{P,n}^*\hat{\phi}_{P,n}$ (cf.~(\ref{phi-definitions})) and note that $\cpsi_{P,n}:= W_{P,n}^*\cphi_{P,n}$ differs at 
 most by a complex factor from $\psi_{P,n}$ appearing in (\ref{Froehlich-form}). 
From these relations we get
$\hat{\phi}_{P,n+1}=\|\hat{\phi}_{P,n+1}\|\wt{W}_{P,n+1}\cpsi_{P,n+1}$. 
Next, we set $\hat{\phi}^*_{P,n+1}:=\hat{\phi}_{P,n+1} / \|\hat{\phi}_{P,n+1}\| $ and compute
\beqa
b(h^{\eps}_{n+1})\hat\phi^*_{P,n+1}= b(h^{\eps}_{n+1}) \wt{W}_{P,n+1}\cpsi_{P,n+1}=
 \wt{W}_{P,n+1}(b(h^{\eps}_{n+1})+ \lan h^{\eps}_{n+1},\ti f^{0}_{n+1}\ran)\cpsi_{P,n+1}.
\label{pull-through}
\eeqa
Making use of {\bcb Lemmas}~\ref{pull-through-lemma}, \ref{vector-shift-Lemma} we estimate the first term on the r.h.s. above  
\beqa
\|b(h^{\eps}_{n+1})\cpsi_{P,n+1}\|\1&\leq &\1 
\|\hint_{\1\si_{n+1}}^{\ka} d^3k\, h^{0}_{n+1}(k)\frac{\lambda}{\sqrt{2|k|}}\frac{1}{E_{P,n+1}-|k|-H_{P-k,n+1}}\cpsi_{P,n+1}\|\non\\
& &+\eps \|\bint_{\1\si_{n+1}}^{\si_n} d^3k\,       \frac{1}{\alpha_{P,n}(\hk)}  \frac{\lambda^2}{2|k|^2}\frac{1}{E_{P,n+1}-|k|-H_{P-k,n+1}}\cpsi_{P,n+1}\| \non\\
\1&\leq&\1c\la^2\bint_{\1\si_{n+1}}^{\ka} d^3k\, \fr{1}{2|k|^3  } \bigg|\frac{1}{\alpha_{P,n+1}(\hk)}
- \fr{1}{\alpha_{P,n}(\hk)}      \bigg|\|\cpsi_{P,n+1}\| + c\eps \la^2(\alf)^{-1} \si_n^{2\alf}  \non\\
\1&=&\1 c\la^2\bint_{\1\si_{n+1}}^{\kappa} d^3k\, \fr{1}{2|k|^3  } \bigg|\frac{\hat k\cdot(\nabla E_{P,n+1}- \nabla E_{P,n} )  }{\alpha_{P,n+1}(\hk)\alpha_{P,n}(\hk)}  
\bigg|  + c\eps \la^2(\alf)^{-1} \si_n^{2\alf} \non\\
\1&\leq&\1 c|\la|^2|\nabla E_{P,n+1}- \nabla E_{P,n}|\, |\ln\si_{n+1}|+ c\eps \la^2(\alf)^{-1} \si_n^{2\alf}, \label{gradient-difference-bound}
\eeqa
{\bcb where we refer to ``Standing assumptions and conventions" in Section \ref{Preliminaries-and-results} for the definitions of $\bint_{\1\si_{n+1}}^{\kappa}d^3k$,  
$\hint_{\1\si_{n+1}}^{\kappa}d^3k$.}
Now we consider the second term on the r.h.s. of (\ref{pull-through}):
\beqa
\lan h^{\eps}_{n+1},\ti f^0_{n+1}\ran\1&=&\1\la^2 \bint_{\!\!\! \si_{n+1}}^{\ka}   d^3k\, \left(\frac{1}{\sqrt{2}|k|^{\frac{3}{2}}\alpha_{P,n}(\hk)}-
\frac{1}{\sqrt{2}|k|^{\frac{3}{2}}\alpha_{P,n+1}(\hk)}\right)\frac{1}{\sqrt{2}|k|^{\frac{3}{2}}\alpha_{P,n}(\hk)}\non\\
 & &+\eps\la^2 \bint_{\!\!\! \si_{n+1}}^{\si_n}   d^3k\, \left(\frac{1}{\sqrt{2}|k|^{\frac{3}{2}}\alpha_{P,n}(\hk)}  \right)\frac{1}{\sqrt{2}|k|^{\frac{3}{2}}\alpha_{P,n}(\hk)}. \label{App-D-interm}
\eeqa
This satisfies again a bound of the form (\ref{gradient-difference-bound}). Finally, we consider the first expression on the
r.h.s. of (\ref{auxiliary-b-formula}):
{\bcb
\beqa
\|h^{\eps}_{n+1}\| \1&\leq&\1 |\la| \bigg( \bint_{\!\!\! \si_{n+1}}^{\ka}  d^3k\,\left|\frac{1}{\sqrt{2}|k|^{\frac{3}{2}}
\alpha_{P,n}(\hk)}-
\frac{1}{\sqrt{2}|k|^{\frac{3}{2}}\alpha_{P,n+1}(\hk)}\right|^2\bigg)^{1/2}+\eps|\la| \bigg( \bint_{\!\!\! \si_{n+1}}^{\si_n}   d^3k\, \left(\frac{1}{\sqrt{2}|k|^{\frac{3}{2}}\alpha_{P,n}(\hk)}  \right)^2\bigg)^{1/2}\non\\
\1&\leq&\1 c|\la| |\nabla E_{P,n+1}- \nabla E_{P,n}| |\ln\si_{n+1}|^{1/2}+c\eps |\la|(\alf)^{-1/2}\si_n^{\alf} 
\eeqa
}
which again satisfies the required bounds. {\bcb Finally, we note that replacing $\hat{\phi}_{P,n+1}$ with 
$\phi_{P,n+1}$  amounts to replacing $\ti f^0_{n+1}$ with $\ti f^1_{n+1}$ on the r.h.s. of (\ref{pull-through}). Consequently, the last term on the r.h.s. of (\ref{App-D-interm})
can be dropped and the region of integration in the first term restricted to $[\si_n, \ka]$. Clearly the resulting expression satisfies again  a bound of the form (\ref{gradient-difference-bound}).} \qed
\bel\cite{Pi03}\label{gradient-shift-lemma} Let $|P|\leq P_{\max}=1/3$ and $|\la|\in (0,\la_0]$. Then 
\beqa
|\nabla E_{P,n-1}-\nabla E_{P,n}|\leq c_{1}[\lambda^{2}\sigma_{n-1}+\|\frac{\hat{\phi}_{P,n}}{\|\hat{\phi}_{P,n}\|}-\frac{\phi_{P,n-1}}{\|\phi_{P,n-1}\|}\|].
\eeqa
\eel
\proof  First, we recall the definitions $\cpsi_{P,n}:=W_{P,n}^*\cphi_{P,n}$, $\cphi_{P,n}:=\phi_{P,n}/\|\phi_{P,n}\|$ 
and $\phi_{P,n}:= W_{P,n} \wt{W}_{P,n}^*\hat{\phi}_{P,n}$ (cf.~(\ref{phi-definitions}) and Definition~\ref{abuse-definition}). Now by a standard computation and definition~(\ref{Pi-formula})
\beqa
W_{P,n} P_{\pho} W_{P,n}^*
\1&=&\1P_{\pho}-\la\hint_{\!\!\!\si_n}^{\ka} d^3k\,  \fr{k}{\sqrt{2}|k|^{3/2}\al_{P,n}(\hk)  }   (b^*(k)+b(k))
+\la^2{\bint}_{\!\!\!\si_n}^{\ka} d^3k\, \fr{k}{2|k|^3 \al_{P,n}^2(\hk) }\non\\
\1&=&\1\Pi_{P,n}+\la^2{\bint}_{\!\!\!\si_n}^{\ka} d^3k\, \fr{k}{2|k|^3 \al_{P,n}^2(\hk) }. \label{gradient-E-definition}
\eeqa
Thus we have by the Hellmann-Feynman theorem and formula~(\ref{gradient-E-definition})
\beqa
\nabla E_{P,n}\1&=&\1\lan \cpsi_{P,n},(P-P_{\pho}) \cpsi_{P,n}\ran=\lan  W_{P,n}\cpsi_{P,n}, W_{P,n}(P-P_{\pho}) W_{P,n}^* 
 W_{P,n}\cpsi_{P,n}\ran\non\\
\1&=&\1-\lan  W_{P,n}\cpsi_{P,n}, \Pi_{P,n}  W_{P,n}\cpsi_{P,n}\ran+P-\la^2    \bint_{\!\!\! \si_{n} }^{\ka}  d^3k\, \fr{k}{2|k|^3 \al_{P,n}^2(\hk) }, \\
\nabla E_{P,n-1}\1&=&\1-\lan  W_{P,n-1}\cpsi_{P,n-1}, \Pi_{P,n-1}  W_{P,n-1}\cpsi_{P,n-1}\ran+P
-\la^2{\bint}_{\!\!\!\si_{n-1}}^{\ka} d^3k\, \fr{k}{2|k|^3 \al_{P,n-1}^2(\hk) }.
\eeqa
Recalling that $\hat{\Pi}_{P,n}:=\wt{W}_{P,n}W^*_{P,n} \Pi_{P,n}W_{P,n}\wt{W}_{P,n}^* $, we can write
\beqa
\lan  W_{P,n}\cpsi_{P,n}, \Pi_{P,n}  W_{P,n}\cpsi_{P,n}\ran=\lan\cpsi_{P,n}, \widetilde{W}_{P,n}^{*}  \hat \Pi_{P,n} \widetilde{W}_{P,n} \cpsi_{P,n}\ran
=\bigg\lan \fr{\hat \phi_{P,n}}{\|\hat \phi_{P,n} \|},   \hat \Pi_{P,n}  \fr{\hat \phi_{P,n}}{\|\hat \phi_{P,n} \|}   \bigg\ran,
\eeqa
where we used $\hat\phi_{P,n}/\|\hat\phi_{P,n}  \|=\widetilde{W}_{P,n} \cpsi_{P,n}$. Denoting for any vector $\psi$ its normalized counterpart by $\psi^*:=\psi/\|\psi\|$, so that $\cphi_{P,n}=\phi^*_{P,n}$, we obtain
\beqa
& &\lan\hat \phi_{P,n}^*,   \hat \Pi_{P,n} \hat\phi_{P,n}^*\ran-\lan\phi^*_{P,n-1},    \Pi_{P,n-1} \phi^*_{P,n-1}\ran\non\\
& &= \lan\hat \phi^*_{P,n},   \hat \Pi_{P,n} (\hat\phi^*_{P,n}-\phi^*_{P,n-1})  \ran+\lan\hat \phi^*_{P,n},   (\hat\Pi_{P,n}-\Pi_{P,n-1}  ) \phi^*_{P,n-1}\ran+
\lan( \hat\phi^*_{P,n}-\phi^*_{P,n-1}),    \Pi_{P,n-1} \phi^*_{P,n-1}\ran. 
\eeqa
Now  we recall formula~(\ref{shift-of-Pi}) 
\beqa
\hat{\Pi}_{P,n}-\Pi_{P,n-1}
=\mathcal{L}|_{n}^{n-1}+\la^2{\bint}_{\!\!\!\si_{n}}^{\si_{n-1}} d^3k\, \fr{k(\al_{P,n}^2(\hk)-\al_{P,n-1}^2(\hk))  }{2|k|^3 \al_{P,n-1}^2(\hk)\al_{P,n}^2(\hk)},
\eeqa
and we get 
\beqa
\nabla E_{P,n-1}-\nabla E_{P,n}\1&=&\1\lan\hat \phi^*_{P,n},   \hat \Pi_{P,n} (\hat\phi^*_{P,n}-\phi^*_{P,n-1})  \ran+\lan( \hat\phi^*_{P,n}- \phi^*_{P,n-1}), \Pi_{P,n-1} \phi^*_{P,n-1}\ran\non\\
& &+\lan \hat \phi^*_{P,n},\phi^*_{P,n-1}\ran   \la^2{\bint}_{\!\!\!\si_{n}}^{\si_{n-1}} d^3k\, \fr{k(\al_{P,n}^2(\hk)-\al^2_{P,n-1}(\hk))  }{2|k|^3 \al_{P,n-1}^2(\hk)\al_{P,n}^2(\hk)} \non\\
& &+\lan \hat \phi^*_{P,n}, \mathcal{L}|_{n}^{n-1} \phi^*_{P,n-1}\ran
-\la^2 {\bint}_{\!\!\!\si_{n-1}}^{\ka} d^3k\, \fr{k(\al_{P,n}^2(\hk)- \al_{P,n-1}^2(\hk) )}{2|k|^3 \al_{P,n-1}^2(\hk)\al_{P,n}^2(\hk) }\non\\
& &
+\la^2 \bint_{\!\!\!\si_{n}}^{\si_{n-1}} d^3k\, \fr{k}{2|k|^3 \al_{P,n}^2(\hk) }.
\eeqa
Consequently,
\beqa
\nabla E_{P,n-1}-\nabla E_{P,n}
\1&+&\1\la^2\bint_{\!\!\!\si_{n-1}}^{\ka} d^3k\, 
\fr{k( -\hat k\cdot \nabla E_{P,n}+\hat k\cdot \nabla E_{P,n-1} )(2-\hat k\cdot \nabla E_{P,n}-\hat k\cdot \nabla E_{P,n-1}    )  }
{2|k|^3 \al_{P,n-1}^2(\hk)\al_{P,n}^2(\hk) }\non\\
\1&=&\1\lan \hat \phi^*_{P,n},\phi^*_{P,n-1}\ran   \la^2\bint_{\!\!\!\si_{n}}^{\si_{n-1}} d^3k\, \fr{k(\al_{P,n}^2(\hk)-\al^2_{P,n-1}(\hk)) }{2|k|^3 \al_{P,n-1}^2(\hk)\al_{P,n}^2(\hk)} \non\\
& &+\lan\hat \phi^*_{P,n},   \hat \Pi_{P,n} (\hat\phi^*_{P,n}-\phi^*_{P,n-1})  \ran+\lan(\hat\phi^*_{P,n}-\phi^*_{P,n-1}),    \Pi_{P,n-1} \phi^*_{P,n-1}\ran\non\\
& &+\lan \hat \phi^*_{P,n},  \mcL|^{n-1}_n  \phi^*_{P,n-1}\ran+\la^2\bint_{\!\!\!\si_{n }}^{\si_{n-1}} d^3k\, \fr{k}{2|k|^3 \al_{P,n}^2(\hk) }. \label{D-long-form}
\eeqa
To conclude, we have to analyze the term involving $\mathcal{L}|_{n}^{n-1}$. We recall that
\beqa
\mcL|^{n-1}_n:= -\la\hint_{\!\!\!\sigma_{n}}^{\sigma_{n-1}}d^{3}k\frac{k(b(k)+b^{*}(k))}{\sqrt{2}|k|^{3/2}\alpha_{P,n-1}(\hat{k})}
\eeqa
and denote by $(\mcL|^{n-1}_n)^{(\pm)}$ its creation (+) and annihilation (-) parts. We note that $(\mcL|^{n-1}_n)^{(-)}\phi^*_{P,n-1}=0$, since 
$\phi^*_{P,n-1}\in \mcF_{n-1}$, thus it suffices to study $\lan \hat \phi^*_{P,n},  (\mcL|^{n-1}_n)^{(+)}  \phi^*_{P,n-1}\ran$. We write
\beqa
 (\mcL|^{n-1}_n)^{(-)}\hat \phi^*_{P,n}\1&=&\1 (\mcL|^{n-1}_n)^{(-)}\wt{W}_{P,n}W_{P,n}^*\phi^*_{P,n}= (\mcL|^{n-1}_n)^{(-)}\wt{W}_{P,n}\cpsi_{P,n} \non\\
\1&=&\1 \wt{W}_{P,n}\bigg\{  (\mcL|^{n-1}_n)^{(-)}-  \la^2\bint_{\!\!\!\si_{n }}^{\si_{n-1}} d^3k\, \fr{k}{2|k|^3 \al_{P,n-1}^2(\hk) } \bigg\}\cpsi_{P,n} \non\\
\1&=&\1\wt{W}_{P,n} (\mcL|^{n-1}_n)^{(-)}\cpsi_{P,n}+\mco(\la^2\si_{n-1}). \label{appD-one}
\eeqa
Now we estimate using Lemmas~ \ref{pull-through-lemma}, \ref{lemma-distr},
\beqa
 (\mcL|^{n-1}_n)^{(-)}\cpsi_{P,n}=-\la^2\bint_{\!\!\!\sigma_{n}}^{\sigma_{n-1}}d^{3}k\frac{k}{2|k|^{2}\alpha_{P,n-1}(\hat{k})}\frac{1}{E_{P,n}-|k|-H_{P-k,n}} \cpsi_{P,n}=\mco(\la^2\si_{n-1}). \label{appD-two}
\eeqa
Now making use of (\ref{appD-two}), (\ref{appD-one}) and (\ref{D-long-form}), and exploiting the fact that 
$ \| \hat \Pi_{P,n}\phi^*_{P,n} \|, \|\Pi_{P,n-1}\phi^*_{P,n-1}\|<\infty$ (which follows from (\ref{gradient-E-definition})) {\color{black}together with $|\nabla E_{P,n}|\leq c$}, we conclude the proof. \qed

\section{{\bcc Convexity of the ground state energy}}\label{convexity}\label{strict}
\setcounter{equation}{0}

The discussion in this appendix is similar to~\cite{FP10} but the proof is streamlined using estimates from the present paper. 
\subsection{Main line of the argument}

In order to show the strict convexity of $E_P:=\lim_{n\to \infty}E_{P,n}$ it is convenient to first state a result concerning $\partial_{|P|}^{2}E_{P,n}$.
\bel \label{telescop} Under the assumptions of Theorem~\ref{main-technical-result} 
\begin{eqnarray}
|\pa_{|P|}^2E_{P,n}-\pa^2_{|P|}E_{P,n-1}|\leq  c|\la|^{1/8 }\sigma_{n-1}^{1-2\delta}. \label{estimate-convex}
\end{eqnarray}
\eel
\proof Starting from formula~(\ref{E-sec-der}), coming back to the discrete cut-off and
 exploiting rotation invariance of the model, we can write 
\begin{equation}
\partial_{|P|}^{2}E_{P,n}  =  1-2\lan\check{\phi}_{P,n}, (\Gamma_{P,n})_{i}\frac{1}{H_{P,n}^{W}-E_{P,n}}\,(\Gamma_{P,n})_{i}\check{\phi}_{P,n}\ran|_{P=|P|\hat{P}_{i}},\label{deriv2-cont}
\end{equation}
where $\hat{P}_i$ is the unit vector in the $i$-th direction for some fixed $i$. {\color{black}In the remaining part of this proof} it is tacitly assumed  that
$P=|P|\hat{P}_i$. Now we define the function
\beqa
f_n(z)\1&:=&\1\lan \check{\phi}_{P,n},(\Gamma_{P,n})_{i}\frac{1}{H_{P,n}^{W}-z}\,(\Gamma_{P,n})_{i}\check{\phi}_{P,n}\ran \non\\
& &-\lan\check{\phi}_{P,n-1},  (\Gamma_{P,n-1})_{i}\frac{1}{H_{P,n-1}^{W+}-z+{\bcb (E_{P,n}-E_{P,n-1})}-\De c_P|^{n-1}_n }\,(\Gamma_{P,n-1})_{i}\check{\phi}_{P,n-1}\ran. 
\label{f-n-w}
\eeqa
 Making use of the maximal modulus principle, {\bcb applied to this function}, and the fact that $\lan \check{\phi}_{P,n},(\Gamma_{P,n})_{i}\check{\phi}_{P,n}\ran=0$, we can write
\beqa
& &|\pa_{|P|}^2E_{P,n}-\pa^2_{|P|}E_{P,n-1}|\non\\
& &\leq \sup_{z_{n+1}\in \ti\ga_{P,n+1}}2\big| \lan \check{\phi}_{P,n},(\Gamma_{P,n})_{i}\frac{1}{H_{P,n}^{W}-z_{n+1}}\,(\Gamma_{P,n})_{i}\check{\phi}_{P,n}\ran 
- \lan\check{\phi}_{P,n-1},  (\Gamma_{P,n-1})_{i}\frac{1}{H_{P,n-1}^{W+}-z_{n+1}}\,(\Gamma_{P,n-1})_{i}\check{\phi}_{P,n-1}\ran \big|\quad\quad\label{convexity-leading}\\
& &\ph{44}+\mco(|\la|^2\si_{n-1}^{1-2\de}), \label{convexity-error}
\eeqa
where the error term  (\ref{convexity-error}) comes from the shift of the resolvent in (\ref{f-n-w}), Theorem~\ref{thm:induction-convergence} and the bounds
$|E_{P,n-1}-E_{P,n}|, |\De c_P|^{n-1}_n|\leq c|\la|^2\si_{n-1}$. Now we rearrange the first term under the absolute value in (\ref{convexity-leading}):
\beqa
F(z_{n+1})\1&:=&\1\lan \check{\phi}_{P,n},  (\Gamma_{P,n})_{i} \frac{1}{H_{P,n}^{W}-z_{n+1}}\,(\Gamma_{P,n})_{i}\check{\phi}_{P,n}\ran=
 \lan \hat{\phi}^*_{P,n}, (\hat\Gamma_{P,n})_{i} \frac{1}{\hat{H}_{P,n}^{W}-z_{n+1}}\,(\hat{\Gamma}_{P,n})_{i}\hat{\phi}^*_{P,n}\ran,
\eeqa
where $\hat{\phi}^*_{P,n}:=\hat{\phi}_{P,n}/ \|  \hat{\phi}_{P,n}\|$. Furthermore, we write
\beqa
F(z_{n+1})\1&=&\1 \lan (\hat{\phi}^*_{P,n}- \check{\phi}_{P,n-1}), (\hat\Gamma_{P,n})_{i} \frac{1}{\hat{H}_{P,n}^{W}-z_{n+1}}\,(\hat{\Gamma}_{P,n})_{i}\hat{\phi}^*_{P,n}\ran
\label{conv-correction-one}\\
& &+\lan \check{\phi}_{P,n-1}, (\hat\Gamma_{P,n})_{i} \frac{1}{\hat{H}_{P,n}^{W}-z_{n+1}}\,(\hat{\Gamma}_{P,n})_{i} (\hat{\phi}^*_{P,n}- \check{\phi}_{P,n-1} )\ran 
\label{conv-correction-two}\\
& &+\lan \check{\phi}_{P,n-1}, (\hat\Gamma_{P,n})_{i} \frac{1}{\hat{H}_{P,n}^{W}-z_{n+1}}\,(\hat{\Gamma}_{P,n})_{i} \check{\phi}_{P,n-1} \ran. \label{conv-leading-term}
\eeqa
 Using Lemmas~\ref{convexity-auxiliary-one}, \ref{convexity-auxiliary-two} 
and the estimate $\|\hat{\phi}^*_{P,n}- \check{\phi}_{P,n-1}\|\leq 4|\la|^{1/4}\si_{n-1}^{1-\de}$ (which follows from Theorem~\ref{thm:induction-convergence}) we can write
\beqa
(\ref{conv-correction-one})+(\ref{conv-correction-two})=\mco(|\la|^{1/8}\si_{n-1}^{1-2\de}),
\eeqa
where we also exploited $|\la|\leq \veps^{8}$.
 As for (\ref{conv-leading-term}), we have 
\beqa
(\ref{conv-leading-term})\1&=&\1\lan \check{\phi}_{P,n-1}, (\De \Ga_{P}|^{n-1}_n)_{i} \frac{1}{\hat{H}_{P,n}^{W}-z_{n+1}}\,(\De \Ga_{P}|^{n-1}_n)_{i} \check{\phi}_{P,n-1}\ran 
\label{conv-second-corrections-one}\\
\1& &\1+ \lan \check{\phi}_{P,n-1}, ( \De \Ga_{P}|^{n-1}_n)_{i} \frac{1}{\hat{H}_{P,n}^{W}-z_{n+1}}\,(\Gamma_{P,n-1})_{i} \check{\phi}_{P,n-1}\ran 
\label{conv-second-corrections-two} \\
\1& &\1+ \lan \check{\phi}_{P,n-1}, (\Gamma_{P,n-1})_{i} \frac{1}{\hat{H}_{P,n}^{W}-z_{n+1}}\,( \De \Ga_{P}|^{n-1}_n  )_{i} \check{\phi}_{P,n-1}\ran 
\label{conv-second-corrections-three} \\
\1& &\1+ \lan \check{\phi}_{P,n-1}, (\Gamma_{P,n-1})_{i} \frac{1}{\hat{H}_{P,n}^{W}-z_{n+1}}\,( \Gamma_{P,n-1})_{i} \check{\phi}_{P,n-1}\ran,  
\eeqa \label{conv-second-leading}
{\color{black}where $\De \Ga_{P}|^{n-1}_n$ is defined in (\ref{De-E-I-0}).}
Making use of Lemma~\ref{convexity-auxiliary-three} and of $\| (\hat{H}_{P,n}^{W}-z_{n+1})^{-1}\|_{\mcF_n}=\mco(\si_{n+1}^{-1})$, we obtain
\beqa
(\ref{conv-second-corrections-one})+(\ref{conv-second-corrections-two})+(\ref{conv-second-corrections-three}) =\mco(|\la|^{1/4}\si_{n-1}^{1-2\de}),
\eeqa
where we also used $|\la|\leq \veps^{8}$.

Thus coming back to (\ref{convexity-leading}) and making use of the expansion  in (\ref{eq:resolvent-expansion}) we can write
\beqa
& &|\pa_{|P|}^2E_{P,n}-\pa^2_{|P|}E_{P,n-1}|\non\\
& &\leq \sup_{z_{n+1}\in \ti\ga_{P,n+1}}2\big|\lan \check{\phi}_{P,n-1}, (\Gamma_{P,n-1})_{i} 
\sum_{j=1}^{\infty}\{\frac{1}{H_{P,{\color{black}n-1}}^{W+}-z_{n+1}} (-H_I^{W}|_{n}^{n-1})\}^j\frac{1}{H_{P,n-1}^{W+}-z_{n+1}} \,( \Gamma_{P,n-1})_{i} \check{\phi}_{P,n-1}\ran |
\label{convex-modulus} \\
& &\ph{44}+\mco(|\la|^{1/8}\si_{n-1}^{1-2\de}).
\eeqa
Adopting the notation of (\ref{R-R'-def}), (\ref{R'''-V-def}), the term under the modulus in (\ref{convex-modulus}) can be rewritten as follows
\beqa
\lan \check{\phi}_{P,n-1}, (\Gamma_{P,n-1})_{i}  \sum_{j=1}^{\infty}\{  RV \}^j R ( \Gamma_{P,n-1})_{i} \check{\phi}_{P,n-1}\ran\1&=&\1
\lan \check{\phi}_{P,n-1}, (\Gamma_{P,n-1})_{i}  \sum_{j=0}^{\infty}\{  RV \}^j R V R ( \Gamma_{P,n-1})_{i} \check{\phi}_{P,n-1}\ran \non\\
\1&=&\1 \lan \check{\phi}_{P,n-1}, (\Gamma_{P,n-1})_{i}  \sum_{j=0}^{\infty}\{  RV \}^j Q^{\bot}_{P,n-1}R V R ( \Gamma_{P,n-1})_{i} \check{\phi}_{P,n-1}\ran \label{appl-b-star}\\
\1& &\1+ \lan \check{\phi}_{P,n-1}, (\Gamma_{P,n-1})_{i}  \sum_{j=0}^{\infty}\{  RV \}^j RV Q_{P,n-1}R V R ( \Gamma_{P,n-1})_{i} \check{\phi}_{P,n-1}\ran,\,.\,\quad\quad
 \label{appl-b-star-next}
 \eeqa
{\bcb where in the last step we used $\lan   \check{\phi}_{P,n-1}, (\Gamma_{P,n-1})_{i} \check{\phi}_{P,n-1}\ran=0$.}
We recall that $\|\sum_{j=0}^{\infty}\{  RV \}^j\|_{\mcF_n}=\mco(1)$ by Lemma~\ref{lem:Basic-estimates} and $\|\Ga_{P,n-1}\check{\phi}_{P,n-1}\|=\mco(1)$. 
Now making use of Lemma~\ref{b-star-lemma}  we write
\beqa
(\ref{appl-b-star})\1&=&\1 \lan \check{\phi}_{P,n-1}, (\Gamma_{P,n-1})_{i}  \sum_{j=1}^{\infty}\{  RV \}^j 
b^*\big(  g_{i}|^{n-1}_n\big) \phi_{P,n-1} \ran +\mco(|\la|^{1/2}\si_{n-1}^{1-2\de})\non\\
\1&=&\1 \lan \check{\phi}_{P,n-1}, (\Gamma_{P,n-1})_{i}  \sum_{j=0}^{\infty}\{  RV \}^j 
RV b^*\big(  g_{i}|^{n-1}_n\big) \phi_{P,n-1} \ran +\mco(|\la|^{1/2}\si_{n-1}^{1-2\de})=\mco(|\la|^{1/2}\si_{n-1}^{1-2\de}),
\eeqa
where in the second step we made use of the fact that $b\big(  g_{i}|^{n-1}_n\big)(\Gamma_{P,n-1})_{i}  \check{\phi}_{P,n-1}=0$
and in the last step we applied Lemma~\ref{interaction-x}. Furthermore, we get by Lemma~\ref{first-case-aux-lemma}, estimate~(\ref{R-prime-V})
and Theorem~\ref{thm:induction-convergence}
\beqa
(\ref{appl-b-star-next})=\mco(|\la| \si_{n-1}^{1-2\de}).
\eeqa
Thus altogether we obtain the bound in (\ref{estimate-convex}). \qed

\begin{thm}\label{hoelder}
Assume $|\lambda_0|$ sufficiently small and $|P|<P_{\max}$. Then $\partial_{|P|}^{2}E_{P,n}$ defined in (\ref{deriv2-cont}) is strictly positive for any $n\in\nat_0$ and converges to a limiting function $\mathcal{E}_{\la}''(|P|)>0$ as $n\to \infty$. The function $\mathcal{E}_{\la}''(|P|)$ is H\"older continuous for some exponent $\eta(>0)$ and 
$\lim_{\la\to 0}\mathcal{E}_{\la}''(|P|)=1$.
\end{thm}
\proof First, we note that the estimate in (\ref{estimate-convex}) yields the existence of $\lim_{n\to \infty} \partial_{|P|}^{2}E_{P,n}<\infty$ by a telescopic argument. 
Next, we observe that $\partial_{|P|}^{2}E_{P,n=0}=1$ (see (\ref{deriv2-cont})), because $(\Gamma_{P,n=0})_{i}\equiv (P_{\pho})_i$ and $\phi_{P,n=0}\equiv \Omega$. 
Consequently, again by a telescopic argument,
\beqa
2|\lan\check{\phi}_{P,n}, (\Gamma_{P,n})_{i}\frac{1}{H_{P,n}^{W}-E_{P,n}}\,(\Gamma_{P,n})_{i}\check{\phi}_{P,n}\ran|_{P=|P|\hat{P}_{i}}|\leq c|\la|^{1/8},
\eeqa
which implies that $\partial_{|P|}^{2}E_{P,n}>0$ and $\mathcal{E}_{\la}''(|P|)>0$ for $P\in S$  and $\la_0$ sufficiently small. This also gives  
$\lim_{\lambda\to0}\mathcal{E}_{\la}''(|P|)=1$.

Starting from the expression in (\ref{deriv2-cont}) and the convergence $\lim_{n \to \infty} \partial_{|P|}^{2}E_{P,n}=\mathcal{E}''(|P|)$, a standard argument ensures that $\mathcal{E}''(|P|)$ is H\"older continuous for some small exponent. Indeed, for $P,P+\Delta P \in S$ it is enough to write 
\begin{eqnarray}
& &\mathcal{E}''(|P+\Delta P|)-\mathcal{E}''(|P|)\non\\
&=&\mathcal{E}''(|P+\Delta P|)-\partial_{|P|}^{2}E_{P+\Delta P,n_*}+\partial_{|P|}^{2}E_{P+\Delta P,n_*}-\partial_{|P|}^{2}E_{P,n_*}+\partial_{|P|}^{2}E_{P,n_*}-\mathcal{E}''(|P|),
\label{Hoelder-convex}
\end{eqnarray}
where {\color{black}the infrared cut-off is $\epsilon^{n_*}=\mathcal{O}(|\Delta P|^{\frac{1}{4}})$} and {\color{black} we} exploit the convergence rate of $\partial_{|P|}^{2}E_{P,n}\to \mathcal{E}''(|P|)$, the estimate on the gap of ${H}^{W}_{P,n}\upharpoonleft\mathcal{F}_{n}$, and the fact that 
$\{\,{H}^{W}_{P,n}\}_{P\in S}$ is an analytic family of type A.  {\bcb(Alternatively, one can use our bound  $|\pa_{P}^{\be_3}E_{P,\si}|\leq c/\si^{\de_{\la_0}} $ and the Taylor theorem  to estimate $ \partial_{|P|}^{2}E_{P+\Delta P,n_*}-\partial_{|P|}^{2}E_{P,n_*}$ )}.\qed
\begin{cor}\label{sconv}
Under the assumptions of Theorem \ref{hoelder}, the limiting function $E_{P}:=\lim_{n\to \infty}E_{P,n}$ is twice {\bcb continuously} differentiable and $\partial_{|P|}^{2}E_{P}\equiv \mathcal{E}_{\lambda}''(|P|)$.
\end{cor}
\proof  Clearly, there exists $C$
such that $|\partial_{|P|}E_{P,n}|\,,\,|\partial^2_{|P|}E_{P,n}|<C$ for all $P\in \PS:=\{\, P\in \real^3\,|\, |P|< P_{\maxi}=\frac{1}{3}\,\}$ and for all $n\in\nat_0$. 
 Thus we can write for any $P$, $P+\Delta P$ in $S$
 \begin{eqnarray}
& &E_{P+\Delta P}-E_{P}=\lim_{n \to \infty}\{E_{P+\Delta P,n}-E_{P,n}\}={\color{black}\lim_{n \to \infty}}\int_{0}^{1}\Delta P\cdot \{\frac{Q}{|Q|}\partial_{|Q|} E_{Q,n}\}_{Q=P+u\Delta P}\,du\\
&=&\int_{0}^{1}\lim_{n\to \infty}\Big\{\Delta P\cdot \{\frac{Q}{|Q|}\partial_{|Q|} E_{Q,n}\}_{Q=P+u\Delta P}\Big\}\,du\,.
\end{eqnarray}
The H\"older continuity of $\lim_{n \to \infty} \partial_{|P|}E_{P,n}$ is shown as in (\ref{Hoelder-convex}), exploiting (\ref{grad-difference}) and Theorem~\ref{thm:induction-convergence} instead of (\ref{estimate-convex}).
 By this H\"older continuity  and the fundamental theorem of calculus we conclude that $E_{P}$ is differentiable and 
 $\nabla E_{P}=\lim_{n\to \infty} \partial_{|P|}E_{P,n}\frac{P}{|P|}$. An analogous argument implies that  the second derivatives of $E_{P}$ exist  with $\partial_{|P|}^{2}E_{P}\equiv \mathcal{E}_{\lambda}''(|P|)$.
\qed

\subsection{Auxiliary lemmas}

\bel\label{convexity-auxiliary-one} Under the assumptions of Theorem~\ref{main-technical-result}
\beqa 
\|(\hat\Gamma_{P,n})_{i} \frac{1}{\hat{H}_{P,n}^{W}-z_{n+1}}\,(\hat{\Gamma}_{P,n})_{i}\hat{\phi}^*_{P,n}\|=\mco(\si_n^{-\de}).
\eeqa
\eel
\proof Clearly we can drop the `hats' and  write
\beqa
\|(\Gamma_{P,n})_{i} \frac{1}{{H}_{P,n}^{W}-z_{n+1}}\,({\Gamma}_{P,n})_{i}{\cphi}_{P,n}\|
\1&\leq &\1 \|(\Gamma_{P,n})_{i}\chi(H^W_{P,n}-E_{P,n}\leq 1) \frac{1}{{H}_{P,n}^{W}-z_{n+1}}\,({\Gamma}_{P,n})_{i}{\cphi}_{P,n}\| \label{second-spectral-projection}\\
& &\1+ 
\|(\Gamma_{P,n})_{i}\chi(H^W_{P,n}-E_{P,n}\geq 1) \frac{1}{{H}_{P,n}^{W}-z_{n+1}}\,({\Gamma}_{P,n})_{i}{\cphi}_{P,n}\|, \label{first-spectral-projection} 
\eeqa
where we inserted spectral projections of $H^W_{P,n}$. Since $(\Gamma_{P,n})_{i}\chi(H^W_{P,n}-E_{P,n}\leq 1)$ is bounded, uniformly in $n$, we have by
Theorem~\ref{thm:induction-convergence} that  $(\ref{second-spectral-projection})=\mco(\si_{n}^{-\de})$.

As for (\ref{first-spectral-projection}),  we note that $\chi(H^W_{P,n}-E_{P,n}\geq 1)$ mollifies the infrared singularity of the resolvent. Hence, 
making use of 
\beqa
\|\chi(H^W_{P,n}-E_{P,n}\geq 1)(i+{H}_{P,n}^{W})^{{\color{black}\frac{1}{2}}}\frac{1}{{H}_{P,n}^{W}-z_{n+1}}\|=\mco(1), \quad
\|(\Gamma_{P,n})_{i}(i+{H}_{P,n}^{W})^{-1/2}\|=\mco(1), \label{energy-bounds-convex}
\eeqa
and of $\|(\Gamma_{P,n})_{i}{\cphi}_{P,n}\|=\mco(1)$ we obtain $(\ref{first-spectral-projection})=\mco(1)$ and conclude the proof. \qed
\bel \label{convexity-auxiliary-two} Under the assumptions of Theorem~\ref{main-technical-result}
\beqa
\|(\hat\Gamma_{P,n})_{i} \frac{1}{\hat{H}_{P,n}^{W}-z_{n+1}}\,(\hat{\Gamma}_{P,n})_{i}{\cphi}_{P,n-1}\|=\mco(\si_n^{-\de}).
\eeqa
\eel
\proof We write
\beqa
\|(\hat\Gamma_{P,n})_{i} \frac{1}{\hat{H}_{P,n}^{W}-z_{n+1}}\,(\hat{\Gamma}_{P,n})_{i}{\cphi}_{P,n-1}\|
\1&=&\1 \|(\Gamma_{P,n})_{i} \frac{1}{{H}_{P,n}^{W}-z_{n+1}}\,({\Gamma}_{P,n})_{i} W_{P,n}\wt{W}^*_{P,n} {\cphi}_{P,n-1}\|\non\\
\1&\leq&\1 \|(\Gamma_{P,n})_{i} \frac{1}{{H}_{P,n}^{W}-z_{n+1}}\,({\Gamma}_{P,n})_{i}  {\cphi}_{P,n}\| \label{convex-aux-two-one}\\
\1& &\1+ \|(\Gamma_{P,n})_{i} \frac{1}{{H}_{P,n}^{W}-z_{n+1}}\,({\Gamma}_{P,n})_{i}\|_{\mcF_n} \| W_{P,n}\wt{W}^*_{P,n} {\cphi}_{P,n-1}-{\cphi}_{P,n}\|. 
\label{second-auxiliary-second-term}
\eeqa
We immediately get from Lemma~\ref{convexity-auxiliary-one} that $(\ref{convex-aux-two-one})=\mco(\si_n^{-\de})$.
Furthermore, we  note that by (\ref{phi-definitions}) and Theorem~\ref{thm:induction-convergence} 
\beqa
\| W_{P,n}\wt{W}^*_{P,n} {\cphi}_{P,n-1}-{\cphi}_{P,n}\|=\|{\cphi}_{P,n-1}-\hat\phi^*_{P,n}\|\leq   4|\la|^{1/4}\si_{n-1}^{1-\de}. \label{second-aux-vector-shift}
\eeqa
Finally, we estimate similarly as in (\ref{second-spectral-projection}), (\ref{first-spectral-projection})
\beqa
\|(\Gamma_{P,n})_{i} \frac{1}{{H}_{P,n}^{W}-z_{n+1}}\,({\Gamma}_{P,n})_{i}\|_{\mcF_n}\1&\leq&\1 \|(\Gamma_{P,n})_{i} \chi(H^W_{P,n}-E_{P,n}\leq 1) 
\frac{1}{{H}_{P,n}^{W}-z_{n+1}}\,({\Gamma}_{P,n})_{i}\|_{\mcF_n} \label{again-spectral-projection-one}\\
\1& &\1+ \|(\Gamma_{P,n})_{i} \chi(H^W_{P,n}-E_{P,n}\geq 1) \frac{1}{{H}_{P,n}^{W}-z_{n+1}}\,({\Gamma}_{P,n})_{i}\|_{\mcF_n}
\label{again-spectral-projection-two}\\
\1&=&\1 \mco(\si_{n+1}^{-1}), \label{again-spectral-three}
\eeqa
where we made use of $\|(\Gamma_{P,n})_{i} \chi(H^W_{P,n}-E_{P,n}\leq 1)\|_{\mcF_n}=\mco(1)$ and $\|({H}_{P,n}^{W}-z_{n+1})^{-1}\|_{\mcF_n}=\mco(\si_{n+1}^{-1})$  
to estimate~(\ref{again-spectral-projection-one}) and applied (\ref{energy-bounds-convex}) to estimate~(\ref{again-spectral-projection-two}). 
From (\ref{again-spectral-three}) and (\ref{second-aux-vector-shift}) we obtain $(\ref{second-auxiliary-second-term})=\mco( \si_{n-1}^{-\de})$,
where we also exploited $|\la|\leq \veps^8$. This concludes the proof. \qed
\bel \label{convexity-auxiliary-three} Under the assumptions of Theorem~\ref{main-technical-result}
\beqa
& &(\De \Ga_{P}|^{n-1}_n)_{i} \check{\phi}_{P,n-1}=\mco(|\la|^{1/4}\si_{n-1}^{1-\de}), \label{convexity-three-one}\\
& & \frac{1}{\hat{H}_{P,n}^{W}-z_{n+1}}\,(\Gamma_{P,n-1})_{i} \check{\phi}_{P,n-1}=\mco(\si_{n-1}^{-\de}).  \label{convexity-three-two}
\eeqa
\eel
\proof As for (\ref{convexity-three-one}), we recall from {\color{black}(\ref{De-E-I-0})} that $\Delta\Gamma_{P}|_{n}^{n-1}=X|^{n-1}_n
+\mathcal{L}|_{n}^{n-1}$, where $X|^{n-1}_n:=-\nabla E_{P,n-1}+\nabla E_{P,n}+\mathcal{I}|_{n}^{n-1}$ is a vector in $\real^3$
satisfying
\beqa
|X|^{n-1}_n|\leq  
c|\la|^{1/4}\si_{n-1}^{1-\de}. \label{standard-bound-X-convex}
\eeqa
Thus it suffices to use that
\beqa
\mathcal{L}|_{n}^{n-1}\check{\phi}_{P,n-1}=(\mathcal{L}|_{n}^{n-1})^{(+)}\check{\phi}_{P,n-1}=\mco(|\la|\si_{n-1}),
\eeqa
which follows from the definition of $\mathcal{L}|_{n}^{n-1}$. 

To prove (\ref{convexity-three-two}), we write, making use of the expansion~(\ref{eq:resolvent-expansion}) and Theorem~\ref{thm:induction-convergence} 
\beqa
\frac{1}{\hat{H}_{P,n}^{W}-z_{n+1}}\,(\Gamma_{P,n-1})_{i} \check{\phi}_{P,n-1}=\sum_{j=0}^{\infty}\{\frac{1}{H_{P,{\color{black}n-1}}^{W+}-z_{n+1}} (-H_I^{W}|_{n}^{n-1})\}^j\frac{1}{H_{P,n-1}^{W+}-z_{n+1}}{\color{black}\,(\Gamma_{P,n-1})_{i}}\check{\phi}_{P,n-1}=\mco(\si_{n-1}^{-\de}),
\eeqa
where we controlled the sum as explained below (\ref{appl-b-star-next}). \qed


\end{document}